%% file: main.tex
\documentclass{houches}
\usepackage{epsfig}
\usepackage{latexsym}
\makeatletter\newcommand\myhline{\old@hline}
\makeatother

\begin{document}
\setcounter{chapter}{0}
\setcounter{page}{1}
\font\sixrm=cmr6
\font\eightrm=cmr8
\def\exercise#1{
\vspace{0pt plus 0.5cm}\penalty-100 \vspace{0pt plus-0.5cm}
\vspace{2ex}
\noindent{\small\sc Exercise:}~{\small#1}
\par\vspace{2ex}}

\input title.tex

\input LH.tex

\end{document}

%% file: title.tex
\author{Rajan Gupta}
\address{Theoretical Division, Group T-8, Mail Stop B-285, \\
	Los Alamos National Laboratory\\
	Los Alamos, NM 87545, U.~S.~A \\
	E-mail: rajan@qcd.lanl.gov}

\editor{}
\title{}
\chapter{INTRODUCTION TO LATTICE QCD}

%% file: LH.tex
\input{rgmac.latex}

\def\gsim{\mathrel{\raise2pt\hbox to 8pt{\raise -5pt\hbox{$\sim$}\hss{$>$}}}}
\def\rsim{\mathrel{\raise2pt\hbox to 8pt{\raise -5pt\hbox{$\sim$}\hss{$>$}}}}
\def\lsim{\mathrel{\raise2pt\hbox to 8pt{\raise -5pt\hbox{$\sim$}\hss{$<$}}}}
\def\half{{\textstyle{1\over2}}} \def\third{{\textstyle{1\over3}}}
\def\sdsdop{{(\bar s\gamma_\mu L d)(\bar s \gamma_\mu L d)}}
\def\gsqeff{{g^2_{\rm eff}}}

\newcommand{\be}{\begin{equation}}
\newcommand{\ee}{\end{equation}}
\newcommand{\bea}{\begin{eqnarray}}
\newcommand{\eea}{\end{eqnarray}}

\newcommand\figcaption[1]{\vskip-0.0truein\caption{#1}\vskip0.0truein}
\newcommand\alphamsbar{\hbox{$\alpha_{\overline{MS}}$}}
\newcommand\alphav{\hbox{$\alpha_v$}}
\newcommand\lili{\hbox{$\{ U_i U_i \}$}}
\newcommand\lilj{\hbox{$\{ U_i U_j \}$}}
\newcommand\sli{\hbox{$ \{ S U_i   \}$}}
\newcommand\ssli{\hbox{$ \{ SS, S U_i   \}$}}
\newcommand\cli{\hbox{$ \{ C U_i   \}$}}
\newcommand\MSbar{\hbox{$\overline{MS}$}}
\newcommand\mmsbar{\hbox{$m_{\overline{MS}}$}}
\newcommand\mbar{\hbox{$\overline{m}$}}
\def\cpt{\hbox{$ \chi $PT}}
\def\trace{{\rm tr}}

\def\vev#1{\langle      #1 \rangle}
\def\VEV#1{{\big\langle #1 \big\rangle}}
\def\me#1#2#3{{\langle    #1\vert    #2\vert    #3\rangle}}
\def\ME#1#2#3{{\big\langle#1\big\vert#2\big\vert#3\big\rangle}}

\input{chap-intro.tex}

\input{chap-connection.tex}

\input{lqcd.tex}

\input{chap-impglue.tex}

\input{chap-impdirac.tex}

\input{chap-confine.tex}

\input{chap-phase.tex}

\input{chap-errors.tex}

\input{chap-corr.tex}

\input{chap-hm.tex}

%
\input{chap-alpha.tex}

\input{chap-mq.tex}

\clearpage
\centerline{\bf Conclusions and Acknowledgements}
\bigskip

I hope that these lectures succeed in providing a pedagogical
introduction to Lattice QCD, and to the analyses of numerical data.
Undoubtedly, some topics have been left out and many are sketchy. My
purpose was to give a broad overview.  I have concentrated on
providing the basics and have illustrated the analyses of Monte Carlo
data by the calculations of the hadron spectrum, strong
coupling constant $\alpha_s$, and quark masses. Prof. Martinelli will,
among other things, discuss the extraction of matrix elements of weak
Hamiltonian, while Prof. L\"uscher will provide a panoramic view of
what future calculations will look like and the improvement possible
with better discretization schemes.  Together, I hope these set of
lectures provide sufficient details and excitement to motivate some of
you to probe deeper into the field.

I express my gratitude to all the lecturers, students, and staff who
made this Les Houches school fun and exciting. It, most certainly, was
a memorable six weeks for me. In preparing these lectures I would like
to thank G. Bali, U.~Heller, C.~Michael, M.~Peardon, the CP-PACS
collaboration, in particular S. Hashimoto and T.~Yoshie, for providing
unpublished data. Special thanks go to my collaborators
T.~Bhattacharya, D. Daniel, G.~Kilcup, A.~Patel, and S.~Sharpe from
whom I have, over the years, learned so much.  Finally, I would like
to thank T. Bhattacharya, D. Daniel, W. Lee, and S.~Sharpe for proof
reading these notes and for their suggestions.

%% file: rgmac.latex
\newskip\defaultbaselineskip\defaultbaselineskip=12pt
\def\linf{\rightarrow \infty}
\def\tr{\mathop{\rm tr}\nolimits}
\def\Tr{\mathop{\rm Tr}\nolimits}
\def\GeV{\mathord{\rm \;GeV}}
\def\MeV{\mathord{\rm \;MeV}}
\def\fermi{\mathop{\rm fm}\nolimits}

\def\cosh{\mathop{\rm cosh}\nolimits}
\def\sinh{\mathop{\rm sinh}\nolimits}
\def\bar{\overline}

\def\cdf{{\hbox{$\chi^2/N_{DF}$}}}

\def\eqinf{\mathrel{\raise2pt\hbox to 24pt{\raise -8pt\hbox{$t \to \infty$}\hss{$=$}}}}

\def\gsim{\mathrel{\raise2pt\hbox to 8pt{\raise -5pt\hbox{$\sim$}\hss{$>$}}}}
\def\rsim{\mathrel{\raise2pt\hbox to 8pt{\raise -5pt\hbox{$\sim$}\hss{$>$}}}}
\def\lsim{\mathrel{\raise2pt\hbox to 8pt{\raise -5pt\hbox{$\sim$}\hss{$<$}}}}
\def\ssqr#1#2{{\vbox{\hrule height.#2pt
      \hbox{\vrule width.#2pt height#1pt \kern#1pt\vrule width.#2pt}
      \hrule height.#2pt}\kern-.#2pt}}
\def\sqr{\mathchoice\ssqr34\ssqr34\ssqr{2.1}3\ssqr{1.5}3}
\def\sqrm{\mathchoice\ssqr44\ssqr44\ssqr{3.1}3\ssqr{2.5}3}
\def\sqrb{\mathchoice\ssqr54\ssqr54\ssqr{4.1}3\ssqr{3.5}3}
\def\sqrB{\mathchoice\ssqr64\ssqr64\ssqr{5.1}3\ssqr{4.5}3}

\def\Dsl{\,\raise.15ex\hbox{$/$}\mkern-13.5mu D} 
\def\dsl{\raise.15ex\hbox{$/$}\kern-.57em\hbox{$\partial$}}
\def\Psl{\dsl}

%
\def\qq{q \bar q}
\def\qbarq{$q \bar q$}
\def\alps{\alpha_s}
\def\coup{{6/g^2}}
\def\greg{${6/g^2}$}

\def\meff{$m_{\rm eff}(t)$}
\def \exppit { {\hbox{$e^{-m_\pi t $} }}}
\def \expkt  { {\hbox{$e^{-m_K   t $} }}}
\def \expNt  { {\hbox{$e^{-m_N   t $} }}}
\def \expmt#1{ {\hbox{$e^{-m_{#1}t $} }}}

\def \gmu {\gamma_\mu}
\def \pl {{\hbox{(1- \gamma_5)}}}
\def \pr {{\hbox{(1+ \gamma_5)}}}
\def\pbarp{{\hbox{$\bar\psi \psi$}}}
\def\xbarx{{\hbox{$\bar\xi \xi$}}}
\def\wl{{\hbox{$\langle L \rangle$}}}

\def\zero{$0^{++}$}
\def\twop{$2^{++}$}
\def\rat{${{m_{0^{++}}} \over {\sqrt\sigma}}$}

\def\deltaE{\delta\! E}
\def\pipi{{\pi\pi}}
\def\deltaEeff{\deltaE_{\scriptstyle\rm eff}}


\def\CA{{\cal A}}  \def\CB{{\cal B}}  \def\CC{{\cal C}} \def\CD{{\cal D}}
\def\CE{{\cal E}}  \def\CF{{\cal F}}  \def\CF{{\cal F}} \def\CH{{\cal H}}
\def\CI{{\cal I}}  \def\CJ{{\cal J}}  \def\CK{{\cal K}} \def\CL{{\cal L}}
\def\CM{{\cal M}}  \def\CN{{\cal N}}  \def\CO{{\cal O}} \def\CP{{\cal P}}
\def\CQ{{\cal Q}}  \def\CR{{\cal R}}  \def\CS{{\cal S}} \def\CT{{\cal T}}
\def\CU{{\cal M}}  \def\CV{{\cal V}}  \def\CW{{\cal W}} \def\CX{{\cal X}}
\def\CY{{\cal Y}}  \def\CZ{{\cal Z}}  

\def\fnsym#1{${}^{#1}$}

\def\fnsym#1{${}^{#1}$}
\def\jrorg{\footnote{\fnsym{\S}}{J. Robert Oppenheimer Fellow.}}
\def\doesrs{\footnote{\fnsym{\ast}}{\absfnt%
     Work supported by the Department of Energy, contract DE-AC03-76SF00515} }

\def\authorrgb{\bigskip\centerline{Ralph G. Brickner}
\smallskip\centerline{\it C-Division,
      Los Alamos National Laboratory, Los Alamos, NM 87545}}

\def\authorrg{\bigskip\centerline{Rajan Gupta} 
    \smallskip\centerline{\it T-8, MS-B285,
                     Los Alamos National Laboratory, Los Alamos, NM 87545}}

\def\authordd{\bigskip\centerline{David Daniel} 
    \smallskip\centerline{\it T-8, MS-B285,
                     Los Alamos National Laboratory, Los Alamos, NM 87545}}

\def\authorrgdd{\bigskip\centerline{Rajan Gupta and David Daniel}
    \smallskip\centerline{\it T-8, MS-B285,
                     Los Alamos National Laboratory, Los Alamos, NM 87545}}

\def\authorrgddjg{\bigskip\centerline{Rajan Gupta, David Daniel, Jeffrey Grandy}
    \smallskip\centerline{\it T-8, MS-B285,
                     Los Alamos National Laboratory, Los Alamos, NM 87545}}

\def\authorddrg{\bigskip\centerline{David Daniel and Rajan Gupta}
    \smallskip\centerline{\it T-8, MS-B285,
                     Los Alamos National Laboratory, Los Alamos, NM 87545}}

\def\authorgg{\bigskip\centerline{Gerald Guralnik}
    \smallskip\centerline{\it Physics Department, Brown University, 
                          Providence, RI 02912}}

\def\authorgk{\bigskip\centerline{Gregory W. Kilcup}
     \smallskip\centerline{\it Physics Department, The Ohio State University,
                               Columbus, OH 43210}}

\def\authoradp{\bigskip\centerline{Apoorva Patel}
     \smallskip\centerline{\it Supercomputer Education and Research Centre
                               and Centre for Theoretical Studies}
     \centerline{\it Indian Institute of Science, Bangalore 560012, India}}

\def\authorsrs{\bigskip\centerline{Stephen R. Sharpe}
     \smallskip\centerline{\it Physics Department, University of Washington,
                          Seattle, WA 98195}}

\def\authorpdf{\bigskip\centerline{Philippe de Forcrand}
     \smallskip\centerline{\it Physics Department, University of Minnesota,
          Minneapolis, MN 55455}
     \centerline{\it and}
     \centerline{\it Cray Research Inc., 1333 Northland Drive, Mendota Heights,
                MN 55120}}


\ifx\href\undefined\def\href#1#2{{#2}}\fi
\def\spireshome{http://www.slac.stanford.edu/cgi-bin/spiface/find/hep/www?FORMAT=WWW&}
{\catcode`\%=12
\xdef\spiresjournal#1#2#3{\noexpand\protect\noexpand\href{\spireshome
                          rawcmd=find+journal+#1%2C+#2%2C+#3}}
\xdef\spireseprint#1#2{\noexpand\protect\noexpand\href{\spireshome rawcmd=find+eprint+#1%2F#2}}
\xdef\spiresreport#1{\noexpand\protect\noexpand\href{\spireshome rawcmd=find+rept+#1}}
\xdef\spireskey#1{\noexpand\protect\noexpand\href{\spireshome key=#1}}
}
\def\eprint#1#2{\spireseprint{#1}{#2}{#1/#2}}
\def\report#1{\spiresreport{#1}{#1}}
\def\nohref{}

\def\putpaper{\edef\refpage{\the\count0}%
              \def\nohref{}%
              {\def\ {+}\def\nohref##1{}\edef\temp{\noexpand\spiresjournal
               {\journalname}{\volume}{\refpage}}\expandafter}\temp
               {\sfcode`\.=1000{\journalname} {\bf \volume} (\refyear)
                \refpage}\egroup}
\def\putpage{\edef\refpage{\the\count0}%
              \def\nohref{}%
              {\def\ {+}\def\nohref##1{}\edef\temp{\noexpand\spiresjournal
               {\journalname}{\volume}{\refpage}}\expandafter}\temp
              {\refpage}\egroup}
\def\dojournal#1#2 (#3){\def\journalname{#1}\def\volume{#2}\def\refyear
                        {#3}\afterassignment\putpaper\bgroup\count0=}
\def\morepage{\afterassignment\putpage\bgroup\count0=}
\def\supresslink{\def\spiresjournal##1##2##3{}}

\def\APNY#1{\dojournal{Ann.\ Phys.\ \nohref{(N.\ Y.)}}{#1}}
\def\CMP#1{\dojournal{Comm.\ Math.\ Phys.}{#1}}
\def\IJMPC#1{\dojournal{Int.\ J.\ Mod.\ Phys.}{C#1}}
\def\IJMPE#1{\dojournal{Int.\ J.\ Mod.\ Phys.}{E#1}}
\def\JAP#1{\dojournal{J.\ App.\ Phys.}{#1}}

\def\MPA#1{\dojournal{Mod.\ Phys.\ Lett.}{A#1}}
\def\MPLA#1{\dojournal{Mod.\ Phys.\ Lett.}{A#1}}
\def\NP#1{\dojournal{Nucl.\ Phys.}{B#1}}
\def\NPA#1{\dojournal{Nucl.\ Phys.}{A#1}}
\def\NPB#1{\dojournal{Nucl.\ Phys.}{B#1}}
\def\NPBPS#1{\dojournal{Nucl.\ Phys.\ \nohref(Proc.\ Suppl.\nohref)}{\nohref B#1}}
\def\NPAPS#1{\dojournal{Nucl.\ Phys.\ \nohref(Proc.\ Suppl.\nohref)}{\nohref A#1}}
\def\NC#1{\dojournal{Nuovo Cimento }{#1}}
\def\PRL#1{\dojournal{Phys.\ Rev.\ Lett.}{#1}}
\def\PR#1{\dojournal{Phys.\ Rev.}{#1}}
\def\PRep#1{\dojournal{Phys.\ Rep.}{#1}}
\def\PRB#1{\dojournal{Phys.\ Rev.}{B#1}}
\def\PRC#1{\dojournal{Phys.\ Rev.}{C#1}}
\def\PRD#1{\dojournal{Phys.\ Rev.}{D#1}}
\def\PRE#1{\dojournal{Phys.\ Rev.}{E#1}}
\def\PL#1{\dojournal{Phys.\ Lett.}{#1B}}
\def\PLA#1{\dojournal{Phys.\ Lett.}{#1A}}
\def\PLB#1{\dojournal{Phys.\ Lett.}{#1B}}
\def\RMP#1{\dojournal{Rev.\ Mod.\ Phys.}{#1}}
\def\PREP#1{\dojournal{Phys.\ Rep.}{#1}}
\def\ZEITC#1{\dojournal{Z.\ Phys.}{C#1}}
\def\ZPC#1{\dojournal{Z.\ Phys.}{C#1}}

\def\ie{{\sl i.e.}}
\def\etal{{\it et al.}}
\def\etc{{\it etc.}}
\def\ibid{{\it ibid}}

\def\wuppertal{in {\it Lattice Gauge Theories: A Challenge in Large Scale 
              Computing}, Wuppertal 1985, Eds. K-H M\"utter and K. Schilling, 
              Plenum Press (1986)}
\def\brookhaven{in {\it ``Lattice Gauge Theory '86''}, 
             Brookhaven, New York, 1986, NATO ASI Series B: Physics Vol. 159}
\def\china{in {\it ``Lattice Gauge Theory Using Parallel Processors''}, 
             CCAST Symposium/Workshop Volume 1, Beijing, China, 1987, 
             Eds. X. Li $et\ al.$, Gordon \& Breach 1987}
\def\seillac{in {\it ``Field theory on the Lattice''}, Proceedings of the 
             International Symposium, Seillac, France, 1987, edited by 
             A. Billoire $et\ al.$,  \NPBPS{4}, (1988)}
\def\tasione{ in {\it ``The Santa Fe TASI-87''}, Proceedings of the 1987 TASI in 
             Elementary Particle Physics, Santa Fe, edited by R. Slansky and G. West, 
             World Scientific 1988}
\def\fermilab{in {\it ``LATTICE 88''},  Proceedings of the 1988 Symposium on Lattice 
             Field Theory, Fermilab, Batavia, Eds. A.S. Kronfeld and P. B. 
             Mackenzie, \NPBPS{9} (1989)}
\def\tasitwo{ in {\it ``From Actions to Answers''}, Proceedings of the 1989 TASI in 
             Elementary Particle Physics, Boulder, Eds. T. DeGrand and D. Toussaint, 
             World Scientific 1990}
\def\capri{ in {\it ``LATTICE 89''}, Proceedings of the 1989 Symposium on Lattice 
             Field Theory, Capri, Italy, 1989, Eds. N. Cabibbo $et\ al.$, 
             \NPBPS{17}, (1990)}
\def\talla{Int. Symp. {\it ``LATTICE 90''}, Proceedings of the International 
             Conference on Lattice Field Theory, Tallahassee, Florida, 1990, 
             Eds. U. M. Heller $et\ al.$, \NPBPS{20}, (1991) }
\def\tsukuba{Int. Symp. {\it ``LATTICE 91''}, Proceedings of the 
             International Symposium on Lattice Field Theory, Tsukuba, Japan, 
             1991, Eds. Fukugita $et\ al.$, \NPBPS{26}, (1992) }
\def\amsterdam{{\it ``LATTICE 92''}, Proceedings of
             the International Symposium on Lattice Field Theory, Amsterdam, 
             The Netherlands, 1992, Eds. J.~Smit $et\ al.$, 
             \NPBPS{30}, (1993) }
\def\dallas{{\it ``LATTICE 93''}, Proceedings of
             the International Symposium on Lattice Field Theory, Dallas, 
             U.S.A., 1993, Eds. T.~Draper  $et\ al.$, 
             \NPBPS{34}, (1994) }
\def\bielefeld{{\it ``LATTICE 94''}, Proceedings of
             the International Symposium on Lattice Field Theory, Bielefeld, 
             Germany, 1994, Eds. F.~Karsch $et\ al.$, 
             \NPBPS{42}, (1995) }

\def\melbourne{{\it ``LATTICE 95''}, To appear in Proceedings of
             the International Symposium on Lattice Field Theory, Melbourne,
             Australia, 1995, Eds. T.~D.~Kieu $et\ al.$, 
             \NPBPS{}, (1996)}

\def\ganga{R. Gupta, G. Kilcup and S. Sharpe}
\def\gangb{R. Gupta, G. Kilcup, A. Patel and S. Sharpe}
\def\gangc{R. Gupta, G. Guralnik, G. Kilcup, A. Patel and S. Sharpe}
\def\ucsb{S. A. Gottlieb, W. Liu, D. Toussaint,R. L. Renken and R. L. Sugar}
\def\fsuhmca{K. Bitar, A.D. Kennedy, R. Horsley, S. Meyer and P. Rossi}

\def\HMCA{S. Duane, A.D. Kennedy, B. Pendleton, D. Roweth, \PL{195} (1987) 216}
\def\swref{B.~Sheikholeslami and R.~Wohlert, \NPB{259} (1985) 572.}

%% file: chap-intro.tex
\section{Introduction}
\label{s:intro}

The goal of the lectures on lattice QCD (LQCD) is to provide an
overview of both the technical issues and the progress made so far in
obtaining phenomenologically useful numbers. The lectures consist of
three parts. My charter is to provide an introduction to LQCD and
outline the scope of LQCD calculations.  In the second set of
lectures, Guido Martinelli will discuss the progress we have made so
far in obtaining results, and their impact on Standard Model
phenomenology. Finally, Martin L\"uscher will discuss the topical
subjects of chiral symmetry, improved formulation of lattice QCD, and
the impact these improvements will have on the quality of results
expected from the next generation of simulations.

QCD is the regnant theory of strong
interactions. It is formulated in terms of quarks and gluons which we
believe are the basic degrees of freedom that make up hadronic matter.
It has been very successful in predicting phenomena involving large
momentum transfer. In this regime the coupling constant is small and
perturbation theory becomes a reliable tool. On the other hand, at the
scale of the hadronic world, $\mu \lsim 1$ GeV, the coupling constant
is of order unity and perturbative methods fail. In this domain
lattice QCD provides a non-perturbative tool for calculating the
hadronic spectrum and the matrix elements of any operator within these
hadronic states from first principles. LQCD can also be used to
address issues like the mechanism for confinement and chiral symmetry
breaking, the role of topology, and the equilibrium properties of QCD at
finite temperature. Unfortunately, these latter topics were not
covered at this school, so I will give appropriate references to compensate 
for their omission.

Lattice QCD is QCD formulated on a discrete Euclidean space time
grid. Since no new parameters or field variables are introduced in
this discretization, LQCD retains the fundamental character of QCD.
Lattice QCD can serve two purposes.  First, the discrete space-time
lattice acts as a non-perturbative regularization scheme. At finite
values of the lattice spacing $a$, which provides an ultraviolet
cutoff at $\pi/a$, there are no infinities. Furthermore, renormalized
physical quantities have a finite well behaved limit as $a \to 0$.
Thus, in principle, one could do all the standard perturbative
calculations using lattice regularization, however, these calculations
are far more complicated and have no advantage over those done in a
continuum scheme.  Second, the pre-eminent use of transcribing QCD on to a
space-time lattice is that LQCD can be simulated on the computer using
methods analogous to those used for Statistical Mechanics systems.  (A
brief review of the connection between Euclidean field theory and
Statistical Mechanics is given in Section~\ref{s:connection}.)  These
simulations allow us to calculate correlation functions of hadronic
operators and matrix elements of any operator between hadronic states
in terms of the fundamental quark and gluon degrees of freedom.

The only tunable input parameters in these simulations are the strong
coupling constant and the bare masses of the quarks.  Our belief is
that these parameters are prescribed by some yet more fundamental
underlying theory, however, within the context of the standard model
they have to be fixed in terms of an equal number of experimental
quantities. This is what is done in LQCD. Thereafter all predictions
of LQCD have to match experimental data if QCD is the correct theory
of strong interactions.  

A very useful feature of LQCD is that one can dial the input
parameters.  Therefore, in addition to testing QCD we can make
detailed predictions of the dependence of quantities on $\alpha_s$ and
the quark masses.  These predictions can then be used to constrain
effective theories like chiral perturbation theory, heavy quark
effective theory, and various phenomenological models.

My first lecture will be devoted to an overview of the scope of LQCD
and to showing that simulations of LQCD are a step by step implementation
of field theory.  The second lecture will be devoted to explaining the
details of how to transcribe the quark and gluon degrees of freedom on
to the lattice, and to construct an action that, in the limit of zero
lattice spacing, gives continuum QCD. I will also spend some time on
issues of gauge invariance, chiral symmetry, fermion doubling problem,
designing improved actions, the measure of integration, and gauge
fixing.

Numerical simulations of LQCD are based on a Monte Carlo integration
of the Euclidean path integral, consequently, the measurements have
statistical errors in addition to the systematic errors due to lattice
discretization. In order to judge the quality of lattice calculations
it is important to understand the origin of these errors, what is
being done to quantify them, and finally what will it take to achieve
results with a given precision.  These issues will be covered in the
third lecture along with an elementary discussion of Monte Carlo
methods.

The fourth lecture will be devoted to the most basic applications of
LQCD -- the calculation of the hadron spectrum and the extraction of
quark masses and $\alpha_s$.  Progress in LQCD has required a
combination of improvements in formulation, numerical techniques, and
in computer technology.  My overall message is that current LQCD
results, obtained from simulations at $a \approx 0.05-0.1$ fermi and
with quark masses roughly equal to $m_s$, have already made an impact on
Standard Model phenomenology. I hope that the three sets of lectures
succeed in communicating the excitement of the practitioners.

In a short course like this I can only give a very brief description
of LQCD.  In preparing these lectures I have found the following books
\cite{Creutz83,Creutz92,Montvay,Rothe}, and reviews
\cite{Sharpe94Book,TASI89,Kogut79RMP,Kogut83RMP} very useful. In addition, 
the proceedings of the yearly conferences on lattice field theory
\cite{LQCDproceedings} are excellent sources of information on 
current activity and on the status of results. I hope that the lectures at
this school and these references will allow any interested reader to
master the subject.

\section{Standard Model of Particle Interactions}
\label{s_SModel}

The Standard Model (SM) of particle interactions is a synthesis 
of three of the four forces of nature.  These forces are described 
by gauge theories, each of which is characterized by a 
coupling constant as listed below. 

\begin{description}
\item[] STRONG INTERACTIONS    \hfill  $\alpha_s \sim 1$ 
\item[] ELECTROMAGNETIC INTERACTIONS    \hfill  $\alpha_{em} \approx 1/137$ 
\item[] WEAK INTERACTIONS    \hfill  $G_F \approx 10^{-5}\ GeV^{-2}$. 
\end{description}

The basic constituents of matter are the six quarks, $u,\ d,\ s,\ c,\
b,\ t$, each of which comes in 3 colors, and the six leptons $e,\
\nu_e,\ \mu,\ \nu_\mu,\ \tau,\ \nu_\tau$.  The quarks and leptons are
classified into 3 generations of families.  The interactions between
these particles is mediated by vector bosons: the 8 gluons mediate
strong interactions, the $W^{\pm}$ and $Z$ mediate weak interactions,
and the electromagnetic interactions are carried by the photon
$\gamma$.  The weak bosons acquire a mass through the Higgs mechanism,
and in the minimal SM formulation there should exist one neutral Higgs
particle.  This has not yet been seen experimentally.  The
mathematical structure that describes the interactions of these
constituents is a local gauge field theory with the gauge group $SU(3)
\times SU(2) \times U(1)$.  The parameters that characterize the SM
are listed below; their origins are in an even more
fundamental but as yet undiscovered theory.

\vfill\newpage
\centerline{\bf Parameters in the Standard Model}
\medskip
\setlength\tabcolsep{0.2cm}
\begin{tabular}{lll}
Parameters            & Number           & Comments          \\
\myhline 
 Masses of quarks          & 6         & $u,\ d,\ s$ light    \\
                           &           & $c,\ b$ heavy        \\
                           &           & $t=175\pm6\ GeV$ \\
&& \\
 Masses of leptons         & 6         & $e,\ \mu,\ \tau$         \\
                           &           & $M_{\nu_e,\ \nu_\mu,\ \nu_\tau}=?$  \\
&& \\
 Mass of $W^{\pm}$         & 1         & 80.3 GeV               \\
 Mass of $Z$               & 1         & 91.2 GeV               \\
 Mass of gluons, $\ \gamma$&           &  0 (Gauge symmetry)  \\
&& \\
 Mass of Higgs             & 1         &  Not yet seen           \\
&& \\
 Coupling $\alpha_s$       & 1         & $\approx 1$ for energies $ \lsim 1$ GeV   \\
 Coupling $\alpha_{em}$    & 1         & 1/137 (=1/128.9 at $M_Z$)          \\
 Coupling $G_F$            & 1         & $10^{-5}\ GeV^{-2}$                \\
&& \\
 Weak Mixing Angles        & 3         & $\theta_1$, $\theta_2$, $\theta_3$ \\
 CP Violating phase        & 1         & $\delta$             \\
&& \\
 Strong CP parameter       & 1         & $\Theta =?$        \\
&& \\
\end{tabular}
\bigskip

The question marks indicate that these values are tiny and not yet
measured.  No mixing angles have been put down for the lepton sector
as I have assumed that the neutrino masses are all zero. A status
report on experimental searches for neutrino masses and their mixing
has been presented by Prof. Sciulli at this school.  The structure of
weak interactions and the origin of the weak mixing angles has been
reviewed by Daniel Treille, while David Kosower covered perturbative
QCD.

In the SM, the charged current interactions of the W-bosons with
the quarks are given by
\be
{\cal H}_W \ \ = \ \ {1 \over 2} (\ J_\mu W^\mu \ + \ h.c.)
\ee
where the current is 
\be
J_\mu \ = \ \ {g_w \over 2\sqrt2}
          \ \ (\bar u, \bar c, \bar t) \ \gamma_\mu (1-\gamma_5)\ V \
              \left(\matrix{d\cr s\cr b\cr}\right)
\ee
and the Fermi coupling constant is related to the $SU(2)$ coupling as
${G_F / \sqrt2} = {g_w^2 / {8 M_W^2}}$. 
$V$ is the $3 \times 3$ Cabibbo-Kobayashi-Maskawa ($CKM$) matrix that 
has a simple representation in terms of the flavor transformation matrix 
\be
V_{CKM} \ \ = \ \ \left(\matrix{
V_{ud}     & V_{us}         & V_{ub}    \cr
V_{cd}     & V_{cs}         & V_{cb}    \cr
V_{td}     & V_{ts}         & V_{tb}    \cr
                         }\right)
\ee
This matrix has off-diagonal entries because the eigenstates of weak
interactions are not the same as those of strong interactions, $i.e.$
the $d,\ s,\ b$ quarks (or equivalently the $u, c, t$ quarks) mix
under weak interactions.

For 3 generations the unitarity condition $V^{-1} = V^\dagger$ imposes
constraints.  As a result there are only four independent parameters
that can be expressed in terms of 4 angles,
$\theta_1,\theta_2,\theta_3$ and the CP violating phase $\delta$. The 
representation originally proposed by Kobayashi and Masakawa is 
\be
V_{CKM} \ \ = \ \ \left(\matrix{
c_1     &-s_1c_3                         &-s_1s_3                     \cr
s_1c_2  &c_1c_2c_3-s_2s_3e^{i\delta}     &c_1c_2s_3+s_2c_3e^{i\delta} \cr
s_1s_2  &c_1s_2c_3+c_2s_3e^{i\delta}     &c_1s_2s_3-c_2c_3e^{i\delta} \cr
                         }\right)
\label{eq:CKMstandard}
\ee
where $ c_i = \cos\theta_i\ $ and $s_i = \sin\theta_i \ $
for $i=1,2,3$. A phenomenologically more useful 
representation of this matrix is the Wolfenstein parameterization 
(correct to $O(\lambda^4)$) 
\be
V_{CKM} \ = \ 
\left(\matrix{
1-\lambda^2/2     & \lambda         & \lambda^3A(\rho-i\eta)   \cr
 -\lambda         & 1-\lambda^2/2   & \lambda^2 A              \cr
\lambda^3A(1-\rho-i\eta) & -\lambda^2 A & 1    \cr
                         }\right)  + O(\lambda^4).
\label{eq:CKMWolfenstein}
\ee
where $\lambda = \sin \theta_c = 0.22$ is known to $1\%$ accuracy, $A
\approx 0.83$ is known only to $10\%$ accuracy, and $\rho$ and $\eta$
are poorly known.  The elements $V_{td}$ and $V_{ub}$ of the CKM
matrix given in Eq.~\ref{eq:CKMWolfenstein} are complex if $\eta \neq
0$, and as a result there exists a natural mechanism for CP-violation
in the SM. The phenomenology of the CKM matrix and the determinations 
of the four parameters have been reviewed by Profs. Buras and 
Richman at this school \cite{LH97Buras,LH97Richman}.

\begin{figure} 
\hbox{\epsfxsize=\hsize\epsfbox{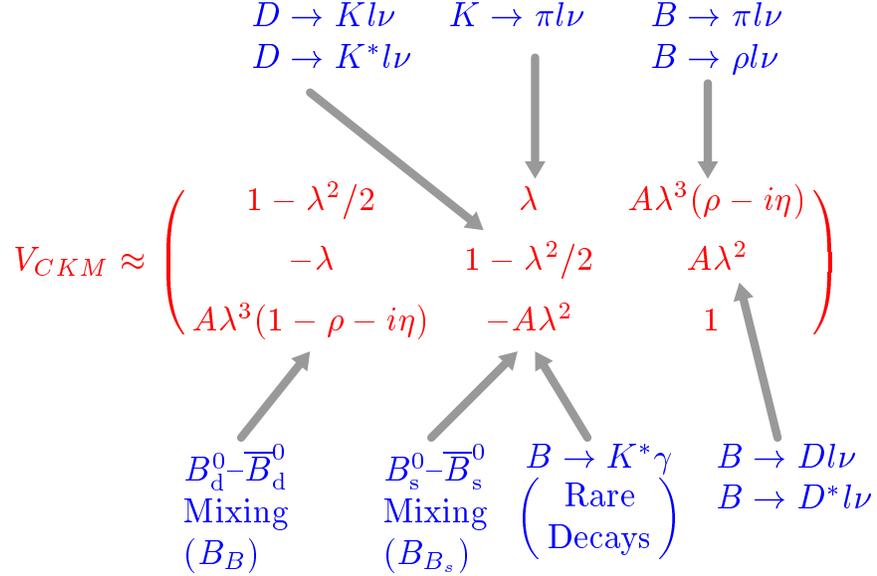}}
\caption{The CKM matrix in the Wolfenstein parameterization. I show
examples of physical processes that need to be measured experimentally
and compared with theoretical predictions via the ``master'' equation
to estimate various elements of the CKM matrix.}
\label{f_CKM1}
\end{figure}

A list of processes that are amongst the most sensitive probes of the
CKM parameters is shown in Fig.~\ref{f_CKM1}.  Lattice QCD
calculations are beginning to provide amongst the most reliable
estimates of the hadronic matrix elements relevant to these
processes.  Details of these calculations and their impact on the
determination of $\rho$ and $\eta$ parameters will be covered by
Prof. Martinelli at this school. My goal is to provide you with the
necessary background.

\section{Scope of Lattice QCD Calculations}
\label{s:scope}

\begin{figure} 
\hbox{\epsfxsize=\hsize\epsfbox{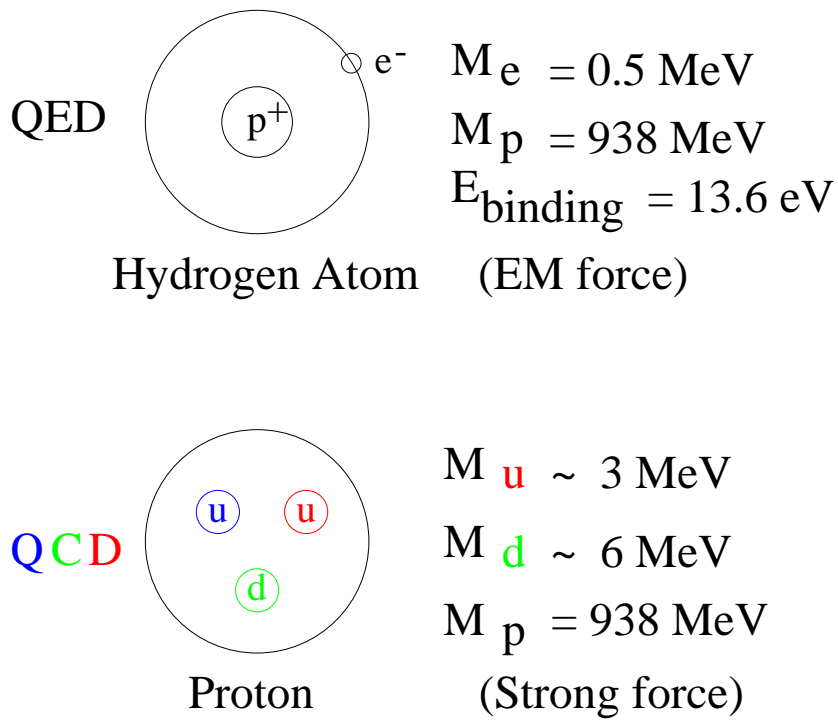}}
\caption{An illustration of the difference in the binding energy
between the electron and the proton in a hydrogen atom (interacting
via electromagnetic forces) and the quarks in a proton (strong
force). }
\label{f_hydrogen}
\end{figure}

The primary goal of LQCD is to test whether QCD is the
correct theory of strong interactions.  Most particle physicists
believe that the answer is yes because of its successes in describing
processes with large momentum transfer, its mathematical elegance, and
uniqueness.  However, quantitative confirmation is lacking.  The
problem with the conventional perturbative approach (expansion in a
small parameter) when analyzing hadronic process at scales $\lsim 1 $
GeV is that the strong coupling constant $\alpha_s \sim 1$.  Thus,
perturbation theory in $\alpha_s$ is not reliable.  As a result we
cannot calculate the masses of mesons and baryons from QCD even if we
are given $\alpha_s$ and the masses of quarks.  This is illustrated in
Fig.~\ref{f_hydrogen} where I show a comparison between the binding
energy of a hydrogen atom and that of the proton. It turns out that
almost all the mass of the proton is attributed to the strong
non-linear interactions of the gluons.

Simulations of lattice QCD have six unknown input parameters. These
are the coupling constant $\alpha_s$ and the masses of the up, down,
strange, charm and bottom quarks (the top quark is too short lived,
$0.4 \times 10^{-24}$ seconds, to form bound states and is thus not
suitable for lattice studies). Once these have been fixed in terms of
six precisely measured masses of hadrons, thereafter, the masses and
properties of the hundreds of other particles made up of these quarks
and gluons have to agree with experiments.  For example, consider the
chiral perturbation theory ($\cpt$) relation for the pion mass
\be
(M_\pi^+)^2 = B_\pi (m_u + m_d) + C_\pi (m_u+m_d)^2 + \ldots .
\label{eq:mpichiral}
\ee
Lattice calculations will provide a check of this relation and fix the
constants $B_\pi, C_\pi, \ldots$, $i.e.$ constrain the chiral lagrangian. In
addition, having measured $M_\pi$ for a number of values of the quark
mass, one can invert Eq.~\ref{eq:mpichiral} to determine the current
quark masses. I shall call such tests of QCD and the determination of
the six unknown parameters of the $SM$ a direct application of lattice
QCD.

\begin{figure} 
\hbox{\epsfxsize=\hsize\epsfbox{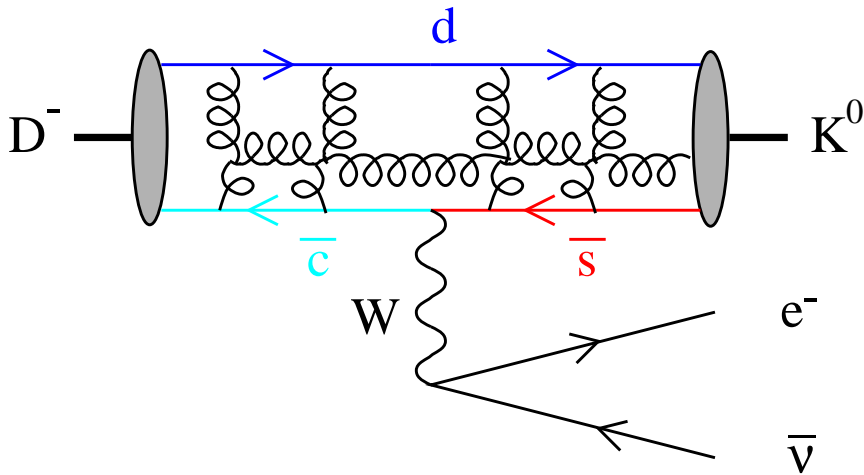}}
\caption{The Feynman diagram for the semi-leptonic decay $D^- \to K^0
e^- \bar \nu_e$.  The QCD corrections are illustrated by the various
gluons being exchanged between the initial and final hadrons. The
leptonic vertex can be calculated reliably using perturbation theory, 
whereas the hadronic vertex requires non-perturbative methods.}
\label{f_semileptonic1}
\end{figure}

The second area in which perturbative estimates are not reliable is in
the calculation of the matrix elements occurring in the weak decays of
hadrons. The reason is that the non-perturbative QCD corrections to
the basic weak process can be large due to the exchange of soft gluons
between the initial and final states.  This is illustrated in
Fig.~\ref{f_semileptonic1} for the case of the semi-leptonic decay $D
\to K l \nu$.  In this case the matrix element of the weak
interactions Hamiltonian between the initial $D$ meson and final kaon 
receives large corrections from QCD which cannot be estimated using PQCD. 

The result of the experiment is a number, $i.e.$ the decay rate.  To
check whether the Standard Model prediction is in agreement with this
number theorists must derive an analytical expression for the same
process.  This expression, in general, consists of a product of three
parts: a combination of the parameters in the $SM$, matrix elements
($ME$) of the interaction Hamiltonian between the initial and final
states, and kinematic factors. (This is the well-known Fermi's golden
rule.)  Schematically one can write this ``master'' equation as
follows.
\be 
Expt. \#  =  \big( {\rm SM\ parameters }    \big) 
             \big( {\rm matrix\ elements }  \big) 
	     \big( {\rm kinematic\ factors }\big) .  
\label{eq:master}
\ee
Thus, for each such experimental number one gets a constraint on one
particular combination of the $SM$ parameters provided the $ME$ are
known. By using several precision experiments and calculating the
necessary $ME$ one can over-constrain the model and therefore test it.
I shall call lattice calculations of $ME$ an ``indirect'' probe of
Standard Model parameters as one usually needs additional
phenomenological/experimental input to extract the unknown parameters
via the master equation.


Physicists have been busy calculating and measuring these parameters
for the last 25 years, so one already has reasonable estimates.  It
is, therefore, useful to keep in mind the level of precision that
lattice calculations have to aim for to make an impact. Non-lattice
estimates, taken from the 1996 Review of Particle Properties and
from Prof. Buras's lectures \cite{LH97Buras}, are

\begin{tabular}{lll}
Parameters            &      Value       & Comments          \\
\hline
 $m_u(\bar{MS}, 2\GeV)$    & 3-4.5 MeV           & $\chi$PT, sum-rules  \\
 $m_d(\bar{MS}, 2\GeV)$    & 5-8 MeV             & $\chi$PT, sum-rules  \\
 $m_s(\bar{MS}, 2\GeV)$    & 100-140 MeV         & $\chi$PT, sum-rules  \\
 $m_c(\bar{MS}, m_c)$      & 1.0-1.6 GeV         & $J/\psi$   sprectra  \\
 $m_b(\bar{MS}, m_b)$      & 4.1-4.6 GeV         & $\Upsilon$ spectra \\
&& \\
 $M_{Higgs}$               &           &  Not Found           \\
&& \\
 $\alpha_s(M_Z)$           & $0.118\pm 0.003$    & World average   \\
&& \\
 $|V_{ud}|$                & $0.9736\pm 0.0010$  & $n \to p e \nu_e$ and \\
                           &                     & $\mu \to e \bar \nu_e \nu_\mu$\\
 $|V_{us}|$                & $0.2196\pm 0.0023$  & $K^+\to \pi^0e^+\nu_e$ and \\
                           &                     & $K_L^0 \to \pi^\pm e^\mp \nu_e$\\
 $|V_{ub}|$                & $(3.3\pm 0.2\pm 0.4\pm 0.7) 10^{-3}$  & $B \to \pi l\nu$; $B \to \rho l\nu $ \\
 $|V_{cb}|$                & $0.040\pm 0.003$    & $B \to X_c l\nu$ and  \\
                           &                     & $B \to D^*l\nu$ \\
&& \\
 $|\epsilon|$              & $2.26 \times 10^{-3}$ &  $K^0 \leftrightarrow \bar{K^0}$ mixing       \\
 $|{\epsilon / \epsilon'}|$ & $0-30 \times 10^{-4}$ &  $K\to\pi\pi$ decays \\
&& \\
\end{tabular}
\bigskip

\noindent Note that only the magnitude of the CKM elements can be
determined from experiments.  Clearly, the most enticing opportunity for
lattice QCD is to help determine the quark masses, $\alpha_s$, and the
essentially unknown Wolfenstein parameters $\rho $ and $\eta$. For
other elements of $V_{CKM}$ one has reasonable estimates, however here
too further progress requires significant improvement in the estimates
of the hadronic matrix elements.  Phenomenologists are relying on LQCD 
to provide improved estimates for these. 

\medskip

Let me end this section by mentioning that in addition to validating
QCD as the correct theory of strong interactions by reproducing the
measured spectrum of mesons, baryons, and glueballs, and the
calculation of weak matrix elements, simulations of LQCD will also
allow us to
\begin{itemize}
\item
Investigate the topological structure of the QCD vacuum, and the
mechanisms responsible for confinement and spontaneously broken chiral
symmetry. For a recent review see ~\cite{TOP97vanbaal}.
\item
Calculate hadronic properties like wavefunctions, decay constants,
form factors for electromagnetic, weak, and rare decays, and structure
functions for deep inelastic scattering. Estimate the pion nucleon
sigma term and the proton spin carried by the quarks by 
calculating  the matrix elements $<N|\overline qq |N>$ and
$<N|\overline q \gamma_\mu\gamma_5 q |N>$.  Some of these topics will 
be covered by Prof. Martinelli, for others a good reference is 
the proceedings of LATTICE conferences~\cite{LQCDproceedings}. 
\item
Analyze the behavior of QCD at high temperatures and address questions
like the nature of the phase transition, the transition temperature
$T_c$, the equation of state, and the collective excitations near and
above $T_c$.  The topic of QCD at finite temperature has been covered
by J-P Blaizot at this school, so I will only list for you my favorite
pedagogical reviews of the subject \cite{finitetemperature}.
The lectures by Profs. Martinelli, L\"uscher, and me will be
confined to understanding and probing the properties of QCD at zero
temperature.
\end{itemize}

\section{Overview of the Lattice Approach}
\label{s:overviewLQCD}

LQCD calculations are a non-perturbative implementation of field theory
using the Feynman path integral approach. The calculations proceed
exactly as if the field theory was being solved analytically had we
the ability to do the calculations.  The starting point is the
partition function in Euclidean space-time
\begin{equation}
Z \ = \
\int {\cal D}A_{\mu} \ {\cal D}\psi \ {\cal D}\bar \psi \ e^{-S}
\end{equation}
where $S$ is the QCD action 
\begin{equation}
\CS \ = \ \int d^4 x \ \big( {1\over 4} F_{\mu\nu}F^{\mu\nu} -
                            \bar \psi M \psi \big) \ .
\end{equation}
and $M$ is the Dirac operator. The fermions are represented by Grassmann 
variables $\psi$ and $\bar \psi$. These can be integrated out exactly with 
the result 
\begin{equation}
Z \ = \
\int {\cal D}A_{\mu} \ {\rm det}M\ e^{\int d^4 x \ (-{1\over 4} F_{\mu\nu}F^{\mu\nu})} .
\label{eq:pathintegral}
\end{equation}
The fermionic contribution is now contained in the highly non-local
term ${\rm det}M $, and the partition function is an integral over
only background gauge configurations.  One can write the action, after
integration over the fermions, as $S \ = {S_{gauge} + S_{quarks}} = \
\int d^4 x \ \big( {1\over 4} F_{\mu\nu}F^{\mu\nu} \big) - \sum_i {\log}
( {\rm Det} M_i) $ where the sum is over the quark flavors,
distinguished by the value of the bare quark mass.

It is expedient to define the ``quenched'' approximation (QQCD) from the 
very start due to the central role it has played in simulations. It
consists of setting ${\rm Det} M = constant$ which corresponds to
removing vacuum polarization effects from the QCD vacuum.  Details and
some consequences of this approximation are discussed in
Section~\ref{ss:QQCD}.

Results for physical
observables are obtained by calculating expectation values
\begin{equation}
\vev{{\cal O}} \ = \ {1 \over Z}
\int {\cal D} A_{\mu} \ {\cal O} \ e^{-S} \ .
\label{eq:expvalue}
\end{equation}
where ${\cal O}$ is any given combination of operators expressed in
terms of time-ordered products of gauge and quark
fields. (Expectation values in the path integral approach correspond 
to time-ordered correlation functions.)  The quarks
fields in $\CO$ are, in practice, re-expressed in terms of quark
propagators using Wick's theorem for contracting fields. In this way
all dependence on quarks as dynamical fields is removed. The basic
building block for fermionic quantities is the Feynman propagator,
\begin{equation}
S_F(y,j,b;x,i,a) \ = \ \big( M^{-1} \big)^{y,j,b}_{x,i,a} \, ,
\end{equation}
where $M^{-1}$ is the inverse of the Dirac operator calculated on a
given background field. A given element of this matrix $\big( M^{-1}
\big)^{y,j,b}_{x,i,a}$ is the amplitude for the propagation of a 
quark from site $x$ with spin-color $i,a$ to site-spin-color $y,j,b$. 

So far all of the above is standard field theory. The problem we face
in QCD is how to actually calculate these expectation values and how
to extract physical observables from these.  I will illustrate the
second part first by using as an example the mass and decay constant
of the pion.

Consider the 2-point correlation function, $\vev{0 | T[\sum_x {\cal
O}_f(\vec x, t) {\cal O}_i(\vec 0,0)] |0 }$ with $t > 0$, where the
operators ${\cal O}$ are chosen to be the fourth component of the
axial current $ {\cal O}_f = {\cal O}_i = A_4 = \bar \psi \gamma_4
\gamma_5 \psi$ as these have a large coupling to the pion. This 2-point
correlation function gives the amplitude for creating a state with the
quantum numbers of the pion out of the vacuum at space-time point $0$
by the ``source'' operator $ {\cal O}_i $; the evolution of this state
to the point $(\vec x,t)$ via the QCD Hamiltonian (or more precisely
the transfer matrix to be discussed in Section~\ref{s:connection});
and finally the annihilation by the ``sink'' operator ${\cal O}_f $ at
$(\vec x,t)$.  The rules of quantum mechanics tell us that ${\cal
O}_i$ will create a state that is a linear combination of all possible
eigenstates of the Hamiltonian that have the same quantum numbers as
the pion, $i.e.$ the pion, radial excitations of the pion, three pions
in $J=0$ state, $\ldots$. The second rule is that on propagating for
Euclidean time $t$, a given eigenstate with energy $E$ picks up a
weight $e^{-Et}$.  Thus, the 2-point function can be written in terms
of a sum over all possible intermediate states
\begin{equation}
\vev{0 | \sum_x {\cal O}_f(\vec x,t) {\cal O}_i(0) |0 } = \sum_n 
{\vev{0 | {\cal O}_f | n} \vev{n| {\cal O}_i | 0} \over 2 E_n} \ e^{-E_n t} \, .
\label{eq:2point}
\end{equation}
To study the properties of the pion at rest we need to isolate this state
from the sum over $n$. To do this, the first simplification is to use the
Fourier projection $\sum_{\vec x}$ as it restricts the sum over states to
just zero-momentum states, so $E_n \to M_n$. (Note that it is
sufficient to make the Fourier projection over either ${\cal O}_i$ or
${\cal O}_f$.)  The second step to isolate the pion, $i.e.$ project in
the energy, consists of a combination of two strategies. One, make a
clever choice of the operators ${\cal O}$ to limit the sum over
states to a single state (the ideal choice is to set ${\cal O}$
equal to the quantum mechanical wave-functional of the pion), and two,
examine the large $t$ behavior of the 2-point function where only the
contribution of the lowest energy state that couples to ${\cal O}_i$
is significant due to the exponential damping. Then 
\begin{equation}
\vev{0 | \sum_x {\cal O}_f(x,t) {\cal O}_i(0) |0 } 
\ \ \eqinf\ \ \ \ 
{\vev{0 | {\cal O}_f | \pi} \vev{\pi| {\cal O}_i | 0} \over 2 M_\pi} \ e^{-M_\pi t} \, .
\label{eq:2point1}
\end{equation}
The right hand side is now a function of the two quantities we want
since $ \vev{0 | A_4(\vec p = 0) | \pi} = M_\pi f_\pi$. In this way, the mass and
the decay constant are extracted from the rate of exponential fall-off
in time and from the amplitude.

\begin{figure}[t] 
\vspace{9pt}
\hbox{\hskip15bp\epsfxsize=0.9\hsize \epsfbox {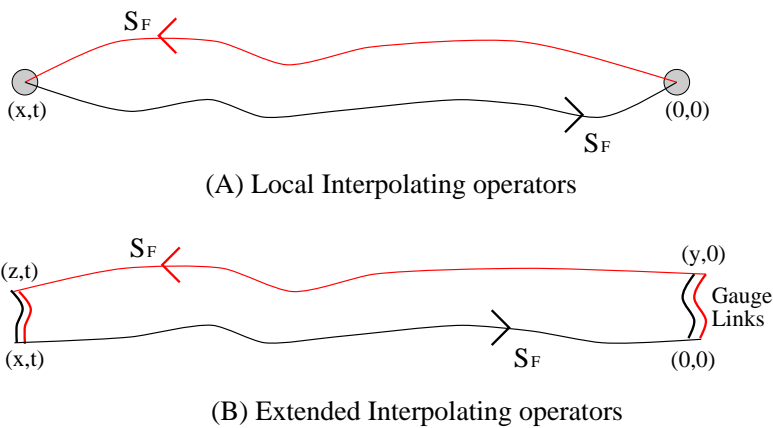}}
\caption{A schematic of the pion 2-point correlation function for (A) local  
and (B) non-local interpolating operators.}
\label{f:pionfig}
\end{figure}

Let me now illustrate how the left hand side of Eq.~\ref{eq:2point1}
is expressed in terms of the two basic quantities we control in the
path integral -- the gauge fields and the quark propagator.  Using
Wick contractions, the correlation function can be written in terms of
a product of two quark propagators $S_F$,
\bea
-\vev{0 | \sum_x \bar \psi (x,t) \gamma_4 \gamma_5 \psi(x,t) &{}& 
                \bar \psi (0,0) \gamma_4 \gamma_5 \psi(0,0) |0 }  \nonumber \\
 &{}& \hskip -0.6 in \equiv \vev{0 | \sum_x S_F(0; \vec x, t) \gamma_4\gamma_5 
                           S_F(\vec x, t; 0) \gamma_4 \gamma_5 | 0  } .
\label{eq:2point2}
\eea
This correlation function is shown in Fig.~\ref{f:pionfig}A. 
To illustrate this Wick contraction
procedure further consider using gauge invariant non-local operators,
for example take $\CO = \bar \psi(x,t) \gamma_4 \gamma_5 (\CP
e^{\int^{y}_{x}dz igA_\mu(z)}) \psi(y,t)$ where $\CP$ stands for
path-ordered.  After Wick contraction the correlation function reads
\begin{equation}
\vev{0 | \sum_x S_F(0; \vec x, t)   \gamma_4\gamma_5 (\CP e^{\int^{x}_{z} igA_\mu})
                S_F ( \vec z, t; \vec y,0) \gamma_4\gamma_5 (\CP e^{\int^{y}_{0} igA_\mu}) | 0  } .
\label{eq:2point3}
\end{equation}
and involves both the gauge fields and quark propagators. This
correlation function would have the same long $t$ behavior as shown in
Eq.~\ref{eq:2point}, however, the amplitude will be different and
consequently its relation to $f_\pi$ will no longer be simple. The
idea behind strategies for improving the projection of $\CO$ on to the
pion is to construct a suitable combination of such operators that
approximates the pion wave-function. For example, if $\phi(x)$ is the
pion wave-functional, then the ``smeared'' operator $\CO = \bar \psi(0)
\gamma_4 \gamma_5 \big( \int d^3x \phi(x) \psi(x) \big) $ is the ideal choice.

Finally, it is important to note that to get the 2-point function
corresponding to the propagation of the physical pion we have to
average the correlation function over gauge configurations as defined
in Eq.~\ref{eq:expvalue}, $i.e.$ do the functional integral.  

An overview of how such calculations of expectation values can be done 
consists of the following steps.


\noindent $\bullet$ {\it Defining a finite dimensional system}: The 
Yang-Mills action for gauge fields and the Dirac operator for fermions
have to be transcribed onto a discrete space-time lattice in such a
way as to preserve all the key properties of QCD -- gauge invariance,
chiral symmetry, topology, and a one-to-one relation between continuum
and lattice fields. This step is the most difficult, and even today we
do not have a really satisfactory lattice formulation that is chirally
symmetric in the $m_q=0$ limit and preserves a one-to-one relation
between continuum and lattice fields, $i.e.$ no doublers.  In fact,
the Nielson-Ninomiya theorem states that for a translationally
invariant, local, hermitian formulation of the lattice theory one
cannot simultaneously have chiral symmetry and no doublers
\cite{NNnogo}. One important consequence of this theorem is that,
in spite of tremendous effort, there is no practical formulation of
chiral fermions on the lattice. For a review of the problems and
attempts to solve them see
\cite{CGT95shamir,CGT97narayanan,CGT97testa} and \cite{LuscherHasenfratz}.

\noindent $\bullet$ {\it Discretization Errors}: This problem is encountered when
approximating derivatives in the action by finite differences, 
\bea 
\partial_x \phi(x) \to \Delta_x \phi(x) &=& {1 \over 2a} (\phi(x+a) - \phi(x-a)) \nonumber \\
                                        &=& {1 \over 2a} \big( e^{a \partial_x} - e^{-a \partial_x} \big) \phi(x) \,.
\eea
As is obvious this introduces discretization errors proportional to
the square of the lattice spacing $a$. These errors can be reduced by
either using higher order difference schemes with coefficients
adjusted to take into account effects of renormalization, or
equivalently, by adding appropriate combinations of irrelevant
operators to the action that cancel the errors order by order in $a$.
The various approaches to improving the gauge and Dirac actions are
discussed in Sections \ref{s:improvedglue} and \ref{s:improveddirac}.  In my
lectures I will concentrate on the three most frequently used
discretizations of the Dirac action -- Wilson \cite{Waction},
Sheikholeslami-Wohlert (clover)
\cite{SWaction}, and staggered \cite{STAGaction}, which have errors of
$O(a)$, $O(\alpha_s a)-O(a^2)$ depending on the value of the
coefficient of the clover term, and $O(a^2)$ respectively.  Reducing
the discretization errors by improving the action and operators
increases the reliability of the results, and for dynamical
simulations may even prove necessary due to limitations of computer
power. At the same time it is important to note that there exists a
well defined procedure for obtaining continuum results with even the
simplest formulation, Wilson's original gauge and fermion action
\cite{Waction}. In this case, however, it is necessary to quantify and
remove the discretization errors by a reliable extrapolation to $a=0$.

\noindent $\bullet$ {\it Generating background gauge fields}: 
The Euclidean action $\CS = \int d^4x (\ {1\over 4}
F_{\mu\nu}F^{\mu\nu} - {\rm Tr}\, \log M) $ for QCD at zero chemical
potential is real and bounded from below.  Thus $e^{-\CS}$ in the path
integral is analogous to the Boltzmann factor in the partition
function for statistical mechanics systems, $i.e.$ it can be regarded
as a probability weight for generating configurations.  Since $\CS$ is
an extensive quantity the range of $\CS$ is enormous, however, the
distribution is very highly peaked about configurations that minimize
the free energy. Configurations close to the peak dominate the
functional integral. Such important configurations can be generated by
setting up a Markov chain in exact analogy to, say, simulations of the
Ising model.  They are called ``importance sampled'' as they are
generated with probability of occurrence given by the weight
$e^{-\CS}$. For a discussion of the methods used to update the
configurations see \cite{Creutz83} or the lectures by Creutz and Sokal
in \cite{Creutz92}.

\noindent $\bullet$ {\it Calculation of the quark propagator on a given background gauge
field}: For a given background gauge configuration, the Feynman quark
propagator $S_F$ is a matrix labeled by three indices -- site, spin,
and color. A given element of this matrix gives the amplitude for the
propagation of a quark with some spin, color, and space-time point to
another space-time point, spin, and color.  Operationally, $S_F$ is
simply the inverse of the Dirac operator $M$.  Once space-time is made
discrete and finite, the Dirac matrix is also finite and its inverse
can be calculated numerically.  Inverting the Dirac operator is the
key computational step in LQCD and consumes over $90\%$ of the CPU
cycles.  To construct the correlation functions requires $S_F$ only 
from a given source point to all other sites of the lattice. This
corresponds to 12 columns (or rows) of $M^{-1}$, one each for the 12
spin and color degrees of freedom.

\noindent $\bullet$ {\it Correlation functions}: The correlation functions are expressed
as path ordered products of quark propagators and gauge fields using
Wick contraction of the operator fields. There are two kinds of
correlation functions that are calculated -- gauge invariant
correlation functions as discussed above for the case of the pion, or
those in a fixed gauge.

\noindent $\bullet$ {\it Expectation values}: The expectation value is simply an average
of the correlation function evaluated over the set of ``importance
sampled'' configurations. Even on a finite lattice the set of
background gauge configurations is infinite as the variables are
continuous valued.  Thus, while it is possible to calculate the
correlation functions on specified background gauge configurations,
doing the functional integral exactly is not feasible. It is,
therefore, done numerically using monte carlo methods.

My recommendation for the simplest way to understand the computational
aspects of LQCD calculations is to first gain familiarity with the
numerical treatment of any simple statistical mechanics system, for
example the Ising model.  The differences are: (i) the degrees of
freedom in LQCD are much more complicated -- SU(3) link matrices
rather than Ising spins, and quark propagators on these background
configurations evaluated by inverting the Dirac operator; (ii) The
action involves the highly nonlocal term ${\rm Ln\ Det\ } M$ which
makes the update of the gauge configurations very expensive; and (iii)
the correlation functions are not simple products of spin variables
like the specific heat or magnetic susceptibility, but complicated
functions of the link variables and quark propagators.

The subtleties arising due to the fact that LQCD is a renormalizable
quantum field theory and not a classical statistical mechanics system
come into play when analyzing the behavior of the correlation
functions as a function of the lattice spacing $a$, and in the quantum
corrections that renormalize the input parameters (quark and gluon fields 
and their masses) and the composite operators used in the
correlation functions. At first glance it might seem that we have
introduced an additional parameter, the lattice spacing $a$, in LQCD,
however, as will be shown in Section~\ref{s:confinement}, the coupling
$g$ and the cutoff $a$ are not independent quantities but are related
by the renormalization group as
\begin{equation}
\Lambda_{QCD} = \lim_{a \to 0} {1\over a} e^{-1/2\beta_0 g^2(a)}\ 
                \big(\beta_0 g^2(a) \big)^{-\beta_1/2\beta_0^2} \ ,
\label{eq:Lambdadef}
\end{equation}
where $\Lambda_{QCD}$ is the non-perturbative scale of QCD, and
$\beta_0 $ and $\beta_1$ are the first two, scheme independent,
coefficients of the $\beta$-function.  In statistical mechanics
systems, the lattice spacing $a$ is a physical quantity -- the
intermolecular separation. In a quantum field theory, it is simply the
ultraviolet regulator that must eventually be taken to zero keeping
physical quantities, like the renormalized couplings ($\Lambda_{QCD}$
and quark masses), spectrum, etc, fixed.

The reason that lattice results are not exact is because in numerical
simulations we have to make a number of approximations as discussed in
Section~\ref{s:syserrors}.  The size of the associated uncertainties
is dictated by the computer power at hand. They are being improved
steadily with computer technology, better numerical algorithms, and
better theoretical understanding. I shall return to the issue of
errors after discussing the formulation of LQCD.

%% file: chap-connection.tex
\section{Connection between Minkowski and Euclidean Field Theory and with Statistical Mechanics}
\label{s:connection}

This section is a brief recapitulation of technical issues regarding the
Euclidean path integral and extracting physics from Euclidean
correlation functions.  For details see \cite{Montvay,Rothe} or the texts 
on statistical field theory, for example~\cite{FT88Parisi,FT90Zinn}. 

A $D$-dimensional Minkowski field theory ($D-1$ spatial dimensions and
one time dimension) is connected to a $D$-dimensional Euclidean
field theory through analytical continuation. Under Wick rotation 
\bea
x_0 &\equiv& t \to -i x_4 \equiv -i \tau \,, \nonumber \\
p_0 &\equiv& E \to i p_4 \,.
\eea
The  Euclidean convention is 
\bea
x_E^2 &=& \sum_{i=1}^4 x_i^2 = \vec{x}\,^2-t^2 = -x_M^2 \,, \nonumber \\
p_E^2 &=& \sum_{i=1}^4 p_i^2 = \vec{p}\,^2-E^2 = -p_M^2 \,.
\eea

The connection between statistical mechanics and a Euclidean field
theory in $D$-dimensions is most transparent when the field theory is
quantized using the Feynman path integral approach. This connection 
forms the basis of numerical simulations using methods common to 
those used to study statistical mechanics systems. A synopsis of the
equivalences between the two is given in Table~\ref{t:connection}. 

\begin{table} 
\begin{center}
\setlength\tabcolsep{0.15cm}
\caption{The equivalences between a Euclidean
field theory and Classical Statistical Mechanics.}
\begin{tabular}{|l|l|}
\myhline
Euclidean Field Theory           &  Classical Statistical Mechanics \\
\myhline
Action                           & Hamiltonian                     \\
unit of action $h$               & units of energy $\beta = 1/kT$  \\
Feynman weight for amplitudes    & Boltzmann factor $e^{-\beta H}$ \\
$\qquad e^{-\CS/h}=e^{-\int\CL dt /h}$   &                                 \\
Vacuum to vacuum amplitude       & Partition function $\sum_{conf.} e^{-\beta H}$  \\
$\qquad$ $\int\CD\phi e^{-\CS/h}$  &                                  \\
Vacuum energy                    & Free Energy                      \\
Vacuum expectation value $\ME{0}{\CO}{0}$ & Canonical ensemble average $\VEV{\CO}$ \\
Time ordered products            & Ordinary products                \\
Green's functions $\ME{0}{T[\CO_1\ldots\CO_n]}{0}$ & Correlation functions $\VEV{\CO_1\ldots\CO_n}$ \\
Mass $M$                         & correlation length $\xi = 1/M$ \\
Mass-gap                         & exponential decrease of correlation functions \\
Mass-less excitations            & spin waves  \\
Regularization: cutoff $\Lambda$ & lattice spacing $a$ \\
Renormalization: $\Lambda \to \infty$ & continuum limit $a \to 0$ \\
Changes in the vacuum            & phase transitions \\
\myhline
\end{tabular}
\label{t:connection}
\end{center}
\end{table}
\smallskip

Let me start with a quick summary of the Euclidean theory.  Consider
the generic 2-point correlation function $\Gamma_M(t,\vec k) = \ME{0}{\int
d^3 x e^{-ik.x} T[ \CF(x,t) \CF^\dagger(0,0) ]}{0}$ in Minkowski
space.  The Heisenberg operator $\CF(x,t)$ can be written as
\be
\CF(x,t) = e^{i\CH t-ipx} \CF e^{-i\CH t+ipx} 
\ee
so that 
\be
\Gamma_M(t,\vec k) = \ME{0}{\int d^3 x e^{-ik.x} 
               \bigg(e^{i\CH t-ipx} \CF e^{-i\CH t+ipx} \CF^\dagger \bigg) }{0}
\ee
where, in dropping the time-ordered product, it is assumed that $t >0$. 
Using the invariance of the vacuum under time and space 
translations and assuming $\CF^\dagger = \CF$ for brevity, this becomes 
\bea
\Gamma_M(t,\vec k) &=& \ME{0}{\int d^3 x e^{-ik.x} \bigg(\CF e^{-i\CH t+ipx} \CF \bigg) }{0} \nonumber \\
                   &=& \ME{0}{\delta(\vec p - \vec k) \bigg(\CF e^{-i\CH t} \CF \bigg) }{0} \,.
\eea
Since the time dependence is explicit, we can Wick rotate to Euclidean space 
using $t \to -i\tau$ and keeping $\CH $ unchanged
\be
\Gamma_E(t) = \ME{0}{\delta(\vec p - \vec k) \bigg(\CF e^{-\CH \tau} \CF \bigg) }{0} \ . 
\ee
To get the path integral representation we divide the time interval into $N$ steps 
$\tau = Na$ and insert a complete set of states $| \phi \rangle \langle \phi |$ at 
each step. 
\bea 
\Gamma_E(t) = \delta(\vec p - \vec k) \sum_{\phi_1 \ldots \phi_N} &{}&
				      \ME{0}{\CF}{\phi_1} 
                                      \ME{\phi_1}{e^{-\CH a}}{\phi_2} \ldots \nonumber \\
                                 &{}& \ldots \ME{\phi_{N-1}}{e^{-\CH a}}{\phi_N} 
                                      \ME{\phi_N}{\CF}{0} 
\label{eq:gen2pointA}
\eea
The discussion of how, by this introduction of the complete set of
states, one trades the operator form of the quantum mechanical
amplitude by the sum over paths is standard and thus not repeated.
On this ``lattice'' the transfer matrix $\CT \equiv \exp(-\CH
a)$ plays the role of the time translation operator, from which
the continuum Hamiltonian is obtained as
\be
\CH = \lim_{a \to 0} \ - {1 \over a } \ \log \CT \,.
\ee
In order for $\CH$ to be a self-adjoint Hamiltonian it is necessary
that $\CT$ is a symmetric, bounded, and positive operator acting on a
Hilbert space of states with a positive norm. The necessary condition
for this, is the Osterwalder-Schrader {\it reflection positivity}
\cite{RP73OS}. That this condition holds for LQCD has been shown by an
explicit construction by M.~L\"uscher~\cite{TM77luscher}. Therefore,
the correlation functions in the desired physical Minkowski theory can
be obtained by analytical continuation of their Euclidean
counterparts.

Assuming that the conditions for a sensible Euclidean theory have been
established, the vacuum to vacuum amplitude (also called the
generating functional or the partition function), in presence of a source
$J$, is 
\be 
Z[J] = \int d \phi e^{-\CS_E + J \phi}
\ee
where $S_E$ is the Euclidean action.  Correlation functions can be
obtained from this in the standard way by differentiating $\log Z[J]$
with respect to $J$, for example $\ME{0}{\phi(x) \phi(0)}{0} = {\delta
\over \delta J(x)} {\delta \over \delta J(0)} \log Z[J] |_{J=0}$.

Lastly, I summarize the relations between Euclidean and Minkowski quantities. 
The basic tool is the repeated insertion of a complete set of momentum states in
Eq.~\ref{eq:gen2pointA}. I use the relativistic normalization of states
\be
\langle k | p \rangle = 2 E L^3 \delta_{k_xp_x} \delta_{k_yp_y} \delta_{k_zp_z} 
		      \ \ { \buildrel {L \to \infty} \over \longrightarrow  } \ \ {}
                        2 E (2 \pi)^3 \delta^3(\vec p - \vec k) \,.
\ee
Then, for a spectrum of stable isolated states $|n\rangle$ with
masses $M_n$ that couple to an interpolating field $\CF$,
\be
\Gamma_E(\tau) = \sum_n {\ME{0}{\CF}{n} \ME{n}{\CF}{0} \over 2 M_n} \ e^{-M_n \tau} \,.
\label{eq:gen2pointB}
\ee
Now it is easy to obtain the connection between these $M_n$ and the
physical spectrum. A fourier transform in Euclidean time followed by a 
rotation to Minkowski space gives 
\bea 
\int d\tau e^{ip_4\tau} {e^{-M_n |\tau|} \over 2 M_n} 
        &=& {1 \over 2M_n(M_n-ip_4)} + {1 \over 2M_n(M_n+ip_4)}  \,, \nonumber \\
        &=& {1 \over M_n^2+p_4^2} \,, \nonumber \\
        &{\longrightarrow \atop p_4 \to -iE}& {1 \over M_n^2-E^2} \,.
\eea
The $M_n$ are precisely the location of the poles in the propagator of
$|n\rangle$. This is why one can simply short step the necessary analytical
continuation as the masses measured from the exponential fall-off are
the pole masses of the physical states. For $\tau \to \infty$ only the
lowest state contributes due to the exponential suppression, and
this provides a way of isolating the lightest mass state.

This simple connection holds also for matrix elements. Consider, for
$\tau_2 >> \tau_1 >>0$, the three-point function
\bea
\int d^3 x d^3 y e^{-ik.x} e^{-iq.y} \ \ME{0}{T[ \CF_f(x,\tau_2) \CO(y,\tau_1) \CF_i(0,0)]}{0} &=& \nonumber \\
{\hskip -1.5 in }
\sum_a \sum_b {\ME{0}{\CF_f}{a} e^{-M_a (\tau_2-\tau_1)} \over 2 M_a}\ \ME{a}{\CO}{b}\ {}
{\ME{b}{\CF_i}{0} \ e^{-M_b \tau_1} \over 2 M_b} \,.
\eea
The two factors on the left and right of $\ME{a}{\CO}{b}$ are each, up to
one decay amplitude, the same as the 2-point function discussed in
Eq.~\ref{eq:gen2pointB}.  They, therefore, provide, for large
$\tau_2-\tau_1$ and $\tau_1$, the isolation of the lowest states in
the sum over states. These can be gotten from a calculation of the
2-point function. Consequently, $\ME{a}{\CO}{b}$ measured in Euclidean
simulations is the desired matrix element between these states and no
analytic continuation to Minkowski space is necessary.

These arguments break down once the sum over states is not just over
stable single particle states.  Consider, for example, an $\CF$ that
is the $\rho$ meson interpolating field. In that case there are also
the $2\pi, 4\pi$ intermediate states with a cut beginning at $E =
\sqrt{2M_\pi}$.  The physical characteristics of $\rho$ meson (mass
and width) can be represented by a complex pole in the propagator
that, in Minkowski space, lies on the second sheet under the cut.
This physical pole can no longer be reproduced by measuring the
``energy'' given by the exponential fall-off of Euclidean
correlation function at large $\tau$ \cite{Unstable91degrand}. It
requires measuring the correlation function for all intermediate
states and momenta (all $\tau$), and then doing the fourier transform
in $\tau$ followed by the analytical continuation $p_4 \to -iE$.

For the same reason there does not exist a simple way for calculating
n-particle scattering amplitudes.  A general 2-particle scattering
amplitude (4-point correlation function) is complex precisely because
the intermediate states involve a sum over resonances and a continuum
with all values of momenta.  In principle this can be obtained from
Euclidean correlation functions but only by measuring them with
essentially infinite precision for all possible $\tau$ and then doing
the analytical continuation. In practice this is a hopeless
possibility.  The one exception is the amplitude at threshold which
is real.  M.~L\"uscher has shown how the 2-particle scattering
amplitudes at threshold can be calculated from the energy shift in a
finite volume, the difference between the 2-particle energy and the
sum of the two particles in isolation~\cite{LH88Luscher,PS88Luscher,PS92LANL,PS95JLQCD}.

%% file: lqcd.tex
\def\opi{\hbox{${1\over  -\Dsl^2  + m^2 }$}}
\def\etoz{\hbox{$\epsilon \to 0$}}
\def\cd{\hbox{\cal D}}
\def\etal{\hbox{\sl et.al.}}
\def\ie{\hbox{\sl i.e.\ }}
\def\otau{\hbox{${\cal O}(\tau)$}}
\def\ozer{\hbox{${\cal O}(0)$}}
\def\otoz{\hbox{${\cal O}(\tau){\cal O}(0)$}}
\def\xx{\hbox{$\langle \bar\chi\chi \rangle$}}
\def\xxw{\hbox{$\langle \bar\psi\psi \rangle$}}
\def\wl{\hbox{$\langle L \rangle$}}
\def\Db{\hbox{$\Delta\beta$} }
\def\Dright#1#2{ \hbox{ ${\buildrel \rightarrow  \over D_#1}  \psi(#2) $}}
\def\Dleft#1#2{ \hbox { ${\bar \psi(#2) {\buildrel \leftarrow  \over D_#1}}$}}

\def\cropen#1{\cr \noalign{\vskip #1}}

\section{Formulation of Lattice Gauge Theories}
\label{s_lqcd}
\bigskip

In 1974 Wilson \cite{Wilson74} formulated Euclidean gauge theories on the
lattice as a means to study confinement and carry out non-perturbative
analysis of QCD. The numerical implementation of the path integral
approach requires the following five steps.
\begin{itemize}
\item Discretization of space-time. 
\item The transcription of the gauge and fermion degrees of freedom. 
\item Construction of the action.
\item Definition of the measure of integration in the path integral.
\item The transcription of the operators used to probe the physics.
\end{itemize}
Of these, the construction of the action and the operators is the most
intricate. Rather than give you the most recent and improved version
of the lattice action, which will be the subject of Martin L\"uscher's
lectures, I have chosen to follow the original path of discovery.

\bigskip

\subsection{Discrete space-time grid}

There are a number of possible ways to discretize space-time in
4 Euclidean dimensions. These include hypercubic, body-centered cubic
\cite{celmaster}, and random \cite{randomlattices} lattices. Of these the
simplest is the isotropic hypercubic grid with spacing $a = a_S = a_T$ and size
$N_S \times N_S \times N_S \times N_T$. Since this grid is used almost
exclusively, it will be the only one discussed in these sets of
lectures.  In the case of the random lattice even the free field
analysis becomes difficult \cite{randomlattices}. In fact very little
exploration of variant grid geometries has actually been done, as
there is little expectation of significant physics gains to offset the
additional computational complexity. Recently, proposals involving
anisotropic lattices -- a hypercubic grid with different spacing in
space and time directions -- have shown promise for reducing the
discretization errors.  I shall come back to this development in
Section~\ref{ss:glueballs} when discussing the glueball spectrum.

\subsection{Lattice transcription of field variables $\psi(x)$ and $A_\mu(x)$:}

The lattice transcription of the matter fields $\psi(x)$ is
straightforward.  The quark field is represented by anticommuting
Grassmann variables defined at each site of the lattice.  They belong
to the fundamental representation of SU(3).  The rules for integration
over a pair of fermion variables $\psi$ and $\bar \psi$ are those
standard for Grassmann variables
\begin{eqnarray}
\label{eq:grassmanrules}
\int d \psi\ d {\bar \psi}\ &=& \ \int d \psi\ d {\bar \psi}\ \psi = \int d \psi\ d {\bar \psi}\ \bar\psi = \ 0 \nonumber \\
\int d \psi\ \psi  \ &=& \int d {\bar \psi}\ {\bar \psi} = \ i \nonumber \\
\int d \psi\ d {\bar \psi}\ \psi\ {\bar \psi}  \ &=& \ 1
\end{eqnarray}
Since the fermion action is linear in both $\psi$ and ${\bar \psi}$,
these rules can be used to integrate over them, and the path integral
reduces to one over only the gauge degrees of freedom.  Thus, it turns
out that in practice one does not have to worry about transcribing
grassmann variables on the lattice and implementing the Pauli exclusion
principle.

As an aside I would like to mention that the rules of integration for
Grassmann variables are sufficient to do the fermion path integral.
Consider $\vev{\bar \psi(x_1) \ldots \bar \psi(x_n) \psi(x_1) \ldots
\psi(x_n)}$.  To get a non-zero expectation value one needs to
saturate each site with a $\bar \psi \psi$. The sites not covered by
the fields in the operator have to come from the $\CS_f^{V-n}$ term in
the expansion $\exp{\CS_f} = 1 + \CS_f + \CS_f^2/2 + \ldots$, where
$V$ is the total number of sites. Since $\CS_f = \sum_x \bar \psi(x)
(\Dsl + m) \psi(x)$ is a sum over sites, the number of terms in
$\CS_f^{V-n}$ poses an exponentially hard combinatorial
problem. Furthermore, each term contributing to the connected diagram
will contain a path-ordered product of gauge links (coming from the
discretized $\Dsl$ term in $\CS_f$. For example, see
Eq.~\ref{eq:Dslnaive}) connecting the sites in the operator, while the
fermionic integration is $\pm m^{V-n} $ with the sign given by the
number of times the $\psi$ and $\bar \psi$ are anti-commuted.  You
should convince yourself that the fermionic functional integral is
the sum over all possible paths touching each site a maximum of one
time. Having enumerated all such paths and their weights, the gauge
functional integral now requires integrating over each link in each
path.  This is clearly a very formidable approach, and in practice
rarely considered.

The construction of gauge fields is less intuitive.  In the continuum,
the gauge fields $A_\mu(x)$ carry 4-vector Lorentz indices, and
mediate interactions between fermions.  To transcribe them Wilson
noted that in the continuum a fermion moving from site $x$ to $y$ in
presence of a gauge field $A_\mu (x)$ picks up a phase factor given by
the path ordered product
\be
\label{eq:13.2}
\psi (y)\ =\ {\cal P}\,e^{\int_x^{y} igA_{\mu} (x) dx_{\mu}}\ \psi (x) \ .
\ee
The eight types of gluons that mediate interactions between quarks are
written in terms of the matrix $A_\mu(x) \equiv\ A^a_\mu (x)
\cdot \lambda_a/2$, where the group generators $\lambda_a$ are normalized to
$Tr \lambda_a \lambda_b = 2 \delta_{ab}$.  Eq.~\ref{eq:13.2} suggested
that gauge fields be associated with links that connect sites on the
lattice.  So, with each link Wilson associated a discrete version of
the path ordered product
\be
\label{eq:13.3}
U(x,x+\hat\mu)\ \equiv\ U_{\mu} (x)\ =\ e^{ iagA_{\mu}(x +{\hat\mu \over 2})} \,,
\ee
where, for concreteness, the average field $A_{\mu}$ is defined at
the midpoint of the link, and $U$ is a $3 \times 3 $ unitary matrix
with unit determinant.  Also, the path ordering in eqn. (2.3)
specifies that
\be
\label{eq:13.4}
U(x,x-{\hat\mu})\ \equiv\ U_{-\mu}(x)\ =\ e^{-iagA_{\mu}(x-{\hat\mu\over2})}\ 
=\ U^\dagger (x-{\hat\mu},x) \ \ .
\ee

\subsection{Discrete Symmetries of the lattice theory} 

The symmetry group of the continuum theory -- Poincar\'e invariance -- 
is reduced to a discrete group.  On a hypercubic lattice 
rotations by only $90^o$ 
are allowed so the continuous rotation group is replaced by the 
discrete hypercubic group \cite{Zweigmandula}.  
Translations have to be by at least one lattice unit, so the allowed momenta 
are discrete 
$$
k \ = \ {{2 \pi n } \over {La} } \quad \quad n=0,1, \ldots L
$$
or equivalently 
$$
k \ = \ \pm \ {{2 \pi n } \over {La} } \quad \quad n=0,1, \ldots L/2 \ .
$$
On the lattice momentum is conserved {\it modulo } $2\pi$. 

In addition to the local gauge symmetry and Poincar\'e invariance, the
lattice action is invariant under parity ($\cal P$), charge
conjugation ($\cal C$) and time reversal ($\cal T$).  The behavior of
the field variables under these discrete
transformations is given in Table~\ref{t:CPTdefs}.

\begin{table} 
\begin{center}
\setlength\tabcolsep{0.32cm}
\caption{The behavior of the gauge and fermion degrees of freedom under the discrete 
transformations of parity, charge-conjugation, and time-reversal. The
charge-conjugation matrix $\CC$ satisfies the relation ${\cal C}
\gamma_\mu {\cal C}^{-1} = - \gamma_\mu^T = -\gamma_\mu^* $, and shall
be represented by $\CC = \gamma_4 \gamma_2$. Note that in the
Euclidean formulation $\CT$ is a linear operator.}
\begin{tabular}{|l|c|c|c|}
\myhline
                           &  $\CP $   & $\CC $   &  $\CT $     \\
\myhline
  &  &  & \\[-7pt]
$U_4 (\vec x, \tau)      $ & $U_4    (-\vec x,  \tau)$
                           & $U^*_4  ( \vec x,  \tau)$
                           & $U_{-4} ( \vec x, -\tau)$    \\[2pt]
$U_i (\vec x, \tau)      $ & $U_{-i} (-\vec x,  \tau)$
                           & $U^*_i  ( \vec x,  \tau)$
                           & $U_i    ( \vec x, -\tau)$    \\[2pt]
\myhline
  &  &  & \\[-7pt]
$\psi(\vec x, \tau)      $ & $\gamma_4 \psi(-\vec x, \tau)      $
                           & $\CC \bar \psi^T(\vec x, \tau)      $
                           & $\gamma_4\gamma_5 \psi(\vec x, -\tau) $     \\[2pt]
$\bar \psi(\vec x, \tau) $ & $\bar \psi(-\vec x, \tau)\gamma_4  $
                           & $- \psi^T(\vec x, \tau) \CC^{-1}  $
                           & $\bar \psi(\vec x, -\tau)\gamma_5\gamma_4$  \\[3pt]
\myhline
\end{tabular}
\label{t:CPTdefs}
\end{center}
\end{table}
\smallskip

\subsection{Local Gauge Symmetry}

\begin{figure} 
\hbox{\epsfxsize=\hsize\epsfbox{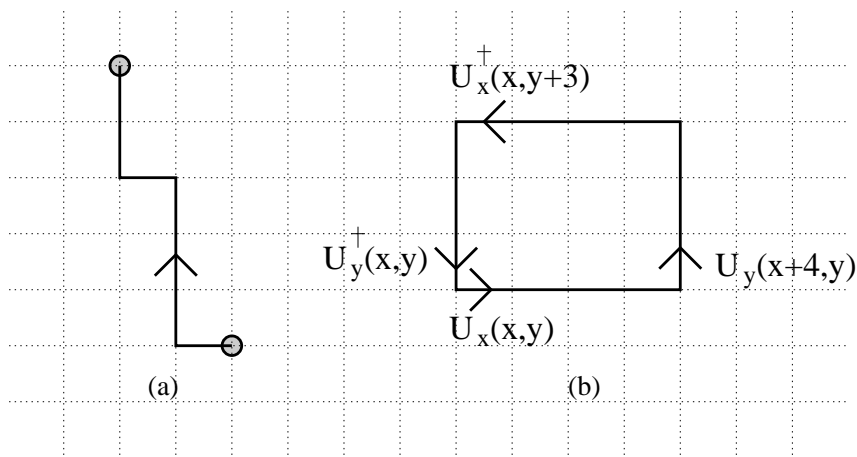}}
\caption{The two gauge invariant quantities. a) An ordered 
string of $U's$ capped by a fermion and an anti-fermion and b) closed 
Wilson loops.}
\label{f_loopline}
\end{figure}

The effect of a local gauge transformation $V(x)$ on the variables
$\psi(x)$ and $U$ is defined to be
\bea
\label{eq:13.5}
\psi (x)     \ &\to& \ V(x)\psi (x)                            \nonumber\\
\bar\psi (x) \ &\to& \ \bar\psi (x) V^\dagger (x)              \nonumber\\
U_{\mu}(x)   \ &\to& \ V(x)U_{\mu}(x) V^\dagger(x+{\hat\mu})            
\eea
where $V(x)$ is in the same representation as the $U$, $i.e.$, it is an
$SU(3)$ matrix.  With these definitions there are two types of gauge
invariant objects that one can construct on the lattice. 
\begin{itemize}
\item
A string consisting of a path-ordered product of links capped by a fermion 
and an antifermion as shown in Fig.~\ref{f_loopline}a.  A simple example is 
\be
\label{eq:13.6}
{\rm Tr}\ \bar\psi (x)\ U_{\mu} (x)\ U_{\nu} (x+{\hat\mu})\ldots 
              U_\rho(y-{\hat\rho}) \ \psi (y) 
\ee
where the trace is over the color indices.  I will use the word string
for a generalized version; a single spin and color trace as defined in
Eq.~\ref{eq:13.6}, or a product of elementary strings that are
path-ordered and do not have a spatial discontinuity. On lattices with
periodic boundary conditions one does not need the $\psi, \bar \psi$
caps if the string stretches across the lattice and is closed by the
periodicity. Such strings are called Wilson/Polyakov lines.
\item
Closed Wilson loops as shown in Fig.~\ref{f_loopline}b. The simplest example 
is the plaquette, a $1 \times 1$ loop, 
\be
\label{eq:13.7}
W_{\mu\nu}^{1 \times 1}\ =\ \Re \ {\rm Tr}\ \big( 
U_{\mu} (x)\ U_{\nu} (x+{\hat\mu})\ 
U_{\mu}^\dagger (x+{\hat\nu})\ U_{\nu}^\dagger (x) \big) \ \ .
\ee
Unless otherwise specified, whenever I use the term Wilson loops I
will assume that the real part of the trace, $\Re\ {\rm Tr}$, has been
taken.  For $SU(N \ge 3)$ the trace of any Wilson loop in the
fundamental representation is complex, with the two possible
path-orderings giving complex conjugate values. Thus, taking the trace
insures gauge invariance, and taking the real part is equivalent to
averaging the loop and its charge conjugate.

\end{itemize}

\medskip
\noindent{\it Exercise: Convince yourself that
there do not exist any other independent gauge invariant quantities.}
\medskip

\noindent{\it Exercise: 
Convince yourself that to preserve gauge invariance one has to formulate 
the lattice theory in terms of link variables $U_\mu$ 
rather than the $A_\mu$ directly.}
\medskip

A gauge invariant action, therefore, has to be built out of loops and
strings.  However, note that these two basic gauge invariant objects
can be of arbitrary size and shape.  Furthermore, they can be taken to
lie in any representation of color $SU(3)$.  The only necessary
constraint is that it give the familiar continuum theory in the limit
$a \to 0$. The reasons for wanting to preserve gauge invariance at all
$a$ is that otherwise one would have many more parameters to tune (the
zero gluon mass, and the equality of the quark-gluon, 3-gluon, and
4-gluon couplings are consequences of gauge symmetry) and there would
arise many more operators at any given order in $a$ (these are
irrelevant operators, but at finite $a$ give rise to discretization
errors as discussed in Section~\ref{ss:symanzikImp}). In fact, one
would like to preserve as many symmetries of the theory as possible at
all values of $a$ and not just in the continuum limit. It turns out to
be relatively simple to maintain local gauge invariance, however,
maintaining chiral symmetry, which has proven so important in the
phenomenology of light quarks, and a one-to-one relation between Dirac
and lattice fermion degrees of freedom has eluded us. This is
discussed in Section~\ref{s:doubling}.  Also, discretization $de\
facto$ breaks continuous rotational, Lorentz, and translational
symmetry.  The key idea of improved discretization schemes, to be
discussed in Sections
\ref{ss:symanzikImp}, \ref{s:improvedglue}, and \ref{s:improveddirac}, 
will be to implement the breaking of these symmetries by operators
that come with sufficiently many powers of $a$ so that their effect is
very local, and thus small on physical (long-distance) quantities.

\section{Simplest Formulation of the Lattice Action}

\subsection{Gauge Action}
\label{s:gaugeaction}

The gauge action can be expressed in terms of closed loops.  I will
outline the construction for the simpler case of the abelian $U(1)$
model in which the link variables are commuting complex numbers
instead of SU(3) matrices.  Consider the simplest Wilson loop, the
$1\times1$ plaquette $W_{\mu\nu}^{1 \times 1}$:
\bea
\label{eq:plaqexpA}
W_{\mu\nu}^{1 \times 1} &=& U_{\mu}(x) U_{\nu} (x+{\hat\mu}) U_{\mu}^\dagger (x+{\hat\nu}) 
U_{\nu}^\dagger (x)  \nonumber \\
&=& e^{iag  \lbrack A_{\mu} (x+{\hat\mu\over 2}) + 
                    A_{\nu} (x+\hat\mu + {\hat\nu\over 2}) -
                    A_{\mu} (x+\hat\nu+{\hat\mu\over 2}) - 
		    A_{\nu} (x+{\hat\nu\over 2}) \rbrack }
\eea
Expanding about $x+{\hat\mu + \hat\nu \over 2} \ $ gives 
\bea
\label{eq:plaqexpB}
&{}& \ \exp \ \lbrack \ ia^2g(\partial_\mu A_\nu - \partial_\nu A_\mu) \ + \ 
    {ia^4g \over 12} (\partial_\mu^3 A_\nu - \partial_\nu^3 A_\mu) \ + 
    \ \ldots\rbrack \nonumber \\
&=& \ 1+ia^2gF_{\mu\nu} \ -\ {a^4g^2\over 2} F_{\mu\nu} F^{\mu\nu} \ +\ 
     O(a^6) \ +\ \ldots 
\eea
The real and imaginary parts of the loop give
\bea
\label{eq:plaqexpC}
\Re {\rm Tr} (1-W_{\mu\nu}^{1 \times 1}) &=&
   {a^4g^2\over 2} F_{\mu\nu} F^{\mu\nu} \ +\ terms \ higher\ order\ in\ a \nonumber \\
\Im(W_{\mu\nu}^{1 \times 1}) &=& a^2gF_{\mu\nu} \ .
\eea
So far the indices $\mu$ and
$\nu$ are uncontracted.  There are $6$ distinct positively oriented
plaquettes, $\{\mu < \nu \}$, associated with each site.  Summing over
$\mu,\nu$, and taking care of the double counting by an extra factor
of $1\over2$, gives
$$
{1\over g^2} \sum_x \sum_{\mu < \nu} \Re {\rm Tr} (1-W_{\mu\nu}^{1 \times 1})\ = \ 
{a^4\over 4} \sum_x \sum_{\mu , \nu} F_{\mu\nu}  F^{\mu\nu} \ \to \ 
{1\over 4} \int d^4 x F_{\mu\nu} F^{\mu\nu} \ \ .
$$

Thus, to lowest order in $a$, the expansion of the plaquette gives the
continuum action.  The steps in the derivation for non-abelian groups
are identical and the final expression (up to numeric factors) is the
same. The result for the gauge action for SU(3), defined in terms of
plaquettes and called the Wilson action, is
\be
S_g = {6\over g^2} \sum_x \sum_{\mu < \nu} \Re {\rm Tr} \ {1 \over 3} (1-W_{\mu\nu}^{1 \times 1}) \,.
\label{eq:Wgaugeaction}
\ee
For historic reasons the lattice calculations are mostly presented in
terms of the coupling $\beta \equiv {6/ g^2}$.  Since $\beta$ is used
for many other quantities like $1/kT$ or the $\beta$-function, I shall
try to be more explicit in the use of notation ($g$ versus $\beta$) if
there is a possibility for confusion.

\medskip
\noindent{\it Exercise: Repeat the Taylor expansion for 
the physical case of SU(3). Show that the 
non-abelian term in $F_{\mu\nu} = \partial_\mu A_\nu - \partial_\nu
A_\mu + g[A_\mu, A_\nu] $ arises due to the non-commuting property of
$\lambda$ matrices.  The details of the derivation for SU(2) are given
in section 3 of Ref.~\cite{Kogut83RMP}.}
\medskip

There are four important points to note based on the above
construction of the lattice action.
\begin{description}
\item{1)}
The leading correction is $O(a^2)$. The term ${a^2 \over 6}
F_{\mu\nu} (\partial_\mu^3 A_\nu - \partial_\nu^3 A_\mu) $ is present
in the expansion of all planar Wilson loops.  Thus at the classical
level it can be gotten rid of by choosing an action that is a linear
combination of say $1\times 1$ and $1\times 2$ Wilson loops with the
appropriate relative strength given by the Taylor expansion (see
Section~\ref{s:improvedglue}).
\item{2)}
Quantum effects will give rise to corrections, $i.e.$
$a^2 \to X(g^2) a^2$ where in perturbation theory $X(g^2) = 1 + c_1 g^2 + \ldots$, and will
bring in additional non-planar loops. Improvement of the action will
consequently require including these additional loops, and adjusting
the relative strengths which become functions of $g^2$.  This is
also discussed in Section~\ref{s:improvedglue}.
\item{3)}
The reason for defining the action in terms of small loops is
computational speed and reducing the size of the discretization
errors. For example the leading correction to $1\times 1$ loops is
proportional to $a^2/6 $ whereas for $1\times 2$ loops it increases
to $ 5 a^2/12$.  Also, the cost of simulation increases by a factor of $2-3$. 
\item{4)}
The electric and magnetic fields $E$ and $B$ are proportional to
$F_{\mu\nu}$.  Eq.~\ref{eq:plaqexpC} shows that these are given in terms 
of the imaginary part of  Wilson loops. 

\end{description}

\subsection{Fermion Action}

To discretize the Dirac action, 
Wilson replaced the derivative with the symmetrized difference 
and included appropriate gauge links to maintain gauge invariance
\begin{equation}
\label{eq:Dslnaive}
\bar \psi \Dsl \psi =\ {1 \over 2a} \bar\psi (x) \ \sum_\mu\ \gamma_{\mu} \lbrack 
            U_{\mu} (x)\psi (x+{\hat\mu}) \ - \ 
            U^\dagger_\mu(x-{\hat\mu})\ \psi (x-{\hat\mu}) \rbrack \ .
\end{equation}
It is easy to see that one recovers the Dirac action in the limit $a
\to 0$ by Taylor expanding the $U_\mu$ and $\psi (x+{\hat\mu})$ in powers of the
lattice spacing $a$.  Keeping only the leading term in $a$, Eq.~\ref{eq:Dslnaive} 
becomes 
\bea
  &{}& {1 \over 2a} \bar\psi (x) \gamma_{\mu} 
       \lbrack \ \big(1+iagA_{\mu} (x+{\hat\mu\over 2}) + \ldots \big)
               \ \big(\psi (x) + a \psi' (x) + \dots \big) - \nonumber \\
  &{}& \hskip 0.72in
             \big(1-iag A_{\mu} (x- {\hat\mu\over 2})) + \ldots \big)
           \ \big(\psi (x) - a \psi' (x) + \dots \big) \ \rbrack \nonumber \\
  &=& \ \bar\psi (x) \gamma_{\mu} (\partial_{\mu} + 
        {a^2 \over 6} \partial_{\mu}^3 + \ldots ) \psi (x)\ \nonumber \\
  &{}&\ + ig\bar\psi (x) \gamma_{\mu} [A_{\mu} + {a^2 \over 2} 
         \big( {1 \over 4} \partial_{\mu}^2 A_\mu  + 
          {(\partial_{\mu}} A_\mu)  \partial_{\mu} + 
          A_\mu  \partial_{\mu}^2 \big) + \ldots  ] \psi (x)\ \nonumber ,
\label{eq:naiveexpansion}
\eea
which, to $O(a^2)$, is the kinetic part of the standard continuum Dirac
action in Euclidean space-time.  Thus one arrives at the simplest
(called ``naive'') lattice action for fermions 
\bea
\label{eq:naiveact}
\CS^N  &=&  m_q\ \sum_x \bar\psi (x) \psi (x)  \nonumber \\
     &{}&\ +\ {1\over 2a}\ \sum_x \bar\psi(x) \gamma_\mu \lbrack 
          U_\mu (x) \psi (x+\hat\mu) - U_\mu^\dagger (x-\hat\mu) 
          \psi (x-\hat\mu) \rbrack \nonumber \\
     &\equiv&  \sum_x \bar\psi (x) M^N_{xy}[U] \psi (y)  
\eea
where the interaction matrix $M^N$ is 
\be
\label{eq:naiveactM}
M^N_{i,j}[U] \ = \  m_q \delta_{ij}\ +\ {1 \over 2a} \sum_\mu\ 
\ \bigl[ \gamma_\mu  U_{i,\mu}\delta_{i,j-\mu}\ -\ 
         \gamma_\mu  U^\dagger_{i-\mu,\mu}\delta_{i,j+\mu}
\bigr]
\ee
The Euclidean $\gamma$ matrices are hermitian, $\gamma_{\mu} =
\gamma_{\mu}^\dagger$, and satisfy $\{ \gamma_{\mu} , \gamma_{\nu} \}
= 2 \delta_{\mu\nu} $.  The representation I shall use is
\be
\label{eq:gammamatdef}
\vec \gamma \ = \ \pmatrix{0&i\vec\sigma \cr  -i\vec\sigma & 0 \cr} \ \ , \ \ 
\gamma_4 \ = \ \pmatrix{1 & 0 \cr  0 & -1 \cr} \ \ , \ \ 
\gamma_5 \ = \ \pmatrix{0&1 \cr  1 & 0 \cr} \ \ ,
\ee
which is related to Bjorken and Drell conventions as follows: 
$ \gamma_{i} =  i \gamma^{i}_{BD}$, 
$ \gamma_{4} =    \gamma^{0}_{BD}$, 
$ \gamma_{5} =    \gamma^{5}_{BD}$.
In this representation $\gamma_1, \gamma_3$ are pure imaginary, 
while $\gamma_2, \gamma_4, \gamma_5$ are real. 

The Taylor expansion showed that the discretization errors 
start at $O(a^2)$. For another simple illustration 
consider the inverse of the free-field propagator $ m + i/a\sum_\mu
\gamma_\mu sin(p_\mu a)$.  Set $\vec p = 0$ and rotate to
Minkowski space ($p_4 \to iE $, $i.e.$ ${\rm sin} p_4 a \to i {\rm
sinh} Ea$). Then, using the forward propagator (upper two
components of $\gamma_4$), gives 
\be
m_q^{pole} a \ = \ {\rm sinh} Ea  
\label{eq:Npolemass}
\ee
for the relation between the pole mass and
the energy. This shows that, even in the free field case, the continuum 
relation $E(\vec p =0) = m$ is violated by corrections that are $O(a^2)$.

\noindent{\bf Symmetries}: The invariance group of the fermion action under
rotations in space and time is the hypercubic group. Full Euclidean
invariance will be recovered only in the continuum limit.  The action
is invariant under translations by $a$ and under $\cal P$, $\cal C$,
and $\cal T$ as can be checked using Table~\ref{t:CPTdefs}.

The $naive$ action $\bar \psi_x M^N_{xy} \psi_y $ has the 
following global symmetry:
\bea
\label{eq:vector_psi}
 \psi(x)\     &\rightarrow&\ e^{i \theta} \psi(x)	\nonumber \\
 \bar\psi(x)\ &\rightarrow&\ \bar\psi(x) e^{-i \theta}   
\eea
where $\theta$ is a continuous parameter. This symmetry is related to 
baryon number conservation and leads to a conserved vector current. 
For $m_q = 0$ the action is also invariant under 
\bea
\label{eq:axial_psi}
 \psi(x)\     &\rightarrow&\ e^{i \theta \gamma_5} \psi(x) \nonumber \\
 \bar\psi(x)\ &\rightarrow&\ \bar\psi(x) e^{i \theta \gamma_5}   
\eea
Having both a vector and an axial symmetry in a hard-cutoff
regularization would imply a violation of the Adler-Bell-Jackiw
theorem. It turns out that while the naive fermion action preserves chiral
symmetry it also has the notorious fermion doubling problem as
explained in Section~\ref{s:doubling}. The chiral charges of these
extra fermions are such as to exactly cancel the anomaly.  In fact the
analysis of $\CS^N$ lead to a no-go theorem by Nielsen-Ninomiya
\cite{NNnogo} that states that it is not possible to define a local,
translationally invariant, hermitian lattice action that preserves
chiral symmetry and does not have doublers.

We will discuss two fixes to the doubling/anomaly/chiral symmetry
problem. First, include an additional term in the action that breaks
chiral symmetry and removes the doublers (Wilson's fix).  Second,
exploit the fact that the naive fermion action has a much larger symmetry
group, $U_V(4) \otimes U_A(4)$, to reduce the doubling problem from
$2^d = 16 \to 16/4$ while maintaining a remnant chiral symmetry
(staggered fermions).  In both cases one regains the correct anomaly
in the continuum limit.  Before diving into these fixes, I would like
to first discuss issues of the measure of integration and gauge fixing, and 
give an overview of Symanzik's improvement program.

\subsection{The Haar Measure} 
\label{ss:haar}

The fourth ingredient needed to complete the formulation of LQCD as a
quantum field theory via the path integral is to define the measure of
integration over the gauge degrees of freedom. Note that, unlike the
continuum fields $A_\mu$, lattice fields are SU(3) matrices with
elements that are bounded in the range $[0,1]$. Therefore, Wilson
proposed an invariant group measure, the Haar measure, for this
integration. This measure is defined such that for any elements $V$
and $W$ of the group
\begin{equation}
\int dU f(U) \ = \ \int dU f(UV) \ = \ \int dU f(WU) 
\end{equation}
where $f(U) $ is an arbitrary function over the group. This
construction is simple, and has the additional advantage that in
non-perturbative studies it avoids the problem of having to include a
gauge fixing term in the path integral. This is because the field
variables are compact.  Hence there are no divergences and one can
normalize the measure by defining
\begin{equation}
\int dU \ = \ 1 .
\end{equation}
For gauge invariant quantities each gauge copy contributes the same
amount.  Consequently the lack of a gauge fixing term just gives an
overall extra numerical factor. This cancels in the calculation of
correlation functions due to the normalization of the path
integral. Note that gauge fixing and the Fadeev-Papov construction
does have to be introduced when doing lattice perturbation theory~\cite{plqcd81kawai}.

The measure and the action are invariant under $\cal C$, $\cal P$,
$\cal T$ transformations, therefore when calculating correlation
functions via the path integral we must average over a configuration
$\cal U$ and the ones obtained from it by the application of $\cal C$,
$\cal P$, $\cal T$.  In most cases this averaging occurs very
naturally, for example, Euclidean translation invariance guarantees
$\cal P$ and $\cal T$ invariance. Similarly, taking the real part of
the trace of a Wilson loop is equivalent to summing over $U$ and its
the charge conjugate configuration.  A cook-book recipe for
evaluating the behavior of correlation functions after average over
configurations related by the lattice discrete symmetries is given in
Section~\ref{s:corrfns}.

One important consequence of these symmetries is that only the real
part of Euclidean correlation functions has a non-zero expectation
value.  This places restriction on the quantities one can calculate
using LQCD. For example, simulations of LQCD have not been very
successful at calculations of scattering amplitudes, which are in
general complex~\cite{testamaiani}. The connection between Minkowski
and Euclidean correlation functions was discussed in
Section~\ref{s:connection}. To get scattering amplitudes what is
required is to first calculate Euclidean amplitudes for all values of
the Euclidean time $\tau$, followed by a fourier transform in $\tau$,
and finally a Wick rotation to Minkowski space. For such an procedure
to work requires data with phenomenal accuracy.  Unfortunately,
simulations give Euclidean correlation functions for a discrete set of
$\tau$ and with errors that get exponentially magnified in the
rotation.  The best that has been done is to use L\"uscher's method
that relates scattering phase shifts to shifts in two particle energy
states in finite volume \cite{LH88Luscher,PS88Luscher}. So far this
method has been tested on models in 2-4 dimensions, and used to
calculate scattering lengths in QCD ~\cite{PS92LANL,PS95JLQCD}.

\subsection{Gauge Fixing and Gribov Copies}

Gauge invariance and the property of group integration, $\int dU \ U =
0$, are sufficient to show that only gauge invariant correlation
functions have non-zero expectation values. This is the celebrated
Elitzur's theorem \cite{Elitzur75}.  It states that a local 
continuous symmetry cannot be spontaneously broken. 

In certain applications (study of quark and gluon
propagators~\cite{GP98Leinweber}, 3-gluon
coupling~\cite{3GC98Parrinello}, hadronic
wavefunctions~\cite{WF93LANL}, and determination of renormalization
constants using quark states~\cite{NPZ98ROME}) it is useful to work in
a fixed gauge. The common gauge choices are axial ($\eta_\mu A_\mu =
0$ where $\eta$ in Minkowski space is a fixed time-like vector. The
common choice is $A_4 = 0$), Coulomb ($\partial_i A_i = 0$), and
Landau ($\partial_\mu A_\mu = 0$).  Their lattice analogues, assuming
periodic boundary conditions, are the following.

{\bf Maximal Axial Gauge}: On a lattice with periodic boundary conditions
one cannot enforce $A_4 =0$ by setting all links in the time direction
to unity as the Wilson line in each direction is an invariant.  The
gauge freedom at all but one site can be removed as follows.  Set all
time-like links $U_4 = 1$ on all but one timeslice. On that time-slice
all $U_4$ links are constrained to be the original Wilson line in
4-direction, however, one can set links in $\hat z$ direction $U_3 =
1$ except on one z-plane. On that z-plane set $U_3$ links to the
Wilson line in that direction and all links in $\hat y$ direction $U_2
= 1$ except on one y-line. On this y-line set all $U_2$ links to the
value of y-lines and links in $\hat x$ direction $U_1 = 1$ except on
one point.  While such a construction is unique, the choice of the
time-slice, z-plane, y-line, and x-point and the labeling of $x,y,z,t$
axis are not. Also, this gauge fixing does not respect Euclidean
translation invariance and it is not smooth.

{\bf Coulomb Gauge}: The lattice condition is given by the gauge
transformation $V(x)$ that maximizes the function
\be 
F[V] = \sum_x \sum_i \Re{\rm Tr}
V(x) \bigg( U_i(x)V^\dagger(x+\hat i) + 
            U_i^{\dagger}(x-\hat i)V^\dagger(x-\hat i) \bigg) 
\label{eq:Coulombgauge}
\ee
separately on each time-slice and $i$ runs over the spatial indices $1-3$.

{\bf Landau Gauge}: This is defined by the same function as in Eq.~\ref{eq:Coulombgauge} 
except that the sum is evaluated on the whole lattice and $i=1-4$.

The maximization condition, Eq.~\ref{eq:Coulombgauge}, corresponds, in
the continuum, to finding the stationary points of $F[V] = ||A^V|| =
\int d^4x {\rm Tr} (A_\mu^V)^2(x)$. What we would like is to
find the global minimum corresponding to the ``smoothest'' fields
$A_\mu$, or the copies in case it is not unique.  For any gauge
configuration there is an infinite set of gauge equivalent
configurations obtained by applying the set of transformations $\{
V(x)\} $. This set is called the gauge orbit. The first question is
whether the gauge fixing procedure gives a unique solution, $i.e.$
whether the gauge-fixing algorithm converges to the same final point
irrespective of the starting point on the gauge orbit. For the maximal
axial gauge, the construction is unique (the configuration $\{U\}$ and
$\{VUV^\dagger\}$ for any $\{V\}$ give the same final link matrices),
so the answer is clearly yes.  For the Coulomb and Landau gauge
choices the answer is no -- there are a number of solutions to the
gauge-fixing condition. These are the famous Gribov copies
~\cite{Gribov}. For a recent theoretical review of this subject see
\cite{Gribov97vanbaal}. 

The algorithms for fixing to either the Coulomb gauge or the Landau
gauge are mostly local. Local algorithms generate, for each site, a
gauge matrix $v_i(x)$ that maximizes the sum, defined in
Eq.~\ref{eq:Coulombgauge}, of the link matrices emerging from that
site. This gauge transformation is then applied to the lattice. These
two steps are iterated until the global maximum is found.  The ordered
product $\prod_i v_i(x)$ gives $V(x)$. For all practical purposes we
shall call this end-point of the gauge fixing algorithm a Gribov copy.
It is, in most cases and especially for the local gauge-fixing
algorithms, the nearest extremum, and not the global minimum of
$||A^V||$.  A set of Gribov copies can be generated by first
performing a different random gauge transformation on the
configuration and then applying the same gauge-fixing algorithm, or by
using different gauge-fixing algorithms.  For recent studies of the
efficacy of different algorithms see \cite{GF97Mendes,GF98forcrand}.

The Monte Carlo update procedure is such that it is highly improbable
that any finite statistical sample contains gauge equivalent
configurations.  So one does not worry about gauge copies.  In the
gauge-fixed path integral one would $a\ priori$ like to average over
only one Gribov copy per configuration and choose this to be the
``smoothest''.  Finding the $V(x)$ which gives the smoothest
configuration has proven to be a non-trivial problem.  The open
questions are: (i) how to define the ``smoothest'' solution of the
gauge-fixing algorithm ($i.e.$ defining the fundamental modular
domain), (ii) how to ascertain that a solution is unique (lies within
or on the boundary of the fundamental modular domain), and (iii) how
to find it.

The ultimate question is -- what is the dependence of the quantities
one wants to calculate on the choice of Gribov copy?  A number of
studies have been carried out
\cite{Gribovtests}, however the results and their interpretation are
not conclusive. Firstly, it is not known whether the smoothest
solution lying within the fundamental modular domain can be found
economically for all configurations and over the range of gauge
coupling being investigated.  In the absence of the smoothest solution
it is not clear whether (i) the residual gauge freedom simply
contributes to noise in the correlation function, and (ii) one should
average over a set of Gribov copies for each configuration to reduce
this noise. In my opinion these remain challenging problems that some
of you may wish to pursue.

The problem of Gribov copies and the Faddeev-Papov construction (a
gauge fixing term and the Faddeev-Papov determinant) resurfaces when
one wants to do lattice perturbation theory or when taking the
Hamiltonian limit of the path integral via the transfer matrix
formalism.  This construction is analogous to that in the
continuum~\cite{plqcd81kawai}.  Finally, let me mention that in a
fixed gauge, the lattice theory has only a BRST invariance
\cite{LH88Luscher}. While BRST invariance is sufficient to prove that
the theory is renormalizable, the set of operators of any given
dimension allowed by BRST invariance is larger than that with full
gauge invariance.  In the analyses of improvement of the action and
operators it is this enlarged set that has to be considered
\cite{Imp97Rome}.

\section{Continuum limit and Symanzik's Improvement program}
\label{ss:symanzikImp}

The Taylor expansion of the simple Wilson action for the gauge and
fermion degrees of freedom showed the presence of higher dimensional
operators.  Since these are suppressed by powers of the lattice
spacing, they are irrelevant operators in the language of the
renormalization group, $i.e.$ they vanish at the fixed point at $a \to
0$.  Another way of expressing the same thing is as follows. The
lattice theory with a hard cut-off at $\pi/a$ can be regarded as an
effective theory \cite{Symanzik83}. Integration of momenta from $\infty \to \pi/a$
generates effective interactions. Thus, in general, we can write the
action as
\begin{equation}
\CS_C = \CS_L(\pi/a) + \sum_n \sum_i a^n C_i^n \CA_i^n(\pi/a)
\end{equation}
where $S_C(S_L)$ is the continuum (lattice) action, and the first sum
is over all operators of a given dimension, and the second over all
dimensions greater than four. Similarly,
the operators used to probe the physics can be written as
\begin{equation}
\CO_C = \CO_L(\pi/a) + \sum_n \sum_i a^n D_i^n \CO_i^n(\pi/a)
\end{equation}
where in this case quantum corrections can induce mixing with lower
dimension operators also.  The coefficients $C_i^n $ and $D_i^n$ are
both functions of the coupling $\alpha_s$. Parenthetically, one
important reason for preserving gauge, Lorentz, and chiral symmetries at finite
$a$ is that they greatly restrict the set of possible operators of a
given dimension.

Now consider the expectation value of a given operator 
\begin{eqnarray}
\langle \CO_C \rangle 
            &=& \int dU \CO_C e^{-\CS_C} \nonumber \\
            &=& \int dU \big\{ \CO_L + \sum_n \sum_i a^n D_i^n \CO_i^n \big\}
                        e^{-\CS_L - \sum_m \sum_i a^m C_i^m \CA_i^m} \nonumber \\
            &=& \int dU \big\{ \CO_L + \sum_n \sum_i a^n D_i^n \CO_i^n \big\}
                        \big\{1 - \sum_m \sum_i a^m C_i^m \CA_i^m + \ldots \big\}
                        e^{-\CS_L } \nonumber \\
            &=& \int dU \CO_L e^{-S_L }  \nonumber \\
	    &+& \int dU \big( \CO_L
                        \big\{\sum_m \sum_i a^m C_i^m \CA_i^m + \ldots \big\} \big) 
                        e^{-\CS_L } \nonumber \\
	    &+& \int dU \big( \sum_n \sum_i a^n D_i^n \CO_i^n 
                        \big\{1 - \sum_m \sum_i a^m C_i^m \CA_i^m + \ldots \big\} \big) 
                        e^{-\CS_L } \,.
\label{eq:symanzikimp}
\end{eqnarray}
Note that the operators $\CA_i^n$, which appear in the action, are
summed over all space-time points.  This leads to contact terms in
Eq.~\ref{eq:symanzikimp} which I shall, for brevity, ignore as they
necessitate a proper definition of the operators but do not change the
conclusions.

The basis for believing that LQCD provides reliable results from
simulations at finite $a$ is that the contributions of the lattice
artifacts (last two terms in the last expression) are small, can be
estimated, and thus removed. At this point I will only make a few
general statements about the calculation of expectation values which
will be embellished on in later lectures.

\begin{itemize}
\item
The contribution of the last two terms in Eq.~\ref{eq:symanzikimp}
vanishes in the limit $a \to 0$ unless there is mixing with operators
of lower dimension.  In such cases the mixing coefficient has to be
determined very accurately (necessitating non-perturbative methods)
otherwise the $a \to 0$ limit is divergent.
\item
Calculations have to be done at values of $a$ at which contributions
of irrelevant terms organized in successive powers in $a$ decrease at
least geometrically.
\item
To achieve continuum results simulations are carried out at a
sufficient number of values of $a$ to allow reliable extrapolation to
$a=0$. The ``fineness'' of the lattice spacing $a$ at which to do the
calculations and the number of points needed depends on the size of
the corrections and how well we can determine the functional form to
use in the extrapolation.
\item
To improve the result to any given order in $a$, one has to improve
both the action and the operators to the same order. Such an approach,
based on removing lattice artifacts organized as a power series
expansion in $a$, is called the Symanzik improvement program.
\end{itemize} 

Eq.~\ref{eq:symanzikimp} shows clearly that there is fair degree of
flexibility in constructing LQCD. We are free to add any irrelevant
operator with a sensible strength and still recover QCD in the
continuum limit. The only difference between various constructions of
the action is the approach to the continuum limit. What we would like
to do is improve the action and operators so as to get continuum
results on as coarse lattices as possible. On the other hand
improvement requires adding irrelevant terms to both the action and
the operators. This increases the complexity of the calculations and
the simulation time. Thus, the strategy for minimizing discretization
errors is a compromise -- optimize between simplicity of the action
and operators (defined in terms of the cost of simulation, code
implementation, and analyses) and the reduction of the discretization
errors evaluated at some given fixed value of the lattice spacing, say
$a \approx 0.1 $ fermi.  The remaining errors, presumably small, can
then be removed by extrapolating the results obtained from simulations
at a few values of $a$ to $a=0$.

The number of floating points operations needed in the simulations of
LQCD scale roughly as $L^6$ in the quenched approximation and $L^{10}$
with dynamical fermions. Current lattice simulations of QCD are done
for lattice spacing $a$ varying in the range $2\ GeV \le a^{-1}
\le 5\ GeV$.  For these parameter values the corrections due to the
$O(a), \ O(ma), O(pa), O(a^2), \ldots $ terms in the Wilson action are
found to be large in many observables.  The importance of improving
the lattice formulation (action and operators) is therefore
self-evident, especially for full QCD. Thus, the search for such
actions is a hot topic right now. Before discussing these let me first
summarize what one hopes to gain by, and the criteria by which to
judge, such an improvement program.
\begin{description}
\item{1.}
Improved scaling, $i.e.$ discretization errors are reduced. Consequently 
one can work on coarser lattices (smaller $L=N_S$) for a given accuracy. 
Even a factor of 2 reduction in $L$ translates into $2^{10} \approx 1000$ in 
computer time!
\item{2.}
Better restoration of rotational and internal symmetries like 
chiral, staggered flavor, $etc.$. 
\item{3.}
The trajectory along which the continuum limit is taken should not
pass close to extraneous phase transitions that are artifacts of the
lattice discretization (see Section~\ref{s:phasetransitions} for a 
discussion of phase transitions in the lattice theory).  The
problem with such transitions is that in their vicinity the desired
continuum scaling behavior can be modified for certain observables,
$i.e.$ the artifact terms in Eq.~\ref{eq:symanzikimp} may not have a
geometric convergence in $a$ and may thus be large.
\item{4.}
The study of topology on the lattice is particularly sensitive to
short distance fluctuations that are artifacts of discretizing the
gauge action.  For the Wilson gauge action one finds that the
topological charge, calculated using L\"uscher's geometric
method~\cite{TOP82Luscher}, jumps from $0 \to 1$ at a core radius of
the instanton of $\rho_c a \approx 0.7$, whereas the action of such
instantons, $S_{\rho_c}$, is significantly less than the classical
value $S^{cl} = 8 \pi^2 / \beta$.  The entropy factor for such small
instantons with $S < S^{cl}$ overwhelms the Boltzmann suppression,
leading to a divergent topological susceptibility in the continuum
limit. These short distance artifacts are called dislocations and it
is not $a\ posteriori$ obvious how to separate the physical from the
unphysical when $\rho \approx \rho_c$. It is, therefore, very
important to start with a lattice formulation in which such artifacts
are excluded, and there exist only physical instantons, $i.e.$ those
with mean core size that has the correct scaling behavior with $g$.
One finds that dislocations are suppressed by improved actions, $i.e.$
improving the gauge action increases $S_{\rho_c}$ \cite{rgrandygupta}
and better definition of the topological charge~\cite{TOP98degrand}.
\item{5.}
Lattice perturbation theory becomes increasingly complicated as more
and more irrelevant operators are added to the action. Even though,
with the advent of non-perturbative methods for determining
renormalization constants, the reliance on lattice perturbation theory
has decreased, perturbative analyses serve as a valuable guide and
check, so we would like to retain this possibility. What one expects
for an improved action is that the lattice and continuum value of
$\alpha_s$, and the coefficients of the series expansion in $\alpha_s$
for a given observable, are much closer.  In that case a much more
reliable matching between the two theories is possible. At present the
1-loop matching between the lattice and continuum theories is one of
the larger sources of uncertainties.
\item{6.} 
The Adler-Bell-Jackiw (ABJ) Anomaly: In the continuum the flavor
singlet axial current has an anomaly.  This shows up as part of the UV
regularization scheme. In the Pauli-Villars scheme one introduces a
heavy mass as the cutoff. This breaks $\gamma_5$ invariance, while in
dimensional regularization, $\gamma_5$ is not well-defined away from
$4$ dimensions. $Naive$ fermions, as shown below, preserve the
$\gamma_5$ invariance but at the expense of flavor doubling. A 
consequence of the doublers is a pair-wise cancellation and no net 
anomaly. A viable lattice theory should reproduce the ABJ anomaly.

\end{description}

Before discussing such ``improved'' actions, it is instructive to
revisit the construction of the naive Dirac action to highlight its
problems.

\section{Fermion Doubling Problem in the ``naive'' action} 
\label{s:doubling}

The problem with the naive discretization of the Dirac action,
specified by the operator $M^N$ in Eq.~\ref{eq:naiveactM}, is that in
the continuum limit it gives rise to $2^d=16$ flavors rather than one.
This doubling problem is readily demonstrated by the inverse of the
free field propagator (obtained by taking the fourier transform of the
action with all $U_\mu(x)=1$).
\be
S^{-1}(p) \ = \ m_q \ + \ {i \over a} \sum_{\mu} \gamma_\mu \ \sin p_\mu a
\ee
which has 16 zeros within the Brillouin cell in the limit $m_q \to 0$.
Defining the momentum range of the Brillouin cell to be
$\{-\pi/2,3\pi/2\}$, the zeros lie at $p_\mu = 0$ and $\pi$. As
discussed later, this proliferation holds under very general
conditions specified by the Nielsen-Ninomiya theorem \cite{NNnogo},
and is intimately related to simultaneously preserving chiral
symmetry. The inclusion of the gauge fields does not solve the 
doubling problem. 

Before discussing partial fixes, it is useful to investigate the
properties of these extra zero modes under chiral transformations.
Let us define a set of 16 4-vectors $\Pi^A = \{ (0,0,0,0),
(\pi,0,0,0), \ldots , (\pi, \pi, \pi, \pi) \}$ with $A = \{1 \ldots
16\}$, and consider the expansion of the massless propagator about
these points. Then
\begin{eqnarray}
S^{-1}(p,m=0) \ &=& \ {i \over a} \sum_{\mu} \gamma_\mu \ \sin p_\mu a \nonumber \\
                &=& \ {i \over a} \sum_{\mu} \gamma_\mu \ \sin (\Pi^A + k)_\mu a \nonumber \\
                &=& \ {i \over a} \sum_{\mu} \gamma_\mu \ \CS^A_\mu \sin k_\mu a \nonumber \\
                &\equiv& \ {i \over a} \sum_{\mu} \tilde\gamma_\mu \ \sin k_\mu a \ ,
\label{eq:NFzeros}
\end{eqnarray}
where $\CS^A_\mu = \{ +1,-1\}$ depending on whether the $\mu$
component of $\Pi^A$ is $\{ 0,\pi \}$.  In the last expression a new
representation of gamma matrices has been defined by the similarity
transformation $\tilde\gamma_\mu = Y_A \gamma_\mu Y_A^{\dagger} $
where
\be
Y_A = \prod_\mu (\gamma_\mu\gamma_5)^{n^A_\mu}
\label{eq:NFYdefn}
\ee
and $n^A_\mu$ are $\{0,1\}$ depending on whether the momentum
expansion in the $\mu$ direction is about $0$ or $\pi$.  Now, $ \tilde
\gamma_5 = Y_A \gamma_5 Y_A^{\dagger} = \CS^A_1 \CS^A_2 \CS^A_3 \CS^A_4 
\gamma_5 \equiv X^A \gamma_5 $, with the signs $X^A$ given in 
table~\ref{t:Chiralsigns}. Thus, the sixteen species break up into two
sets of 8 with chiral charge $\pm 1$, and render the theory anomaly-free. 
Consequently, this ``naive'' discretization is phenomenologically not 
acceptable. The presence of doublers can, in fact, be traced to a larger 
flavor symmetry of the naive fermion action as I now discuss. 

\medskip
\begin{table} 
\begin{center}
\setlength\tabcolsep{1cm}
\caption{The chiral sign and degeneracy factor for the doublers.}
\begin{tabular}{|l|l|l|}
\myhline
                 &                 &       \\[-7pt]
$A$              & Degeneracy      & $X^A$ \\[2pt]
\myhline
$(0,0,0,0)$      & 1               & $+1$    \\[2pt]
$(1,0,0,0)$      & 4               & $-1$    \\[2pt]
$(1,1,0,0)$      & 6               & $+1$    \\[2pt]
$(1,1,1,0)$      & 4               & $-1$    \\[2pt]
$(1,1,1,1)$      & 1               & $+1$    \\[2pt]
\myhline
\end{tabular}
\label{t:Chiralsigns}
\end{center}
\end{table}
\smallskip
\medskip

In the continuum the mass-less Dirac action is invariant under the $U(1)_V
\otimes U(1)_A$ flavor transformations 
\bea 
\psi_L &\to& V_L \psi_L, \hskip 0.8in \bar \psi_L \to \bar \psi_L V_L^\dagger \nonumber \\
\psi_R &\to& V_R \psi_R, \hskip 0.8in \bar \psi_R \to \bar \psi_R V_R^\dagger \,.
\eea
Since the left and right handed fields are only coupled through the mass term, 
the vector symmetry $V_L = V_R = e^{i\theta}$ holds for all $m$, whereas 
the axial symmetry $V_L = V_R^\dagger = e^{i\theta}$ (or equivalently 
$V_L = V_R = e^{i\theta \gamma_5}$) holds only for $m=0$. 
The naive fermion action $\CS_N$ for a single flavor has a much larger
symmetry group, $U(4)_V \otimes U(4)_A$, under 
\bea
\psi(x) &\to& \Gamma_x \bigg( {1 \over 2 } \big(1-\epsilon(x) \gamma_5 \big) V_L + 
                              {1 \over 2 } \big(1+\epsilon(x) \gamma_5 \big) V_R 
		       \bigg)  \Gamma_x^\dagger \psi(x) \nonumber \\
\bar \psi(x) &\to& \bar \psi(x) \Gamma_x \bigg( 
			 V_L^\dagger  {1 \over 2 } \big(1+\epsilon(x) \gamma_5 \big) + 
                         V_R^\dagger  {1 \over 2 } \big(1-\epsilon(x) \gamma_5 \big)
 		                          \bigg)  \Gamma_x^\dagger 
\label{eq:NFsymm}
\eea
if one considers translations by $2a$. $\Gamma_x$ is given in
Eq.~\ref{eq:STAGGdef}, the phase $\epsilon(x)$ in
Eq.~\ref{eq:SFphases}, and $\Gamma_x \gamma_5 \Gamma_x^\dagger =
\epsilon(x)\gamma_5$. Here $V_{L,R}$ are general $4 \times 4$ unitary
matrices acting on the spinor index of the fermions. Even though there
is, in Eq.~\ref{eq:NFsymm}, a dependence on $x$ through $\epsilon(x)$,
this is a global symmetry as specifying it at one point fixes it at
all points.

To see how doubling arises, consider the vector ($V_L = V_R$) discrete
subgroup with the thirty two elements $\{\pm Y_A\}$ 
defined in Eq.~\ref{eq:NFYdefn}. Eq.~\ref{eq:NFzeros} shows 
that the $Y_A$ shift the momenta by $\pi/a$ along the direction 
specified by $A$, 
\be
Y_A S_F(k) Y_A^\dagger = S_F(k+{\pi \over a} A) \,.
\ee
Since this is a similarity transform, the sixteen regions of the
Brillouin zone related by this transormation are physically
equivalent. This analysis, which relied only on the spinor structure
of the action, makes it easy to see why the doublers exist even for
the interacting theory. In Section~\ref{s:staggered} this symmetry
will be exploited to construct staggered fermions. In this formulation
the sixteen doublers are reduced to four, and at the same time a
$U(1)_V \otimes U(1)_\epsilon$ symmetry (corresponding to when
$V_{L,R}$ are just phases) is retained. The $U(1)_\epsilon$ plays the
role of a chiral symmetry at finite $a$ and in the continuum limit
becomes the $U(1)_A$.

\bigskip
\subsection{Generalities of Chiral Symmetry}

In the massless limit of the continuum theory, the flavor symmetry at
the classical level is $U_V(1) \times U_A(1) \times SU_V(n_f) \times
SU_A(n_f)$, where $n_f$ is the number of light fermions.  The unbroken
$U_V(1)$ gives rise to baryon number conservation.  The $U_A(1)$ is
broken by instanton contributions and the flavor singlet axial current
is anomalous.  The $SU_A(n_f)$ is spontaneously broken and the quark
condensates acquire a non-zero expectation value in the QCD
vacuum. Associated with this spontaneously broken symmetry are
$(n_f^2-1)$ goldstone bosons, $i.e.$ the pions. Chiral symmetry has
played a very useful role in the continuum theory. For example, one
can (i) classify operators under distinct representations of
$SU_L(n_f) \times SU_R(n_f)$, and chiral symmetry prohibits mixing
between operators in different representations, (ii) derive relations
between matrix elements of a given operator between states with
different number of pions in the final state (for example $K \to \pi
\pi \leftrightarrow K \to \pi \pi\pi$), (iii) exploit the operator
Ward identity, and (iv) derive relations between correlation
functions, $i.e.$ the Ward-Takahashi identities.  It is therefore
desirable to preserve chiral symmetry at finite $a$ and not just in
the continuum limit. To determine if this is possible one needs to
understand the relationship between doublers and chiral symmetry in
the lattice theory.

The chiral symmetry defined in Eq.~\ref{eq:axial_psi} is realized 
by the lattice theory provided $\gamma_5 M + M \gamma_5 = 0$. It is easy 
to verify that the n\"aive lattice action satisfies the hermiticity property 
\begin{equation}
\gamma_5 M \gamma_5 = M^\dagger
\label{eq:hermiticity}
\end{equation}
Thus, in the massless limit, the Euclidean lattice action has to be
anti-hermitian for chiral symmetry, Eq.~\ref{eq:axial_psi}, to hold.
This is true of $M^N$ given in Eq.~\ref{eq:naiveactM}, and was
accomplished by simply taking the symmetric difference for the derivative. The
important question is -- what are the general conditions under which a
lattice theory with this realization of chiral symmetry exhibits
doublers. The answer is given by the Nielsen-Ninomiya theorem
\cite{NNnogo}. Consider a generalized action such that $S^{-1}(p,m=0)
= i F(p)$.  Then, if

\begin{description}
\item{$\bullet$}
the function $F(p)$ is periodic in momentum space with period $2 \pi /a$. 
(This is a consequence of {\it translation invariance}.) 
\item{$\bullet$}
the lattice momenta are continuous in the range $\{0,2\pi\}$. This
is true in the $L \to \infty$ limit at all values of the lattice spacing.
\item{$\bullet$}
$F(p)$ is continuous in momentum space. This is guaranteed if the 
interactions are local.
\item{$\bullet$}
$F(p) \to p_\mu \gamma_\mu$ for small $p_\mu$ and as $a \to 0$ to
match the continuum theory.
\item{$\bullet$}
the action has chiral symmetry, $i.e.$ it satisfies the hermiticity
property $\gamma_5 M \gamma_5 = M^\dagger$ and the massless Dirac action
is anti-hermitian.
\end{description}

\begin{figure} 
\hbox{\epsfxsize=\hsize\epsfbox{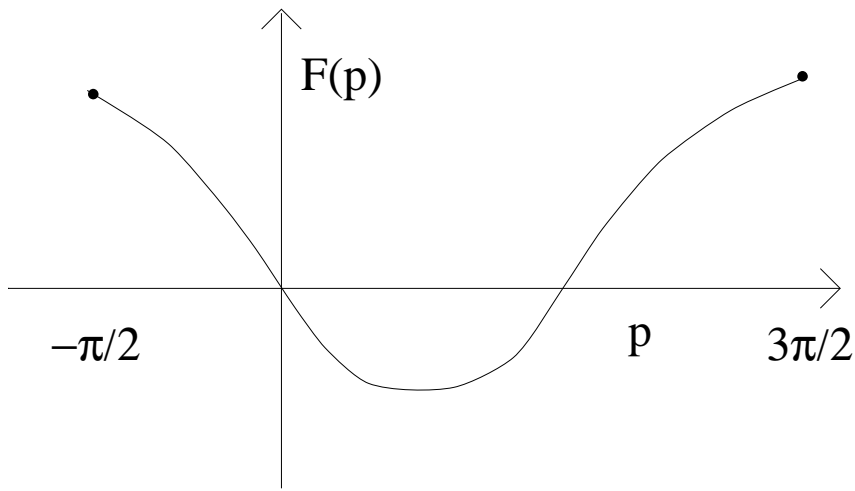}}
\caption{A schematic 1-d representation of the massless inverse propagator $F(p)$.}
\label{f:Fp}
\end{figure}

\medskip
\noindent the theory will have doublers. It is easy to understand this 
from the schematic one dimensional plot of $F(p)$ shown in
Fig.~\ref{f:Fp} which highlights the fact that $F(p)$ has to have an
even number of zeros under the above conditions. The $\pm$ slopes
imply that one state is left moving while the other is right
moving. In 1-dimension the direction is the same as handedness,
consequently left-handed and right-handed fields come in inseparable
pairs, and together form a Dirac fermion.

The key consequence of the Nielsen-Ninomiya theorem is that one cannot
define a lattice theory with chiral fermions. For example if one
writes down $M^N$ for a left-handed Dirac fermion, then its doubler
will be right-handed but have the same transformation under $L \otimes
R$.  Consequently, the two ``states'' coexist to form a Dirac fermion.
Assuming that the Nielsen-Ninomiya theorem is air-tight, then how does
one proceed to (i) define a lattice theory without doublers and which
recovers chiral symmetry in the continuum limit, and (ii)
define chiral gauge theories on the lattice? 

The answer to the first question will be given by explicit
reconstruction of Wilson's and staggered fixes. An alternate very
exciting new development, which I will not do justice to, is the
resurrection \cite{CS98hasenfratz} of the Ginsparg-Wilson condition for 
a chirally symmetric formulation \cite{CS82ginspargwilson}
\be
\gamma_5 D + D \gamma_5 = a D \gamma_5 D \,.
\ee
L\"uscher has recently presented a clear analysis of how this
condition allows the construction of a lattice theory without doublers
and with chiral symmetry, avoids the NN theorem, and gives the ABJ
anomaly \cite{CS98luscher}. Unfortunately, the bottleneck with this
alternative is the lack of a closed form expression for the lattice
Dirac action that is sufficiently local to make simulations feasible.

The second question -- construction of chiral gauge theories
-- is still unresolved. For those interested in this very important
unsolved problem, a good starting point are the reviews 
\cite{CGT95shamir,CGT97narayanan,CGT97testa} which contain 
some of the recent efforts.

\section{Wilson Fermions (WF)}
\label{s:wilsonfermions}

In Section~\ref{ss:symanzikImp}, it was shown that the lattice action
is not unique.  One has the freedom to add an arbitrary number of
irrelevant operators to the action, as these do not change the
continuum limit.  Wilson's solution to the doubling problem was to add
a dimension five operator $a r \bar \psi \,\sqrB\, \psi$, whereby the
extra fifteen species at $p_\mu = \pi$ get a mass proportional to $r /
a$ \cite{kgwf77}.  The Wilson action is
\bea
\label{eq:wilsonacta}
A^W\ &=& \  m_q\ \sum_x \bar\psi (x) \psi (x)  \nonumber \\
     &+& \ {1\over2a}\ \sum_{x,\mu} \bar\psi(x) \gamma_\mu \lbrack 
     U_\mu (x) \psi (x+\hat\mu) \ - \ U_\mu^\dagger (x-\hat\mu) 
     \psi (x-\hat\mu) \rbrack \nonumber \\
     &-&\ {r\over2a}\ \sum_{x,\mu} \bar\psi (x) \lbrack 
     U_\mu (x) \psi (x+\hat\mu) \ - 2 \psi(x) + \ U_\mu^\dagger (x-\hat\mu) 
     \psi (x-\hat\mu) \rbrack  \nonumber \\
     &=& \  {(m_q a \ + 4r) \over a} \sum_x \bar\psi (x) \psi (x)  \nonumber \\
     &+& \ {1\over2a}\ \sum_x \bar\psi \lbrack 
	  (\gamma_\mu-r) U_\mu (x) \psi (x+\hat\mu) \ - 
     (\gamma_\mu + r)\ U_\mu^\dagger (x-\hat\mu) \psi (x-\hat\mu) \rbrack  \nonumber \\
     &\equiv&\  \sum_{x,y} \bar\psi^L_x M_{xy}^W \psi^L_y  
\eea
where the interaction matrix $M^W$ is usually written as 
\be
\label{eq:wilsonactc}
M^W_{x,y}[U] a = \delta_{xy}\ -\ \kappa \sum_\mu
 \bigl[(r - \gamma_\mu ) U_{x,\mu}\delta_{x,y-\mu} + 
        (r + \gamma_\mu ) U^\dagger_{x-\mu,\mu}\delta_{x,y+\mu} \bigr]
\ee
with the rescaling 
\begin{eqnarray}
\kappa &=& 1/(2m_q a + 8r) \nonumber \\
\psi^L &=& \sqrt{m_q a + 4r}\ \psi = \psi/\sqrt{2\kappa} \ .
\label{eq:WFrescaling}
\end{eqnarray}
In this form the Wilson action has the following interpretation as a
Statistical Mechanics model. The local term acts like an inertia
keeping the fermion at the same site while the non-local term makes
the fermion hop to the nearest neighbor site with strength $\kappa$.
During this hop the fermion acquires a twist by $(\gamma_\mu - r) $ in
spin space and by $U_\mu$ in color space. The 
case $r=1$ is special as the spin twist $1 \pm \gamma_\mu$ is 
a projection operator of rank 2, 
\bea
\label{eq:spinprojectors}
\big( { {1\pm\gamma_\mu} \over 2 } \big)^2  \ &=& \ \big( {1\pm\gamma_\mu \over 2 }\big) \nonumber \\
{\rm Tr}  \big( { {1\pm\gamma_\mu} \over 2 } \big)  \ &=& \ 2 \ . 
\eea
Later we shall see that for $r=1$ the dispersion relation has a single branch, $i.e.$ there 
are no doublers in the free field limit. 

From Eq.~\ref{eq:WFrescaling}, one notes that for the free theory the
quark mass is given in terms of the lattice parameters $\kappa$ and
$r$ as
\be
\label{eq:wilsonmass}
m_q a\ = \ {1\over { 2 \kappa}} - {4r }
    \equiv \ {1 \over 2 \kappa } - { 1 \over 2\kappa_c} 
\ee
with a zero at $\kappa = \kappa_c \equiv 1/8r$.  For the interacting theory 
($U_\mu(x) \neq 1$) we will continue to define $m_q a = 1 /2\kappa -
1/2\kappa_c$ with the proviso that $\kappa_c$ depends on $a$.  The
renormalization of $\kappa_c $ independent of $\kappa$ implies that
the quark mass has both multiplicative and additive renormalizations.
This is a consequence of the explicit breaking of chiral symmetry by
the irrelevant term proportional to $r$. Wilson's fix for doublers
comes with a high price tag -- a hard breaking of chiral symmetry
at $O(a)$. 

The free propagator in momentum space for Wilson fermions is 
\be
\label{eq:wilsonfreeprop}
S_F(p) = M_W^{-1}(p) \ = {a \over {\ 1 \ - \ 2\kappa\ \sum_{\mu} 
         \big( r \ \cos p_\mu a- i \gamma_\mu \ \sin p_\mu a\big) }} \ .
\ee
Due to the dimension 5 operator (terms proportional to $r$) the 15
extra states at $p_\mu a = \pi$ get masses of order $2rn/a$,
where $n$ is the number of components of $p_\mu a = \pi$.

The operator $M$ satisfies the following relations:
\bea
\label{eq:WFhermiticity}
\gamma_5 M_W^\dagger \gamma_5 \ &=& \ M_W , \nonumber \\
\gamma_5 S_F^\dagger (x,y) \gamma_5 \ &=& \ S_F (y,x) \ , \nonumber \\
M_W^\dagger (\kappa, r) \ &=& \ M_W (-\kappa, -r)  \ . 
\eea
The first two state that the ``hermiticity'' property of $M^N$ is
preserved.  The second equation relates a quark propagator from $x \to
y$ to the antiquark propagator from $y \to x$.  This important
identity, called hermiticity or $\gamma_5$ invariance, leads to a
significant simplification in numerical calculations as discussed in
Section~\ref{ss:corr2pt}. The adjoint in $S_F^\dagger$ is with respect to the
spin and color indices at each site. The last relation shows that
$M^W$ is not anti-hermitian due to the Wilson $r$ term.

An analysis similar to Eq.~\ref{eq:Npolemass} for the naive fermion action,
shows that the pole mass derived from Eq.~\ref{eq:wilsonfreeprop} is
different from the bare mass and given by
\be
m_q^{pole} a = r(\cosh Ea -1) + \sinh Ea \,.
\label{eq:WFfreemass}
\ee
This shows, as expected, that the discretization corrections in
spectral quantities occur at $O(a)$.

\subsection{Properties of Wilson Fermions}

I give a brief summary of the important properties of Wilson fermions.
\begin{description}
\item{$\bullet$}
The doublers are given a heavy mass, $2 r n /a$, and decouple in 
the continuum limit. 

\item{$\bullet$}
Chiral symmetry is broken explicitly. The derivation of axial Ward
identities, using the invariance under the transformation given in
Eq.~\ref{eq:axial_psi}, have the generic form
\be
\VEV{{\partial \delta\CS \over \partial \theta} \CO } = \VEV{{\partial \delta\CO \over \partial \theta} } \,.
\ee
For WF, the variation of the action under an axial transformation is 
\be
{\partial \delta\CS \over \partial \theta} = \partial_\mu A_\mu - 2 m P + r a X,
\ee
where $X$ is an additional term coming from the variation of the
Wilson $r$ term \cite{WI85Bochicchio}. Thus, in general, all relations
based on axial WI will have corrections at order $ra$ and involve
mixing with operators that would normally be absent due to the chiral
symmetry.
\item{$\bullet$}
Calculation of the matrix element $\ME{0}{T[A_\mu V_\nu V_\rho]}{0}$
shows that the ABJ anomaly is correctly reproduced in the continuum
limit~\cite{KarstenSmit81}.  Karsten and Smit show, by an explicit
1-loop calculation, that while each of the extra 15 states contribute
terms proportional to $r$, the total contribution from all sixteen
states is independent of $r$ and equals the ABJ anomaly.
\item{$\bullet$}
The quark mass gets both additive and multiplicative renormalization. 
\item{$\bullet$}
The zero of the quark mass is set by $\kappa_c$ in
Eq.~\ref{eq:wilsonmass}.  There are two ways to calculate $\kappa_c$
at any given $a$. (i) Assume the chiral relation $M_\pi^2 \propto
m_q$, calculate the pion mass as a function of $1/2\kappa$, and
extrapolate it to zero.  The value of $\kappa$ at which the pion
becomes massless is, by definition, $\kappa_c$.  (ii) Calculate the
quark mass through the ratio $\VEV{\partial_\mu A_\mu(x)
P(0)}/\VEV{P(x) P(0)}$ (based on the axial Ward identity) as a
function of $1/2\kappa$ and extrapolate to zero.  The two estimates
can differ by corrections of $O(a)$. Both these calculations requires
a statistical average over gauge configurations and an extrapolation
in $1/2\kappa$.  Consequently, on an individual configuration the zero
mode of the matrix $M$ will not occur at $\kappa_c$ but will have a
distribution about $\kappa_c$.  

In the full theory ${\rm det} M$ suppresses configurations with zero
modes, however, this protection is lost in QQCD.  Nevertheless, the
problem of zero modes at $\kappa < \kappa_c$ on certain
configurations, called ``exceptional'' configurations, needs to be
addressed in both theories.

Near these zero-modes the quark propagator becomes singular even
though one is in the physical region $\kappa < \kappa_c$.
Consequently, in a large enough statistical sample one will hit
``exceptional'' configurations, and hadronic correlations functions on
these will show fluctuations by orders of magnitude (depending on
small the quark mass is) from the rest. Since this pathology is a
feature of Wilson-like actions and not of the gauge configurations
themselves, one cannot simply remove these configurations from the
statistical sample. The problem of such zero-modes increases with
decreasing quark mass and $\beta$.  Two practical approaches used so
far have been to (i) simulate at $m_q > m_q^{*}$ where $m_q^{*}$ is
sufficiently large such that the spread in the position of the zero
modes does not give rise to large fluctuations in the correlation
function, and (ii) for fixed $m_q$ increase the lattice volume as it
decreases the probability of hitting exceptional configurations.
These fixes do not help if one is interested in simulating QCD near
the chiral limit, $i.e.$, for physical light quarks. For this new
ideas are needed. A more detailed analysis of this problem and a
possible solution is given in
\cite{exceptional97estia,exceptional98estia}.
\item{$\bullet$}
The spin and flavor degrees of freedom are in one-to-one
correspondence with continuum Dirac fermions.  Thus, the construction
of interpolating field operators is straightforward, for example,
$\bar \psi \gamma_5 \psi$ and $\bar \psi \gamma_i \psi$ are
interpolating operators for pseudoscalar and vector mesons just as in
the continuum (see Section~\ref{s:corrfns}).
\item{$\bullet$}
The Wilson term changes the discretization errors to $O(a)$. 
\item{$\bullet$}
{\bf Feynman Rules} and details of gauge fixing for the plaquette gauge
action and Wilson fermions are given in \cite{plqcd81kawai}.
\item{$\bullet$}
{\bf Symmetries of $S_F$}: The transformation of the propagator 
under the discrete symmetries is the same as for naive fermions. For a
given background configuration $U$ these are
\bea
\label{eq:fermionpct}
{\cal P}: \qquad  S_F(x,y,[U]) &\to& \gamma_4\         S_F(x^P,y^P,[U^P])\          
				     \gamma_4 \hfil \nonumber \\
{\cal T}: \qquad  S_F(x,y,[U]) &\to& \gamma_4\gamma_5\ S_F(x^T,y^T,[U^T])\ 
				     \gamma_5\gamma_4 \hfil \nonumber \\
{\cal C}: \qquad  S_F(x,y,[U]) &\to& \gamma_4\gamma_2\ S_F^{\rm T}(x,y,[U^C])\ 
				     \gamma_2\gamma_4 \hfil \nonumber \\
{\cal H}: \qquad  S_F(x,y,[U]) &\to& \gamma_5\         S_F^\dagger(y,x,[U])\      
				     \gamma_5  \hfil \nonumber \\
{\cal CPH}: \quad  S_F(x,y,[U]) &\to& \CC \gamma_4\    S_F^\dagger(y^P,x^P,[U^C])\   
					  \gamma_4 \CC^{-1} \hfil 
\eea
These relations are very useful in determining the properties of
correlation functions (real versus imaginary, even or odd in $\vec p$, 
\etc) as discussed in Section~\ref{s:corrfns}.

\item{$\bullet$}
The dispersion relation, Eq.~\ref{eq:WFfreemass}, has the following 
two solutions
\be
e^{Ea} = {(ma+r) \pm \sqrt{(1 + 2mar + m^2a^2)} \over 1+r} 
\ee
For $r = 0$ and $m$ small (naive fermion action), the two solutions are
\bea
E &=&  m + O(m^2a^2) \nonumber \\
E &=& -m + {i\pi \over a} + O(m^2a^2) \,,
\eea
where the second solution corresponds to the pole in the Euclidean
propagator at $p_4 = \pi$ and is the temporal doubler. For
$r=1$ this doubler is removed, as expected, and the single root $Ea =
\log(1+ma)$ is associated with the physical particle.  For a general
$r \ne 1$ the second root is given by
\be
E = -m + {1 \over a} \bigg(i\pi + \log {(1-r) \over 1+r} \bigg) + \ldots \ .
\label{eq:Eroot}
\ee
Such branches which are not associated with the physical state or with
doublers are called ghost branch.  I use a very restrictive definition
of a doubler $--$ a solution which has $|E|=0$ for $m = \vec p =
0$. Again from Eq.~\ref{eq:Eroot} it is clear that this ghost branch
is pushed to infinity for $r=1$.  In general, if the action couples
$(n+1)$ time slices, then the dispersion relation has $n$
branches. The roots can be real, complex-conjugate pairs, and/or have
an imaginary part $i \pi /a$ as illustrated above. Such additional
branches can be a nuisance for improved actions, for example the
$D234$ action discussed in Section~\ref{ss:D234}. The goal is to
obtain a doubler and ghost free theory.  This can be done by a
combination of tuning the parameters (for example choosing $r=1$ in
the Wilson case) and/or going to anisotropic lattices $a_t \ll a_s$ as
advocated by Lepage and collaborators \cite{IDA97D234c}.

\item{$\bullet$}
{\bf Conserved Vector Current}: The Wilson action is invariant under
the global transformation defined in Eq.~\ref{eq:vector_psi}. To
derive the associated conserved current (for degenerate masses) we use
the standard trick of calculating the variation of the action under a
space-time dependent $\theta(x)$ and associating the coefficient of
$\theta(x)$ in $\delta S$ with $\partial_\mu J_\mu$.  The change in
the action under such an infinitesimal transformation is
\bea
\label{eq:vectorcurrenta}
\delta S &=&
             \kappa \sum_{x,\mu} \bar\psi(x) (\gamma_\mu-r) U_\mu (x) 
              \psi (x+\hat\mu) \exp(i\theta(x) - i\theta(x+\mu)) \nonumber \\
        &-& 
     \kappa \sum_{x,\mu} \bar\psi(x+\mu) (\gamma_\mu + r)\ U_\mu^\dagger (x) 
        \psi (x)  \exp(i\theta(x+\mu) - i\theta(x)) \nonumber 
\eea
which, to first order in $\theta$, is 
\bea
\label{eq:vectorcurrentb}
-\sum_{x,\mu} \ \big\lbrack \bar\psi(x) (\gamma_\mu-r) U_\mu (x) \psi (x+\hat\mu) &+& \nonumber \\
 &{}& \hskip -1.8in 
   \bar\psi(x+\mu) (\gamma_\mu + r)\ U_\mu^\dagger (x) \psi (x) \big\rbrack 
   \big\lbrack i {{ \partial \theta} \over {\partial x_\mu}} \big\rbrack . \nonumber 
\eea
The conserved current, obtained after integration by parts, is 
\bea
\label{eq:vectorcurrentc}
V^{c}_\mu = \bar\psi(x) (\gamma_\mu-r) U_\mu (x) \psi (x+\hat\mu) + 
               \bar\psi(x+\mu) (\gamma_\mu + r)\ U_\mu^\dagger (x) \psi (x) \,. \nonumber \\
\eea
$V^{c}_\mu$ is hermitian and reduces to the symmetrized version of the
1-link vector current for $r=0$. In many applications like decay
constants one uses the local (flavor) currents defined by
\bea
2V_\mu(x) &=& \bar\psi(x)         \gamma_\mu {\lambda^a \over 2} U_\mu (x) \psi (x+\hat\mu) + h.c. \nonumber \\
2A_\mu(x) &=& \bar\psi(x) \gamma_5\gamma_\mu {\lambda^a \over 2} U_\mu (x) \psi (x+\hat\mu) + h.c. \,,
\eea
which are not conserved, and consequently have associated non-trivial
renormalization factors $Z_V$ and $Z_A$ which have to included when 
calculating matrix elements.

\item{$\bullet$}
{\bf The chiral condensate}: As explained in \cite{WI85Bochicchio}, the 
chiral condensate $\vev{\bar \psi \psi}$ is not simply related to the 
trace of the Wilson quark propagator $\vev{S_F(0,0)}$. The breaking of 
chiral symmetry by the $r$ term introduces contact terms that need to be 
subtracted non-perturbatively from $S_F(0,0)$.  This has not proven 
practical. The methods of choice are to use either the continuum 
Ward Identity
\be
\label{eq:WIcondensate}
\vev{\bar\psi\psi}^{\rm WI} \equiv \vev{0|S_F(0,0)|0}
        = \lim_{m_q\to 0} m_q \int d^4x \vev{0|P(x)P(0)|0},
\ee
or the lattice version of the Gell-Mann, Oakes, Renner
relation \cite{GMOR68} 
\be
\label{eq:GMOR}
\vev{\bar\psi\psi}^{\rm GMOR}
        = \lim_{m_q\to 0} -{ f_\pi^2 M_\pi^2 \over {4 m_q}} \,.
\ee
A comparison of the efficacy of the two methods is discussed in \cite{HM91LANL}.  

\item{$\bullet$}
{\bf Operator Mixing}: There is, in general, a mixing of operators
with different chirality in the Callen Symanzik equations.  This is
true even at $\kappa_c$ and is a consequence of the explicit breaking
of chiral symmetry.  This mixing of operators poses serious problems
in the the calculation of matrix elements of the weak interactions
Hamiltonian, $i.e.$, the lattice analyses of Ward identities and
obtaining amplitudes with the correct chiral behavior is more
involved. For example the chiral behavior of matrix elements of the
tree-level 4-fermion operators (relevant to the extraction of $B_K$,
$B_7$, and $B_8$) is not the same as in the continuum theory. The
reason being that the Ward Identities associated with the
spontaneously broken $SU_A(n_f)$ receive corrections at $O(a)$ in the
lattice theory.  This issue will be discussed in more detail by
Prof. Martinelli.

\end{description}

In Section~\ref{s:improveddirac} I will discuss how the $O(a)$ errors
in $WF$ can, with very little extra effort, be reduced to $O(a^2)$ or
higher. For the time being let me use this drawback of Wilson fermions
to motivate the formulation of staggered fermions. This formulation
has $O(a^2)$ errors, retains enough chiral symmetry to give the
correct chiral behavior for amplitudes involving Goldstone pions, but
at the expense of a 4-fold increase in the number of flavors.

\section{Staggered fermions ($SF$)}
\label{s:staggered}

The 16-fold doubling problem of the naive fermion action given in Eq.~\ref{eq:naiveact}
can be reduced to 4 by the trick of spin-diagonalization \cite{STAGaction}.  
Staggered fermions $\chi$ are defined by the transformation 
\be
\psi(x) = \Gamma_x \chi(x)  \qquad
\bar \psi(x) = \bar \chi(x)  \Gamma_x^\dagger \qquad
\Gamma_x = \gamma_1^{x_1}\gamma_2^{x_2}\gamma_4^{x_3}\gamma_4^{x_4}. 
\label{eq:STAGGdef}
\ee
In terms of $\chi$ the action can be written as \cite{STAGaction}
\bea
\label{eq:SFaction}
\CS_S &= m_q \sum_x \bar\chi (x) \chi (x)
       + {1 \over 2} \sum_{x,\mu} \bar\chi_x {\eta_{x,\mu} }
                     \big( U_{\mu,x} \chi_{x+\hat\mu} -
                     U^\dagger_{\mu,x-\hat\mu} \chi_{x-\hat\mu} \big) \nonumber \\
    &\equiv  \sum_{x,y} \bar\chi (x) M_{xy}^S \chi (y)  
\eea
with the matrix $M^S$ given by 
\be
M^{S}[U]_{x,y} = m_q\delta_{xy}\ +\ {1\over 2}\ \sum_\mu\ \eta_{x,\mu}
\ \bigl[U_{x,\mu}\delta_{x,y-\mu}\ -\ U^\dagger_{x-\mu,\mu}\delta_{x,y+\mu}
\bigr] \,.
\label{eq:SFactionM}
\ee
The $\gamma$ matrices are replaced by the phases $\eta_{x,\mu}$. 
It is convenient to define the following phase factors 
\bea
\eta_{x,\mu}  &=& (-)^{\sum_{\nu<\mu} x_\nu} \nonumber \\
\zeta_{x,\mu} &=& (-)^{\sum_{\nu>\mu} x_\nu} \nonumber \\
\epsilon_x    &=& (-)^{x_1+x_2+x_3+x_4}      \nonumber \\
S_R(x)        &=& {1\over 2}\big(1+\zeta_\mu\zeta_\nu-\eta_\mu\eta_\nu+
                              \zeta_\mu\zeta_\nu\eta_\mu\eta_\nu \big)   . 
\label{eq:SFphases}
\eea
From Eqs.~\ref{eq:SFaction},\ref{eq:SFactionM} it is easy to see that
the different spin components of $\chi$ are decoupled as the phase
factor $\eta_{x,\mu}$ depends only on the site index and direction and
do not have a spinor index.  One can therefore drop the spin index on
$\chi$ leaving only color degrees of freedom at each site. This
reduces the original $2^d$-fold degeneracy of naive fermions by a factor
of four. The mass term in Eq.~\ref{eq:SFactionM} is hermition, while
the $\Dsl$ term is anti-hermitian, which as I discussed before is
required to realize chiral symmetry.  

The $SF$ action has translation invariance under shifts by $2a$ due to
the phase factors $\eta_{x,\mu}$. Thus, in the continuum limit, a
$2^4$ hypercube is mapped to a single point, and the 16 degrees of
freedom reduce to 4 copies of Dirac fermions. That is, for every physical
flavor, the staggered discretization produces a 4-fold degeneracy
denoted as the staggered flavor.  At finite $a$ the gauge interactions
break this flavor symmetry and the 16 degrees of freedom in the
hypercube are a mixture of spin and flavor. This is one of the major
drawbacks of staggered fermions. Construction of operators and their
interpretation in terms of spin and flavor is non-trivial as I show
below for the case of bilinears. I begin the discussion of staggered
fermions by listing its symmetries. I will then outline the
construction of bilinear operators and end with a discussion of Ward
identities.

\noindent{\bf Symmetries of SF}: The analogue of $\gamma_5$ invariance for $SF$ is
\bea
\label{eq:SFhermiticity}
\epsilon_x M_S^\dagger(x,y)  \epsilon_y \ &= \ M_S(x,y) , \nonumber \\
\epsilon_x S_F^\dagger (x,y) \epsilon_y \ &= \ S_F(y,x) \ .
\eea
and the symmetries of this action are \cite{STAGgkss}
$$
\tabskip 1 truecm \halign{&$\displaystyle#$\cr
{\rm Translations} \hfil & S_\mu \hfil : &  \chi(x) \to \zeta_\mu(x) \chi(x+\mu) \hfil \cropen{0.1 truecm}
{\rm Rotations} \hfil & R_{\mu\nu} \hfil : &  \chi(x) \to S_R(R^{-1}x)\chi(R^{-1}x)\hfil \cropen{0.1 truecm}
{\rm Inversion} \hfil & I \hfil : &  \chi(x) \to \eta_4(x) \chi(I^{-1}x) \hfil \cropen{0.1 truecm}
{\rm Charge\ Conjugation} \hfil & C \hfil : &  \left\{ \matrix{
          \chi(x) \to \hphantom{-} \epsilon(x) \bar \chi(x), \cr
          \bar \chi(x) \to - \epsilon(x) \chi(x) \cr} \right. \hfil \cropen{0.1 truecm}
{\rm Vector (Baryon \#)} \hfil & U(1)_V \hfil : &  \left\{ \matrix{
                \chi(x) \to e^{i \theta} \chi(x), \cr
                \bar \chi(x) \to e^{-i \theta} \bar \chi(x)  \cr} \right.  \hfil \cropen{0.1 truecm}
{\rm Axial} \hfil & U(1)_A \hfil : &  \left\{ \matrix{
                \chi(x) \to e^{i \theta \epsilon(x)} \chi(x), \cr
                \bar \chi(x) \to e^{i \theta \epsilon(x)} \bar \chi(x) \cr} \right. \hfil \cropen{0.1 truecm}
}
$$
where the phases $\eta, \zeta, \epsilon$ and $S_R$ are defined in
Eq.~\ref{eq:SFphases}.  It is the last invariance, the $U(1)_A$, or
also called the $U(1)_\epsilon$ invariance, that becomes a flavor
non-singlet axial symmetry in the continuum limit, and is the
main advantage of staggered fermions.

\noindent{\bf Operators with Staggered fermions:} 
There are two equivalent ways of understanding flavor identification
of staggered fermions: (i) momentum space \cite{STAGmom}, and (ii)
position space \cite{STAGkmnp,STAGposition}. Here I review the position
space approach.

For the purposes of constructing hadronic operators it is convenient
to represent the 16 degrees of freedom in a hypercube by a hypercube
field $Q$, which in the continuum limit represents the four spin
components of four degenerate flavors ~\cite{STAG86DanielKieu}.  This
is done by dividing the lattice into hypercubes identified by a
4-vector $y$, $i.e.$ $y$ is defined on a lattice of spacing $2a$, and
the set of 16 vectors $\{A\}$ to points within the hypercube. These
are called hypercube vectors, and addition of two such vectors is
defined modulo 2, $C_\mu = (A_\mu + B_\mu)_{mod 2}$. Then, in the free
field limit (setting gauge links $U_{x,\mu}=1$ with the understanding
that for the interacting theory non-local operators are to be
connected by a path ordered product of links that makes them gauge
invariant, or these operators are to be evaluated in a fixed gauge)
\be
Q(y) \ =\ {1 \over N_f } \sum_{\{A\}} \Gamma_A \ \chi(y+A)
\label{eq:stag16field}
\ee
where $N_f = 4$ and 
\be
\Gamma_A \ = \ \gamma_1^{A_1} \gamma_2^{A_2} 
               \gamma_3^{A_3} \gamma_4^{A_4} .
\ee
A general bilinear operators can be written as 
\bea
{\cal O}_{SF} \ &=& \ {\rm Tr} \big( \bar Q \gamma_S Q \gamma_F^\dagger \big) \nonumber \\
                &=& \sum_{A,B} \bar \chi(y+A) {\rm Tr}\big(\Gamma_A^\dagger \gamma_S \Gamma_B \xi_F \big) \chi(y+B) \nonumber \\
                &=& \sum_{A,B} \bar \chi(y+A) (\gamma_S \otimes \xi_F)_{AB} \chi(y+B) \nonumber \\
		&=& \bar Q (\gamma_S \otimes \xi_F) Q .
\label{eq:STAGop2}
\eea
where the matrices $\gamma_S$ determines the spin and $\xi_F$ the
flavor of the bilinear. Each of these can be one of the standard
sixteen elements of the Clifford algebra.  In the last expression, the
field $Q$ is expressed as a 16 component vector.  It is easy to see
from Eq.~\ref{eq:STAGop2} that local operators are given by  $\gamma_S
= \xi_F$.  The identification of meson operators is as follows
~\cite{STAGkmnp,STAGmesons,STAGhm91LANL}.
\bea
\label{eq:SFoperators}
V_\mu &= \bar Q (\gamma_\mu         \otimes \xi_F) Q \nonumber \\
A_\mu &= \bar Q (\gamma_\mu\gamma_5 \otimes \xi_F) Q \nonumber \\
P     &= \bar Q (          \gamma_5 \otimes \xi_F) Q \nonumber \\
S     &= \bar Q (1                  \otimes \xi_F) Q \,.
\eea
In the continuum the mesons lie in the flavor SU(4) representations {\bf 1}
and {\bf 15}, while on the lattice these representations break into
many smaller ones.  For example, the continuum {\bf 15}-plet of pions
breaks into seven lattice representations (four 3-d and three 1-d).
\bea
\label{eq:STAGpions}
{\bf 15} \ \to\ &{}& (\gamma_5\otimes\gamma_{i})                  \oplus
                     (\gamma_5\otimes\gamma_{i}\gamma_4)          \oplus
                     (\gamma_5\otimes\gamma_{i}\gamma_5)          \oplus
                     (\gamma_5\otimes\gamma_{i}\gamma_4\gamma_5)  \oplus \nonumber \\
                &{}& (\gamma_5\otimes\gamma_4)                    \oplus
                     (\gamma_5\otimes\gamma_5)                    \oplus
                     (\gamma_5\otimes\gamma_4\gamma_5)   \ , \nonumber \\
{\bf 1}  \ \to\ &{}& (\gamma_5\otimes1)        \ .
\eea
In the first line, $i=1,2,3$, so the representations are
3-dimensional.  The other representations are 1-dimensional.
Similarly, the continuum rho {\bf 15}-plet (times three spin
components) breaks into eleven representations (four 6-d and seven
3-d).
\bea
\label{eq:STAGrhoreps}
{\bf 15}\ \to\ &{}& (\gamma_{i}\otimes\gamma_{j})                 \oplus
                    (\gamma_{i}\otimes\gamma_{j}\gamma_4)         \oplus
                    (\gamma_{i}\otimes\gamma_{j}\gamma_5)         \oplus
                    (\gamma_{i}\otimes\gamma_{j}\gamma_4\gamma_5) \oplus \nonumber \\
               &{}& (\gamma_{i}\otimes\gamma_{i})                 \oplus
                    (\gamma_{i}\otimes\gamma_{i}\gamma_4)         \oplus
                    (\gamma_{i}\otimes\gamma_{i}\gamma_5)         \oplus
                    (\gamma_{i}\otimes\gamma_{i}\gamma_4\gamma_5) \oplus \nonumber \\
               &{}& (\gamma_{i}\otimes\gamma_4)                   \oplus
                    (\gamma_{i}\otimes\gamma_5)                   \oplus
                    (\gamma_{i}\otimes\gamma_4\gamma_5)   \ , \nonumber \\
{\bf 1}\  \to\ &{}& (\gamma_{i}\otimes1)        \ .
\eea
Of the sixteen pions, the Goldstone pion has flavor $\xi_5$.  The
remaining 15 pions are given by the other 15 members of the Clifford
algebra and are heavier due to the staggered flavor symmetry breaking,
and their masses do not vanish at $m_q=0$. In fact the mass for each
representation can be different. The degree to which the staggered
flavor symmetry is broken can be quantified, for example, by comparing
the mass of the Goldstone pion versus those in the other
representations.

The above examples make it clear that the mixing of spin and flavor
complicates the analysis and is a major disadvantage of staggered
fermions.  I will not go into details of calculations using staggered
fermions.  A useful list of references is as follows.  The
construction of baryon operators is given in \cite{STAGbaryons}.  For
an illustrative calculations of the spectrum see \cite{STAGhm91LANL}.
The transcription of 4-fermion operators arising in effective weak
Hamiltonian and their matrix elements within hadronic states are
discussed in \cite{STAGgkss,STAG98sharpe}.

\noindent{\bf Chiral Symmetry and Ward Identities:} The staggered fermion action can be 
written in terms of the 16 component fields $Q$, defined in
Eq.~\ref{eq:stag16field}, as \cite{Rothe}
\bea 
\CS_S &=& m_q \sum_{y} \bar Q(y) ({\bf 1} \otimes {\bf 1}) Q(y)  \nonumber \\
      &+& \hskip 0.2in \sum_{y,\mu} \bar Q(y) \big[ (\gamma_\mu \otimes {\bf 1}) \partial_x + 
				       (\gamma_5 \otimes \xi_\mu \xi_5) \sqrB_\mu \big] Q(y) \,.
\eea
This action, for $m_q=0$, is invariant under the following abelian transformation 
\bea
Q(y) \to e^{i \theta(\gamma_5 \otimes \xi_5)} Q(y) \nonumber \\
\bar Q(y) \to \bar Q(y) e^{i \theta(\gamma_5 \otimes \xi_5)} \,.
\eea
Thus one preserves a non-trivial piece of the original $U(4) \times U(4)$
symmetry (called $U(1)_\epsilon$ or $U(1)_A$) whose generator is $\gamma_5 \otimes \xi_5$ and
the associated Goldstone boson operator is $\bar Q (\gamma_5 \otimes
\xi_5) Q$. This axial $U(1)_A$ symmetry, along with the $U(1)_V$,
allows, for appropriate transcription of operators, Ward Identities
for lattice amplitudes similar to those in the continuum.  The best
studied example of this is the lattice calculation of the $K^0 \bar
K^0$ mixing parameter $B_K$ \cite{BkstagLANL}.

The basic tools for constructing the Ward Identities are exemplified by 
the following two relations involving quark propagators and 
the zero 4-momentum insertion of the pseudo-Goldstone operator 
$\bar \chi(x) \chi(x) (-1)^x$ and the scalar density $\bar \chi(x) \chi(x)$ 
\cite{STAGgkss}
\bea
\label{eq:stagWIa}
(-1)^{x_f} G_2(x_f, x_i) + G_1(x_f, x_i) (-1)^{x_i} &=& \nonumber \\
  &{}& \hskip -1.0in (m_1 + m_2) \sum_x G_1(x_f, x)  (-1)^{x} G_2(x, x_i)  \\
\label{eq:stagWIb}
G_2(x_f, x_i) - G_1(x_f, x_i) &=& \nonumber \\
  &{}& \hskip -1.0in (m_1 - m_2) \sum_x G_1(x_f, x)  G_2(x, x_i)  
\eea
where $G_i(x_f, x_i)$ is the propagator from $x_i $ to $x_f$ for a
quark of mass of mass $m_i$.  The derivation of these identities, using
the hopping parameter expansion, is given in Ref.~\cite{STAGgkss}. Note
that these relations are valid for all $a$ and at finite $m$. An
example of the usefulness of these relations is in extracting the
chiral condensate as shown below. Eq.~\ref{eq:stagWIa} gives
\bea
\label{eq:stagXX}
\vev{\bar \chi \chi} = G(x_i, x_i) &=& 
       m \sum_x (-1)^{-x_i} G(x_i, x)  (-1)^{x} G(x, x_i)  \nonumber \\
  &=& m \sum_x |G(x_i, x)|^2 \,,
\eea
where the last term on the right is the zero 4-momentum sum of the pion 
correlator.  Similarly Eq.~\ref{eq:stagWIb} gives 
\be
\label{eq:stagXXb}
{\partial \vev{\bar \chi \chi} \over \partial m} = 
 \sum_x G(x_i, x)  G(x, x_i)  = 
 \sum_x |G(x_i, x)|^2 (-1)^{x-x_i} \ .
\ee
Combining these two equations one can extract the value of the 
condensate in the chiral limit
\be
\label{eq:stagXXc}
\vev{\bar \chi \chi}(m=0) = \big( 1 - m {\partial \over \partial m} \big) 
\vev{\bar \chi \chi} =  \sum_{even} |G_1(x_i, x)|^2 
\ee
by summing the Goldstone pion's 2-point function over sites with even $(x-x_i)$. 

\subsection{Summary}

Even though most of the simulations done until recently have used
either the staggered or Wilson formulation of fermions, neither is
satisfactory.  They have their relative advantages and disadvantages
which have been evaluated on a case by case basis.  The rule of thumb
is that in the extraction of observables where chiral symmetry plays
an essential role and the external states are Goldstone bosons,
staggered fermions do better.  Otherwise Wilson fermions are preferred
due to their correspondence with Dirac fermions in terms of spin and
flavor.  The focus of current research is to improve the Wilson
formulation and thus have the best of both worlds. This is being done
in a variety of ways as discussed in Section~\ref{s:improveddirac}.

%% file: chap-impglue.tex
\section{Improved gauge actions}
\label{s:improvedglue}

There have been a number of methods proposed to improve the gauge
action beyond $O(a^2)$.  These include the Symanzik approach,
mean-field improvement, and renormalization group inspired perfect
actions.  Let me begin this discussion by reviewing the Symanzik
program.

\subsection{L\"uscher-Weisz Action}

\begin{figure} 
\hbox{\epsfxsize=\hsize\epsfbox{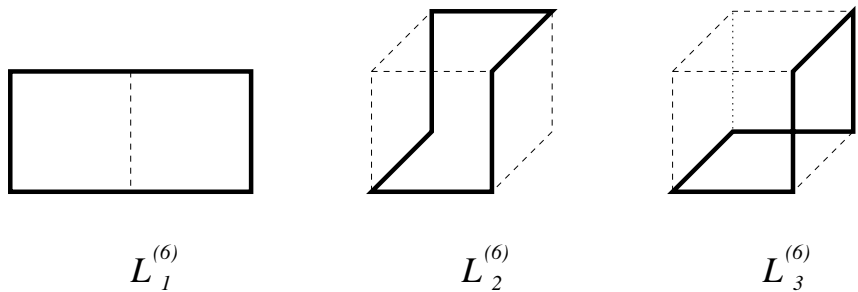}}
\caption{The three types of six link loops.}
\label{f:loop6ops}
\end{figure}

The leading order term in the expansion of all Wilson loops is ${\cal
O}^{(4)} = \sum_{\mu\nu} F_{\mu\nu}F_{\mu\nu}$ and corrections begin
at $O(a^2)$ as there are no dimension 5 operators.  Thus, any lattice
action written as a linear combination of Wilson loops will have the
correct continuum limit with corrections at $O(a^2)$.  There are three
dimension 6 operators, which in continuum notation are:
\bea
\label{gaugedimsix}
{\cal O}_1^{(6)}\ &=&\ \sum_{\mu,\nu} {\rm Tr} 
                     \bigg(D_\mu F_{\mu\nu} D_\mu F_{\mu\nu} \bigg)   \ , \nonumber \\
{\cal O}_2^{(6)}\ &=&\ \sum_{\mu,\nu,\rho} {\rm Tr} 
                     \bigg(D_\mu F_{\nu\rho} D_\mu F_{\nu\rho} \bigg) \ , \nonumber \\
{\cal O}_3^{(6)}\ &=&\ \sum_{\mu,\nu,\rho} {\rm Tr} 
                     \bigg(D_\mu F_{\mu\rho} D_\nu F_{\nu\rho} \bigg) \ . 
\eea
Also there are only 3 six-link loops that one can draw on the lattice.
These, as shown in Fig.~\ref{f:loop6ops}, are the planar ${\cal L}_1^{(6)}$, twisted ${\cal L}_2^{(6)}$
and the L shaped ${\cal L}_3^{(6)}$ respectively.  Thus, classical
improvement of the lattice action, $i.e.$ removing the $O(a^2)$ term,
can be achieved by taking a linear combination of the plaquette
and these three six-link loops.  Each of these loops has the
expansion
\be
\label{eq:gaugeimpa}
{\cal L} \ = \ {r}^{(4)}\ {\cal O}^{(4)}
+ {r}_1^{(6)}\ {\cal O}_1^{(6)} 
+ {r}_2^{(6)}\ {\cal O}_2^{(6)}  
+ {r}_3^{(6)}\ {\cal O}_3^{(6)} + \dots \ ,
\ee
and L\"uscher and Weisz have shown an elegent way of calculating the expansion 
coefficients ${r}_\alpha^{(d)}$ \cite{IGA85luscherweiszA}. 
Their results are summarized in Table~\ref{t:IGAclassical}. 

\begin{table} %
\begin{center}
\setlength\tabcolsep{0.25cm}
\def\Ofour{\hbox {${\cal L}^{(4)}$ }}
\def\Osixa{\hbox {${\cal L}_1^{(6)}$ }}
\def\Osixb{\hbox {${\cal L}_2^{(6)}$ }}
\def\Osixc{\hbox {${\cal L}_3^{(6)}$ }}
\caption{The coefficients of the continuum operators of dimension 4 and 6 in the 
classical expansion of Wilson loops with 4 and 6 links. }
\smallskip
\begin{tabular}{|l|c|c|c|c|}
\myhline
$Loop$  & ${r}^{(4)}$    & ${r}_1^{(6)}$  & ${r}_2^{(6)}$ & ${r}_3^{(6)}$   \\
\myhline
\Ofour  & $-{1 \over 4}$ & ${1 \over 24}$ &    0          &      0         \\
\Osixa  & $-2$           & ${5 \over 6} $ &    0          &      0         \\
\Osixb  & $-2$           & $-{1 \over 6}$ & ${1 \over 6}$ & $ {1 \over 6}$ \\
\Osixc  & $-4$           & ${1 \over  6}$ &    0          & $ {1 \over 2}$ \\
\myhline
\end{tabular}
\label{t:IGAclassical}
\end{center}
\end{table}
\smallskip

The lattice gauge action can be written as 
\be
\label{eq:gaugeimpb}
\CS_g \ = \ {6 \over g^2} \bigg\{ c^{(4)}(g^2)\ {\cal L}^{(4)} 
                   + \sum_{i=1,3} {c}_i^{(6)}(g^2) \ {\cal L}_i^{(6)}  \bigg\}
\ee
in terms of the plaquette and these three 6-link loops.  Using the
results in Table~\ref{t:IGAclassical} one gets the normalization
condition
\be
\label{eq:gaugeimpc}
 c^{(4)}(g^2) + 8{c}_1^{(6)}(g^2) + 8{c}_2^{(6)}(g^2) + 16{c}_3^{(6)}(g^2) = 1
\ee
such that in the continuum limit the action reduces to
$F_{\mu\nu}F_{\mu\nu}/4$.  Also, from
Table~\ref{t:IGAclassical} one can see that tree level improvement can
be obtained by the choice
\be
\label{eq:gaugeclassicalimp}
c^{(4)} + 20 c_1^{(6)} \ = \ 0; \quad c_2^{(6)} = 0;\quad c_3^{(6)}=0 \ .
\ee
This removes ${\cal O}_1^{(6)}$ by cancelling the contribution from
the two planar loops.  Of the three dimension six operators, ${\cal
O}_1^{(6)}$ alone breaks Lorentz (rotational) invariance.  Thus,
rotational invariance of the heavy $q \bar q$ potential can be used to
non-perturbatively tune $c_1^{(6)}$ and to compare the various
improved actions as discussed later.

To include quantum effects, Curci, Menotti, and Paffuti did a 1-loop
calculation.  They imposed the criteria that all Wilson loops be
improved at the leading logarithm
level~\cite{IGA83curcimenotti,IGA84Weisz}.  This corresponds to
improving the string tension.  (To kill the $O(a^2)$ terms at all
length scales instead of only in the long distance heavy quark
potential requires including the finite pieces in the 1-loop
calculations. The fully improved action, as we shall see in the final
L\"uscher-Weisz form, will lead to $O(g^2)$ corrections to $ c^{(4)}$
and $c_1^{(6)}$ and the need for ${\cal O}_2^{(6)}$.)  The result of
their calculation required the improved action to have the same form
as given in Eq.~\ref{eq:gaugeclassicalimp}.  This agreement with the
classical result is fortuitous. Symanzik's on-shell improvement
program, in general, prescribes that one has to derive three relations
to fix the $c_i^{(6)}$ such that on-shell quantities do not have $O(a^2)$
corrections.  L\"uscher and Weisz, in fact, found that improving all
large loops produced only two constraints at tree-level
\cite{IGA85MLPW}
\bea
\label{eq:gaugeimpd}
c^{(4)} + 20 c_1^{(6)} - 4c_2^{(6)} + 4c_3^{(6)} &=& 0 \nonumber \\
c_2^{(6)}  &=& 0 \ . 
\eea
Thus, they were left with a 1-parameter family of improved actions.
The reason for this freedom is the presence of a redundant operator
${\cal O}_3^{(6)}$ generated by the field redefinition $A_\mu \to
A_\mu + a^2 g^2 d(g) D_\nu F_{\nu\mu}$
\cite{IGA85luscherweiszA,IGA85MLPW}.  Since one has the liberty to choose 
the coefficient of an interaction that is part of a redundant operator
without affecting on-shell improvement, the convenient choice
$c_3^{(6)}=0$ simplifies the action and again leads us back to
Eq.~\ref{eq:gaugeclassicalimp}. To summarize, Eqs.~\ref{eq:gaugeimpd}
are the most general set of constraints that can be derived based on
improving large Wilson loops. The criteria employed by Curci \etal,
improving the large distance part of $q \bar q$ potential, is a
subset. Since this involves a spectral quantity one can use the
redundant operator to get the simpler form
Eq.~\ref{eq:gaugeclassicalimp}.


The relation $c_3^{(6)}=0$ holds both at tree and 1-loop level as 
it comes from exploiting the redundant operator. Combining this with 
the first constraint in Eq.~\ref{eq:gaugeimpd} and the normalization condition 
Eq.~\ref{eq:gaugeimpc} leads to a single condition, which at tree-level is 
\be
c_1^{(6)} - c_2^{(6)} = - {1 \over 12} \,,
\ee
and is all that can be gotten from Wilson loops. To get a second 
condition L\"uscher and Weisz had to set up 
other spectral quantities to be improved. Failing to find a suitable 
condition in the real theory, they looked at a situation in which 
the gluons are massive and appear as asymptotic states. Using the mass 
and scattering amplitude of these states they derived the 1-loop 
improved action~\cite{IGA85MLPW}, which goes under the name
L\"uscher-Weisz action and has leading correction at $O(g^4 a^4)$, 
\bea
c_0^{(4)} (g^2) &=&  {5 \over 3}   + 0.2370  g^2 \nonumber \\
c_1^{(6)} (g^2) &=& -{1 \over 12}  - 0.02521 g^2 \nonumber \\
c_2^{(6)} (g^2) &=&                - 0.00441 g^2 \nonumber \\
c_3^{(6)} (g^2) &=&  0 \,.
\eea
In this form for the $c_i$, the normalization condition,
Eq.~\ref{eq:gaugeimpc}, is already incorporated.  

The bottom line on the L\"uscher-Weisz action is that since the
coefficient $c_2^{(6)}$ is small, the action, to a very good
approximation, can be improved by keeping just ${\cal L}^{(4)}$ and
$\CL_1^{(6)}$ as in the classical case.  This 1-loop result can be
mean-field improved without much effort as discussed next.

\subsection{Tadpole/Mean-field Improvement}

The representation of gauge fields by the exponentiated form 
\be
U_\mu(x) = e^{iagA_\mu (x)} = 1 + iagA_\mu (x) - {a^2 g^2 \over 2} A_\mu^2(x) + \ldots
\ee
gives rise to local quark-gluon vertices with one, two, $\ldots$
gluons. Of these vertices, all but the lowest are lattice artifacts.
Contracting the two gluons in ${a^2 g^2 } A_\mu^2(x)/2$ gives rise to
tadpole diagrams. Parisi, and Lepage and
Mackenzie~\cite{TI80Parisi,TI93lepagemackenzie} recognized that these
artifacts, though n\"aively suppressed by powers of $a$, are in fact
only suppressed by powers of $g^2$ due to the ultraviolet divergences
generated by the tadpole loops.  They reasoned that these tadpole
contributions were responsible for (i) the poor match between short
distance quantities (like the link in Landau gauge, small Wilson
loops, $\kappa_c$, $\ldots$) and their perturbative estimates, and
(ii) large coefficients in the perturbative lattice expansions. They,
therefore, proposed a mean-field improvement (also called tadpole
improvement(TI)) to get rid of these artifacts.

The mean-field improvement starts by assuming that the lattice fields
can be split into a UV and IR parts, and that the UV part should be
integrated out~\cite{TI80Parisi,TI93lepagemackenzie}, $i.e.$
\be
e^{iagA_\mu (x)} = e^{iag(A^{IR}_\mu (x)+A^{UV}_\mu (x))} = u_0
e^{iagA^{IR}_\mu (x)} \equiv u_0 \tilde U_\mu(x) \ .
\ee
The rescaling of links by an overall constant factor $u_0 < 1$ keeps
the theory gauge invariant.  Under this rescaling (each $U \to u_0
\tilde U$) the Wilson gauge and fermion action becomes
\be
\CS^W \to \sum_x \bigg[ \beta u_0^4\ \CW_{11} + \bar{\tilde \psi} {\tilde \psi} 
     - \kappa u_0 \bar{\tilde \psi} \tilde D {\tilde \psi}  \bigg] \,.
\ee
Thus, a simplified recipe for mean-field improvement is to use the
better expansion parameters $ \tilde g^2 = g^2 / u_0^4$, $\tilde
\kappa = \kappa u_0$, $\tilde U = U / u_0$, and ${\tilde \psi} =
\psi^{cont}/\sqrt{2\kappa u_0}$. Note that in terms of the tilde 
variables the theory looks the same as the original theory except that
all links are scaled by $u_0$. Lepage and
Mackenzie~\cite{TI93lepagemackenzie}, and subsequent work, have shown
that mean-field improvement of 1-loop estimates gives, in many cases,
results that are very close to non-perturbative estimates, and the
expansion coefficients are much smaller. Tadpole improvement is,
therefore, a useful thing to do, especially since it requires
negligible extra effort once the 1-loop calculation has been done.
The only limitation is that it does not work equally well in all cases
and there is no real understanding of why it does not work well in
some cases. To illustration tadpole improvement I will apply it
to the L\"uscher-Weisz action.

There are two common choices of the tadpole factor $u_0$: (i) 
the fourth root of the plaquette, and (ii) the expectation value of the 
link in Landau gauge. So far most calculation have used the first choice, 
whereas Lepage and collaborators now advocate the link in Landau gauge as 
a better choice ~\cite{IDA98D234c}. To implement tadpole improvement requires 
both the 1-loop expressions for these variables and their 
non-perturbative values. For the Wilson action 
\bea
u_0 &=& \vev{\CW_{11}}^{1/4} = 1 - {1 \over 12} g^2 \nonumber \\
u_0 &=& \vev{U}_{LG}       = 1 - {0.0772} g^2 
\eea
and for the L\"uscher-Weisz action, Eq.~\ref{eq:gaugeclassicalimp}, 
it is \cite{IGA84PWRW}
\be
u_0 = \vev{\CW_{11}}^{1/4} = 1 - ({1 \over 2})({4 \over 3})( {0.366262 \over 4}) g^2 
			   = 1 - {0.061044} g^2 \,.
\ee

Denoting the 1-loop tadpole factor as $u_0 = 1 - X g^2$, the
tadpole-improved L\"uscher-Weisz (TILW) action is \cite{IGA93lepage}
\bea
\CS_{TILW} = \beta_{eff} \bigg( 
{\cal L}^{(4)} &-& {1 \over 20 u_0^2} \big(1 + ( 0.1602-2X)\ g^2 \big) \CL_1^{(6)} \nonumber \\
               &-& {1 \over    u_0^2}  0.00264\ g^2 \CL_2^{(6)} \bigg) 
\label{eq:LWTIaction}
\eea
The non-perturbative value of $u_0$ has to be determined
self-consistently as the relative co-efficients in the action
themselves depend on $u_0$.  Note that $\CL_2^{(6)}$, which comes in
only as a quantum effect, has, after TI, a coefficient is reduced to a
negligible value.

An important point to note about the TILW action is the strength of
$\CL_1^{(6)}$.  Since $u_0 = 0.85-0.90$ for lattice parameters
currently in use, the strength of $\CL_1^{(6)}$ is larger than the
classical value $1/20$, $i.e$ tadpole improvement suggests an
over-correction.  Later on we shall see that this over-correction is
also characteristic of improved actions based on renormalization group
methods.

A simplified description of TI applied to the above example is that
the coefficient of $\CL_1^{(6)}$ was multiplied by $u_0^2(pert) /
u_0^2(non-pert)$, the relative factor of two extra links between the
plaquette and $\CL_1^{(6)}$. The factor $u_0^2(pert)$, which differs
from $u_0^2(non-pert)$ at $O(g^4)$, is absorbed in the 1-loop
expansion. This, as is obvious, makes the coefficient of $g^2$ small.
The correction is, therefore, dominated by $u_0(non-pert)$ which 
can be measured with very high precision. There, however, is the residual
uncertainty due to the freedom in the choice of $u_0$.  To summarize,
the underlying philosophy of TI is that the large tadpole
contributions are present in the perturbative expansion of all
quantities, and can be absorbed by factoring $u_0(pert)$ raised to the
appropriate power depending on the number of links. This makes the
expansion coefficients small.  The modified perturbative expansion is
then in terms of an improved $\tilde g^2 = g^2/u_0^4$, and the
correction is dominated by overall factors of $u_0(non-pert)$, which
should be chosen to be the expectation value of the link in Landau gauge. 

One of the first tests to convincingly demonstrate improvement using the 
TILW action was restoration of rotational invariance in the $q \bar q$ 
potential. The results from \cite{POT95lepage} are reproduced 
in Fig.~\ref{f:POT95lepage}. Since then tests of improvement using 
the scaling of the finite temperature transition, $T_c$, and topological 
susceptibility have been carried out. These results, as will be discussed 
later, all make a strong case for improvement. 

\input{f-POT95lepage.tex}

\subsection{The Renormalization Group and the Renormalized Trajectory}

This is a brief summary of Wilson's formulation of the renormalization
group and the concept of the renormalized trajectory.  For details see
\cite{Wilsonkogut74,Wilson79,Fisher83,RG85Wuppertal,IGA87guptapatel,TASI89rajan,rrajancreutz,RG98hasenfratz}.
I will start with a discussion in momentum space. 
Consider the effective theory formulated at some scale $Q$ in terms of
the generalized action
\be
\CS(Q) = \sum_\alpha \ K_\alpha^{(0)} (Q) \ S_\alpha
\ee
where $S_\alpha$ are all possible operators (interactions) and
$K_\alpha^{(0)}$ are the corresponding coupling constants. Now
consider an exact renormalization group transformation (RGT) $\CT_b$ that
integrates all momenta between $Q$ and $Q/b \equiv Q/2$. Here $b=2$ is
the scale factor of the transformation. The renormalized theory at
scale $Q/2$ is
\be
\CS (Q/2) = \sum_\alpha \ K_\alpha^{(1)} (Q/2) \ S_\alpha
\ee
where $K_\alpha^{(1)}$ are analytical functions of $K_\alpha^{(0)}$, 
\be
K_\alpha^{(1)} = \CT_b \big[ K_\alpha^{(0)} \big] \ .
\ee
The sequence of theories defined by repeated use of $\CT_b$ defines a
flow trajectory in the space of couplings.  The physical theory is
preserved under each $\CT_b$, and thus along the whole
trajectory. Correlation lengths, measured in lattice units, change
trivially $\xi/a \to \xi/ab$.  Fixed points of this transformation are
defined by
\be
K_\alpha^{(*)} = \CT_b \big[ K_\alpha^{(*)} \big] \ , 
\ee
$i.e.$ points at which the theory reproduces itself under
$\CT_b$. Since all correlation lengths (momenta) scale by $b$ under
$\CT_b$, therefore at the fixed points
\be
\xi \big|_{\rm fixed\ point} \ = \ 0 \quad {\rm or} \quad \infty \,.
\ee
Fixed points with $\xi = 0$ are trivial fixed points. Examples of
these are the $T=0$ and $T=\infty$ limit of most statistical mechanics
models.  We are interested in non-trival fixed points with
$\xi=\infty$.  At these points the correlation length $\xi / a =
1/Ma$, measured in units of lattice spacing, is infinite. This
allows us to take the continuum limit of lattice field theories $--$
set $a=0$ holding $M $ fixed.

Now consider a point $\{K_\beta\}$ close to the fixed point
$\{K_\beta^*\}$ and a RGT with an infinitesimal $b$.  Under the
assumption that there are no singularities in the space of coupling
constants, the new couplings $K_\alpha^\prime (K_\beta)$ generated by
$\CT_b$ can be written as a Taylor expansion about the fixed point
\be
\label{ltma}
K_\alpha^\prime(K_\beta)\ =\ K_\alpha^*)\ +\ 
    {\partial K_\alpha^\prime \over \partial K_\beta} \Big|_{K^*}\ 
    \Delta K_\beta \ + \ \dots \ \ .
\ee
In that case one can define a ``linear region'' in the neighborhood 
of $\{K_\beta^*\}$ in which
\be
\label{ltmb}
    \Delta K_\alpha^\prime \ = \ 
    {\partial K_\alpha^\prime \over \partial K_\beta} \Big|_{K^*}\ 
    \Delta K_\beta 
\ee
where $ \Delta K_\beta  =  K_\beta - K^*_\beta $.  
The linearized transformation matrix (also called the stability matrix) 
\be
\label{ltmc}
\CT_{\alpha \beta}\ \equiv \ {\partial K_\alpha^\prime \over 
                            \partial K_\beta} \Big|_{K^*} 
\ee
controls the RG flows locally. In field theory discussions, the
stability matrix $\CT_{\alpha \beta}$ is called the $\beta$-function.
Note that, because of the truncation of Eq.~\ref{ltma} all statements
based on the analysis of $\CT_{\alpha \beta}$ have the implicit
assumption that they are approximately true in the linear region close
to the fixed point and for small $b$, and become exact only as
$\{K_\beta\} \to \{K_\beta^*\}$.

Wilson formulated the real space renormalization group as an alternate
to the momentum space renormalization group discussed above.  This
approach is characterized by a constrained integral over the position
dependent original degrees of freedom $s$ to produce an effective
theory,
\be
e^{K_\alpha^\prime S_\alpha(s^\prime)} = \int \CD s \CT(s^\prime,s) e^{K_\alpha S_\alpha(s)} \,.
\ee
Here $\CT(s^\prime,s)$ is the constraint that specifies how the $s$ in
some local neighborhood (called the block cell) are to be averaged and
replaced by an effective $s^\prime$. A necessary condition on $
\CT(s^\prime,s)$, that preserves the partition function, is
\be
\int  \CD {s^\prime}\ \CT(s^\prime,s) = 1 \,.
\label{eq:blockcondition}
\ee
One can see that if the block cell in $D-$dimensions is $b^D$, and the
constraint $\CT(s^\prime,s)$ is equivalent to integrating momenta in
the range $Q/2 - Q$, then the two approaches are equivalent. In
general, an exact integration of the upper half momenta is not
feasible and $\CT(s^\prime,s)$ is constructed as some sensible
``average'' of the original spins in the block cell.  Other than the
condition Eq.~\ref{eq:blockcondition}, the definition of a sensible
``average'' is that it preserves the physical properties of the
theory.  Because of this dual description, momentum versus real space
formulations, the RGT is also called a blocking transformation $--$ a
usage more common in statistical mechanics and numerical
implementations.

A brief summary of the properties of the matrix $\CT_{\alpha \beta}$, 
evaluated using either the momentum or real space formulations, 
and its associated eigenvalues $\Lambda_i$ and eigenvectors $\Phi_i$
obtained by solving $\CT \Phi_i = \Lambda_i \Phi_i$ is as follows.

\begin{description}
\item{1.}
$\CT_{\alpha \beta}$ is, in general, not a symmetric matrix. Its 
eigenvalues are, however, real. 
\item{2.}
The eigenvalues depend on the scale factor $b$.  Since the theory
obtained after two successive transformations, $b \otimes b$, should
be the same as the one obtained after a single scale change by $b^2$,
the general form of the eigenvalues is $\Lambda_i = b^{\lambda_i}$
where the $\lambda_i$ are independent of $\CT_b$ and are called the
critical exponents.
\item{3.}
The eigenvectors $\Phi_i$ are by definition the linear scaling fields.
Consider a small deviation of size $u$ from the fixed point along a
given eigenvector $\Phi$. Then under a RGT, the renormalized value of
the coupling, $u'$, is given by
\be
\label{ltmd}
u'\Phi = \CT_b u\Phi = b^{\lambda_\Phi} u \Phi \quad . 
\ee
Note the crucial use of the assumption of linearity in 
deducing this simple scaling of $u$. The eigenvectors 
depend on the precise form of $\CT_b$. 
\item{4.}
Eigenvalues $\Lambda_i > 1$ lead to flows away from the fixed point as
deviations along these $\Phi_i$ grow under $\CT_b$.  These eigenvalues
and the corresponding eigenvectors are called relevant.
\item{5.}
Operators corresponding to eigenvalues $\Lambda_i < 1$ die out
geometrically with the number of blocking steps.  These
operators do not contribute to the long distance properties of the
system and are consequently called irrelevant.  The associated
exponents $\lambda_i$ control corrections to scaling.
\item{6.}
Eigenvalues that are exactly unity are called marginal.  Deviations
along marginal directions do not change under RGT. To ascertain that an
eigenvalue is truly marginal, one has to go beyond the linear
approximation.  Typically marginal operators develop logarithmic
corrections at higher order.  A quick examination of the $\beta$-function of
asymptotically free field theories should convince you that $g$ is a marginal
coupling  at tree-level and becomes relevant due to quantum corrections.
\item{7.}
There is an additional class of eigenvectors, called redundant
operators, that are not physical.  Redundant operators are
manifestation of the invariance of the action under a redefinition of
the fields, $i.e.$, they define relations between the different
operators such as the equations of motion.  The corresponding
eigenvalues can be $<1,\ =1,$ or $ >1$.  These eigenvalues depend on
the choice of the $RGT$, $i.e.$ different RGT with the same or
different scale factor $b$ will give different $\lambda_i$.  Since
different non-linear RGT correspond to different rescaling of the
fields, redundant operators are specific to the transformation.  In
fact, one way to isolate them is to repeat the calculation of
$\CT_{\alpha \beta}$ with a different $RGT$; the exponents $\lambda_i$
that change with the RGT correspond to redundant operators.  Even
though they carry no physics information it is important to isolate
these operators, especially in numerical calculations. For example,
when improving the action one wants to use all possible redundant
operators to simplify the action and thus minimize the computational
cost.
\item{8.}
Under certain conditions $\CT_{\alpha \beta}$ is block diagonal,
$i.e.$ the couplings break up into sub-sets that are closed under the
$RGT$.  For example, the spin flip symmetry $s \to -s$ of the Ising
Hamiltonian at zero magnetic field causes the interactions to break up into
odd and even sectors. One can, therefore, calculate
$\CT_{\alpha\beta}$ for each sector separately as the eigenvalue
spectrum is independent.

\end{description}

\medskip

In this generalized space of couplings $\{ K_\alpha \}$, there exist
points, called critical points $\{ K_\alpha^c \}$, with $\xi =
\infty$. Starting from a critical point, a RGT produces another since
$\xi^\prime = \xi/b = \infty$.  Thus the set of critical points define
a hypersurface in the infinite dimensional space $\{ K_\alpha \}$.
The RG flows on this surface can (a) meander randomly, (b) go to some
limit cycle or a strange attractor, or (c) converge to a fixed point
$\{ K_\alpha^* \}$.  We shall only be concerned with possibility
(c). At the fixed points the renormalized couplings are exactly equal
to the original couplings, $K_\alpha' = K_\alpha $, and the theory
reproduces itself at all length scales. This is in distinction to
critical points where only the long distance behavior is reproduced.
The location of the fixed point depends on the RGT and in general a
given RGT will have more than one fixed point with $\xi =
\infty$.  Each such fixed point has a basin of attraction $i.e.$ the
set of $\{ K_\alpha^c \}$ that converge to it under the RGT.  This
hypersurface is thus orthogonal to all relevant eigenvectors of
$\CT_b$. The basin of attraction defines the universality class since
the long distance behavior of all theories is the same and governed by
the fixed point theory.

\begin{figure} 
\hbox{\epsfxsize=\hsize\epsfbox{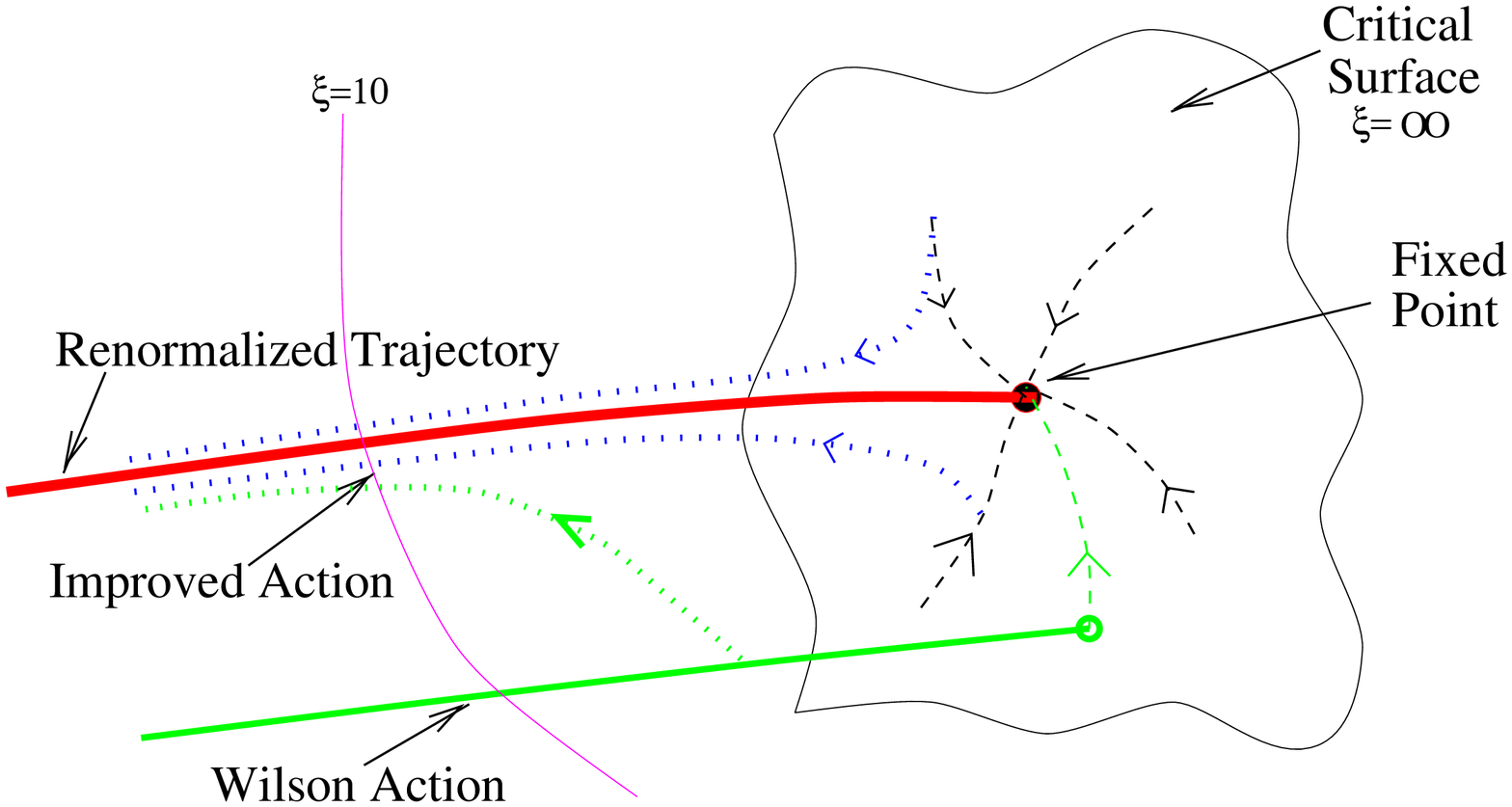}}
\caption{A schematic of the critical hypersurface ($\xi=\infty$), the fixed 
point, the renormalized trajectory (RT) flowing towards the infrared
under blocking transformations, and flows attracted by them. Also
shown is the simple Wilson action and flows starting from it.}
\label{f:RTflow}
\end{figure}

The flow along a relevant scaling field, whose end-point is the fixed
point is called the renormalized trajectory (RT). Just as the fixed
point is an attractor of all critical points in its universality
class, the RT is the attractor of all flows terminating in the
critical points that lie in the basin of attraction of the fixed
point. This is illustrated in Fig.~\ref{f:RTflow}.  For simplicity I
have assumed that there is only one relevant coupling, as is the case
for QCD. Along the RT there are no scaling violations as, by
construction, all the irrelevant scaling fields are zero.  Thus,
simulations done using an action along the exact RT, will reproduce
the continuum physics without discretization errors.  The RT is thus
the holy grail of attempts to improve the lattice formulation, $i.e.$
the perfect action.

Simulations on finite lattices are done in a region of coupling space
where all $\xi \ll L$, and not at the fixed point or at the critical
points. Current typical values lie in the range $\xi/a = 1/\sqrt{\sigma}a = 5-15$. 
Even for quenched simulations this range will not change in the near future. 
Points in this region lie on flows that terminate at critical
points that may lie quite far from the fixed point. Corrections to
scaling along these flows at $\xi \approx 10$ may, therefore, be
large. One can reduce these corrections by adjusting the action to lie
on another flow that starts closer to the fixed point and at $\xi
\approx 10$ is closer to the RT. Thus, to improve the action a 
non-perturbative approach is to estimate the RT.  Since flows are
attracted by the RT, its location can be estimated by starting with
even the simple plaquette action and following the flows generated
under some blocking transformation.  Evaluating the blocked action
along the flow is therefore the basis of all renormalization group
methods for improving the action.  Methods used for determining the
blocked action are reviewed in \cite{RG85Wuppertal}.

A standard assumption underlying the validity of the renormalization
group is that the fixed point action is local (the strength of the
couplings fall off exponentially with the size of the Wilson loops and
the separation between points that are coupled by the Dirac operator),
nevertheless, a local action still involves an ``infinite'' number of
couplings. Keeping all these when evaluating the action generated
under blocking, and in simulations, is impractical. So one is forced
to to truncate the action to a finite (small) number of
interactions. Also, the RT at $\xi \approx 10$ may be sufficiently
different in form from that near the fixed point.  Consequently, the
questions one has to face when designing improved actions are (i) what
is the minimal set of terms to keep, (ii) are all the redundant
interactions removed from this set, and (iii) how to compare such
improved actions evaluated using different RGT. Also note that the
projection of the RT on to a truncated subspace may not be the
action with the smallest scaling violations that one can design in that space. 
I shall explore attempts to answer these questions next.

I will summarize current activity in designing improved gauge actions
by reviewing those that are being used to calculate phenomenologically
relevant quantities.  Unfortunately, there is no standard notation
used by different collaborations. To facilitate comparison I will
therefore try to write them in a common form as well as maintain
contact with the original notation.  Wilson loops will be denoted by
$W^i_{\mu\nu}$ where the index $i$ will be used for both the type of
loop and its representation. Unless otherwise stated, it will be
implicit that the real part of the trace of Wilson loops is to be
taken.  The $W^i_{\mu\nu}$ are normalized such that the maximum value
of the trace is unity. As before $\beta $ is the overall coupling and the $c_i$
give the relative strengths of different loops.

\subsection{Iwasaki Action}

Iwasaki studied block transformations in perturbation theory. His
criterion was to obtain a gauge action that after a few blockings comes
close to the RT. His {\it Ans\"atz} is \cite{IGA85Iwasaki}
\be
\label{eq:iwasakiaction}
\CS_{Iwasaki} = \beta\ \sum_{\mu,\nu}\ \sum_x\ 
             \bigg(W^{1 \times 1}_{\mu\nu} - {0.0907} W^{1 \times 2}_{\mu\nu} \bigg)
\ee
where $W^{1 \times 2}$ is the rectangular loop $\CL_1^{(6)}$ shown in
Fig.~\ref{f:loop6ops}.  (Iwasaki used the notation $(1-8c_0)W^{1
\times 1}_{\mu\nu} + c_0 W^{1 \times 2}_{\mu\nu}$ with $c_0 = -0.331$
which is the same as Eq.~\ref{eq:iwasakiaction} with a rescaled
$\beta$.)  This action shows better rotational invariance and scaling
of $T_c/\sqrt{\sigma}$. It is being pursued by the JLQCD collaboration
\cite{FT98Kanaya}.

\subsection{$b=\sqrt{3}$ blocking transformation}

The $\sqrt{3}$ blocking transformation has certain very nice
properties for 4-dimensional gauge theories in addition to being a
smaller scaling factor than the conventional $b=2$
\cite{rrajancreutz}. The block cell consists of a site and its eight
neighbors. As a result the transformation has much higher rotational
symmetry for both fermions and gauge fields \cite{rrajancreutz}. Gupta
and Patel have studied the flow under blocking using Monte Carlo
Renormalization Group methods.  They investigated the effect of (i)
truncation and (ii) variation in the $c_i$ as a function of
$\beta$~\cite{IGA87guptapatel}.  For phenomenological studies they
approximate the RT by a linear trajectory in a four parameter space
\bea
\label{eq:rootthreedef}
\CS_{\sqrt{3}}\ = \ \beta\ \Re \lbrace\ 
		&{}&\ c_F^{1 \times 1} \sum \Tr W^{(1 \times 1)} \nonumber \\
              \ &+&\ c_F^{1 \times 2} \sum \Tr W^{(1 \times 2)} \nonumber \\
              \ &+&\ c_6^{1 \times 1} \sum [ {3 \over 2} (\Tr W^{(1 \times 1)} )^2
                                      - {1 \over 2} \Tr W^{(1 \times 1)} ] \ \nonumber \\
                &+&\ c_8^{1 \times 1} \sum [ {9 \over 8} |\Tr W^{(1 \times 1)} |^2
                         - {1 \over 8} ] \ \ \rbrace \ , 
\eea
with
\be
\label{eq:rootthreerat}
c_6^{1 \times 1}/c_F^{1 \times 1} = -0.12  ; \quad
c_8^{1 \times 1}/c_F^{1 \times 1} = -0.12  ; \quad
c_F^{1 \times 2}/c_F^{1 \times 1} = -0.04  . 
\ee
This particular choice of couplings (truncation of the action) was
made for the following reasons.  The coupling for the plaquette in the
adjoint representation, $c_8^{1 \times 1}$, was included to avoid the
phase structure in the fundamental-adjoint plane (see
Section~\ref{s:phasetransitions}); that in the 6-dimensional
representation, $c_6^{1 \times 1}$, since it differs from $c_8^{1
\times 1}$ only at $O(a^4)$ and higher, and the ${1 \times 2}$
rectangle in the fundamental representation, $c_F^{1 \times 2}$, as it
is the simplest loop that kills the $O(a^2)$ terms $a\ la$ Symanzik.

Classical expansion, see Eq.~\ref{eq:plaqexpC}, of the 
loops gives the normalization condition
\be
\bigg[ c_F^{1 \times 1} + {5 \over 2} c_6^{1 \times 1} + {9 \over 4} 
c_8^{1 \times 1} \bigg] + 8 c_F^{1 \times 2}  = 1 \,,
\ee
where I have lumped together the first three terms coming from $W^{(1
\times 1)}$ as the effective contribution of the plaquette. Using
the ratios given in Eq.~\ref{eq:rootthreerat}, one gets a very high weight 
$c^{(4)} = 1/0.11$, which as shown in \cite{rgrandygupta} is correlated with 
a suppression of dislocations versus physical instantons. 
Similarly, the coefficient of the leading discretization correction 
$\CO^{(6)}_1$, using Table~\ref{t:IGAclassical}, is 
\be
{1 \over 24} \bigg[ c_F^{1 \times 1} + {5 \over 2} c_6^{1 \times 1} + {9 \over 4} 
c_8^{1 \times 1} \bigg] + {5 \over 6} c_F^{1 \times 2} \,.
\ee
For the ratios given in Eq.~\ref{eq:rootthreerat} this is not zero,
but over-corrected by the proposed strength of $W^{(1 \times 2)}$. At
$1/a \approx 2$ GeV, this over-correction is larger than that for the
tadpole improved L\"uscher-Weisz {\it Ans\"atz}, similar to
$\CS_{Iwasaki}$, and smaller than that in $\CS_{QCDTARO}$.

Early spectrum studies and measurement of the glueball spectrum
\cite{rrajanglueball} to test improvement were inconclusive.  In
hindsight the failure to quantify improvement is due to poor
statistics and not looking at quantities like rotational invariance at
strong coupling, where the lattice artifacts show up clearly for the
simple Wilson action.  Recently this action has also been used to
study lattice topology and the results show that short distance
artifacts are suppressed \cite{rgrandygupta}.

\subsection{$b=2$ blocking transformation}

The QCDTARO collaboration has also investigated the flow of the action
under blocking~\cite{IGA96qcdtaro,IGA97qcdtaro}.  They use a $b=2$
blocking transformation and evaluate the couplings on blocked lattices
using Schwinger-Dyson equations. Their overall conclusion is that the
blocked action, to a very good approximation, can be represented by
the two couplings $W^{1 \times 1}$ and $W^{1 \times 2}$. Their
estimate of the RT cannot be approximated by a linear relation between
the coefficients. An approximate {\it Ans\"atz} in the range of
couplings simulated is
\be
\CS_{QCDTARO} = \beta\ \sum_{i}\ \sum_{\mu,\nu}\ \sum_x\ 
             \bigg(W^{1 \times 1}_{\mu\nu} - {0.11} W^{1 \times 2}_{\mu\nu} \bigg)
\ee
Note that the coefficient of $W^{1 \times 2}$ is larger than the
tadpole improved Luscher-Weisz, Iwasaki, or the $b=\sqrt{3}$ actions.

\subsection{Classically Perfect Action}

Hasenfratz and Niedermayer showed that one can derive the classically
perfect action for asymptotically free-field theories using a
saddle-point integration of the RGT about $g=0$
~\cite{IGA94perfectaction,IMP97hasenfratz}.  They use the $b=2$
blocking transformation proposed by Swendsen ~\cite{BT81swendsen}.
The practical issue here is whether the saddle-point integration gives
an action that can be approximated by a few local terms.  The
application of this method to SU(3) and the tuning of the truncated
version is still under investigation 
\cite{IGA96perfectaction}. The current version of the truncated
classical fixed-point action (CFP) is
\cite{IGA98perfectaction}
\be
\label{eq:Perfectaction}
\CS_{CFP}\ =\ \beta \sum_{i,x} \lbrace c_1^i (1- \Re  \Tr W^{i}) + \ldots + 
                               c_4^i (1- \Re  \Tr W^{i})^4  \rbrace.
\ee
where the sum over $i$ involves two kinds of loops, the plaquette and
the 6-link loop that twists around a 3-cube ($\CL_2^{(6)}$ in
Fig.~\ref{f:loop6ops}). Higher powers of the loops arise naturally in
the saddle-point integration. Since the strengths decrease rapidly,
the authors choose to truncate at the fourth power.  The choice of the
particular 6-link loop is motivated by the desire to keep the action
local in time and to help improve rotational invariance.  The values
of coefficients $c_1 - c_4$ for the two types of loops are given in
Table~\ref{t:perfectaction}.

At this point full control over stability with respect to truncation
has not been achieved. The second issue is whether this {\it Ans\"atz}, 
obtained at $a=0$, continues to be a useful parameterization at $1/a
\sim 2$ GeV where simulations are done, $i.e.$ is the renormalized trajectory
approximately linear.  To test this requires a non-perturbative tuning
about $1/a \sim 2$ GeV using the same $b=2$ blocking transformation
used in the saddle-point integration. The calculations necessary to do
this tuning have not yet been initiated. As a result this {\it
Ans\"atz} is considered as work in progress.  Nevertheless, the
tests for improvement in the scaling of $T_c$, string tension,
topological susceptibility, and in the restoration of rotational
symmetry are already very encouraging
\cite{IGA98perfectaction,TOP98degrand}.

\begin{table} %
\begin{center}
\setlength\tabcolsep{0.35cm}
\caption{The current coefficients of the truncated 
         ``perfect'' action with 4 and 6 link loops. }
\begin{tabular}{|l|c|c|c|c|}
\myhline
$Loop$        & $c_1$    & $c_2$   & $c_3$  & $c_4$     \\
\myhline
  &  &  & & \\[-7pt]
$1 \times 1$  &  3.248   & -1.580  & 0.1257 & 0.0576    \\[2pt]
Twisted 6 link&  -0.2810 &  0.0051 & 0.0049 & -0.0096   \\[2pt]
\myhline
\end{tabular}
\label{t:perfectaction}
\end{center}
\end{table}
\smallskip

\subsection{Summary}

The question -- which is the most improved gauge action -- is still
far from settled. As should be obvious, the drawback of the
renormalization group improved actions is that the expansion is in the
size of the loops or the dimensionality of the representation and not
in the dimensionality of the irrelevant operators. Since each loop
contributes to operators of all dimensions it is not clear how to
parameterize the residual error ($O(a^2) + O(a^4)$, or $O(a^4)$, or
higher). Also, implementing the equations of motions (removing the
freedom associated with redundant directions) has not been easy. So
there is not yet a consensus on the obvious minimal set of operators
that should be kept. Different operators have been added based on (i)
their importance in perturbative expansions, (ii) size of the
coefficients in blocked actions (keep in mind that the freedom of the
choice of the blocking transformation and consequently the associated
RT effects these coefficients), and (iii) specific incorporation of
non-perturbative effects; for example the inclusion of a negative
adjoint coupling helps avoid the phase structure in the
fundamental-adjoint plane (see Section~\ref{s:phasetransitions}), and
adding loops mimicking the $F^{\mu\nu} {\tilde F}_{\mu\nu}$ operator
help give scale invariant instantons
\cite{IGA98perfectaction,IGA98FPAtop}. One hopes that a more precise
method for tuning the gauge action will emerge in the
future. Fortunately, in spite of this ambiguity, the situation is very
encouraging. The improvement seen in the restoration of rotational
invariance (measured using $ q \bar q$ potential as shown in
Fig.~\ref{f:POT95lepage}), glueball spectrum, and finite temperature
studies is remarkable. To a good approximation, most of these actions
are very well represented by adding just the $W^{(1 \times 2)}$ term
with a coefficient larger than the value suggested by tadpole
improvement, and keeping a few higher representations of the
plaquette. Such a truncated space keeps the computational problem
tractable.  

The bottom line is that even though there is not a commonly
accepted improved gauge action, the improvements seen with any of the
above actions are significant. At present the more pressing challenge,
in terms of the size of the errors and the computational cost, is
improving the Dirac discretization.  This is discussed next.

%% file: f-POT95lepage.tex
\begin{figure}
\begin{center}
\setlength{\unitlength}{0.240900pt}
\ifx\plotpoint\undefined\newsavebox{\plotpoint}\fi
\sbox{\plotpoint}{\rule[-0.175pt]{0.350pt}{0.350pt}}%
\begin{picture}(1200,900)(0,0)
\sbox{\plotpoint}{\rule[-0.175pt]{0.350pt}{0.350pt}}%
\put(264,158){\rule[-0.175pt]{210.065pt}{0.350pt}}
\put(264,158){\rule[-0.175pt]{0.350pt}{151.526pt}}
\put(264,284){\rule[-0.175pt]{4.818pt}{0.350pt}}
\put(242,284){\makebox(0,0)[r]{$1$}}
\put(1116,284){\rule[-0.175pt]{4.818pt}{0.350pt}}
\put(264,410){\rule[-0.175pt]{4.818pt}{0.350pt}}
\put(242,410){\makebox(0,0)[r]{$2$}}
\put(1116,410){\rule[-0.175pt]{4.818pt}{0.350pt}}
\put(264,535){\rule[-0.175pt]{4.818pt}{0.350pt}}
\put(242,535){\makebox(0,0)[r]{$3$}}
\put(1116,535){\rule[-0.175pt]{4.818pt}{0.350pt}}
\put(264,661){\rule[-0.175pt]{4.818pt}{0.350pt}}
\put(242,661){\makebox(0,0)[r]{$4$}}
\put(1116,661){\rule[-0.175pt]{4.818pt}{0.350pt}}
\put(438,158){\rule[-0.175pt]{0.350pt}{4.818pt}}
\put(438,113){\makebox(0,0){$1$}}
\put(438,767){\rule[-0.175pt]{0.350pt}{4.818pt}}
\put(613,158){\rule[-0.175pt]{0.350pt}{4.818pt}}
\put(613,113){\makebox(0,0){$2$}}
\put(613,767){\rule[-0.175pt]{0.350pt}{4.818pt}}
\put(787,158){\rule[-0.175pt]{0.350pt}{4.818pt}}
\put(787,113){\makebox(0,0){$3$}}
\put(787,767){\rule[-0.175pt]{0.350pt}{4.818pt}}
\put(962,158){\rule[-0.175pt]{0.350pt}{4.818pt}}
\put(962,113){\makebox(0,0){$4$}}
\put(962,767){\rule[-0.175pt]{0.350pt}{4.818pt}}
\put(264,158){\rule[-0.175pt]{210.065pt}{0.350pt}}
\put(1136,158){\rule[-0.175pt]{0.350pt}{151.526pt}}
\put(264,787){\rule[-0.175pt]{210.065pt}{0.350pt}}
\put(45,472){\makebox(0,0)[l]{\shortstack{$a\,V(r)$}}}
\put(700,68){\makebox(0,0){$r/a$}}
\put(395,693){\makebox(0,0)[l]{a) Wilson Action}}
\put(264,158){\rule[-0.175pt]{0.350pt}{151.526pt}}
\put(438,293){\circle{12}}
\put(511,392){\circle{12}}
\put(566,478){\circle{12}}
\put(613,427){\circle{12}}
\put(654,510){\circle{12}}
\put(787,553){\circle{12}}
\put(962,636){\circle{12}}
\put(438,293){\usebox{\plotpoint}}
\put(428,293){\rule[-0.175pt]{4.818pt}{0.350pt}}
\put(428,293){\rule[-0.175pt]{4.818pt}{0.350pt}}
\put(511,388){\rule[-0.175pt]{0.350pt}{1.927pt}}
\put(501,388){\rule[-0.175pt]{4.818pt}{0.350pt}}
\put(501,396){\rule[-0.175pt]{4.818pt}{0.350pt}}
\put(566,473){\rule[-0.175pt]{0.350pt}{2.409pt}}
\put(556,473){\rule[-0.175pt]{4.818pt}{0.350pt}}
\put(556,483){\rule[-0.175pt]{4.818pt}{0.350pt}}
\put(613,426){\rule[-0.175pt]{0.350pt}{0.482pt}}
\put(603,426){\rule[-0.175pt]{4.818pt}{0.350pt}}
\put(603,428){\rule[-0.175pt]{4.818pt}{0.350pt}}
\put(654,508){\rule[-0.175pt]{0.350pt}{1.204pt}}
\put(644,508){\rule[-0.175pt]{4.818pt}{0.350pt}}
\put(644,513){\rule[-0.175pt]{4.818pt}{0.350pt}}
\put(787,545){\rule[-0.175pt]{0.350pt}{3.854pt}}
\put(777,545){\rule[-0.175pt]{4.818pt}{0.350pt}}
\put(777,561){\rule[-0.175pt]{4.818pt}{0.350pt}}
\put(962,586){\rule[-0.175pt]{0.350pt}{24.090pt}}
\put(952,586){\rule[-0.175pt]{4.818pt}{0.350pt}}
\put(952,686){\rule[-0.175pt]{4.818pt}{0.350pt}}
\sbox{\plotpoint}{\rule[-0.250pt]{0.500pt}{0.500pt}}%
\put(326,158){\usebox{\plotpoint}}
\put(335,176){\usebox{\plotpoint}}
\put(346,194){\usebox{\plotpoint}}
\put(358,210){\usebox{\plotpoint}}
\put(371,226){\usebox{\plotpoint}}
\put(385,242){\usebox{\plotpoint}}
\put(399,257){\usebox{\plotpoint}}
\put(414,271){\usebox{\plotpoint}}
\put(429,285){\usebox{\plotpoint}}
\put(445,299){\usebox{\plotpoint}}
\put(461,312){\usebox{\plotpoint}}
\put(477,325){\usebox{\plotpoint}}
\put(493,338){\usebox{\plotpoint}}
\put(510,351){\usebox{\plotpoint}}
\put(526,364){\usebox{\plotpoint}}
\put(543,376){\usebox{\plotpoint}}
\put(560,388){\usebox{\plotpoint}}
\put(576,401){\usebox{\plotpoint}}
\put(593,413){\usebox{\plotpoint}}
\put(610,425){\usebox{\plotpoint}}
\put(627,437){\usebox{\plotpoint}}
\put(644,449){\usebox{\plotpoint}}
\put(660,461){\usebox{\plotpoint}}
\put(677,473){\usebox{\plotpoint}}
\put(695,485){\usebox{\plotpoint}}
\put(711,497){\usebox{\plotpoint}}
\put(728,509){\usebox{\plotpoint}}
\put(745,521){\usebox{\plotpoint}}
\put(762,533){\usebox{\plotpoint}}
\put(779,545){\usebox{\plotpoint}}
\put(796,557){\usebox{\plotpoint}}
\put(813,568){\usebox{\plotpoint}}
\put(830,580){\usebox{\plotpoint}}
\put(847,592){\usebox{\plotpoint}}
\put(864,604){\usebox{\plotpoint}}
\put(882,615){\usebox{\plotpoint}}
\put(899,627){\usebox{\plotpoint}}
\put(916,639){\usebox{\plotpoint}}
\put(932,651){\usebox{\plotpoint}}
\put(950,663){\usebox{\plotpoint}}
\put(967,674){\usebox{\plotpoint}}
\put(984,686){\usebox{\plotpoint}}
\put(1001,698){\usebox{\plotpoint}}
\put(1018,710){\usebox{\plotpoint}}
\put(1035,721){\usebox{\plotpoint}}
\put(1053,733){\usebox{\plotpoint}}
\put(1070,745){\usebox{\plotpoint}}
\put(1087,756){\usebox{\plotpoint}}
\put(1104,768){\usebox{\plotpoint}}
\put(1121,780){\usebox{\plotpoint}}
\put(1131,787){\usebox{\plotpoint}}
\end{picture}
\setlength{\unitlength}{0.240900pt}
\ifx\plotpoint\undefined\newsavebox{\plotpoint}\fi
\sbox{\plotpoint}{\rule[-0.175pt]{0.350pt}{0.350pt}}%
\begin{picture}(1200,900)(0,0)
\sbox{\plotpoint}{\rule[-0.175pt]{0.350pt}{0.350pt}}%
\put(264,158){\rule[-0.175pt]{210.065pt}{0.350pt}}
\put(264,158){\rule[-0.175pt]{0.350pt}{151.526pt}}
\put(264,284){\rule[-0.175pt]{4.818pt}{0.350pt}}
\put(242,284){\makebox(0,0)[r]{$1$}}
\put(1116,284){\rule[-0.175pt]{4.818pt}{0.350pt}}
\put(264,410){\rule[-0.175pt]{4.818pt}{0.350pt}}
\put(242,410){\makebox(0,0)[r]{$2$}}
\put(1116,410){\rule[-0.175pt]{4.818pt}{0.350pt}}
\put(264,535){\rule[-0.175pt]{4.818pt}{0.350pt}}
\put(242,535){\makebox(0,0)[r]{$3$}}
\put(1116,535){\rule[-0.175pt]{4.818pt}{0.350pt}}
\put(264,661){\rule[-0.175pt]{4.818pt}{0.350pt}}
\put(242,661){\makebox(0,0)[r]{$4$}}
\put(1116,661){\rule[-0.175pt]{4.818pt}{0.350pt}}
\put(438,158){\rule[-0.175pt]{0.350pt}{4.818pt}}
\put(438,113){\makebox(0,0){$1$}}
\put(438,767){\rule[-0.175pt]{0.350pt}{4.818pt}}
\put(613,158){\rule[-0.175pt]{0.350pt}{4.818pt}}
\put(613,113){\makebox(0,0){$2$}}
\put(613,767){\rule[-0.175pt]{0.350pt}{4.818pt}}
\put(787,158){\rule[-0.175pt]{0.350pt}{4.818pt}}
\put(787,113){\makebox(0,0){$3$}}
\put(787,767){\rule[-0.175pt]{0.350pt}{4.818pt}}
\put(962,158){\rule[-0.175pt]{0.350pt}{4.818pt}}
\put(962,113){\makebox(0,0){$4$}}
\put(962,767){\rule[-0.175pt]{0.350pt}{4.818pt}}
\put(264,158){\rule[-0.175pt]{210.065pt}{0.350pt}}
\put(1136,158){\rule[-0.175pt]{0.350pt}{151.526pt}}
\put(264,787){\rule[-0.175pt]{210.065pt}{0.350pt}}
\put(45,472){\makebox(0,0)[l]{\shortstack{$a\,V(r)$}}}
\put(700,68){\makebox(0,0){$r/a$}}
\put(395,693){\makebox(0,0)[l]{b) Improved Action}}
\put(264,158){\rule[-0.175pt]{0.350pt}{151.526pt}}
\put(438,270){\circle{12}}
\put(511,332){\circle{12}}
\put(566,377){\circle{12}}
\put(613,402){\circle{12}}
\put(654,429){\circle{12}}
\put(757,496){\circle{12}}
\put(787,530){\circle{12}}
\put(962,669){\circle{12}}
\put(438,269){\usebox{\plotpoint}}
\put(428,269){\rule[-0.175pt]{4.818pt}{0.350pt}}
\put(428,270){\rule[-0.175pt]{4.818pt}{0.350pt}}
\put(511,332){\usebox{\plotpoint}}
\put(501,332){\rule[-0.175pt]{4.818pt}{0.350pt}}
\put(501,332){\rule[-0.175pt]{4.818pt}{0.350pt}}
\put(566,376){\rule[-0.175pt]{0.350pt}{0.482pt}}
\put(556,376){\rule[-0.175pt]{4.818pt}{0.350pt}}
\put(556,378){\rule[-0.175pt]{4.818pt}{0.350pt}}
\put(613,401){\rule[-0.175pt]{0.350pt}{0.482pt}}
\put(603,401){\rule[-0.175pt]{4.818pt}{0.350pt}}
\put(603,403){\rule[-0.175pt]{4.818pt}{0.350pt}}
\put(654,427){\rule[-0.175pt]{0.350pt}{0.964pt}}
\put(644,427){\rule[-0.175pt]{4.818pt}{0.350pt}}
\put(644,431){\rule[-0.175pt]{4.818pt}{0.350pt}}
\put(757,495){\rule[-0.175pt]{0.350pt}{0.482pt}}
\put(747,495){\rule[-0.175pt]{4.818pt}{0.350pt}}
\put(747,497){\rule[-0.175pt]{4.818pt}{0.350pt}}
\put(787,524){\rule[-0.175pt]{0.350pt}{3.132pt}}
\put(777,524){\rule[-0.175pt]{4.818pt}{0.350pt}}
\put(777,537){\rule[-0.175pt]{4.818pt}{0.350pt}}
\put(962,656){\rule[-0.175pt]{0.350pt}{6.022pt}}
\put(952,656){\rule[-0.175pt]{4.818pt}{0.350pt}}
\put(952,681){\rule[-0.175pt]{4.818pt}{0.350pt}}
\sbox{\plotpoint}{\rule[-0.250pt]{0.500pt}{0.500pt}}%
\put(339,158){\usebox{\plotpoint}}
\put(349,175){\usebox{\plotpoint}}
\put(362,192){\usebox{\plotpoint}}
\put(375,208){\usebox{\plotpoint}}
\put(388,223){\usebox{\plotpoint}}
\put(404,238){\usebox{\plotpoint}}
\put(419,252){\usebox{\plotpoint}}
\put(434,266){\usebox{\plotpoint}}
\put(450,279){\usebox{\plotpoint}}
\put(466,292){\usebox{\plotpoint}}
\put(482,305){\usebox{\plotpoint}}
\put(498,318){\usebox{\plotpoint}}
\put(515,331){\usebox{\plotpoint}}
\put(531,344){\usebox{\plotpoint}}
\put(548,356){\usebox{\plotpoint}}
\put(565,368){\usebox{\plotpoint}}
\put(582,380){\usebox{\plotpoint}}
\put(598,392){\usebox{\plotpoint}}
\put(615,405){\usebox{\plotpoint}}
\put(632,417){\usebox{\plotpoint}}
\put(649,429){\usebox{\plotpoint}}
\put(666,441){\usebox{\plotpoint}}
\put(683,453){\usebox{\plotpoint}}
\put(700,465){\usebox{\plotpoint}}
\put(717,477){\usebox{\plotpoint}}
\put(734,489){\usebox{\plotpoint}}
\put(751,501){\usebox{\plotpoint}}
\put(768,512){\usebox{\plotpoint}}
\put(785,524){\usebox{\plotpoint}}
\put(802,536){\usebox{\plotpoint}}
\put(819,548){\usebox{\plotpoint}}
\put(836,559){\usebox{\plotpoint}}
\put(853,571){\usebox{\plotpoint}}
\put(870,583){\usebox{\plotpoint}}
\put(887,595){\usebox{\plotpoint}}
\put(904,607){\usebox{\plotpoint}}
\put(921,618){\usebox{\plotpoint}}
\put(938,630){\usebox{\plotpoint}}
\put(956,642){\usebox{\plotpoint}}
\put(973,654){\usebox{\plotpoint}}
\put(990,665){\usebox{\plotpoint}}
\put(1007,677){\usebox{\plotpoint}}
\put(1024,689){\usebox{\plotpoint}}
\put(1041,700){\usebox{\plotpoint}}
\put(1059,712){\usebox{\plotpoint}}
\put(1075,724){\usebox{\plotpoint}}
\put(1093,735){\usebox{\plotpoint}}
\put(1110,747){\usebox{\plotpoint}}
\put(1127,759){\usebox{\plotpoint}}
\put(1136,765){\usebox{\plotpoint}}
\end{picture}
\end{center}
\caption{Static-quark potential computed on $6^4$ lattices with $a\approx
0.4$\,fm using the Wilson action and the
TILW action. The dotted line
is the standard infrared parameterization for the continuum potential,
$V(r)=Kr-\pi/12r + c$, adjusted to fit the on-axis values of the
potential. Figure reproduced from ~\cite{POT95lepage}.}
\label{f:POT95lepage}
\end{figure}

%% file: chap-impdirac.tex
\section{$O(a)$ improvement in the Dirac Action}
\label{s:improveddirac}

The first attempt to improve the Wilson fermion action by removing the
$O(a)$ corrections was by Hamber and Wu \cite{hamberwuaction}.  They
added a two link term to the Wilson action
\bea
\label{eq:HWaction}
\CS_{HW}\ &=& \ A_W  \ + \ {r \kappa \over 4a}\ \sum_{x,\mu}
     \bar\psi^L(x) \bigg[ U_\mu (x) U_\mu (x+\hat \mu) \psi^L
     (x+2\hat\mu) \ + \nonumber \\
     &{}& \hskip 1.4in U_\mu^\dagger
     (x-\hat\mu) U_\mu^\dagger (x-2\hat\mu) \psi^L (x-2\hat\mu) \bigg] \nonumber \\ 
       &\equiv& \sum_{x,y} \bar\psi_L (x) M_{xy}^{HW} \psi_L (y) .
\eea
It is easy to check by Taylor expansion that, at the classical level,
the $O(a)$ artifact in $\CS_W$ is cancelled by the HW term.  One can
include quantum corrections by calculating the relative coefficient in
perturbation theory. It turns out that this action has not been
pursued -- in the 1980's due to the lack of computer power, and in the
1990's due to the merits of the Sheikholeslami-Wohlert (called SW or 
the clover) action \cite{SWaction}.

\subsection{The Sheikholeslami-Wohlert (clover) Action}

SW proposed adding a second dimension five operator to $\CS_W$, the
magnetic moment term, to remove the $O(a)$ artifacts without
spoiling Wilson's fix for doublers. The resulting SW action is
\be
\label{eq:SWact}
\CS_{SW} \ = \ \CS_W -
    {ia C_{SW} \kappa r\over 4}\ \bar\psi(x) \sigma_{\mu\nu} F_{\mu\nu} \psi(x)  \ .
\ee
The fact that this action is $O(a)$ improved for on-shell quantities
cannot be deduced from a Taylor expansion. To show improvement
requires an enumeration of all dimension 5 operators, the use of
equations of motion to remove some of them, and incorporating the
remaining two ($m^2 \bar \psi \psi$ and $m F^{\mu \nu} F_{\mu \nu}$)
by renormalizing $g^2 \to g^2(1+b_g ma) $ and $m \to m(1+b_m ma)$. The
construction of the SW action has been explained in detail by Prof.
L\"uscher at this school, so instead of repeating it, I direct you to
his lectures. I will simply summarize results of tuning $C_{SW}$ using
both perturbative and non-perturbative methods. A list of
perturbatively improved values of $C_{SW}$ that have been used in
simulations is as follows.

\medskip
\begin{tabular}{|l|l|}
\myhline
$C_{SW} = 1                          $&  tree-level improvement \\
$C_{SW} = 1 + 0.2659(1) g_0^2        $&  1-loop improved \cite{SW87Wohlert}  \\
$C_{SW} = 1/u_0^3                    $&  tadpole improved \\
$C_{SW} = 1/u_0^3(1 + 0.0159 g_I^2 ) $&  1-loop tadpole improved \\
\myhline
\end{tabular}
\medskip

In the 1-loop TI result $u_0$ was chosen to be the fourth root of the
plaquette. As an exercise convince yourself that using the expectation
value of the link in Landau gauge gives a value closer ($1.6$)
to the non-perturbatively improved result ($1.77$ at $g=1$) given
below. For each of the above values of $C_{SW}$, the errors are reduced
from $O(a)$ (Wilson action) to at least $O(\alpha_s a)$.

The recent exciting development, also reviewed by L\"uscher, is that
the axial Ward identity $C_{SW}$ can be used to non-perturbatively
tune $C_{SW}$ and remove all $O(a)$ artifacts provided one
simultaneously improves the coupling $g^2$, the quark mass $m_q$, and
the currents ~\cite{IDA96ALPHA}. Estimates of $C_{SW}$ have been made
at a number of values of $g$ for the quenched theory with Wilson's
gauge action \cite{IDA96ALPHA,Csw97SCRI}. These data can be fit by \cite{Csw97SCRI} 
\be
C_{SW}= {{1 - 0.6084g^2 - 0.2015 g^4 + 0.03075 g^6} \over {1 - 0.8743 g^2}} . 
\label{eq:CSWnonpert}
\ee
for $6/g^2 > 5.7$. 

The advantages of the SW action is that it is local and leaves
perturbation theory tractable \cite{rSWactROME}. Adding the clover
term is only a $\sim 15\%$ overhead on Wilson fermion simulations.  An
example of improvement obtained using the SW versus Wilson action 
for fixed quark mass ($M_\pi/M_\rho = 0.7$) is that the deviations 
from the $a=0$ values  in $M_N /\sqrt{\sigma}$ and $M_{vector} /\sqrt{\sigma}$ at
$a=0.17$ fermi are reduced from $30-40\% \to \sim 3-5\%$
\cite{Csw97SCRI}. Further tests and evaluation of the improvement are 
being done by a number of collaborations. I expect a number of
quenched results to be reported at the LATTICE 98 conference, and
these should appear on the e-print archive by late 1998.

\subsection{D234c action} 
\label{ss:D234}

As mentioned before, there are two sources of discretization errors: 
(i) approximating derivatives by finite differences, and (ii) quantum 
corrections, for example those due to modes with momentum $ > \pi/a$ 
that are absent from the lattice theory. 
Alford, Klassen, and Lepage have proposed improved gauge 
and Dirac actions that reduce (i) by classical improvement and (ii) 
by tadpole improvement~\cite{IDA97D234c,IDA98D234c}. Their current 
proposal for isotropic lattices, called the {\bf D234c action} as 
it involves second, third, and fourth order derivatives, is 
\bea
M_{D234c} &=& m(1+0.5ram) + \sum_\mu \bigg\{ \gamma_\mu \Delta_\mu^{(1)} 
                                    - {C_3 \over 6} a^2 \gamma_\mu \Delta_\mu^{(3)} \bigg\} \nonumber \\
          &+& r \sum_\mu \bigg\{ - {1 \over 2} a \Delta_\mu^{(2)}
                                 - {C_F \over 4} a \sum_\nu \sigma_{\mu\nu} F_{\mu\nu}
				 + {C_4 \over 24} a^2 \Delta_\mu^{(4)} \bigg\} \,.
\label{eq:D234caction}
\eea
Here $\Delta_\mu^{(n)}$ is the $n^{th}$ order lattice covariant
derivative (symmetric and of shortest extent).  The term proportional
to $C_3$ is precisely what is needed to kill the leading
discretization error in $\Delta_\mu^{(1)}$ as shown in
Eq.~\ref{eq:naiveexpansion}. The terms proportional to $r$ are
generated by a field redefinition and thus represent a redundant
operator. They, therefore leave unalterated the $O(a)$ improvement of
the naive action for on-shell quantities. The three terms in it are
respectively the Wilson term, that is again used to remove the
doublers, clover term (for $F_{\mu\nu}$ they use an $O(a^2)$ improved
version) as in the SW action needed to remove the $O(a)$ error
introduced by the Wilson fix, and an $a^3$ correction.  To include
quantum corrections to the classical values $C_3 = C_F = C_4 = 1$ they
resort to tadpole improvement, $i.e.$, they divide all links appearing
in $M_{D234c}$ by $u_0$, which is taken to be the mean link in Landau
gauge. 

Quantum corrections generally also induce additional operators,
however, in perturbation theory these start at $O(\alpha_s)$. After
tadpole improvement the coefficients of $\alpha_s$ for such operators
are all expected to be small, so one hopes that their contributions
can be neglected. The philosophy/motivation is analogous to dropping the small
$\CL_2^{(6)}$ term in the tadpole improved Luscher-Weisz pure gauge
action given in Eq.~\ref{eq:LWTIaction}.

For two of the coefficients, $C_F $ and $C_3$, the authors have carried 
out non-perturbative tests of the accuracy of tadpole improvement
\cite{IDA98D234c}. Since only the operator $\Delta_\mu^{(3)}$ breaks
rotational symmetry at $O(a^2)$, therefore $C_3$ can be tuned
non-perturbatively by studying rotational invariance of dispersion
relations.  Similary, the clover term and the Wilson term break chiral
symmetry at $O(a)$, thus their relative coefficient $C_F$ can be tuned
by studying the axial Ward identity. Results of their tests show that
errors in the spectrum and dispersion relations for momenta as high as
$p = 1.5/a$ are $\approx 10\%$ even at $a = 0.4 $ fermi!  The authors
are extending their analyses of the $D234$ action to anisotropic
lattices which, as will be discussed in Section~\ref{ss:glueballanisotropic}, 
have already yielded rich dividends in the calculation of glueball
masses.

\subsection{NRQCD action} 

\newcommand{\del}{{\bf \Delta}}
\newcommand{\dl}{{\delta{\cal L}}}
\newcommand{\delv}{{\bf \del}}
\newcommand{\delvc}{{\bf D}}
\newcommand{\delfour}{{\Delta^{(4)}}}
\newcommand{\delsq}{\Delta^{(2)}}
\newcommand{\delsqc}{D^{(2)}}
\newcommand{\Mbz}{{M_0}}
\newcommand{\Dv}{{\bf D}}
\newcommand{\Ev}{{\bf E}}
\newcommand{\Bv}{{\bf B}}
\newcommand{\psid}{\bar{\psi}}
\newcommand{\sigmav}{\mbox{\boldmath$\sigma$}}

In the simulation of heavy quarks one has to worry about $O(m_H a)$
errors in addition to $O(pa)$ and $O(\Lambda_{QCD} a)$.  For charm and
bottom quarks these errors are $O(1)$ in the range of lattice spacings
one can reasonably simulate. Lepage and collaborators, recognizing
that the heavy quark in heavy-heavy or heavy-light systems is, to a
good approximation, non-relativistic, developed a non-relativistic
formulation (NRQCD) valid for $m_H a > 1$ \cite{NRQCD91LepageThacker}. 
In this approach, the action for the two-component Pauli spinors 
is obtained by a Foldy-Wouthuysen transformation, and relativistic and 
discretization corrections are added as a systematic expansion in $1/M$ 
and $p/M$. 
One version of the NRQCD action is \cite{fB98NRQCD}
\bea 
\label{eq:nrqcdact}
\CL &=&  \overline{\psi}_t \psi_t \\
    &-&  \overline{\psi}_t
 \left(1 \!-\!\frac{a \delta H}{2}\right)_t
 \left(1\!-\!\frac{a H_0}{2n}\right)^{n}_t
 U^\dagger_4
 \left(1\!-\!\frac{a H_0}{2n}\right)^{n}_{t-1}
 \left(1\!-\!\frac{a\delta H}{2}\right)_{t-1} \psi_{t-1}, \nonumber
\eea
where $\psi$ is the two-component Pauli spinor and 
$H_0$ is the nonrelativistic kinetic energy operator,
\be
H_0 = - {\delsq\over2\Mbz} \,.
\ee
$\delta H$ includes relativistic and finite-lattice-spacing
corrections,
\begin{eqnarray}
\delta H
&=& - \frac{g  c_1 }{2\Mbz}\,\sigmav\cdot\Bv \nonumber \\
& & + \frac{ig c_2 }{8(\Mbz)^2}\left(\delv\cdot\Ev - \Ev\cdot\delv\right)
    - \frac{g  c_3 }{8(\Mbz)^2} \sigmav\cdot(\delv\times\Ev - \Ev\times\delv) \nonumber \\
& & - \frac{   c_4 (\delsq)^2}{8(\Mbz)^3} 
    + \frac{   c_5 a^2\delfour}{24\Mbz}  
    - \frac{   c_6 a(\delsq)^2}{16n(\Mbz)^2} .
\label{eq:NRQCDdeltaH}
\end{eqnarray}
Here $\Ev$ and $\Bv$ are the electric and magnetic fields, $\delv$ and
$\delsq$ are the spatial gauge-covariant lattice derivative and
laplacian, and $\delfour$ is a lattice version of the continuum
operator $\sum D_i^4$. The parameter $n$ is introduced to remove
instabilities in the heavy quark propagator due to the highest
momentum modes \cite{NRQCD91LepageThacker}. Again, the tree-level
coefficients $c_i = 1$ can be tadpole-improved by rescaling all the 
link matrices by $u_0$.

This formulation has been used very successfully to calculate
$\alpha_s$ from the upsilon spectrum \cite{alpha97NRQCD} as discussed
in Section~\ref{s:alpha}, the heavy-heavy and heavy-light spectrum
\cite{HM98NRQCD}, and the heavy-light decay constants $f_B$ and $f_{B_s}$
\cite{fB98NRQCD}.  The drawback of NRQCD is that it is an effective theory 
and one cannot take $a \to 0$. In fact, to simulate charm and bottom
quarks, one has to work in a narrow window $ 1 \lsim 1/a \lsim 2.5$
GeV to keep $m_H a > 1$. The good news is that current data 
\cite{alpha97NRQCD,HM98NRQCD,fB98NRQCD} show that
the discretization and relativistic corrections are small and under
control with the terms included in Eq.~\ref{eq:NRQCDdeltaH}.

\subsection{Perfect action}

The development of a perfect Dirac action is not complete.  The method 
is same as used for the gauge action and discussed in ~\cite{RG98hasenfratz}. 
The bottleneck is that the saddle point integration gives an action that even
in the free field limit involves a number of sites, and the gauge
connections between any pair of sites is the average of a large number
of paths \cite{PA97weise}. First tests of such an action truncated to
a $3^4$ hypercube have been carried out by Degrand and collaborators
\cite{IGA98perfectaction}.  From these it is not clear whether the
improvement, compared to other discretization schemes, justifies the
extra cost of simulating such an action.  Hopefully better ways to tune 
and test such actions will be formulated soon. 

\subsection{Fermilab action}

The Fermilab group has proposed a ``Heavy Wilson'' fermion action
\cite{IDA97Fermilab}. This action is meant to interpolate between the 
light $O(a)$ improved SW-clover action and the NRQCD action for heavy
quarks.  The goal is to remove not only $O(a)$ discretization
corrections but also take into account all powers of $ma$ errors so
that charmonium and bottomonium systems can be simulated with the same
reliability as light-light physics. This improvement is especially
targeted towards improving heavy-light matrix elements.  So far no
tests have been reported with the proposed action, therefore I will
not reproduce the action here but refer interested readers to
\cite{IDA97Fermilab} for details.

\subsection{ Summary}

In analogy to discussions of improved gauge actions, having seen
tangible evidence for improvement in quenched lattice data with each of the
above formulations, the question is, which action is the most
efficient for a comprehensive study?  The deciding factor, in my
opinion, will be dynamical simulations since the prefactor and the
scaling exponent ($L^{\sim 10}$) makes the cost of update of
background configurations prohibitive without improvement on today's
computers.  Dynamical simulations using improved actions have just
begun, so it will be a few years before enough data can be generated
to determine the advantages and disadvantages of different approaches.

Finally, I should mention a recent novel approach $--$ domain wall
fermions $--$ for improving the chiral properties of Wilson fermions
\cite{DW92kaplanshamir}.  In this approach the system (domain) is 
5-dimensional and the fermions live on 4-dimensional surfaces that
make up the opposite boundaries (walls).  The 4-dimensional gauge
fields on the two walls are identical, whereas the fermion zero-modes
have opposite chirality.  The overlap between these modes is
exponentially suppressed by the separation in the fifth
dimension. Consequently, the chiral symmetry breaking contributions are
also supressed, and results with good chiral properties for vector
like theories like QCD can be obtained by adding the contribution from
the two domain walls.  First simulations of weak matrix elements show
results with good chiral properties and in general look extremely
promising \cite{DW97blumsoni}. Since I do not do justice to this
exciting approach in these lectures, interested readers should consult
\cite{DW97blumsoni} and references therein.

%% file: chap-confine.tex
\section{Confinement and Asymptotic Freedom}
\label{s:confinement}

For any theory to provide a successful description of strong
interactions it should simultaneously exhibit the phenomena of
confinement at large distances and asymptotic freedom at short
distances.  Lattice calculations support the hypothesis that for
non-abelian gauge theories the two domains are analytically connected,
and confinement and asymptotic freedom co-exist.  Similary, one way to
show that QCD is the correct theory of strong interactions is that the
coupling extracted at various scales (using experimental data or
lattice simulations) is unique in the sense that its variation with
scale is given by the renormalization group. The data for $\alpha_s$
is reviewed in Section~\ref{s:alpha}.  In this section I will discuss 
what these statements mean and imply.

\subsection{Asymptotic Freedom}

The term asymptotic freedom implies that the running coupling $g \to 0$ 
as the momentum scale of the probe $\mu \to \infty$. This is characterized 
by the $\beta$-function 
\be
\mu {\partial g \over \partial \mu} = - \beta(g) = 
          - (\beta_0 g^3  + \beta_1 g^5 + \ldots )
\label{eq:betafndef}
\ee
where $\beta(g)$ is positive for non-abelian gauge groups. In this
case $g = 0$ is a UV stable fixed point of the theory, and about this
point the theory can be analyzed using standard perturbative
expansions. This novel behavior was confirmed by 't Hooft, Politzer,
and by Gross and Wilczek who showed that the perturbative
$\beta$-function satisfies Eq.~\ref{eq:betafndef}
\cite{AF73pgw}, $i.e.$, for $N$ colors and $n_f$ active flavors
\bea
\label{eq:betafncoeffs}
\beta_0 &=& ({{11N - 2n_f} \over 3})/16\pi^2 \,, \nonumber \\
\beta_1 &=& ({34N^2\over 3} - {10Nn_f \over 3} - {n_f(N^2-1) \over N} )
/(16\pi^2)^2 \,.
\eea
These two leading terms in the expansion of $\beta(g)$ are gauge and
regularization scheme invariant. From these it is easy to see that
$\beta(g)$ is positive for $n_f < 8$.  This result was essential in
establishing QCD as the theory of strong interactions for it explained
existing experimental data which showed that the strength of strong
interactions decreases as the momentum exchanged in a process
increases.

Asymptotic freedom, Eq.~\ref{eq:betafndef}, implies that QCD
dynamically generates a mass scale.  Integrating
Eq.~\ref{eq:betafndef} from momentum scale $\mu_1$ to $\mu_2$ with
$\mu_2 > \mu_1$ and keeping only the $\beta_0$ term to simplify the
standard calculation, gives
\be
{ 1 \over { 2 \beta_0 g^2(\mu_2)}}  \ - \ { 1 \over { 2 \beta_0 g^2(\mu_1)} }
\ = \ \log {\mu_2 \over \mu_1} \ , 
\label{eq:betafnint} 
\ee
$i.e.$ the coupling constant of non-abelian 
gauge theories depends logarithmically on the momentum scale of the 
process. Rewriting Eq.~\ref{eq:betafnint} 
\bea
\label{eq:asfrunning}
{ 1 \over { 2 \beta_0 g^2(\mu)}}  \ - \  \log {\mu} \ &=& \ constant \nonumber \\
\Longrightarrow \quad {\exp \big\{ {1 \over 2 \beta_0 g^2(\mu)} \big\} } \ &=& \ 
           {\mu \over \Lambda_{QCD}}  \nonumber \\
\Longrightarrow \quad \alpha_s(\mu) \ = \ {g^2(\mu) \over 4\pi }  \ &=& \ 
           {1 \over {8 \pi \beta_0 \log {\mu \over \Lambda_{QCD}}}}  
\eea
introduces $\Lambda_{QCD}$, the invariant scale of the theory with
dimensions of mass.  Thus QCD, a theory with a dimensionless coupling
constant and no intrinsic mass scale in the absence of quark masses,
dynamically generates a mass scale.  This happens because in order to
specify $g$ one has to also specify a momentum scale at which it is
defined.  Extending the above analysis to include $\beta_1$ in
Eq.~\ref{eq:betafndef} gives
\be
\label{eq:asftwoloop}
\Lambda_{QCD} \ = \ \lim_{\mu \to \infty} \ \mu \  
        { \big( {1\over \beta_0 g^2(\mu)} \big) }^{{\beta_1\over 2\beta_0^2}} 
        \ \exp \lbrack -\ {1\over 2\beta_0 g^2(\mu)} \rbrack \ \equiv \ 
           \mu\ f_p \big( g(\mu) \big) \ \ .
\ee
This 2-loop definition of $\Lambda_{QCD}$ is not unique; the value of
$\Lambda_{QCD}$ depends on the the precise relation between $g$ and
$\mu$.  However, once the value of $\Lambda$ is determined in one
scheme it can be related to that in any other
perturbative scheme.  For example, in the lattice
regularized theory $\Lambda_{lattice}$ is also defined by
Eq.~\ref{eq:asftwoloop} but with $\mu$ replaced by $1/a$.  Then to
1-loop
\be
\label{eq:asflama}
{\Lambda_{QCD} \over \Lambda_{lattice} }\ = \ \mu a\  
         \exp \bigg\{ -\ {1\over 2\beta_0 } 
             \bigg[ {1 \over g^2(\mu)}  - {1 \over g^2(a)} \bigg] \bigg\} \ .
\ee
In perturbation theory the two coupling constants are related as 
\be
\label{eq:asflamb}
g^2(\mu) \ = \ g^2(a) \bigg\{1 - \beta_0 g^2(a) 
                       \bigg( \log(\mu a)^2 - \log C^2 \bigg) + O(g^4) \bigg\}
\ee
By substituting Eq.~\ref{eq:asflamb} into Eq.~\ref{eq:asflama} one finds 
\be
\label{eq:asflamc}
\Lambda_{QCD} \ = \ C\ \Lambda_{lattice} 
\ee
$i.e.$ the two constants, $\Lambda_{QCD}$ and $\Lambda_{lattice}$, are related 
by a multiplicative constant.  To calculate $C$ requires knowing the 
finite part of the coupling constant renormalization to 1-loop
in both the lattice and continuum  regularization schemes~\cite{celmaster79,plqcd81kawai}. 
The results are listed in Table~\ref{t:Lambdaconn} for 
$\Lambda_{MOM}$ and $\Lambda_{\MSbar}$.  It is important
to note that even though the above definition and estimate of
$\Lambda_{QCD}$ is made using perturbation theory, it is intrinsically
a non-perturbative quantity.

\begin{table} 
\begin{center}
\setlength\tabcolsep{0.32cm}
\caption{The relation between $\Lambda_{MOM}$ and $\Lambda_{\MSbar}$ and 
$\Lambda_{latt}$ as a function of the number of active flavors. The
results for $\Lambda_{MOM} / \Lambda_{latt}$ with $g$ defined by the
triple gluon vertex are in Feynman gauge. }
\begin{tabular}{|l|c|c|c|c|c|}
\myhline
                                  &          &          &          &          &           \\[-7pt]
$n_f$                             &    0     &    1     &    2     &    3     &    4      \\[2pt]
\myhline
                                  &          &          &          &          &           \\[-7pt]
$\Lambda_{\MSbar}/\Lambda_{latt}$ &  28.8    &  34.0    &  41.1    &   51.0   &   65.5    \\[2pt]
\myhline
                                  &          &          &          &          &           \\[-7pt]
$\Lambda_{MOM}/\Lambda_{latt}$    &  83.4    &  89.4    &  96.7    &  105.8   &  117.4    \\[2pt]
\myhline
\end{tabular}
\label{t:Lambdaconn}
\end{center}
\end{table}

All dimensionful quantities in lattice simulations are measured in 
units of the lattice spacing, for example one measures $Ma$  and not $M$. 
Such dimensionless quantities vary with $g$ as 
\be
\label{eq:asfmscaling}
M_i a\ = \ c_i f_i(g) \ \equiv \ c_i \Lambda^i_{non-pert} a \ .  
\ee
where the $c_i$ are constants representing the continuum value and
$f_i$ are scaling functions. Lattice correlation lengths $\xi_i \equiv
1/M_i a $ diverge as the fixed point at $g = 0$ is approached. This
divergence is an artifact -- it is the unit of measurement $a$ that is
going to zero and not $M_i$. This is precisely what one wants to have
happen to get rid of the scaffold (lattice) erected to regularize the
theory. Non-perturbative renormalization consists of taking the
continuum limit holding some physical quantity $M_i$ fixed while
allowing $a \to 0$ according to $f_i(g)$. The particular quantity
$M_i$ (string tension, or nucleon mass, $etc.$) held fixed defines the
renormalization condition, and the constraint between $a$ and $g$ is
the scaling relation.

\begin{itemize}
\item
{\bf Scaling}: The hypothesis of scaling is that all $f_i(g)$ converge
to a universal scaling function $f(g) = \Lambda_{non-pert}a$.  The
renormalization condition is then independent of the state used to
define it.  For $ g < g_{scaling}$ each lattice quantity is
characterized by the non-perturbative number $c_i$ and all
dimensionless ratios of physical quantities are independent of $g$.

\item
{\bf Asymptotic Scaling}: Close to the fixed point at $g=0$, 
perturbation theory is apposite and all $f_i(g) \to f_{pert}(g) \to
\Lambda_{QCD} a$.
\end{itemize}

Lattice simulations, therefore, provide continuum results to within
some error tolerance from simulations done in a neighborhood of
$g=0$. In this region scaling holds. For a fixed error criteria, the
extent of this scaling window can be enlarged by improving the action
and operators as discussed in Section~\ref{ss:symanzikImp}. For
simulations done outside the scaling window it is imperative that a
reliable extrapolation to $a=0$ be done. Obviously, one way to get an
upper bound on the errors is to compare the measured scaling function
with $f_{pert}(g)$.  The good news is that simulations in the past few
years show that corrections to scaling can be reduced to a few percent
level for lattice spacings with $5 \lsim \xi/a \lsim 10$ ( where
$\xi/a = 1/\sqrt{\sigma}a$). This corresponds to $2 \lsim 1/a \lsim 4$
GeV. Consequently, realistic results with physical light quarks can
be obtained on lattices of side $\sim 75-150$, and for many
quantities, like those involving the heavy quarks, much smaller
lattices are sufficient.  The goal of the lattice approach is
twofold. First, improve the discretization of the action and operators
to extend the scaling region to stronger coupling, and second do
simulations at a number of points within this region and remove the
remaining errors by extrapolation to $a=0$.

An implication of the above discussion is that if all states in QCD
have zero mass in the continuum limit then $g=0$ would be a trivial
fixed point as the renormalized coupling $g_R$ would also be zero.  On
the other hand one can ask whether some of the states of QCD can be
massless while the others are massive?  It is interesting to note that
we can actually say something about this aspect of the spectrum of QCD
once the scaling behavior is fixed by Eq.~\ref{eq:asfmscaling}.  In
fact, under the following two assumptions, the pure gauge sector of QCD will 
have a mass gap:
\begin{itemize}
\item
There are no zero mass states at any finite non-zero value of $g$,
$i.e.$ the dimensionless quantity $Ma$ does not vanish at any non-zero
value of the lattice spacing $a$ in a region that lies in the same
thermodynamic phase as the continuum limit at $a=0$.
\item
There exists only one relevant coupling $g$ in QCD, and corresponding
to it a single universal scaling function defined by say
Eq.~\ref{eq:asftwoloop}. In this case the scaling behavior of all
observables is fixed and all mass-ratios have to stay finite as the
continuum limit is taken.
\end{itemize}
\medskip\noindent
These two assumptions are borne out by present numerical data, $i.e.$
there is no evidence for a massless state in the pure gauge sector.
Also there is no indication of a phase boundary separating the region
where simulations have been done and $g=0$.  Therefore, combining
lattice data with renormalization group scaling, QCD predicts that the
lightest glueball state is massive.

This prediction does not change with the introduction of $n_f$ flavors
of massless quarks into the theory.  Spontaneously broken chiral
symmetry will give rise to $n_f^2 -1$ massless pions in the spectrum.
To take the chiral limit one has to tune the quark masses to zero to
define the physical world. This tuning has to be done at each value of
$a$. Thus one cannot form mass-ratios with respect to these Goldstone
states as the corresponding $c_i(g)$ are tuned to zero.

To conclude, once we have verified that scaling (perturbative or
non-perturbative) exists then the only unknowns needed to predict the
spectrum of QCD are the constants $c_i$.  Asymptotic freedom
simplifies QCD by providing an analytical prediction of the universal
scaling behavior as $g \to 0$.  Current data show that the
coefficients $c_i$, which are intrinsically non-perturbative, can be
extracted using LQCD from simulations done at $a \sim 0.1$ fermi.

\subsection{Confinement}

There are two ways one can test whether a theory confines. One,
demonstrate that the free energy of an isolated charge is infinite;
two, show that the potential energy between two charges grows
with distance. The first test shows that it takes infinite energy to
add an isolated test charge to the system, while the second shows that
it requires infinite energy to separate two charges by infinite
distance. I shall show that the first is probed by the expectation
value of the Wilson line $\vev{\CL}$ in the pure gauge theory, while
the second by the expectation value of Wilson loops $\vev{\CW}$ or by
the correlation function $\vev{\CL{(\tau)}\CL^\dagger{(0)}}$.

\subsubsection{Wilson Loops}
\label{ss:Wloops}

\begin{figure} 
\hbox{\epsfxsize=\hsize\epsfbox{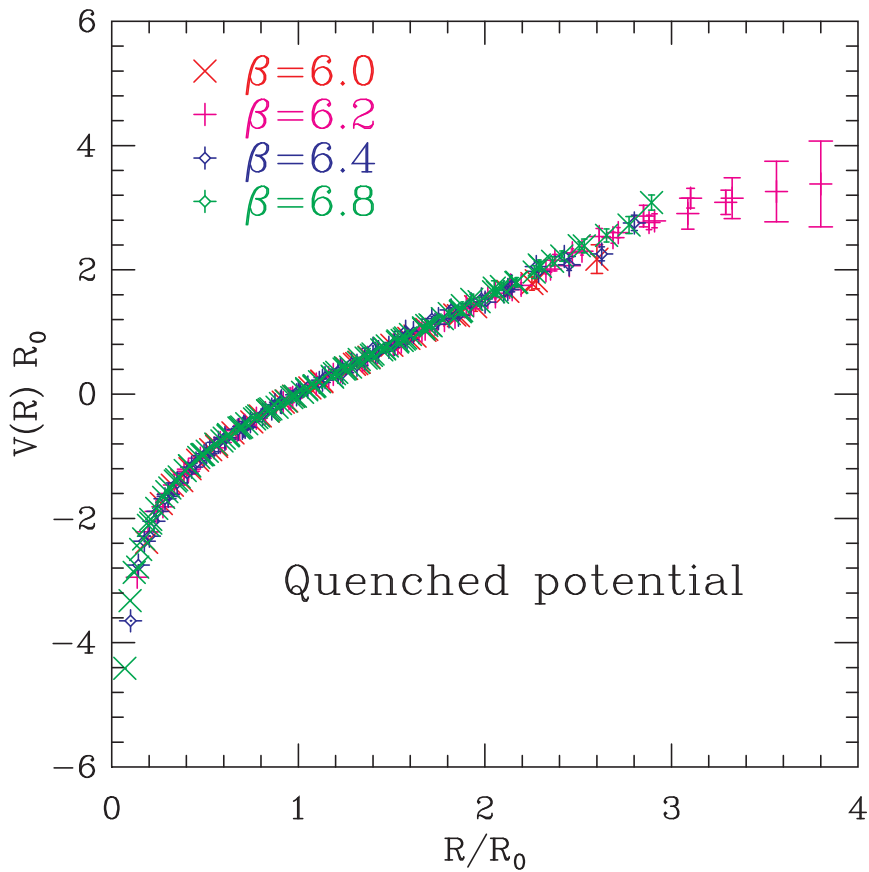}}
\caption{ The static $q \bar q$ potential in the quenched approximation 
obtained by the Wuppertal collaboration
{\protect{\cite{POT93Wuppertal}}}. The data at $\beta=6.0,\ 6.2,\ 6.4$
and $6.8$ has been scaled by $R_0$ defined in Eq.~\ref{eq:sommerR0},
and normalized such that $V(R_0) = 0$.  The collapse of the different
sets of data on to a single curve after the rescaling by $R_0$ is
evidence for scaling. The linear rise at large $R$ implies
confinement. }
\label{f:qqpotentialSU3}
\end{figure}

The extra action associated with an external charge placed in a gauge
field is $\CS_J = \int d^4 x J_\mu^a A_\mu^ a$. For a point charge $J(x) =
\delta^4(x)$, $S_J$ is given by the path ordered integral of the gauge
field along the world line of the charge. Thus
\be
\VEV{\CW} = e^{-\CS_J} \sim e^{-V(R) T} 
\label{eq:wloops1}
\ee
for a $R \times T$ rectangular Wilson loop. One can regard the
expectation value of a Wilson loop as the creation of a $q \bar q$
pair at time $T=0$ at point $R/2$, separated instantaneously to $R$
and $0$, allowed to evolve for time $T$, and then allowed to
annihilate. Amongst the many terms (instanteous creation,
annihilation, $\ldots$) that contribute to the action for this
procedure is that due to the potential between two static charges
separated by distance $R$ and integrated for time $T$, $i.e.$ $V(R)
T$.  Thus
\be 
V(R) = \lim_{T\rightarrow\infty} - {1 \over T} \log \vev{\CW(R,T)} \,.
\ee
The simplest {\it Ans\"atz} for a confining potential is the Cornell potential
\be 
V(R) = V_0 + \sigma R - {\alpha \over R} \,,
\label{eq:wloops4}
\ee
where $\sigma$ is the string tension. At large distances $\sigma R$
dominates, while at short distances it is the Coulomb term $\alpha /
R$ that is relevant. Such a potential, therefore, simultaneously
exhibits confinement and asymptotic freedom. For such linear
potentials large Wilson loops show an area law, $\log \vev{\CW(R,T)}
\propto RT$, and thus probe confinement. A state-of-the-art calculation
by the Wuppertal collaboration \cite{POT93Wuppertal} of $V(R)$ for
SU(3) gauge theory is shown in Fig.~\ref{f:qqpotentialSU3}. It shows
the expected behavior -- a linear rise at large $R$ and a Coulomb
behavior at small $R$.

One can also extract the potential from the correlation function
$\vev{\CL{(R)}\CL^\dagger{(0)}}$ of Wilson lines, defined in
Section~\ref{ss:Wilsonline}, pointing in the $T$ direction and
separated by distance $R$. It is easy to see that this correlation
function is equivalent to a Wilson loop with the sides at $T=0,N_T$
removed.  The Wilson lines are closed due to the periodic boundary
conditions on the gauge links and thus individually gauge
invariant. This correlation function measures the action due to two
static test charges separated by distance $R$.

\begin{figure} 
\hbox{\epsfxsize=\hsize\epsfbox{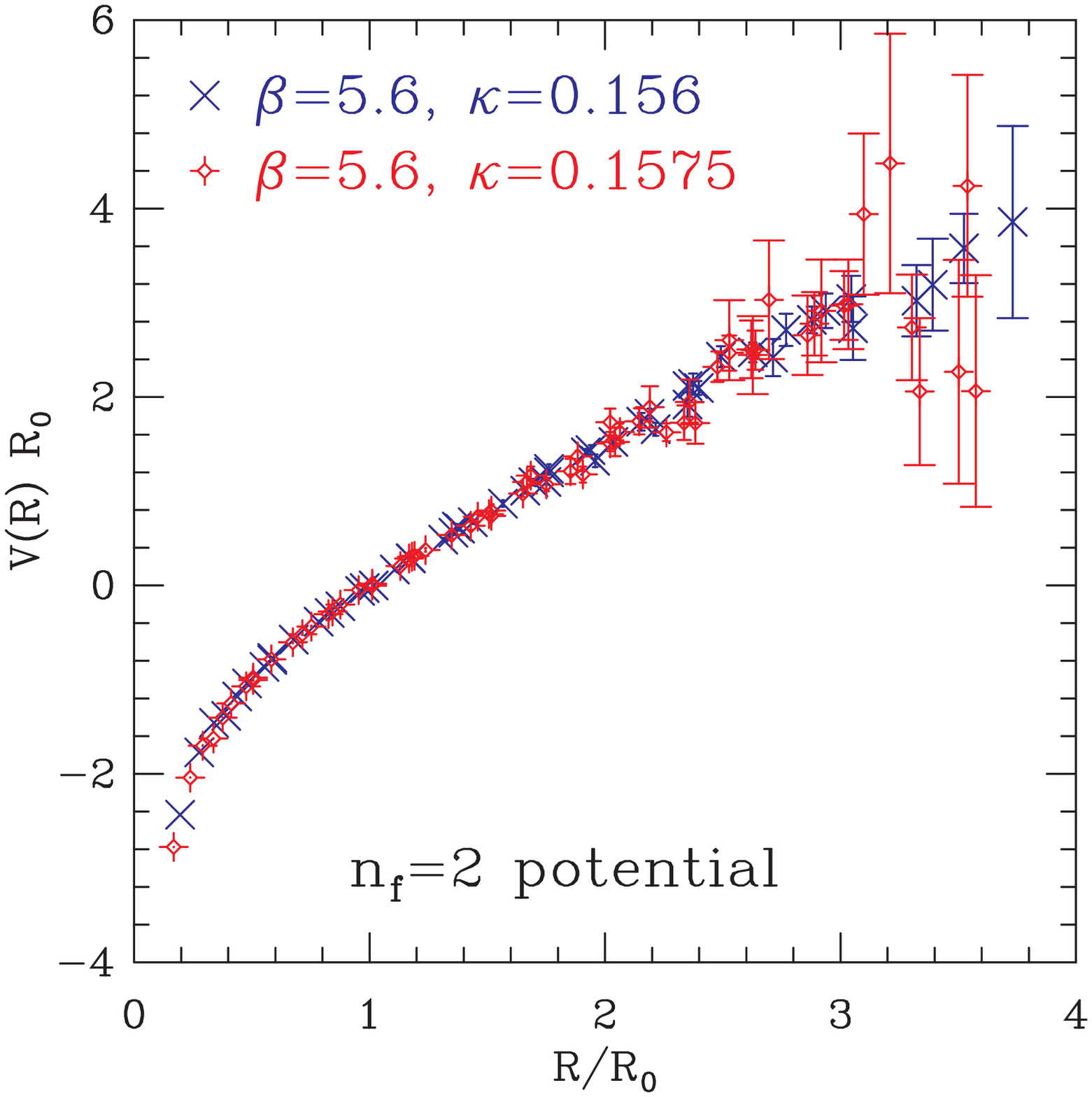}}
\caption{ The static $q \bar q$ potential with two flavors of dynamical 
fermions obtained by the Wuppertal collaboration
{\protect{\cite{SP97Wuppertal}}}. The dynamical quark mass in the
$\beta=6.0,\ \kappa=0.156$ simulation is $\sim 2 m_s$ while that at
$\kappa=0.1575$ is $\sim m_s$. The rest is same as in
Fig.~\ref{f:qqpotentialSU3}. These data lie lower than the quenched
points, however, for these dynamical quark masses there is no evidence
yet of string breaking at large $R$.}
\label{f:qqpotential}
\end{figure}

Measurements of the potential provide a way to set the scale of
lattice calculations. One approach is to use $\sigma$. 
The string tension can be extracted from the slope of $V(R)$ at large $R$ 
or from the asymptotic behavior of Creutz ratios \cite{ST80Creutz}
\be
\sigma = \lim_{{T\rightarrow\infty \atop R\rightarrow\infty}}
        - \log \bigg( {\vev{\CW(R,T)}  \vev{\CW(R+1,T+1)} \over 
                       \vev{\CW(R+1,T)}\vev{\CW(R,T+1)}}  \bigg) \,.
\ee
It can be measured very accurately in lattice simulations, however,
its physical value is either inferred from the Regge slope or from
potential models. Regge phenomenology gives the relation $M_l^2(l) =
l/\alpha' + c$ where $M_l$ is the mass of a state on the Regge
trajectory with orbital angular momentum $l$. To get a relation
between $\sigma$ and $\alpha'$ two approaches are used: (1) string
models which give $\sigma = 1/2\pi\alpha'$, and (ii) potential models
which give $\sigma = 1/8\alpha'$ ~\cite{regge94Dubin}. The value of
$\alpha'$ depends on the quark mass. For example $\alpha'$ from the
light hadrons, $c \bar c$, and $b \bar b$ is $0.85 \to 0.85/2 \to
\approx 0.85/4.5$ GeV${}^{-2}$.  It is customary to choose the value
$\alpha'=0.85$ GeV${}^{-2}$ from the light sector as it is expected to
be most sensitive to the large distance part of the potential.
Combining this with the string {\it Ans\"atz} $\sigma = 1/2\pi\alpha'$
yields $\sqrt{\sigma} \approx 440$ MeV
\cite{regge91lucha,regge98burakovsky}.  A survey of results from 
potential models can be found in \cite{regge96Veseli} and one finds 
$\sigma$ varies in the range $0.18 - 0.22$ GeV${}^{2}$.  In most lattice 
calculations the value $\sqrt{\sigma} = 440$ MeV is used, however, one 
should be aware of the significant uncertainty in its extraction. 

\begin{figure} 
\hbox{\epsfxsize=\hsize\epsfbox{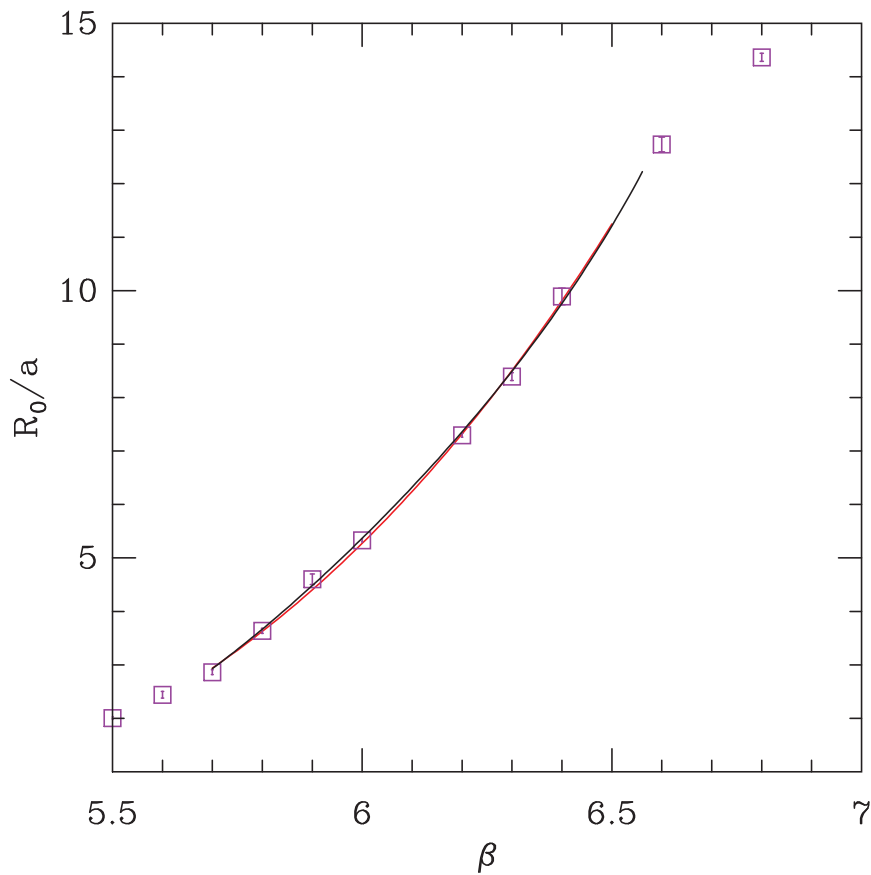}}
\caption{ The scale $R_0$ extracted from the static $q \bar q$ potential. The data in 
purple are from ~\cite{POT93Wuppertal,SP97Wuppertal,R094UKQCD} and the red line is a fit 
to these data.  The black line is the fit 
$\log(a/R_0) = -1.6805 - 1.7139(\beta-6) + 0.8155(\beta-6.0)^2 - 0.6667(\beta-6)^3$
to the data by the Alpha collaboration \cite{R098alpha}. }
\label{f:R0}
\end{figure}

An even better quantity, proposed
by Sommer~\cite{POT94Sommer}, is the distance scale $R_0$ defined
through the force between two static charges
\be
R^2 {\partial V(R) \over \partial R} \bigg|_{R=R_0} = 1.65
\label{eq:sommerR0}
\ee
The Cornell and Richardson potentials, which have been very successful in 
predicting levels of charmonium and Upsilon system, yield 
\be
R_0 \approx 0.49 \ {\rm fermi}
\ee
Setting the scale using $R_0$ has the following advantages: (i) its
value is not very sensitive to the choice of the phenomenological
potential, (ii) it is defined at an intermediate distance where
phenomenological potentials are highly constrained by the $c \bar c$
and $b \bar b$ spectra, (iii) lattice measurements of $V(R)$ at these
intermediate distances can be obtained with high precision, and (iv)
this construction is valid for both full and quenched QCD. The status
of the current quenched data from three collaborations is shown in
Fig.~\ref{f:R0}.  The data agree, and it turns out that over this same
range of $\beta$ the relation $R_0 = 1.18 / \sqrt{\sigma}$ holds to a
very good approximation. (The $\sigma$ is extracted from the large $R$
behavior of the $q \bar q$ potential as discussed above.) Thus, in
subsequent analyses of lattice data I will assume this relation when
using either $\sigma$ or $R_0$ to set the scale.

For full QCD, the linear rising potential is screened. The string
between the $q \bar q$ breaks when the energy is large enough to pop
an additional $q \bar q$ out of the vacuum and create two mesons. This
phenomena of screening can be tested by plotting $V(R)$ for different
values of the dynamical quark mass and observing the flattening of the
linear rise. The value of $R/R_0$ at which the flattening should occur
should decrease with $m_q$. A recent analyses, again by the Wuppertal
collaboration, is shown in Fig.~\ref{f:qqpotential}
\cite{SP97Wuppertal}. The data show no flattening -- presumably 
because the sea quark masses used in the update are not light enough
to cause string breaking at these $R/R_0$.

\subsubsection{Strong Coupling Expansion}
\label{ss:Scoupling}

Confinement, for $\beta \to 0$, can be demonstrated  using strong
coupling expansions. I will illustrate this technique by two examples
-- string tension and the $0^{++}$ glueball mass.

The strong coupling expansion for SU(N) gauge theory depends on the 
following identities for integration over link matrices
\bea
\int dg\         &=& 1 \nonumber \\
\int dg\  U_{ij} &=& 0 \nonumber \\
\int dg\  U^\dagger_{ij} &=& 0 \nonumber \\
\int dg\  U_{ij} U_{kl} &=& 0 \nonumber \\
\int dg\  U^\dagger_{ij} U^\dagger_{kl} &=& 0 \nonumber \\
\int dg\  U_{ij} U^\dagger_{kl} &=& {1 \over N}\ \delta_{il}\delta_{jk} \nonumber \\
\int dg\  U_{i_1j_1}  \ldots U_{i_Nj_N} &=& {1 \over N}\ \epsilon_{i_1 \ldots i_N}\ \epsilon_{j_1 \ldots j_N} \,.
\label{eq:SCintegrals} 
\eea
The essential point is that the group integration gives a
non-zero result only if each link occurs in a combination 
from which a color singlet can be
formed. Eq.~\ref{eq:SCintegrals} shows this for the identity,
``meson'' and ``baryon'' configuration of links. 

\begin{figure} 
\hbox{\epsfxsize=\hsize\epsfbox{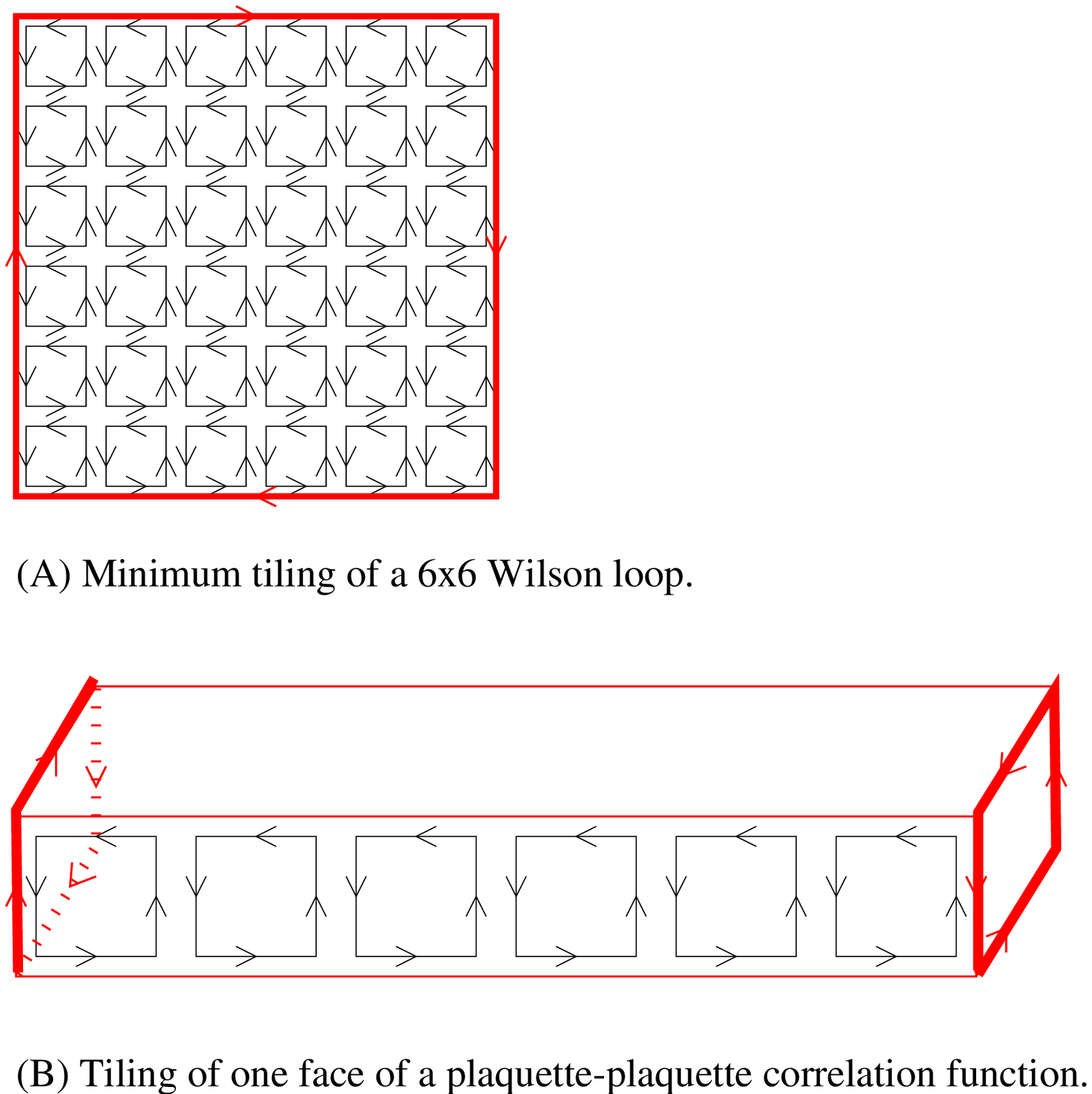}}
\caption{Examples of minimum tiling of (A) a $6 \times 6 $ Wilson loop, and 
(B) the plaquette-plaquette correlation function. }
\label{f:strongcoupling}
\end{figure}

The expectation value of the Wilson loop, at small $\beta$ (large $g$)
for the plaquette action, can be expanded as follows.
\bea
\VEV{\CW_{RT}} &=& {1 \over Z} \int dU \CW_{RT} e^{ {\beta/2N} (\CW^+_{11} + \CW^-_{11})} \nonumber \\
               &=& {1 \over Z} \int dU \CW_{RT} \big( 1 + {\beta/2N} (\CW^+_{11} + \CW^-_{11}) + \ldots \big)
\eea
where $\CW^+_{11} ,\ \CW^-_{11}$ are the two orientations of the
plaquette and the trace over color incides in each loop is implicit.
An inspection of Fig.~ \ref{f:strongcoupling}A shows that
the first non-zero contribution to the integral occurs when the loop
$\CW_{RT}$ is tiled by elementary plaquettes with the right
orientation. Each such plaquette brings a factor of ${\beta / 2N}$
from the expansion and another factor of $1/N$ from the
integration. (Note that SU(2) is special and the combined factor is
${\beta / N^2}$ as the two orientations of the loop have the same
value, $i.e.$ the trace is real.) Thus
\be
\VEV{\CW_{RT}} = {\beta^{RT} \over 2N^2} \bigg( 1 + \ldots \bigg)
\ee
where the leading corrections come from replacing any given tile with a 
pillbox. Now using Eqs.~\ref{eq:wloops1} and \ref{eq:wloops4} gives 
\be
\sigma = -\log {\beta \over 2N^2} + O(\beta) .
\ee
Thus all gauge theories, including U(1), confine in the strong
coupling limit.  What distinguishes U(1) from non-abelian gauge
theories is that the U(1) gauge theory has a phase transition at $g^2
\sim 1$ (see Section ~\ref{s:phasetransitions}), and the physical
theory (electrodynamics) lies in the weak coupling region which is
analytically disconnected from the confining strong coupling limit.

As a second application of the strong coupling expansion consider the
calculation of $0^{++}$ glueball mass.  The minimum tiling of the
connected plaquette-plaquette correlation function is the rectangular
tube consisting of four sides of size $1 \times T$ as shown in 
Fig.~\ref{f:strongcoupling}B. Thus
\be
\VEV{\CW_{11}(T) \CW_{11}(0)} \sim e^{-M_{0^{++}}T} = 
  \bigg( {\beta \over 2N^2} \bigg)^{4T} \bigg(1 + \ldots \bigg) \,,
\label{eq:SCglueball}
\ee
and 
\be
M_{0^{++}}  = - 4 \log {\beta \over 2N^2} + \ldots \,.
\ee
The fact that the lowest order result $M_{0^{++}} / \sqrt{\sigma} = 4$ is very close to 
present day estimates (see Section~\ref{ss:glueballs}) is fortuitous. 

\medskip
\noindent{\it Exercise: What is the minimal tiling for the ${0^{++}}$ 
correlation function built out of a $R \times R$ Wilson loop. Show
that Eq.~\ref{eq:SCglueball} holds independent of the size of the loop. }


\subsubsection{Wilson/Polyakov Line}
\label{ss:Wilsonline}

The Wilson line, pointing in say the time direction, is defined by the
path ordered product
\be 
\vev{\CL}(x,y,z) = {\rm Tr}\ \CP \prod_{t=1}^{N_T} U_4(x,y,z,t)
\ee
It is gauge invariant due to periodic boundary conditions on gauge
fields, $U_4(x,y,z,N_T+1) = U_4(x,y,z,1) $.  Repeating the above
arguments made for the Wilson loop one finds that the free energy of
an isolated quark is given by the Wilson/Polyakov line
\be
\vev{\CL} \sim e^{- F_q N_T} \,.
\ee
Thus, the possibility of having isolated charges in the theory 
requires $F_q$ be finite. 

In addition to gauge invariance, SU(N) gauge theories have a global
Z(N) invariance. It consists of multiplying all links on a given time
slice by the same element $z$ of Z(N) without changing the action.
Under this transformation $\vev{\CL} \to z \vev{\CL} $. Such a global
symmetry can be spontaneously broken. Thus we have two possibilities
\bea
Broken:   \qquad \vev{\CL} &\ne& 0  \qquad {\rm Deconfined}\ (F_q\ {\rm finite}) \nonumber \\
Unbroken: \qquad \vev{\CL} &=& 0    \qquad {\rm Confined}\phantom{de}\   (F_q\ {\rm infinite})
\eea
Thus, $\vev{\CL}$ is an order parameter, and tests for confinement in
pure gauge theories. It is also used to probe the location and order
of the finite temperature transition -- a discontinuity in $\vev{\CL}$
signals a first order transition, whereas a continuous change implies
second order. 

Svetitsky and Yaffe \cite{FTsvetitskyYaffe} argued that for the
purposes of determining the order of the finite temperature transition
the relevant effective theory is a 3-dimensional spin model of
short-ranged interactions (between Wilson lines reduced to spin
variables) with a Z(N) symmetry. Then using the universality classes
known from renormalization group analyses, they predicted that the
transition for SU(2) should be second order (Ising class), whereas for
SU(N$ \ge 3$) it should be first order (P\"otts class). Numerical
results are consistent with these predictions, in particular for SU(2)
the critical exponents have been determined using standard techniques
of Statistical Mechanics and found to agree with those of the
3-dimensional Ising model~\cite{FTpurereviews}.

In the presence of dynamical quarks, it is easy to see that the
discretized Dirac action is not invariant under the Z(N)
symmetry. Consequently, $\vev{\CL} \ne 0$ for all values of $\beta$,
so $\vev{\CL}$ is no longer an order parameter. Nevertheless, changes
(discontinuous or continuous) in it are used to search for the
location of the finite temperature phase transition and its order.

%% file: chap-phase.tex
\section{Phase Transitions in the Lattice Theory}
\label{s:phasetransitions}
\bigskip

The lattice theory with a finite cut-off $\mu = \pi/a$ can be
regarded as an effective theory.  The integration over momenta in the
range $\{\pi/a,\infty\}$ renormalizes the couplings and generates
additional effective interactions (see Section~\ref{ss:symanzikImp}
and the lectures by M.~L\"uscher for a detailed discussion of this
approach). Thus, one can regard the lattice theory as a point in an
infinite dimensional space of couplings, and taking the continuum
limit as a flow in this space to the critical point at $a=0$ that defines
the physical theory. In this way of thinking about LQCD as a
Statistical Mechanics system one is automatically lead to the
questions
\begin{itemize}
\item
What is the phase diagram in this extended coupling constant space?
\item
Are there other fixed points, and if so what is the nature of the
theory at those points?
\item
What are the order parameters that characterize different phases?
\end{itemize}

The starting point of my discussion of these questions will be the
pure gauge theory.  As before, the generalized form of the gauge action is
\be
\label{eq:Ggaugeaction}
\CS_g = \beta\ \sum_{i,j}\ \sum_{\mu,\nu}\ \sum_x\ c_{i,j} \ %
{\Re}\,{\rm Tr}\, W^{i,j}_{\mu\nu} \nonumber
\ee
where the sum over $i$ in $W^{i,j}_{\mu\nu}$ is over all possible
Wilson loops and $j$ is over all representations of the loop.  $\beta
$ is the overall coupling and the $c_{i,j}$ give the relative strengths of
the different loops. 

The three simplest gauge invariant probes of the phase diagram are
(gauge non-invariant quantities have zero expectation values as shown
by Elitzur~\cite{Elitzur75})
\begin{itemize}
\item
Wilson loops: These, as discussed in Section~\ref{ss:Wloops}, contain 
information on the potential between static quarks. 
\item
Wilson/Polyakov lines. These measure the free energy of an isolated quark. 
\item
The chiral condensate $\VEV{\bar \psi \psi}$. This provides information on 
vacuum alignment under spontaneous chiral symmetry breaking. 
\end{itemize}

LQCD, in the limit $g \to \infty$, can be analyzed using strong
coupling expansions as discussed in Section~\ref{ss:Scoupling}.  Using
these it was shown that, in the limit $g \to \infty$, the expectation
value for Wilson loops has an area law for all gauge groups (U(1) and
SU(N)).  Since electrodynamics (U(1)) does not confine, this region of
coupling space cannot correspond to the physical theory. In other
words, there should exist a phase transition separating the strong
coupling and weak coupling phases of non-confining theories like U(1).
This has been established for U(1), both analytically \cite{U1phasesA}
and by Monte Carlo simulations \cite{U1phasesMC}. For a theory like
QCD, with both confinement and asymptotic freedom, we need to know if
the two regions of parameter space are analytically
connected. Analytical methods like strong and weak coupling expansions
can be used provided the range of validity of the expansions gives a
sufficiently large overlap, otherwise non-perturbative methods are
necessary.

\begin{figure}[thbp]
\hbox{\hskip15bp\epsfxsize=0.9\hsize\epsfbox{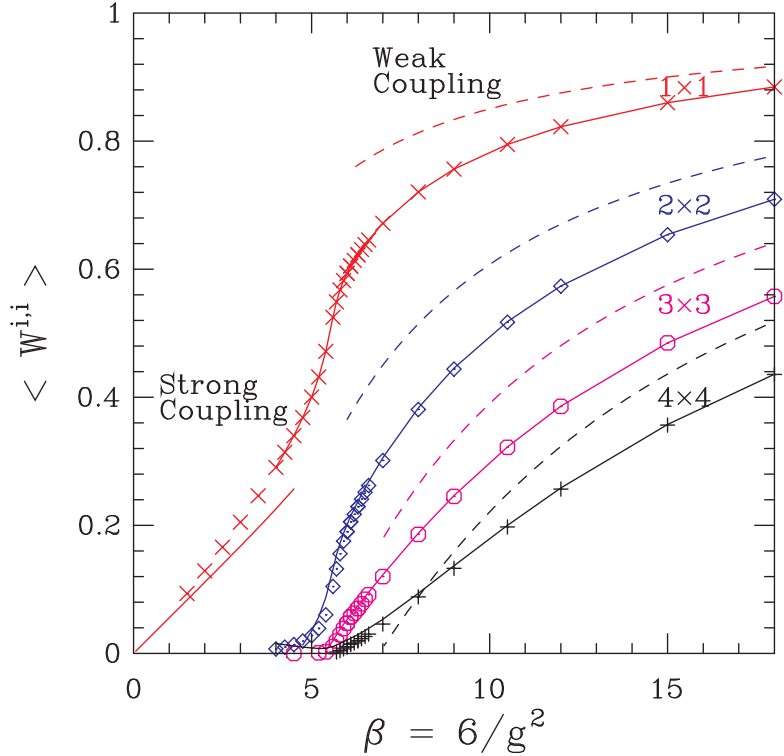}}
\vskip \baselineskip
\figcaption{The behavior of the expectation value of square Wilson loops 
as a function of coupling $\beta$. The data, show a sharp crossover
between $5 \lsim \beta \lsim 6.0$.  Result of the strong coupling
expansion to $O(\beta^5)$ and weak coupling expansion to $O(g^4)$ are
shown by the dashed line. Results of tadpole improved weak coupling
expansion (TIPT) are shown by the solid lines. Since the improved
coupling is defined in terms of the $1 \times 1$ loop, $g^2_I =
g^2/<W^{1 \times 1}>$, the agreement between the data and TIPT is a
trivial result of the construction as explained in the text.  The
improvement in larger loops, however, demonstrates the efficacy of
TIPT.}
\label{f:wilsonloops}
\end{figure}

To address these questions for SU(3) I show, in Fig.~\ref{f:wilsonloops}, the
behavior of square loops, $\lbrace {\Re}\,{\rm Tr}\, W^{i \times i}
\rbrace$, as a function of the coupling $\beta$ for the simplest
action -- Wilson's plaquette action. Also shown for comparison are the
strong coupling expansion to $O(\beta^5)$ \cite{WLSC80munsterweisz},
the weak coupling expansion to $O(g^4)$ \cite{WLPT85hellerkarsch} and
its tadpole improved (TIPT) version. These numerical data show that there is
a smooth though sharp transition between the weak coupling and strong
coupling phases, $i.e.$ no phase transitions.  The crossover takes
place in the window $5 \lsim \beta \sim 6$, with the larger loops
becoming perturbative at slightly higher $\beta$. Both strong coupling
and weak coupling expansions fail in this region (for TIPT this is
evident only for large loops), so numerical methods are
essential. Such an analysis was the basis of the pioneering work of
Creutz, and Creutz, Jacobs, and Rebbi~\cite{Creutz79}, who showed that
the string tension changes from its strong to weak coupling behavior
without a phase transition.  Their calculations yielded the first clue
that it may be possible to get continuum physics with
$O(20\%)$ errors from simulations at $\beta \gsim 6.0$.

The TIPT result shown in Fig.~\ref{f:wilsonloops} is obtained as
follows. The mean-field improved Wilson loop expectation value is 
defined to be 
\be
\VEV{W^{i \times i }}  \ = \ \VEV{{\widetilde W}^{i \times i }}\  u_0^{4i} \,,
\ee
where for $u_0$ I use the non-perturbative value obtained from the plaquette as it is 
more readily available.  The 
improved expansion for $\VEV{\widetilde W^{i \times i }}$, after
absorbing the tadpoles using the plaquette, is given by
\be
\VEV{\widetilde W^{i \times i }} = {\VEV{W^{i \times i }} \over 
		   		    \VEV{W^{1 \times 1 }}^i} \,,
\ee
where both terms on the right are first expanded in terms of $\tilde
g^2 = g^2/u_0^4(pert) = g^2/\VEV{W^{1 \times 1 }}_{pert}$ to order
$\tilde g^4$, and then the ratio is truncated at order $\tilde
g^4$. It is obvious that using $u_0$ from the plaquette guarantees a
perfect fit for $ \VEV{W^{1 \times 1 }}$, however the improvement in
even the $ \VEV{W^{4 \times 4 }}$, is remarkable.

The first study of the phase diagram in the generalized coupling
constant space was done using an action constructed from the plaquette
in both the fundamental and adjoint representations of the SU(N) gauge
theory \cite{FA81BhanotCreutz}
\begin{equation}
S = \beta_F \sum {1 \over N} \Re {\rm Tr} W_{1 \times 1} \ + \ 
    \beta_A \sum {1 \over N^2} |{\rm Tr}  W_{1 \times 1} |^2 \ .
\label{eq:Sfundadj}
\end{equation}
For this action, the effective bare coupling, obtained using a Taylor expansion, is 
\be
\beta_{eff} \equiv {2N \over g_{eff}^2}  = {\beta_F } + {2\beta_A }  \,.
\ee
Thus simulations can be done in the negative $\beta_A$ region as long
as $\beta_F > 2 |\beta_A|$. This is a classical condition that 
avoids the weak coupling singularity.

\begin{figure}[thbp]
\hbox{\hskip15bp\epsfxsize=0.9\hsize\epsfbox{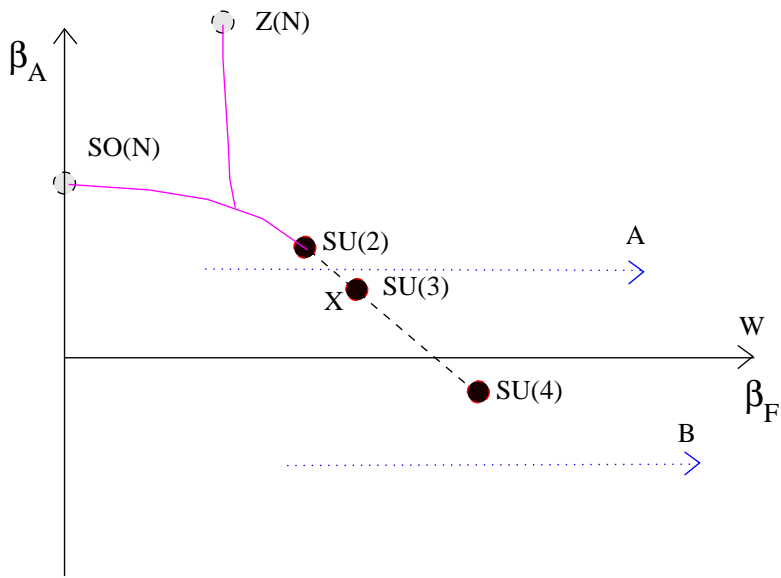}}
\vskip \baselineskip
\figcaption{The phase diagram of SU(N) gauge theories in the fundamental-adjoint 
coupling constant space.}
\label{f:fundadj}
\end{figure}

The resulting lines of first order bulk transitions, along with
the location of the end point for three gauge groups, are show in
Fig.~\ref{f:fundadj}. The point at $\beta_A= \infty$ is the first
order transition in the $Z(N)$ gauge theory, while that at $\beta_F =
0$ corresponds to that in the $SO(N)$ theory.  The location of the third
end-point with respect to the $\beta_F$ axis depends on $N$. Current
estimates of the location are
\cite{FA96Gavai,FA95Heller,FA81SU4}
\bea
\label{eq:FAendpoint}
\beta_F = 1.22    \quad\qquad \beta_A &=& 1.25    \quad\qquad {\rm SU(2)} \nonumber \\
\beta_F = 4.00(7) \qquad      \beta_A &=& 2.06(8) \qquad {\rm SU(3)} \nonumber \\
\beta_F \sim 12-15   \qquad   \beta_A &=& (-1) - (-5) \qquad {\rm SU(4)} 
\eea
The fact that for $N \le 3$ this point lies above the $\beta_F$ axis
is consistent with the observation that there is no discontinuity in
the behavior of Wilson loop data as show in Fig.~\ref{f:wilsonloops}.
If one were to take the continuum limit along a line crossing the
transition line above this point, for example along the dotted line
$A$ in Fig.~\ref{f:fundadj}, then there would be discontinuities in
Wilson loops (and consequently in the string tension), specific heat,
and glueball masses at the point of intersection with the transition
line. At the end-point $X$ the specific heat diverges and consequently
the $0^{++}$ glueball mass goes to zero. (Note that the specific heat
is the volume integral of the $0^{++}$ glueball correlation function
with the plaquette as the interpolating operator. Thus the
divergence in the specific heat implies a zero in the $0^{++}$
glueball mass.)  The string tension and masses of glueball states
with other spin-parity quantum numbers are non-zero and continuous at
this point.  Such phase transitions are lattice artifacts. For
example, taking the continuum limit at $X$, $i.e.$ setting $a=0$,
would give a free theory as all states other than the $0^{++}$
glueball become infinitely heavy.

One can take the continuum limit for the $SU(3)$ gauge theory along
any trajectory like $A$ or $W$ or $B$.  In case $A$ it is necessary to
keep $\beta_F > \beta_F^*$ to avoid the lattice artifacts.  Along $W$,
while there are no singularities, the artifact $X$ could, around
$\beta_F = 5.7$, cause significant deviations from the physics of
the relevant fixed point at $g=0$.  Lattice data verify this scenario
-- the non-perturbative $\beta$-function measured from observables
other than $M_{0^{++}}$ shows a large dip around $\beta_F =
5.7$~\cite{rrajancreutz}.  Thus, to avoid these artifacts requires
simulations for $\beta_F \gsim 6.0$.  On a trajectory like $B$ the
influence of $X$ is expected to be less than along $W$, and it is
possible that contact with continuum physics can be made on coarser
lattices.  If so then the trajectory $B$ corresponds to an ``improved
action'' in accordance with the third criteria enumerated in
Section~\ref{ss:symanzikImp}.

This qualitative argument for taking the continuum limit along a
trajectory like $B$ is supported by Monte Carlo renormalization group
(MCRG) estimates of the renormalized
trajectory~\cite{IGA87guptapatel}.  Such calculations of the RT yield
a negative value for $\beta_A$.  As explained above, the most
sensitive probe of the expected improvement on including a $\beta_A$ term
in the action, is the behavior of $M_{0^{++}}$.  Until recently tests
of this conjecture were limited by statistical accuracy
\cite{gballs91LANL}. The recent calculations on anisotropic lattices
of glueball spectra by Morningstar and Peardon~\cite{GB98Morningstar}
validate this scenario. I shall discuss their results in
Section~\ref{ss:glueballs}.

It is clear that there are additional phase transitions in the
generalized lattice theory. For example one would get a similar
diagram to that in Fig.~\ref{f:fundadj} if the action was defined in
terms of the $1 \times 2$ loop in the fundamental and adjoint
representations.  The question, therefore, is whether there exist
other second order transition points at which all correlation lengths
(measured in lattice units) diverge $i.e.$ where a non-trivial
continuum limit exists. The commonly held belief, supported by
numerical data, is that all such points lie on the hypersurface
defined by $g_{eff} = 0$. The long distance behavior of these theories
is controlled by the fixed point defined with respect to a suitable
renormalization group transformation.  Taking the continuum limit of
the lattice theory on this hypersurface provides a non-perturbative
definition of QCD.  This is the standard scenario.

An alternative scenario is the presence of a non-trivial fixed point
at $g_{eff} \ne 0$~\cite{PatSeiler}. If this unlikely possibility
turns out to be true then the perturbative relation between $g_{eff}$
and $a$, based on asymptotic freedom, would change -- the coupling
would no longer run to zero as the momentum scale $Q \to \infty$. This
could be exposed by precise measurements of $\alpha_s$ as a function of
$Q^2$, or by a detailed search for non-trivial second order phase
transitions in the lattice theory.

In either case the non-perturbative lattice analyses of quantities
such as the spectrum and weak matrix elements does not change. To get
to the continuum limit we would still extrapolate the lattice data in
$a$ to $a=0$.  Only the PQCD relation between $a$ and $g$ would be
invalid, consequently extrapolations using the perturbative scaling
relation could not be used. My bottom line on this issue, having
mentioned the heretical view, is to proceed assuming that the standard
scenario is correct.

%% file: chap-errors.tex
\section{Errors in Lattice Results}
\label{s:syserrors}

Numerical simulations of lattice QCD introduce both statistical and a
number of systematic errors. I have collected together a brief
description of the various errors here, and in the remainder of the
article I will only give quantitative estimates. I will label a 
particular systematic uncertainty as negligible/important if it is
negligible/comparable in magnitude to the statistical error for that
observable as determined in a state-of-the-art calculation.

\subsection {Statistical errors} 

The monte carlo method for doing the functional integral employs
statistical sampling. The results, therefore, have statistical errors.
The current understanding, based on agreement of results from
ensembles generated using different update algorithms, different
initial starting configuration, and different random number generators
in the Markov process, is that the functional integral for QQCD is
dominated by a single region.  Second, we find that configurations
with non-trivial topology are properly represented in an ensemble
generated using a Markov chain based on small changes to link
variables.  Based on these observations we conclude that the energy
landscape is simple, consequently, the statistical errors should fall
as $1/\sqrt{N}$, where $N$ is the number of independent measurements.
Tests like binning of the data and the calculation of auto-correlation
times for various observables are used to determine the monte-carlo
time between independent configurations.

Estimates of decorrrelation times in dynamical simulations are just
becoming available and the understanding of decorrelation times for
various update algorithms and different fermion formulations is not
complete.  A recent study of decorrelations for update of Wilson fermions 
\cite{HMCA98Sesam} using the hybrid Monte Carlo algorithm \cite{HMCA87} (the
algorithm of choice) presented very encouraging results.  They found
that decorrelation time based on tunneling rate between different
topological sectors (expected to be amongst the slowest modes) was
within a factor of two of standard probes like large Wilson loops and
hadron correlators, and $\approx 80$ units of molecular dynamics time
for $m_\pi / M_\rho \ge 0.56 $.  Thus trajectory lengths of 5000-10000
units, which are feasible on todays computers, provide a reasonable
statistical sample. Even though details like how this decorrelation
time grows with decreasing quark mass still need to be filled in, it
is clear that current Monte Carlo algorithms do provide a reliable way
of doing the functional integral for both QQCD and full QCD.

\subsection {Finite Volume}

This is the simplest error to visualize as it is associated with using
finite lattices to represent an infinite system.  L\"uscher has shown
that for sufficiently large lattices of side $L$, the finite volume
corrections in the mass $M$ of a given state fall off as $e^{-M L}$ 
\cite{LH88Luscher}.  This anaylsis assumes that the interactions with 
the mirror states on a periodic lattice are small.  On small lattices
the wave-function of the states is squeezed, resulting in an much more
rapid increase of the mass with decreasing lattice size.  This effect
is seen clearly for $L < 1.5$ fermi, where the finite size behavior
fits a power-law, $\sim 1/L^3$ \cite{FSE93JAPAN}.  The goal,
therefore, is to work on lattices large enough such that the finite
size effects are exponentially suppressed.

To determine the lattice size needed for observables involving quark
propagation I choose the pion as the test state as it the lightest and
thus has the largest correlation length. Results of quenched
simulations show that for $M_\pi L \ge 5$ the exponential suppression
applies and the corrections are negligible. For physical value of
$M_\pi$ this translates into $ L \ge 7$ fermi. Another way of stating
the same criteria is that the quantum mechanical properties of
hadrons, typically of size $\le 1$ fermi, are unaltered if the box
size is larger than $7$ fermi.  Current simulations explore quark
masses in the range $m_q \ge m_s/4$ or equivalently $M_\pi / M_\rho
\ge 0.4$.  For these ``heavier masses'' the criteria $M_\pi L \ge 5$
translates into $ L \ge 3$ fermi. The lattices used in the most recent
calculations, for example the CP-PACS runs, discussed later, satisfy
this condition.

Another consequence of finite $L$ is momentum resolution. For example,
the lowest non-zero momentum, and spacing, is $2 \pi /La = 393$ MeV
for $L=32$ and $1/a=2$ GeV. One would like a much finer resolution
when investigating the dispersion relation for mesons and baryons or
when calculating matrix elements required to study form-factors in 
semi-leptonic and rare decays. This can only be done by increasing
either $L$ or $a$. Increasing $L$ is limited by computer resources, while 
increasing $a$ increases discretization errors which are discussed next.

\subsection {Finite lattice spacing errors}

The lattice is merely a technical scaffold and the physical results
are obtained by taking the $a \to 0$ limit.  At finite $a$, lattice
results have discretization errors whose size depend on the degree to
which the lattice action and operators have been improved (see
Sections~\ref{ss:symanzikImp}, \ref{s:improvedglue}, and
\ref{s:improveddirac}). As discussed in 
Section~\ref{ss:symanzikImp}, we employ two strategies
for removing these errors. First, improve the lattice action and
operators so that the errors at fixed $a$ are small, and secondly,
repeat the simulations at a number of values of $a$ and extrapolate to
$a=0$. Since the functional form used to characterize the errors is
usually truncated to just the leading term in $a$, the extrapolation
has an associated systematic error.  To get a handle on this
uncertainty, extrapolations have been done for three different
formulations (Wilson, clover, staggered) for which the leading
corrections are different ($O(a)$, $O(\alpha a)-O(a^2)$, $O(a^2)$).
The difference between the three results in the $a=0$ limit is a
measure of the residual uncertainty.  I shall illustrate this
procedure for controlling the discretization errors in
Section~\ref{s:mq} where the calculation of quark masses is discussed.

\subsection {Chiral Extrapolations in the light quark masses}

The physical $u$ and $d$ quark masses are too light to simulate on
current lattices. For $1/a = 2$ GeV, realistic simulations require
$L/a \gsim 70$ to avoid finite volume effects, $i.e.$ keeping $M_\pi L
\ge 5$. Current biggest lattice sizes are $L/a=32$ for quenched and
$L/a=24$ for unquenched. Thus, to get results for quantities involving
light quarks, one typically extrapolates in $m_u = m_d$ from the range
$ m_s/4 - 2 m_s$ using functional forms derived using chiral
perturbation theory but truncated to the first couple of terms.  Since
the range of the chiral extrapolation is large, it is important to
quantify the size of the higher order chiral corrections. This
requires precise data at a large number of values of quark masses.
Most of the current fits are based on keeping just the lowest order
terms as the quality of the data is not good enough to quantify the
various higher order corrections, and/or to identify the
quenched artifacts discussed in Section~\ref{ss:QQCD}. The status of
the current data with respect to resolving these higher order
corrections is discussed in Section~\ref{s:hadronspectrum} where an
analysis of SU(3) flavor symmetry breaking, for example the mass
differences between mesons and baryons within SU(3) multiplets like
the baryon octet and the decuplet, is presented.

Current simulations also neglect isospin breaking effects and
electromagnetic contributions as these are a few MeV in nature, and
smaller than the present numerical resolution. The iso-spin
symmetric mass is defined by $\bar m = (m_u + m_d)/2$.  For an exploratory
analysis of iso-spin breaking effects in the presence of a strong U(1)
field see Ref.~\cite{isospin96Estia}. 

I expect that our understanding of the chiral corrections will change
significantly in the next two years due to the factor of 10-100
increase in the computational power made available to LQCD. To study
iso-spin breaking effects properly requires dynamical lattices with
roughly physical values for $m_u$ and $m_d$. Such simulations are
still a few years away.

\subsection {Discretization of Heavy Quarks}

Simulations of heavy quarks ($c$ and $b$) have large discretization
errors of $O(ma)$ in addition to the $O(\Lambda_{QCD} a)$ and
$O(pa)$. This is because quark masses measured in lattice units, $m_c
a$ and $m_b a$, are of order unity for $2 \GeV \le 1/a \le 5 \GeV$.
Data show that these discretization errors are large even for $m_c$ in
the case of Wilson and clover actions.  (Staggered fermions do not
have any advantage over Wilson-like discretizations for heavy quarks
and are hence not used to study heavy quarks.) A number of alternate
approaches are being investigated to address this issue (see
Section~\ref{s:improveddirac}).  These include the non-relativistic QCD
(NRQCD) formulation
\cite{NRQCD91LepageThacker}, lattice versions of the heavy quark
effective theory \cite{HQETrefs}, a variant of the Dirac
discretization that interpolates between NRQCD and clover
discretizations \cite{IDA97Fermilab}, and non-perturbatively $O(a)$
improved SW-clover fermions \cite{IDA96ALPHA}.  Many collaborations
are testing these formulations, however, it is still too early to
judge which approach will provide the best solution to the simulations
of heavy quarks.

\subsection { Matching between lattice and continuum scheme (renormalization
constants)} 

Experimental data are analyzed using some continuum renormalization
scheme like $\overline{MS}$ so, to make contact with phenomenology, 
results in the lattice regularization scheme have to be converted to
the continuum scheme.  The perturbative relation between renormalized
quantities in $\overline{MS}$ and the lattice scheme are, in
almost all cases, known only to 1-loop.  Data show that the
$O(\alpha_s)$ corrections can be large, $\sim 10-50\%$ depending on
the quantity at hand.  However, as explained by Parisi
\cite{TI80Parisi} and by Lepage and Mackenzie \cite{TI93lepagemackenzie},
these large corrections are mostly due to lattice artifacts which,
once identified as coming from tadpole diagrams, can be removed.  The
Lepage-Mackenzie prescription for reorganizing the lattice
perturbation theory to remove these artifacts has significantly
reduced the associated uncertainty. Nevertheless, the improvement is
not uniform and in many cases the corrections are still large, and the
residual errors in TI 1-loop perturbative estimates are hard to
ascertain. The final step in improving the reliability of 
the matching factors is to develop non-perturbative
methods~\cite{ALPHAZs,ROMEZs,APETOZs}. Significant progress has been
made in setting up these calculations, and once they are complete the
uncertainties in existing results due to using 1-loop $Z$'s will be
removed. An illustration of the effect of using perturbative versus 
non-perturbative $Z$'s is presented in Section~\ref{ss:mqWI}. 

\subsection {Operator mixing} 

As mentioned in Section~\ref{s:scope}, a major application of LQCD is
to calculate the matrix elements of operators that arise in the
effective weak Hamiltonian between hadronic states. In addition to the
question of the overall normalization of these operators, the
operators can, in general, mix with operators of the same, higher, and
{\it lower} dimensions. On the lattice this mixing arises due to
quantum corrections and discretization errors. The set of lattice
operators one has to consider is usually larger than those in the
continuum theory because at finite $a$ the symmetries of the lattice
theory are smaller than those of the continuum theory, for example the
hard breaking of chiral symmetry in Wilson fermions.  In cases where
there is mixing with lower dimensional operators, the mixing
coefficients have to be known very accurately otherwise the power
divergences overwhelm the signal as $a \to 0$.  Similarly, in cases
where there is mixing with operators of the same dimension but with
different tensor structures, as in Wilson-like actions due to the
explicit breaking of chiral symmetry, the chiral behavior may again be
completely overwhelmed by these lattice artifacts if the coefficients
are not known precisely. Examples where such mixing has posed serious
computational challenges are the matrix elements of 4-fermion
operators needed in the calculation of the $\Delta I = 1/2$ rule,
$B_K$, $B_6$, and $B_8$. In these cases also non-perturbative methods
for calculating the mixing coefficients are essential. For a
discussion of these methods see Prof. Martinelli's lectures.

\subsection {Quenched Approximation (QQCD)}
\label{ss:QQCD}

This approximation is computationally the hardest to shed and
quantify.  Since it plays a key role in today's simulations, I will
discuss it in some detail.  The approximation consists of ignoring the
fermion contribution to the path integral, $i.e.$ setting ${\rm det}\
M = constant$ in Eq.~\ref{eq:pathintegral}. The sole reason for this 
approximation is limitations of computer power. With known algorithms
the computational requirements of full QCD simulations go up by
factors of $10^3 - 10^5$ depending on the quark mass.  Since QQCD is
confining, asymptotically free, and shows spontaneous chiral symmetry
breaking ($\VEV{\bar \psi \psi}_{QQCD} \neq 0$), and differs from full
QCD only in the relative weighting of the background gauge
configurations, the physics analyses are identical in most cases. It is
therefore considered reasonable to do detailed quenched simulations in
order to understand and control all other sources of errors while
waiting for better algorithms or the computer technology
needed to provide the required additional factor of $10^3 - 10^5$,
$i.e.$ 10-1000 teraflop capability.

\begin{figure} 
\hbox{\epsfxsize=\hsize\epsfbox{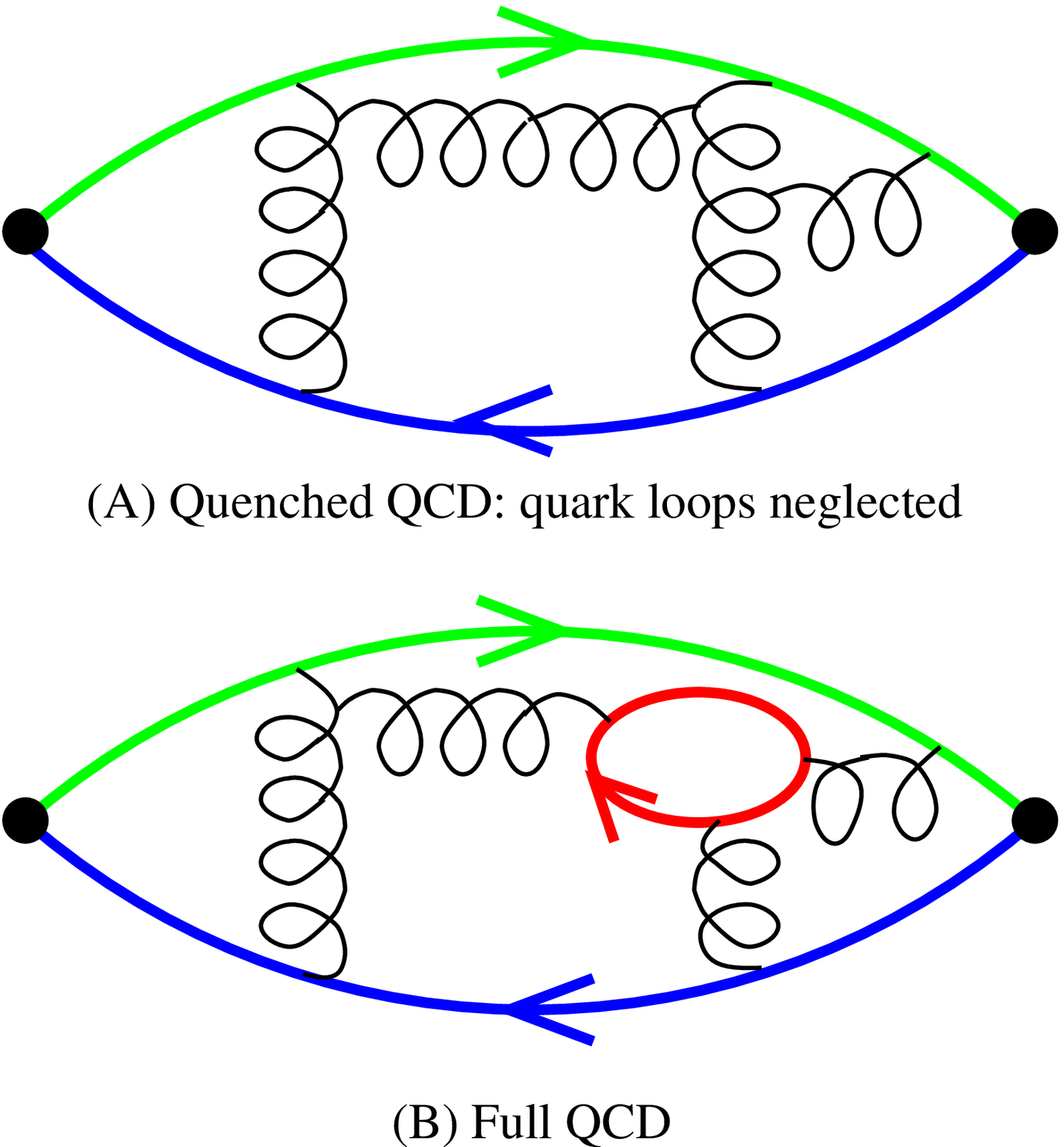}}
\caption{An illustration of the difference between quenched QCD and QCD. QQCD has 
no vacuum polarization loops on virtual gluon lines.}
\label{f:qqcd}
\end{figure}

Physically, the quenched approximation corresponds to turning off
vacuum polarization effects of quark loops. This is illustrated in
Fig.~\ref{f:qqcd} for the pion correlator. One important consequence
of neglecting vacuum polarization effects is the behavior of the
potential between a $q \bar q$ pair as a function of the separation.
In the full theory the string breaks by the creation of a $q \bar q$
pair out of the vacuum, therefore at large distances there is a
screened potential between two mesons. For such screened potential 
large Wilson loops should show a perimeter law. In QQCD the
string does not break and large Wilson loops show an area law. Thus
the long distance behavior of the two theories is very different and
one might be led to believe that QQCD is a bad approximation.
However, because of confinement, the long distance scale that is
relevant to hadronic physics has a natural cutoff of a few
fermi. Thus, if one can match the potential between some fine scale ($
\sim 0.01 $ fermi, below which asymptotic freedom makes the
calculations perturbative) and the typical size of hadrons ($ \sim 1 $
fermi) then QQCD could give reasonable estimates for the hadronic
spectrum.  This argument is really only applicable to heavy onia,
where potential models work well. For light quarks one simply has to
do the simulations to estimate the size of the distortions.  In any
case the community has proceeded by assuming that the two theories can
be roughly matched by adjusting the quenched coupling constant at some
scale like $0.5$ fermi, or equivalently by adjusting the overall scale
$\Lambda_{QQCD}$, and hopes that the quenching corrections are small,
$10-20\%$, for many cases.  If this bears out then even QQCD would
yield many phenomenologically useful predictions.

In spite of the optimism, the fact remains that QQCD is not a unitary
theory, so it is important to understand how it fails. I would like 
to illustrate its shortcomings with the following examples.

\begin{figure} 
\hbox{\epsfxsize=\hsize\epsfbox{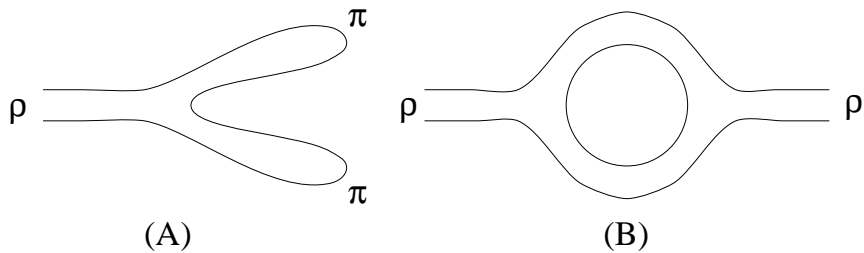}}
\caption{The $\rho \pi \pi$ coupling from the $\rho \to \pi \pi$ 
decay and (B) the $\rho$ correlator in full QCD.}
\label{f:rhopipi}
\end{figure}

We can calculate the $\rho \pi \pi $ coupling in both QCD and QQCD by
measuring the three point function $\VEV{T[\rho \pi \pi]}$ illustrated
in Fig~\ref{f:rhopipi}A.  The difference between the two values is
a measure of the validity of QQCD.  On the other hand the
diagram shown in Fig~\ref{f:rhopipi}B is absent in QQCD, and the
$\rho$ spectral function does not have a cut beginning at $2M_\pi$. 
Thus, the $\rho \pi \pi $ coupling {\it cannot} be extracted
from an analysis of the $\rho$ 2-point function in QQCD.

\begin{figure} 
\hbox{\epsfxsize=\hsize\epsfbox{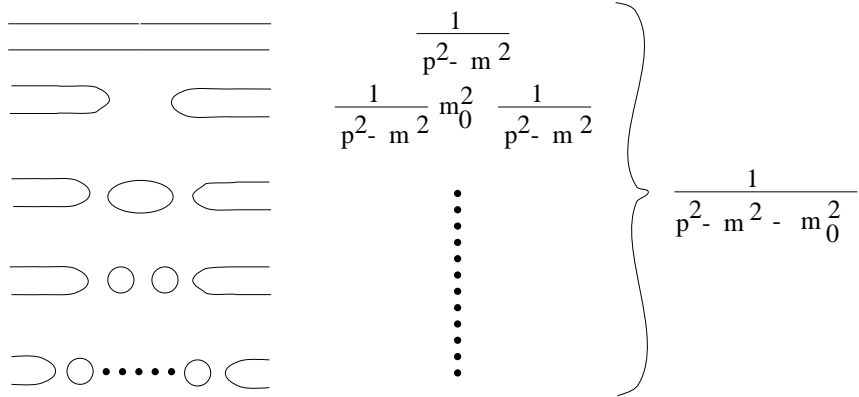}}
\caption{The iteration of the $\eta'$ propagator in full QCD. Only the first two diagrams 
survive in QQCD.}
\label{f:etaprime}
\end{figure}

A very important difference between full QCD and QQCD is the behavior
of the $\eta'$. In full QCD the singlet axial current is anomalous and
the $\eta'$ acquires a large mass due to the summation of vacuum
polarization graphs shown in Fig~\ref{f:etaprime},
\begin{equation}
{1 \over p^2 - M^2} + {1 \over p^2 - M^2} M_0^2 {1 \over p^2 - M^2} + \ldots = 
	{1 \over p^2 - M^2 -M_0^2} 
\end{equation}
where, for simplicity I have approximated the gluonic coupling by a
momentum independent constant $M_0^2$. This is related to the
topological susceptibility via the Witten-Veneziano relation
$ M_0^2 = 2 n_f \chi_t / f_\pi^2 = M_{\eta'}^2 + M_\eta^2 - 2M_K^2$ 
\cite{WittenVeneziano}. In QQCD the absence of the vacuum polarization
diagrams means that the $\eta'$ propagator has a single and a double
pole, $i.e.$ ${(p^2 - M^2)^{-1}}$ and the ``hairpin'' term ${(p^2 -
M^2)^{-1}} M_0^2 {(p^2 - M^2)^{-1}}$, where $M^2$ is the Goldstone
pion mass $M_\pi^2$.  The fact that the $\eta'$, in the limit $m_q \to
0$ of QQCD, is massless has important consequences. Two groups, Sharpe
and collaborators \cite{SRScpt}, and Bernard and Golterman
\cite{CbMg}, have investigated these by formulating a chiral
perturbation theory for QQCD which retains $\eta'$ as an additional
Goldstone boson.  I illustrate the differences between $\chi$PT and
Q$\chi$PT by considering the behavior of the 2-point correlation
function of the pion.

\begin{figure} 
\hbox{\epsfxsize=\hsize\epsfbox{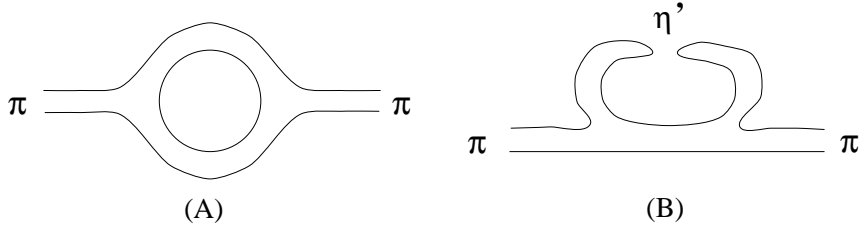}}
\caption{The 1-loop correction to the pion propagator in (A) full QCD (B) QQCD.}
\label{f:QQCDpi}
\end{figure}

At 1-loop full QCD has the diagram shown in Fig~\ref{f:QQCDpi}A which
is absent in QQCD. On the other hand, in QQCD, the hairpin term gives
a contribution via the diagram shown in Fig~\ref{f:QQCDpi}B. Such
diagrams are neglected in full QCD as they iterate to produce a
massive $\eta'$.  The chiral expansions for $M_\pi$ and $f_\pi$ in
full QCD have been calculated by Gasser and Leutwyler
\cite{GasserLeutwyler}, 
\begin{eqnarray}
M_\pi^2 &=& 2 m B \big[ 1 + X_\pi - {1 \over 3} X_{\eta_8} + 
	  {16 \over f_0^2} (2M_K^2+ M_\pi^2)(2L_6 - L_4) \nonumber \\
        &{}& \hskip 1.35in + 
	  {16 \over f_0^2} M_\pi^2 (2L_8 - L_5)  + \dots \big] \nonumber \\
f_\pi^2 &=& f_0 \big[ 1 - 4 X_\pi - 2 X_K + 
	  {16 \over f_0^2} (2M_K^2+ M_\pi^2) L_4 + 
	  {16 \over f_0^2} M_\pi^2  L_5  + \dots \big] 
\end{eqnarray}
where $L_i$ are the $O(p^4)$ coefficients, $X_i = M_i^2 {\log}
({M_i^2 \over \Lambda_{\chi SB}^2}) /(4 \pi f_0)^2$ are the chiral
logs, and $f_\pi = 131$ MeV. In QQCD, Sharpe and Bernard and Golterman
\cite{SRScpt,CbMg} find that the leading terms are 
\begin{eqnarray}
M_\pi^{2\ Q} &=& 2 m B_Q \big[ 1  - \delta {\log} {M_\pi^2 \over \Lambda_{\chi SB}^2} + 
		 \dots \big] \nonumber \\
f_\pi^Q      &=& f_Q \big[ 1 - 
         	  {16 \over f_Q^2} M_\pi^2  \tilde L_5  + \dots \big] 
\end{eqnarray}
where $\delta = M_0^2 / 24 \pi^2 f_\pi^2$ and $\tilde L_i$ are the
quenched $O(p^4)$ coefficients. The term proportional to $\delta$ is a 
quenched chiral log and is singular in the limit $m_q = 0$. 

The differences, in general, between quenched
and unquenched expressions, as illustrated by the above expressions, are
\begin{description}
\item{$\bullet$}
All the chiral constants are different in the two theories. For
example, $B \ne B_Q$ and $f_0 \ne f_Q$.
\item{$\bullet$}
The normal chiral logs are missing in QQCD.
\item{$\bullet$}
The quenched expressions can be singular in the chiral limit due to
the goldstone $\eta'$, $i.e.$ through terms proportional to
$\delta$. These artifacts become dominant as $m_q \to 0$.
\end{description}

The first question raised by the analyses of Sharpe and Bernard and
Golterman is -- are the results of quenched $\chi PT$, an expansion
about $m_q = 0$, meaningful if they are singular in that limit? We do
not have a formal answer to this question. The approach adopted is to
do the simulations and look for such chiral logs by doing fits with and 
without these terms. The status of this search has been reviewed in
\cite{QCPT94rev,QCPT96rev}.  The data, while not conclusive, does suggest 
that $\delta \approx 0.15$, consistent with the value obtained by
using phenomenological values for $M_0 $ and $f_\pi$.

The second question, assuming that predictions of quenched $\chi PT$
make sense, is how does one extract phenomenologically useful results
for quantities that require a chiral extrapolation to $\bar m$? One
approach is to fit the quenched data using the full QCD chiral
expansions in a region of quark masses where the quenched artifacts
are small. Current data suggest that this region is $m_q \gsim
m_s/4$. In this case quenching errors are, by definition, a
combination of those due to the absence of quark loops and those due
to fits being made at heavier quark masses.  The second possibility is
to fit using the QQCD expression but extrapolate keeping only the
terms in the full QCD expressions. This approach has the disadvantage
of increasing the number of free parameters. In either case the hope
is that there exist quantities for which the QQCD artifacts are small
and the differences between QQCD and QCD coefficients are small; in
such cases quenched estimates are useful phenomenological
predictions. One such example is the $K^0 \bar{K^0}$ mixing parameter
$B_K$ \cite{STAGBk}.

\bigskip

In view of these various sources of systematic error, any quantity
measured on the lattice depends not only on the input parameters,
$i.e.$ the quark masses and the coupling $\alpha_s$, but also on the
lattice size $L$, $a$ (discretization errors are $\propto g^m a^n$
with the powers depending on the order of improvement in the action and
operators), the method used to determine the renormalization constants,
and the number of dynamical quark flavors $n_f$.  While the
theoretical understanding of these errors is, in general, not
complete, nevertheless, a lot is known and one can make consistency
checks on the data.  For example, the infinite volume continuum limit
results for fixed $n_f$ simulations should be independent of the
discretization scheme, the numerical approach used, and the definition
of the renormalization constants. As of this writing, results for a
number of observables show stability once extrapolated to the $L \to
\infty$ and $a \to 0$ limits, albeit in the quenched approximation
($n_f=0$).  We regard this consistency check as the first important
step towards providing precise results. Our optimism stems from
knowing that for many phenomenologically important quantities ($B_K,
f_D, f_{D_s}, f_B, f_{B_s}, B_B, \alpha_s$, and the quark masses)
quenched lattice QCD results are already competitive with the best
non-lattice methods, and future enhancements in methodology and
computer technology will systematically improve their precision.

%% file: chap-corr.tex
\section{Lattice Correlators}
\label{s:corrfns}

\begin{figure} 
\hbox{\epsfxsize=\hsize\epsfbox{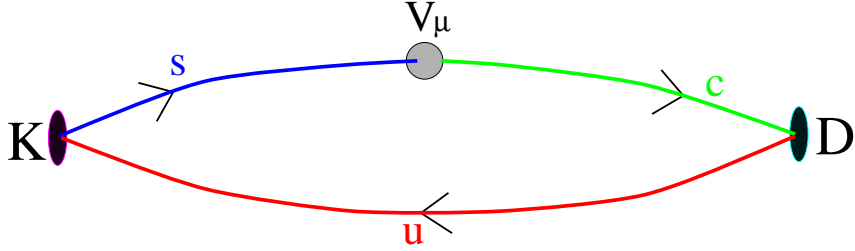}}
\caption{Schematic of Feynman diagram for $\VEV{K V_\mu D}$ in terms of 
$c$, $s$, and $u$ quark propagators.}
\label{f:mfffig}
\end{figure}

In Section~\ref{s:overviewLQCD} we discussed the reduction of the pion
2-point correlation function to a product of quark propagators and
possibly gauge links, and its relation to sum over states from which
$M_\pi$ and $f_\pi$ can be extracted. Here I enlarge the discussion to
general 2-point and 3-point functions.  Consider the calculation of
the matrix element $\VEV{K^+ | V_\mu | D^0}$ which arises in the
extraction of semi-leptonic form factors. The 3-point correlation
function needed to calculate this ME is
\begin{eqnarray}
  C^{PVP}_\mu(t_x, \vec p; t_y, \vec q) &=& 
	\sum_{\vec x, \vec y} e^{-i ( q \cdot y + p \cdot x )} \VEV{T[ K^+ V_\mu D^0 ] }  \\
    &=& \sum_{\vec x, \vec y} e^{-i ( q \cdot y + p \cdot x )}
     \VEV{ \bar s(0) \gamma_5 u(0)
           \bar c(y) \gamma_\mu s(y)
           \bar u(x) \gamma_5 c(x) } \nonumber \\
    &=& \sum_{\vec x, \vec y} e^{-i (q \cdot  y + p \cdot x )}
     \VEV{ \gamma_5 S_F^u(x,0) \gamma_5 S_F^s(0,y) \gamma_\mu S_F^u(y,x) } \nonumber \\
    &\sim&
  { {\exp( -E_K(\vec p + \vec q)t_y -E_D(\vec p)(t_x-t_y)) \over 
      ( 4 E_K(\vec p + \vec q) E_D(\vec p) ) } } \times \nonumber \\
  &{}& \hskip -0.3in
          \me{0}{\bar u \gamma_5 s}{K(\vec p + \vec q)}
          \me{K(\vec p + \vec q)}{\bar s \gamma_\mu c}{D(\vec p)}
          \me{D(\vec p)}{\bar c \gamma_5 u}{0}  \ . \nonumber 
\label{eq:mffexample}
\end{eqnarray}
I have used (i) Wick contractions to express the correlation function
in terms of quark propagators $S_F$ (see Fig.~\ref{f:mfffig}); (ii)
local bilinears for the source and sink operators and the vector
current (to improve the signal one usually uses smeared operators and
an improved current); and (3) the interpolating operator for the $D$
meson and the current insertion are at definite momentum. Momentum
conservation then fixes $\vec p_k$. The final form in
Eq.~\ref{eq:mffexample} shows the behavior for large Euclidean time
separations ($t_x \ll t_y \ll t_0$).  Of the many terms in the right
hand side, we are only interested in the matrix element $\me{K(\vec p
+ \vec q)}{\bar c \gamma_\mu s}{D(\vec p)}$.  The remaining factors
can all be extracted from 2-point correlation functions. Thus, one can
make individual fits to all the required correlators, or design a
ratio in which the maximum number of unwanted factors cancel.  In
practice, to improve the numerical signal, a single fit is made to a
ratio of 3-point to a combination of 2-point functions, $ \VEV{T[ K
V_\mu D ]} / \VEV{T[KK]} \VEV{T[DD]}$, that gets rid of the
exponential in time behavior.

To make fits to such correlation functions projected on to definite
spatial momenta, we have to address three questions:
\begin{itemize}
\item Is the correlator even or odd in its time variables?
\item Is the correlator even or odd in its momenta?
\item Is the correlator real or imaginary? As mentioned before only the 
      real part of Euclidean correlation functions has a signal. In
      practice we usually work in terms of bilinear operators given in 
      Table~\ref{t:wilsonops} where factors of $i$ have been dropped 
      for brevity. In this basis of operators one needs
      to know whether a given n-point function has a signal in the
      real or imaginary part.
\end{itemize}

\noindent To answer these questions  we use the discrete symmetries, $\CT$,
$\CP$, and $\CC\CP\CH$ discussed in Table~\ref{t:CPTdefs} and in
Eq.~\ref{eq:fermionpct}.  For each element of the Dirac algebra,
$\Gamma_A$ define three signs, $\tau_A$, $\pi_A$, and $\chi_A$,
associated with $\CT$, $\CP$, and $\CC\CP\CH$:
\begin{eqnarray}
  \gamma_4 \gamma_5 \Gamma_A \gamma_5 \gamma_4
    &=& \tau_A \Gamma_A; \nonumber \\
  \gamma_4 \Gamma_A \gamma_4
    &=& \pi_A \Gamma_A; \nonumber \\
  C \gamma_4  \gamma_5 \Gamma_A \gamma_5 \gamma_4 C^{-1}
    &=&\chi_A \Gamma_A^*. 
\label{eq:cfsyms}
\end{eqnarray}
These signs are given in Table~\ref{t:CPTcfsigns} and 
the application to 2-point and 3-point correlators is illustrated in the 
next two sub-sections. 

\begin{table} 
\begin{center}
\setlength\tabcolsep{0.32cm}
\caption{The signs for bilinear currents under $\CT$, $\CP$ and $\CC\CP\CH$ transformations.}
\begin{tabular}{|c|c|c|c|c|c|c|c|c|}
\myhline
          &    &    &    &    &    &    &    &    \\[-7pt]
          & 1 & $\gamma_i$ & $\gamma_4$ & $\gamma_i\gamma_j$ & $\gamma_i\gamma_4$
               & $\gamma_i\gamma_5$ & $\gamma_4\gamma_5$ & $\gamma_5$  \\
          &    &    &    &    &    &    &    &    \\[-7pt]
\myhline
$\tau_A $ &  + &  + &  - &  + &  - &  - &  + &  - \\[2pt]
\myhline
$ \pi_A $ &  + &  - &  + &  + &  - &  + &  - &  - \\[2pt]
\myhline
$\chi_A $ &  + &  - &  + &  + &  - &  + &  - &  - \\[2pt]
\myhline
\end{tabular}
\label{t:CPTcfsigns}
\end{center}
\end{table}

\subsection{Two-point meson correlators}
\label{ss:corr2pt}

Consider the general two-point correlator
\begin{eqnarray}
  C^{BA}(\vec p, t)
  =& \sum_{\vec x} e^{ -i p \cdot x}
    \VEV{ \bar\psi_2(x) \Gamma_B \psi_1(x)
          \bar\psi_1(0) \Gamma_A \psi_2(0) }, \nonumber \\
  =& - \sum_{\vec x} e^{ -i p \cdot x}
    \VEV{
      {\rm Tr}\,\left( S_2(0,x) \Gamma_B S_1(x,0) \Gamma_A \right)
    }, 
\label{eq:CPT2cfsigns}
\end{eqnarray}
where ${\rm Tr}$ is trace over spin and color indices and $t \equiv
x_4$.  $\CT$, $\CP$ and $\CC\CP\CH$ then imply the following:
\begin{itemize}
\item
  If $\tau_A\tau_B=\pm1$, $C^{BA}(\vec p, t)$ is even (odd) in $t$.
\item
  If $\pi_A\pi_B=\pm1$, $C^{BA}(\vec p, t)$ is even (odd) in $\vec p$.
\item
  If $\chi_A\chi_B=\pm1$, $C^{BA}(\vec p, t)$ is real (imaginary).
\end{itemize}

Note that in Eq.~\ref{eq:CPT2cfsigns} the second quark propagator is
from $x \to 0$. Using the hermiticity property $S(x,0) = \gamma_5
S(0,x)^\dagger \gamma_5$ we can write this in terms of
$S(x,0)$. This simple property, the relation between the quark
propagator from a given point and the anti-quark propagator to that
point, leads to a huge computational saving. For example, if the
momentum projection is done by summing over the $L^3$ sink points
$\vec x$ with weight $\exp(i \vec p \cdot \vec x)$, then, because of
the hermiticity property we need only one propagator inversion instead
of $L^3 + 1$.

\subsection{Interpolating operators}

For Wilson-like fermions the interpolating field operators for mesons
and baryons at zero-momentum are given in Table~\ref{t:wilsonops}. The 
three flavors are labeled $u,d,s$ for brevity. The
baryon operators need further clarification as discussed in 
\cite{HM96LANL} and reproduced below.

\begin{table}
\begin{center}
\setlength\tabcolsep{1cm}
\caption{The local interpolating field operators for
mesons and baryons in Wilson-like theories. Projection to zero
momentum states is obtained by summing over the points ${\vec x}$ on a
time slice. The C-parity is only relevant for flavor degenerate meson
states. The $()$ in baryon operators denote spin trace. $\CC =
\gamma_2\gamma_4$ and the symmetry properties of flavor indices for
nucleons are discussed in the text. The decuplet baryon operator is
completely symmetric in flavor index }
\begin{tabular}{|l|c|c|}
\myhline 
                   &               &         \\[-4pt]
State              & $I^G(J^{PC}) $& Operator \\[2pt]
\myhline 
                   &               &         \\[-4pt]
Scalar($\sigma$)   & $1^-(0^{++})$ & $\bar u (x) d (x)$ \\[2pt]
                   & $1^-(0^{++})$ & $\bar u (x) \gamma_4 d (x)$ \\[2pt]
Pseudoscalar       & $1^-(0^{-+})$ & $\bar u (x) \gamma_5 d (x)$ \\[2pt]
                   & $1^-(0^{-+})$ & $\bar u (x) \gamma_4 \gamma_5 d (x)$ \\[2pt]
Vector             & $1^+(1^{--})$ & $\bar u (x) \gamma_i d (x)$ \\[2pt]
                   & $1^+(1^{--})$ & $\bar u (x) \gamma_i \gamma_4 d (x)$ \\[2pt]
Axial ($a_1$)      & $1^-(1^{++})$ & $\bar u (x) \gamma_i \gamma_5 d (x)$ \\[2pt]
Tensor($b_1$)      & $1^+(1^{+-})$ & $\bar u (x) \gamma_i \gamma_j d (x)$ \\[2pt]
\myhline
                   &               &         \\[-4pt]
Nucleon octet      & ${1\over 2}({1\over 2} ^-)$ & $(u^T_a \CC d_b) \gamma_5 s_c \epsilon^{abc} $ \\[3pt]
                   & ${1\over 2}({1\over 2} ^-)$ & $(u^T_a \CC \gamma_5 d_b) s_c \epsilon^{abc} $ \\[3pt]
Delta decuplet     & ${3\over 2}({3\over 2} ^+)$ & $(u^T_a \CC \gamma_i d_b) s_c \epsilon^{abc} $ \\[3pt]
\myhline
\end{tabular}
\label{t:wilsonops}
\end{center}
\end{table}

The spin-1/2 baryons are created by the interpolating operators
\be
\label{eq:ebarydef}
{\cal O}_{(ij)k} = 
(\psi_{a,i}^T C\gamma_5 \psi_{b,j})\psi_{c,k} \epsilon^{abc} 
\,, 
\ee
where $a$, $b$ and $c$ label color, while $i$, $j$ and $k$ label flavor.
It is simple to show that ${\cal O}_{(ij)k} = -{\cal O}_{(ji)k}$,
so that there are only nine independent operators---eight $SU(3)$ octets
and the singlet $\sum_{ijk} \epsilon^{ijk} {\cal O}_{ijk}$.
One way to project against the singlet is to form
\be
\label{ebarydefn}
{\cal B}_{ijk} = {\cal O}_{(ij)k} + {\cal O}_{(ik)j} = {\cal B}_{ikj} \,.
\ee
There are eight independent $B_{ijk}$'s, the relation to the
usual states being exemplified by
\bea
\label{ebaryrel}
\sqrt{2} p	  &=& -{\cal B}_{duu}=2{\cal B}_{uud}= 2{\cal B}_{udu} \,,\nonumber \\
\sqrt{2}\Sigma^+  &=& -{\cal B}_{suu}=2{\cal B}_{uus}= 2{\cal B}_{usu} \,,\nonumber \\
\Sigma^0          &=& {\cal B}_{uds}+{\cal B}_{dsu} = - {\cal B}_{sud} \,.\nonumber \\
\sqrt{3}\Lambda^0 &=& {\cal B}_{uds}-{\cal B}_{dsu} \,. 
\eea
The overall factor in these equations is arbitrary,
while the relative normalization is fixed by $SU(3)$ symmetry.

\begin{figure} 
\hbox{\epsfxsize=\hsize\epsfbox{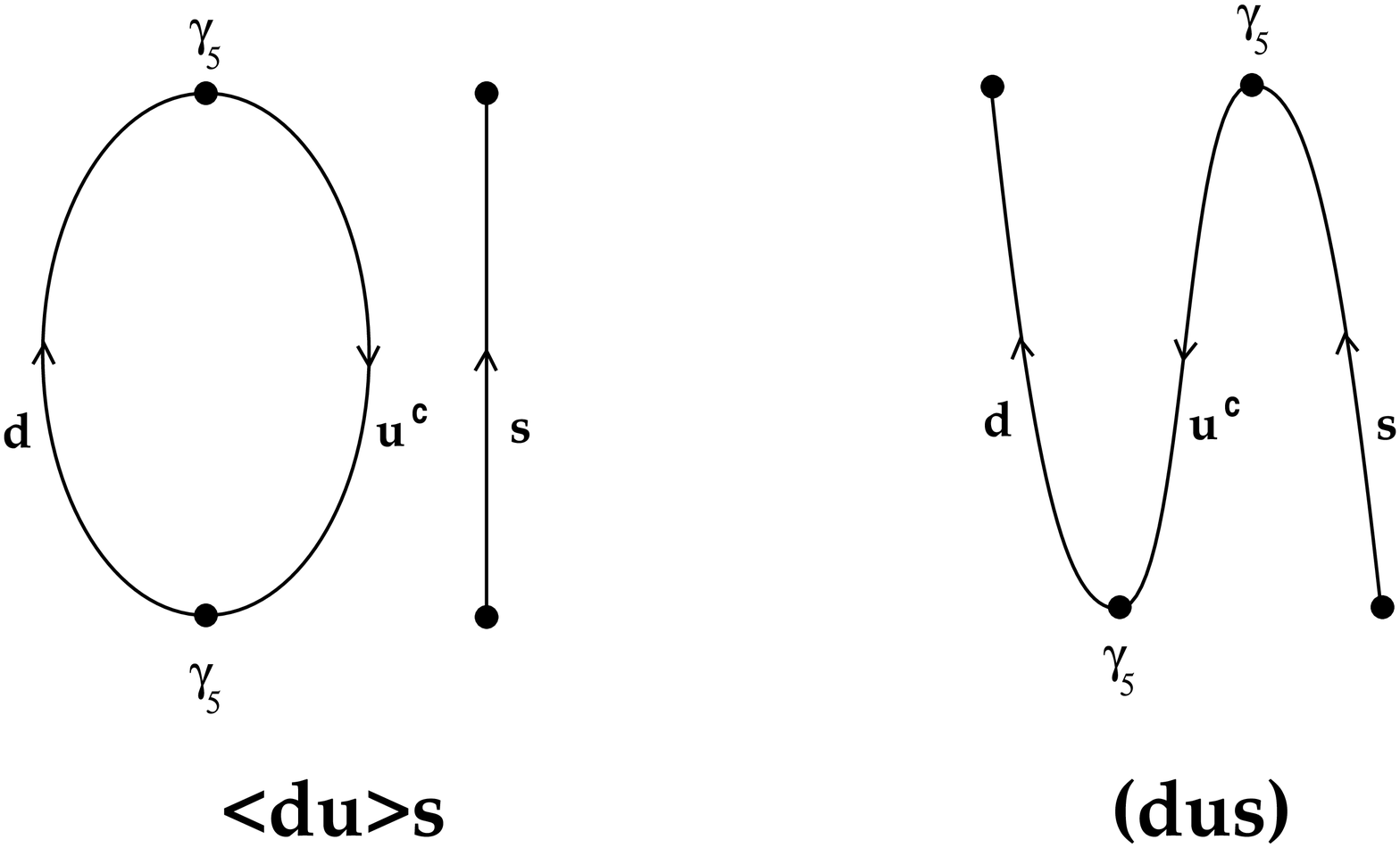}}
\caption{The two different types of contractions for the baryon states.}
\label{f:Bcontractions}
\end{figure}

All spin-1/2 baryon correlators are built out of the two contractions
shown in Fig.~\ref{f:Bcontractions}.  The notation $\langle DU \rangle
S = \langle UD \rangle S$ corresponds to quarks of flavors $U$ and $D$
contracted into a closed loop, while the propagator for $S$ carries
the spin quantum numbers of the baryon.  The notation $(DUS)$
corresponds to a single ordered contraction of the three quarks.  We
consider two types of correlator, ``$\Sigma$-like'' and
``$\Lambda$-like''.  The former is exemplified by that of the
$\Sigma^0$
\bea
\label{ebarysig}
S\{UD\} = S \{DU\} &\equiv& 
          \vev{{\cal B}_{sud}(x) \ \overline{{\cal B}_{sud}(0)}} \nonumber \\
        &=& \langle US \rangle D + \langle DS \rangle U + (USD) + (DSU) \,. 
\eea
This equation defines the sign conventions for the contractions.
The proton, neutron, $\Sigma^+$, $\Sigma^-$, $\Xi^0$ and $\Xi^-$
correlators are also of this type:
they are, respectively, $D\{UU\}$, $U\{DD\}$, $S\{UU\}$, $S\{DD\}$,
$U\{SS\}$ and $D\{SS\}$.
The second type of correlator is that of the $\Lambda^0$
\bea
\label{ebarylam}
S[UD] = S[DU] &\equiv& {1 \over 3} \bigg[ \langle US \rangle D 
                        + \langle DS \rangle U 
                        + 4 \langle UD \rangle S   \nonumber \\
                        &{}& \qquad -  (USD) - (DSU)  + 2(SUD)  \nonumber \\
                        &{}& \qquad  + 2(SDU) + 2(UDS) + 2(DUS) \bigg] \,. 
\eea
When $m_u\ne m_d$, there is also a non-vanishing
$\Lambda^0-\Sigma^0$ cross correlator, for which 
no useful results have been found ~\cite{HM96LANL}. 

Correlators of the form $A[AB]$ and $A\{AB\}$ 
are not independent---they are related by
\bea
\label{ebarynotindep} 
A[AB] 	&= {1\over4} \left( 3 B[AA] + B\{AA\} \right) \,, \nonumber \\
A\{AB\}	&= {1\over4} \left( B[AA] + 3 B\{AA\} \right) \,. 
\eea
One can think of the results for the $A\{BB\}$ and $A[BB]$ masses
as being those for the $\Sigma$ and $\Lambda$, respectively,
with $m_s=m_A$ and $m_u=m_d=m_B$.
Unlike in the real world, there is nothing to stop $m_A=m_B$. 
Note, however, that in this case the $\Sigma$ and
$\Lambda$ are also degenerate, i.e. $M(A\{AA\})=M(A[AA])$.
Indeed, the contractions in the two cases are identical.

The interpretation of the results for the completely non-degenerate
correlators, $A[BC]$ and $A\{BC\}$, is more complicated.  Because
isospin is broken, the $\Sigma^0$- and $\Lambda$-like states mix, with
both correlators containing contributions from both physical states.
Let $M_+$ and $M_-$ be the masses of the heavier and lighter states,
respectively, and $\delta\!M$ the mass difference.  At long times, the
effective mass for both correlators will asymptote to $M_-$.  However,
at times short compared to the inverse mass difference,
i.e. $\delta\!M t_{\rm max} \ll 1$, there will be an approximate
plateau at a value which is a weighted average of the two masses.  
The surprising result, derived at lowest order in the chiral 
expansion in ~\cite{HM96LANL}, is that 
the masses extracted from these short time plateau are insensitive 
to the isospin breaking and can be interpreted as those for 
$A[DD]$ and $A\{DD\}$ where $D = (B+C)/2$ is the mean mass for the two 
quarks. 

\subsection{Three-point correlators}
\label{ss:corr3pt}

A general three-point correlation function of meson operators and
bilinear currents has the form
\begin{eqnarray}
  C^{CBA}(\vec q, t_y; \vec p, t_x)  &{}& \nonumber \\
  &{}&\hskip -1.1in = \ \sum_{\vec x, \vec y} e^{-i ( q \cdot y + p \cdot x )}
    \VEV{ \bar\psi_3(y) \Gamma_C \psi_2(y)
          \bar\psi_2(x) \Gamma_B \psi_1(x)
          \bar\psi_1(0) \Gamma_A \psi_3(0) }, \nonumber \\
  &{}&\hskip -1.1in = - \sum_{\vec x, \vec y} e^{-i ( q \cdot y + p \cdot x )}
    \VEV{ {\rm Tr}\,\left( S_3(0,y) \Gamma_C S_2(y,x) \Gamma_B 
                     S_1(x,0) \Gamma_A  \right)    }, 
\label{eq:CPT3cfsigns}
\end{eqnarray}
where $t_x \equiv  x_4$   and   $t_y \equiv  y_4$. $\CT$,  $\CP$   and
$\CC\CP\CH$ imply 
\begin{itemize}
\item
  If $\tau_A\tau_B\tau_C=\pm1$, $C^{CBA}(\vec q, t_y; \vec p, t_x)$
  is even (odd) under the simultaneous inversion of $t_x$ and $t_y$.
\item
  If $\pi_A\pi_B\pi_C=\pm1$, $C^{CBA}(\vec q, t_y; \vec p, t_x)$ is
  even (odd) under the simultaneous inversion of $\vec p$ and
  $\vec q$.
\item
  If $\chi_A\chi_B\chi_C=\pm1$, $C^{CBA}(\vec q, t_y; \vec p, t_x)$
  is real (imaginary).
\end{itemize}

\subsection{$\CC \CP \CS $ Symmetry}
\label{ss:corrCPS}

The theory has an additional discrete symmetry, $\CC\CP\CS$, where
$\CS$ is the symmetry between switching $s$ and $d$ quarks
\cite{CPS85Bernard}. This symmetry holds both on the lattice and in
the continuum for $m_d = m_s$. For $m_d \ne m_s$ the symmetry is
softly broken, $i.e.$, violations are proportional to $m_s -
m_d$. This symmetry plays a very useful role in the calculation of
weak matrix elements for it (i) restricts the set of possible Wick
contractions, and (ii) restricts the set of operators that mix with
each other. 

For a simple illustration consider $(\bar s \gamma_\mu L u)( \bar u
\gamma_\mu L d)$, a local 4-quark operator which is even under
$\CC\CP\CS$
\bea
(\bar s \gamma_\mu L u)( \bar u \gamma_\mu L d )
	&{\stackrel{\CP}{\rightarrow}}& 
(\bar s \gamma_4 \gamma_\mu L \gamma_4 u)( \bar u \gamma_4 \gamma_\mu L \gamma_4 d ) \nonumber \\
	&{\stackrel{\CC}{\rightarrow}}&
(s^T \CC^{-1} \gamma_4 \gamma_\mu L \gamma_4 \CC \bar u^T)
(u \CC^{-1} \gamma_4 \gamma_\mu L \gamma_4 \CC \bar d^T ) \nonumber \\
&=& (s^T \gamma_4 \gamma_\mu^T L \gamma_4 \bar u^T)
    (u^T \gamma_4 \gamma_\mu^T L \gamma_4 \bar d^T ) \nonumber \\
&=& (s^T \gamma_\mu^T R^T \bar u^T)(u \gamma_\mu^T R^T \bar d^T ) \nonumber \\
&=& (\bar u R \gamma_\mu s)(\bar d  R \gamma_\mu u ) \nonumber \\
&=& (\bar u \gamma_\mu L s)(\bar d   \gamma_\mu L u ) \nonumber \\
	&{\stackrel{\CS}{\rightarrow}}&
(\bar u \gamma_\mu L d)(\bar s  \gamma_\mu L u )
\label{eq:CPSillustration}
\eea
The brackets denote trace over spin and color and the final reordering
of the two traced terms is allowed as time-ordering is implicit when
calculating expectation values.

This symmetry has a further generalization as discussed by Bernard and
Soni in \cite{Creutz92,TASI89}.  For 4-fermion operators of
the form $(\bar \psi_1 \Gamma_A \psi_2)( \bar \psi_3 \Gamma_B
\psi_4)$, there are two general switchings: (i) $ \CC\CP\CS'$ corresponding
to $1 \leftrightarrow 2$ and $3 \leftrightarrow 4$, and (ii) $
\CC\CP\CS''$ corresponding to $1 \leftrightarrow 4$ and $2
\leftrightarrow 3$. Bernard and Soni show in \cite{Creutz92,TASI89}
how these switching
symmetries are used to restrict (i) the possible contractions in the
decay $K \to \pi\pi$ via the operator $(\bar s \gamma_\mu L u)( \bar u
\gamma_\mu L d)$, and (ii) the mixing of 
$\CO_{\pm} \equiv (\bar s \gamma_\mu L d)( \bar u \gamma_\mu L u) \pm 
                  (\bar s \gamma_\mu L u)( \bar u \gamma_\mu L d) - (u \to c)$ 
with other operators. I leave the details of these calculations 
as an exercise for the reader.

%% file: chap-hm.tex
\section{Lattice Calculations of the Hadron Spectrum}
\label{s:hadronspectrum}

Successful calculations of the spectrum will test QCD.  The basic
steps in the lattice analysis are discussed in
Sections~\ref{s:overviewLQCD} and \ref{s:corrfns}.  For a given set of
input parameters, the mass of a given state is determined from the
rate of exponential fall-off of the connected 2-point correlation
function
\begin{equation}
\Gamma(\tau) \equiv \langle O_f(\tau) O_i(0) \rangle - 
                         \langle  O_f \rangle \ \langle  O_i \rangle
 = \ \sum_n {\ME{0}{O_f}{n} \ME{n}{O_i}{0}  \over 2 M_n} e^{-M_n \tau} .
\label{eq:corrfuncdef}
\end{equation}
We shall assume that the sum over $n$ is restricted to the desired
state by (i) optimizing $O_i$ and $O_j$ to get a large overlap with
the wave function, $i.e.$ make $\ME{desired\ state}{O}{0}$ large and
the rest small, (ii) increasing the statistics so that the signal
extends to large enough $\tau$ at which any remaining contamination
from higher states is negligible; and (3) making a zero-momentum
projection on either $O_i$ or $O_f$. I shall further assume that the
spectrum of possible states is discrete and has a mass-gap. Then, in
order to extract $M$ it is useful to define an effective mass
\begin{equation}
M_{eff}(\tau) \ = \ \log \big( {\Gamma(\tau-1) \over \Gamma(\tau) } \big)
\label{eq:defmeff}
\end{equation}
which, in the limit $\tau \to \infty$, converges to the desired value.
The anatomy of the behavior of $M_{eff}(\tau)$ for two different
states (pion and nucleon) and for two different choices of $O_i$ and
$O_f$ are shown in Figs.~\ref{f:pionmeff} and \ref{f:nucleonmeff}. The
points to note are:

\begin{figure} 
\hbox{\epsfxsize=\hsize\epsfbox{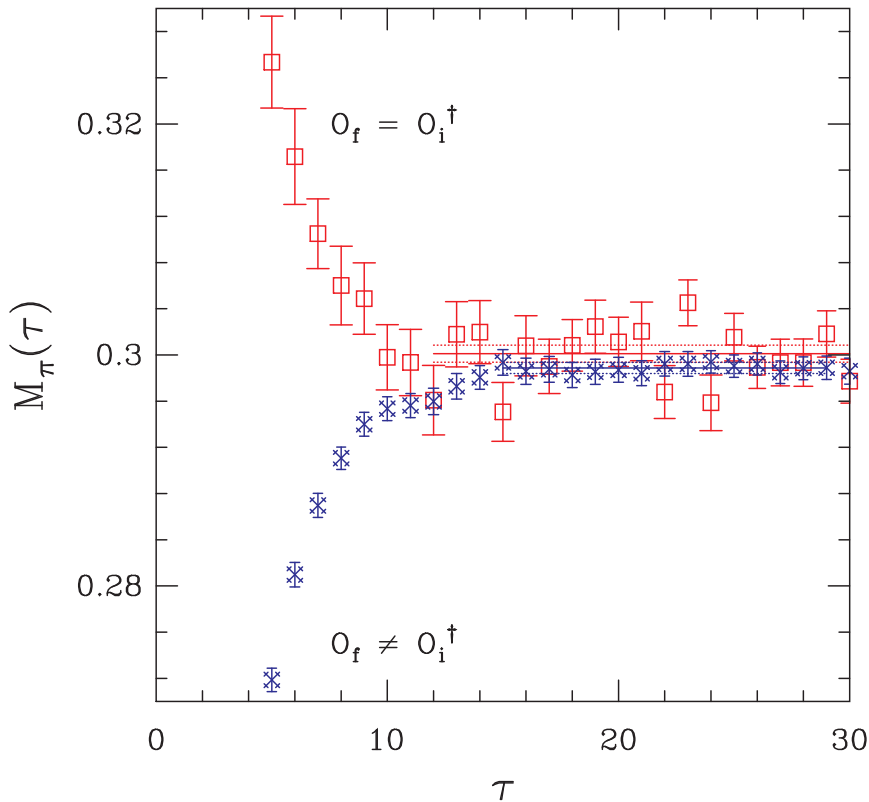}}
\caption{Effective mass plots for $M_\pi$ versus $\tau$. The
horizontal lines denote the estimate of the asymptotic value of $M$
and the range of the plateau used in the fits, while the dotted lines show
the error. The squares denote data for smeared-smeared
interpolating operators ($\CO_f = \CO_i^\dagger$), while fancy crosses
show data with wall-local operators. }
\label{f:pionmeff}
\end{figure}

\begin{figure} 
\hbox{\epsfxsize=\hsize\epsfbox{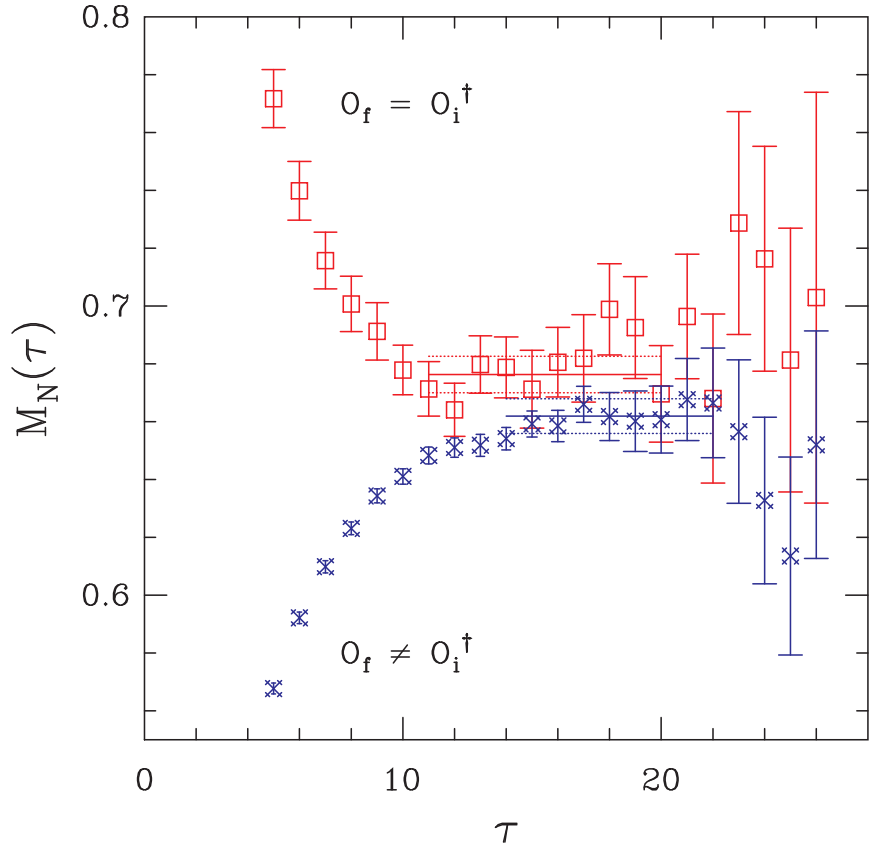}}
\caption{Effective mass plots for $M_N$ versus $\tau$. The rest is 
same as in Fig.~\ref{f:pionmeff}. }
\label{f:nucleonmeff}
\end{figure}

\begin{figure}[htbp] 
\hbox{\epsfxsize=\hsize\epsfbox{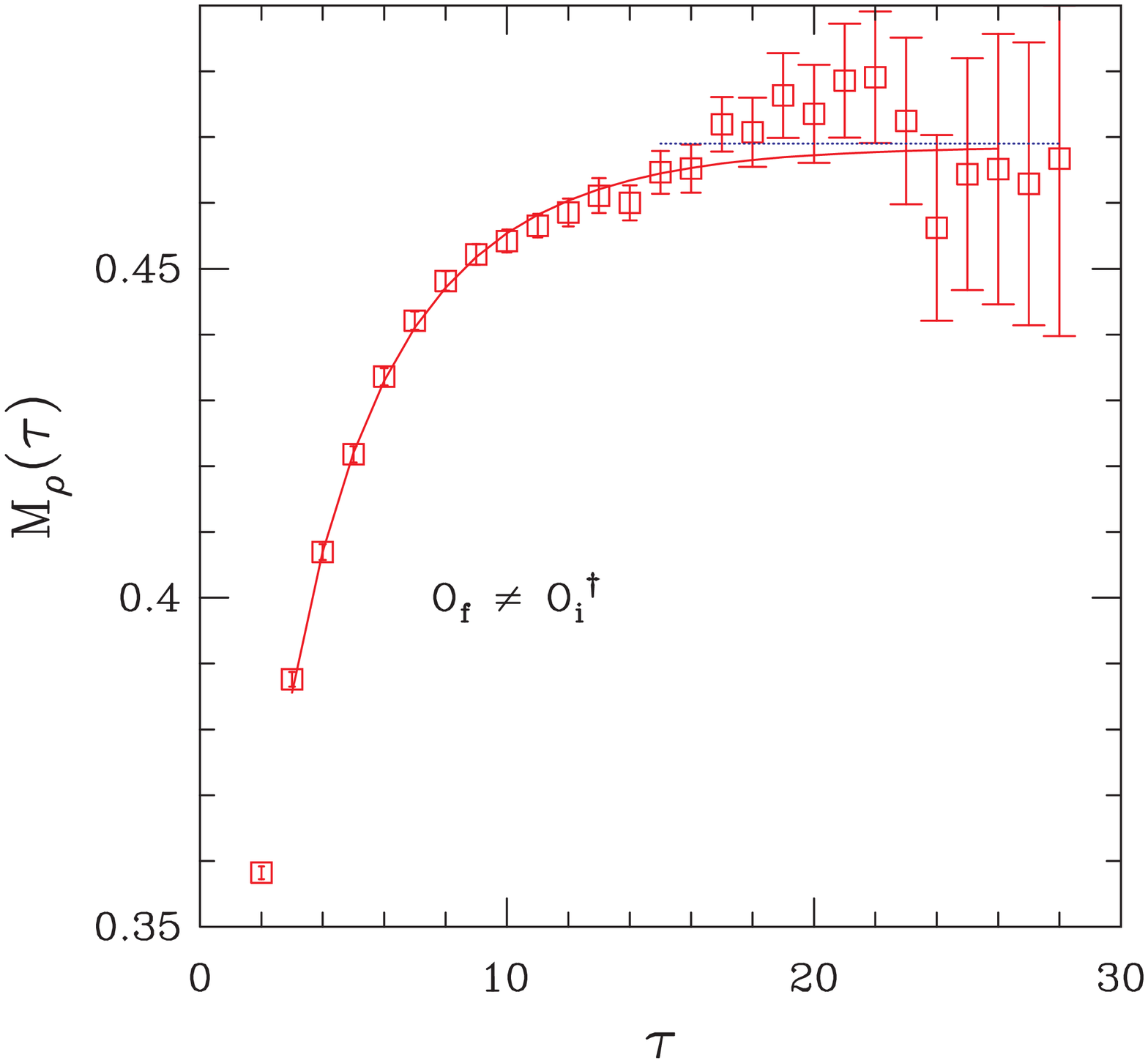}}
\caption{Effective mass plots for $M_\rho$ versus $\tau$ for the wall-local 
source. Since there is no clear plateau, a fit including two states is
made to $3 \le \tau \le 26$. From such fits the masses of $\rho$ and
$\rho^*$ are obtained. The mass of the $\rho$ from the 2-state fit is
consistent with that from a single state fit to data at $ \tau > 15$
(blue dotted line).}
\label{f:rhomeff}
\end{figure}

\begin{itemize}
\item
The convergence to the asymptotic value $M$ can be from above or below
depending on the choice of $O$. Only for $O_f = O_i^\dagger$ is the
correlation function positive definite and the convergence is
monotonic and from above.

\item
For $O_f \ne O_i^\dagger$, the convergence depends on subtle
cancellations between contributions of various states. The relative 
signs are given by the product of the two amplitudes 
$\ME{n}{O_{f}}{0}\ME{n}{O_{i}}{0}$. Since this sign for 
state $|n\rangle$ can be plus or minus, the convergence
is neither guaranteed to be monotonic nor from above.

\item
The contribution of excited states is most obvious at small $\tau$,
where $M_{eff}$ varies rapidly before flattening out. For $\tau$ in
this flat (called plateau) region one assumes that, within the 
numerical precision, only the lowest state contribution is significant. 

\item
The location in $\tau$ of the onset and the length of the plateau region
depends on the interpolating operators.

\item
For a finite statistical sample the plateau region is not flat but the
data usually exhibit correlated fluctuations \cite{JAPANcorrelations}.
Therefore, unless the plateau is long enough that the fit can average
over a few cycles, the extracted $M_{eff}$ can be biased by these statistical
fluctuations.

\item
The statistical errors grow with $\tau$, except for 
the case of the pion. The reason for this was elucidated by Lepage in 
\cite{Lepagetasi90} and a synopsis of his argument is as follows. For a 
state with $N$ valence quark lines, where $N=2(3)$ for mesons
(baryons), the errors are controlled by the square of the correlator
which has $2N$ lines. The state of lowest energy in the squared
correlator is, therefore, two pions for mesons and three pions for
baryons.  Thus, while the signal decreases as $\exp{(-M \tau)}$, the
noise only decreases as $\exp{(-M_\pi \tau \times N/2)}$.
Consequently, all states for which $ M > N M_\pi /2$ the errors will
increase with $\tau$. The only exception to this condition is the pion
for which the signal and noise have the same $\tau$ dependence.

\item
In cases where there is no clear plateau (the signal does not extend
far enough in $\tau$), one can make a fit to a larger range of $\tau$
by including the contribution of higher (radial) states in the sum in
Eq.~\ref{eq:corrfuncdef}. This is illustrated in Fig.~\ref{f:rhomeff}
for a two state fit which involves four free parameters, two
amplitudes and two masses.  The amplitude and mass of the excited
state are less well constrained as they are trying to mimic the effect
of a number of states, consequently such fits may show significant
dependence on the range of $\tau$ selected. One can refine the excited
state parameters by iteratively improving single state fits to
$\Gamma(\tau) - \Gamma_0(\tau)$ at short $\tau$ and to the ground
state $\Gamma_0(\tau)$ at large $\tau$ until the whole range is
covered.  A better approach to gain confidence in ground state
parameters is to make simultaneous fits to a matrix of correlation
functions created with different source and sink operators with the
same quantum numbers. For a matrix with $m$ ($n$) distinct sources
(sinks) the simultaneous fit is more constraining as it involves
$m+n$ amplitudes but a single common mass.

\end{itemize}

To summarize, we mostly extract $M$ from fits to an intermediate range
of $\tau$, the ``plateau'' region, where $M_{eff}$ is essentially
independent of $\tau$. At small $\tau$ the influence of excited states
is large, while at large $\tau$ the statistical noise overwhelms the
signal. Accuracy of the result depends on the presence of a plateau
extending over a large range of $\tau$ that averages a few cycles of
correlated fluctuations.  The data in Figs.~\ref{f:pionmeff},
\ref{f:nucleonmeff}, and \ref{f:rhomeff} show that, with 170 lattices 
of size $32^3 \times 64$ at $1/a =2.3$ GeV, this is true for the pion
but marginal for other states like the vector mesons and baryons.  Our
present estimate is that roughly $1000$ independent lattices are
necessary to get a reliable signal that leads to $0.1-0.4 \%$ accuracy
in baryon masses with quark masses in the range $m_s - m_s/4$
respectively.

Assuming that the above steps have been carried out and we have
reliable estimates for hadron masses as a function of quark masses,
$a$, $L$, and the number $n_f$ of dynamical flavors, we are finally in
a position to compare them to experimental data.

\subsection{Status of the Light Meson and Baryon Spectrum}
\label{ss:hm}

The first test of QCD is to reproduce the masses of pseudoscalar
mesons ($\CP\CS = \{\pi, K, \eta \}$), vector mesons ($\CV=\{\rho,
K^*, \omega, \phi \}$), octet baryons ($\CB_8= \{ N, \Sigma, \Xi, \Lambda
\}$), and decuplet baryons ($\CB_{10} = \{ \Delta, \Sigma^*, \Xi^*, \Omega \}$)
from simulations with four input parameters $m_u, m_d, m_s, $ and
$\alpha_s$. I will assume that the lattices are large enough that finite
size effects can be neglected. The remaining important issues are
\begin{itemize}
\item
Simulations have mainly been done in the quenched
approximation. Rather than testing QCD, these calculations 
quantify the accuracy of QQCD.
\item
Isospin breaking effects and electromagnetic contributions are
neglected. 
\item
Simulations have been done with unphysical values of quark masses,
typically in the range $[2m_s - m_s/4]$. To get physical results the
data are extrapolated to $\bar m = (m_u+m_d)/2$ using functional forms
predicted by (quenched) chiral perturbation theory.
\end{itemize}

The first step in the analyses is to extrapolate the data to physical
values of $\mbar$ and $m_s$. The simplest {\it Ans\"atze} are the lowest order
chiral expressions
\begin{eqnarray}
M_{\CP\CS}^2 a^2 &=& \qquad\ \ B_{\CP\CS}\ a^2 (m_1 + m_2)/2  + \ldots , \nonumber \\ 
M_{\CV}      a   &=& A_{\CV} + B_{\CV}\ a (m_1 + m_2)/2  + \ldots , \nonumber \\ 
M_{\Sigma}   a   &=& A_{N} + 4 F\ \mbar a + 2(F-D)\ m_s a  + \ldots , \nonumber \\ 
M_{\Lambda}  a   &=& A_{N} + 4(F-{2D \over 3})\ \mbar a + 2(F+{D \over 3})\ m_s a + \ldots , \nonumber \\ 
M_{10}       a   &=& A_{10}+ B_{10}\ a (m_1 + m_2 + m_3)/3  + \ldots \ .
\label{eq:chiralforms}
\end{eqnarray}
From these fits we determine the constants $A_{\CV}, A_N, A_{10}$ and
$B_{\CP\CS}$, $B_{\CV}$, $F, D, B_{10}$.  Then using a suitable set, for
example $A_{\CV}, B_{\CP\CS}, B_{\CV}$ or $A_{10}$, $B_{\CP\CS}$,
$B_{10}$, we fix the three inputs parameters $1/a, \mbar, m_s$.  Unless
otherwise stated the following discussion will assume that $\mbar$,
$a$, and $m_s$ have been fixed using $M_\pi, M_\rho$, and $M_\phi$ (or
equivalently $M_{K^*}$) respectively. The question then is -- how well do
the masses of the remaining states compare with experimental numbers?

This simple analysis to test QCD (or to quantify the quenched
approximation) can fall short in two ways. First, the data could show
significant deviations from the linear relations
Eq.~\ref{eq:chiralforms}. In that case the analysis becomes much more
elaborate due to the increase in the number of free parameters; the 
possible leading non-linear terms include $O(m^2)$ corrections and
chiral logarithms. To disentangle these requires very precise data at a
number of values of $m_q $ in the range $m_s/4 - 3 m_s$, however, such
a possibility of physics richer than that predicted by the ``naive''
quark model is exciting. The second possible failing is that different
sets of states used to fix $a, \mbar, m_s$ give different
results. There could be three reasons for this failure: (i) The
coefficients extracted from simulations at the significantly heavier
values of $m_q$ change as the quark masses are decreased; (ii) the
differences are artifacts of discretization errors and vanish in the
limit $a = 0$; (3) quenching errors. The current data show evidence
for non-linearities in fits in
$m_q$~\cite{HM96LANL,HM96gottlieb}. Also, the $a, \mbar, m_s$
extracted from different sets are significantly different, and 
all three causes listed above contribute.  As a result the 
focus of current analyses has been to first disentangle and quantify
the contributions of the various artifacts.  I will briefly illustrate
these effects.

\begin{figure} 
\hbox{\epsfxsize=\hsize\epsfbox{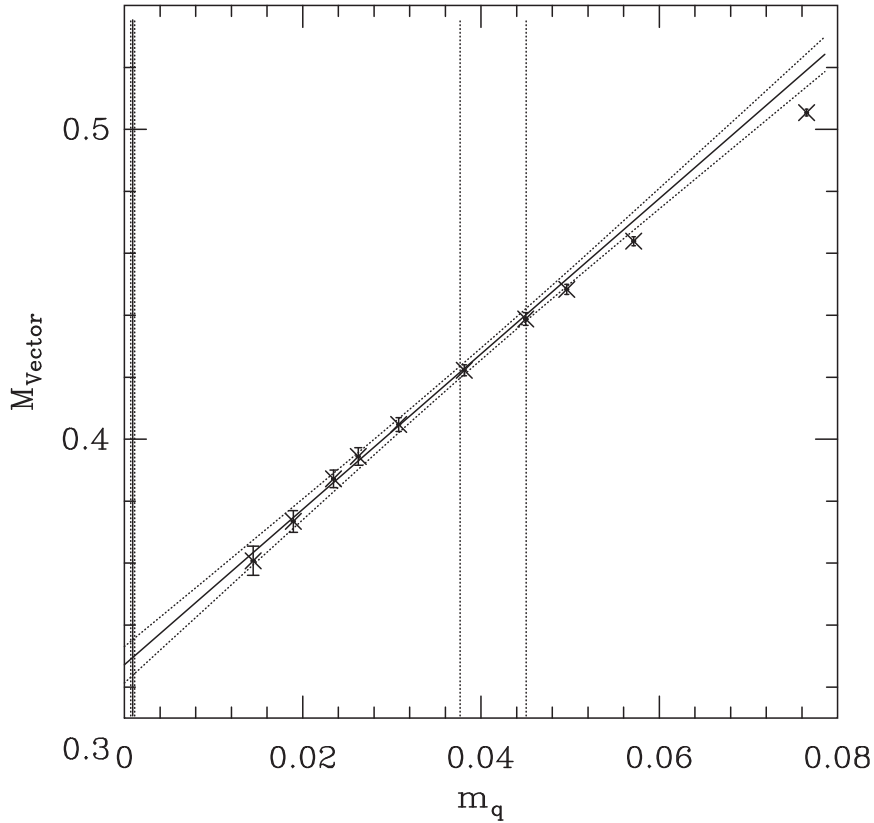}}
\caption{Plot of $M_{vector}$ versus the quark mass $m$.  The data, reproduced
from \cite{HM96LANL}, were obtained on $32^3 \times 64$ quenched
lattices at $\beta\equiv 6/g^2=6$ corresponding to $1/a=2.33(4)$
GeV. The linear fit is to the lightest six points. The vertical lines
show \mbar\ and the range of estimates of $m_s$.}
\label{f:hmrhofit}
\end{figure}

Let me start with evidence for non-linear terms in the chiral fits.
Fig~\ref{f:hmrhofit} is an example of the behavior of the vector meson
mass as a function of the quark mass \cite{HM96LANL}.  The plot makes
it clear that in order to quantify the deviations one needs very
precise data over a large range of $m_q$.  With a statistical accuracy
of $\sim 0.5\%$, a linear fit works for $0.02 \le \mbar a \le 0.04$
corresponding to $50 \le \mbar \le 100$ MeV, but not for the full
range.  The leading chiral corrections are proportional to $\mbar^{3/2}$
from chiral loops, and $\mbar^2$. Including either term allows fits over
the full range, and the data are not precise enough to distinguish
between them.  The effect of these non-linearities seems to be small,
for example the change in $M_{\rho}$, after extrapolation to
$\mbar$, is $\lsim 1\%$ on including either of the higher order
terms in the fit.

\begin{figure}[t*] 
\hbox{\epsfxsize=\hsize\epsfbox{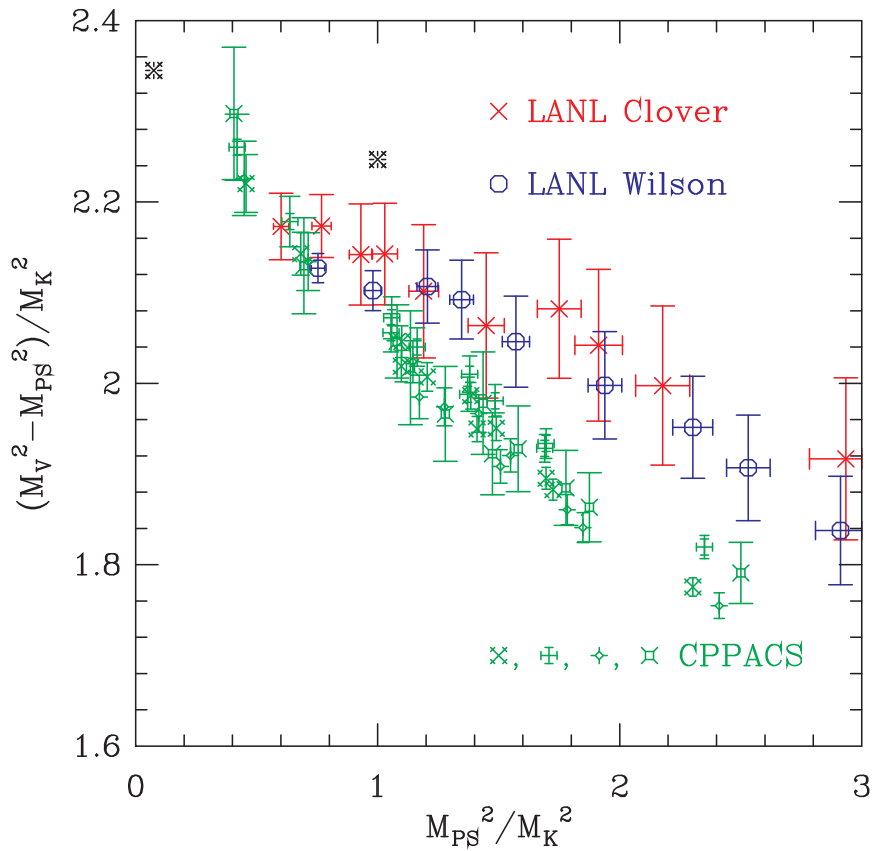}}
\caption{Hyperfine splittings in the meson sector. Quenched data 
from (i) LANL collaboration with Wilson fermions (blue octagons)
\cite{HM96LANL}, (ii) LANL collaboration with tadpole improved clover 
fermions (red crosses)~\cite{HM98LANL}, and (iii) CP-PACS
collaboration Wilson runs at $\beta=5.9, 6.1, 6.25, 6.47$ (green fancy
crosses, pluses, diamonds, and squares) \cite{HM97yoshie}. The
experimental points are shown by symbol burst.}
\label{f:hyperfine}
\end{figure}

The first test of the quenched approximation is whether it can
reproduce the hyperfine splittings in the meson sector. Figure
\ref{f:hyperfine} shows some current data. 
The only input quantity is the scale $1/a$ which is used to convert
lattice masses to physical units. If $M_\pi$ and $M_\rho$ are used to
set $\mbar$ and $1/a$ respectively, then the curves, by
construction, have to pass through the experimental point at
$M_{\CP\CS} = M_\pi$.  Different choices for the scale setting
quantity will change $1/a$ and shift the curves as a whole. On the
other hand, the slope at any given point is independent of the lattice
scale, and is therefore the more robust lattice prediction.  For this
reason the quantity $J \equiv M_V
\partial M_V/\partial M_{PS}^2$ is commonly used as a test of the
hyperfine splittings. Lattice data (for a review see
\cite{HM97yoshie}) gives values for $J$ that are $\sim 20\%$ too small. 
Another manifestation of this discrepancy will appear in
Section~\ref{s:mq} where I show that the estimates of $m_s$ fixed
using $M_K$ versus $M_{K^*}$ (or $M_\phi$) differ by $\sim 15\%$.

The data shown in Fig.~\ref{f:hyperfine} allow two
comparisons. First, the octagens (blue points) and crosses (red) are
LANL data from the same 170 quenched $32^3 \times 64$ lattices at
$\beta=6.0$ with the Wilson and tadpole improved clover Dirac
actions~\cite{HM96LANL,HM98LANL}. One does not find any significant
improvement with clover fermions. Second, the fancy crosses, pluses,
diamonds, and squares (green points) are Wilson fermion data from
CP-PACS collaboration at $\beta=5.9, 6.1, 6.25, 6.47$
\cite{HM97yoshie}. The data at the four $\beta$ values are
consistent. Both comparisons imply that the smaller splittings are not
due to discretization errors.

The data in Fig.~\ref{f:hyperfine} also show a difference between the
CP-PACS and LANL Wilson fermion results at quark masses heavier than
$m_s$.  This is surprising since in this regime both collaborations
have reliable measurements of the vector and pseudoscalar masses.  The
culprit is the lattice scale. It turns out that the estimate of $1/a$
from LANL data is $3\%$ larger compared to the value obtained by an
interpolation of the CP-PACS data.  Adjusting the LANL scale shifts 
their data as a whole and makes all the results overlap.

The conclusion, from this and other quenched data, is that the
splittings come out too small.  Even though the size of discretization
errors is, in my opinion, not fully resolved, the deviations are
thought to be mostly due to the use of the quenched approximation.

\begin{figure}[t*] 
\hbox{\epsfxsize=\hsize\epsfbox{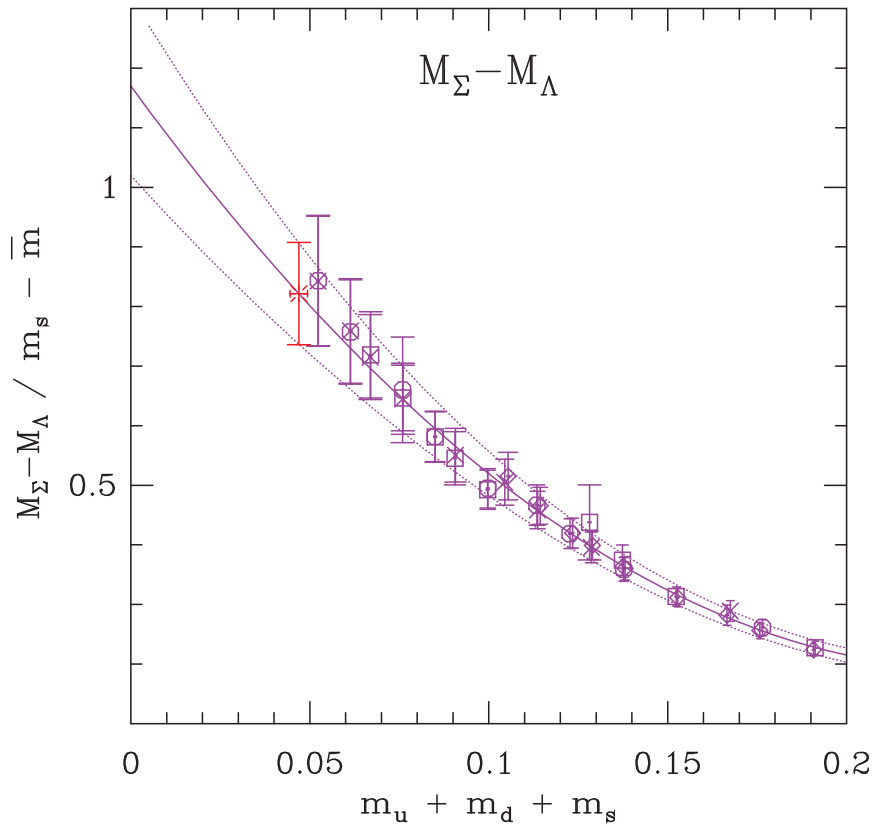}}
\caption{Quadratic fit to $M_\Sigma - M_\Lambda / (\mbar - m_s) $,
including baryons composed of completely non-degenerate quarks
\cite{HM96LANL}.  The value extrapolated to the physical point is
shown by the burst symbol at the extreme left. The linear relations
shown in Eq.~\ref{eq:chiralforms} would predict a constant value.}
\label{f:sigmalambda}
\end{figure}

The most extensive discussion of the mass-splittings in the baryon
octet and decuplet using the Wilson action is given in
\cite{HM96LANL}.  Note that since this calculation was done at only one $1/a \approx 2.3$
GeV, therefore discretization errors are not resolved.  There are two
important observations. (i) The mass splittings in the baryon octet
show significant non-linearities, for example the dependence of
$M_\Sigma - M_\Lambda$ on the quark mass is shown in
Fig.~\ref{f:sigmalambda}. The non-linear terms significantly increase
the splittings, and extrapolations to $\mbar$ give results roughly
consistent with experimental values. (ii) The decuplet baryon masses
show a behavior linear in $m_1+m_2+m_3$, and the splittings come out
too small for either choice $m_s(M_K)$ or $m_s(M_\phi)$.

\begin{figure}[t*] 
\hbox{\epsfxsize=\hsize\epsfbox{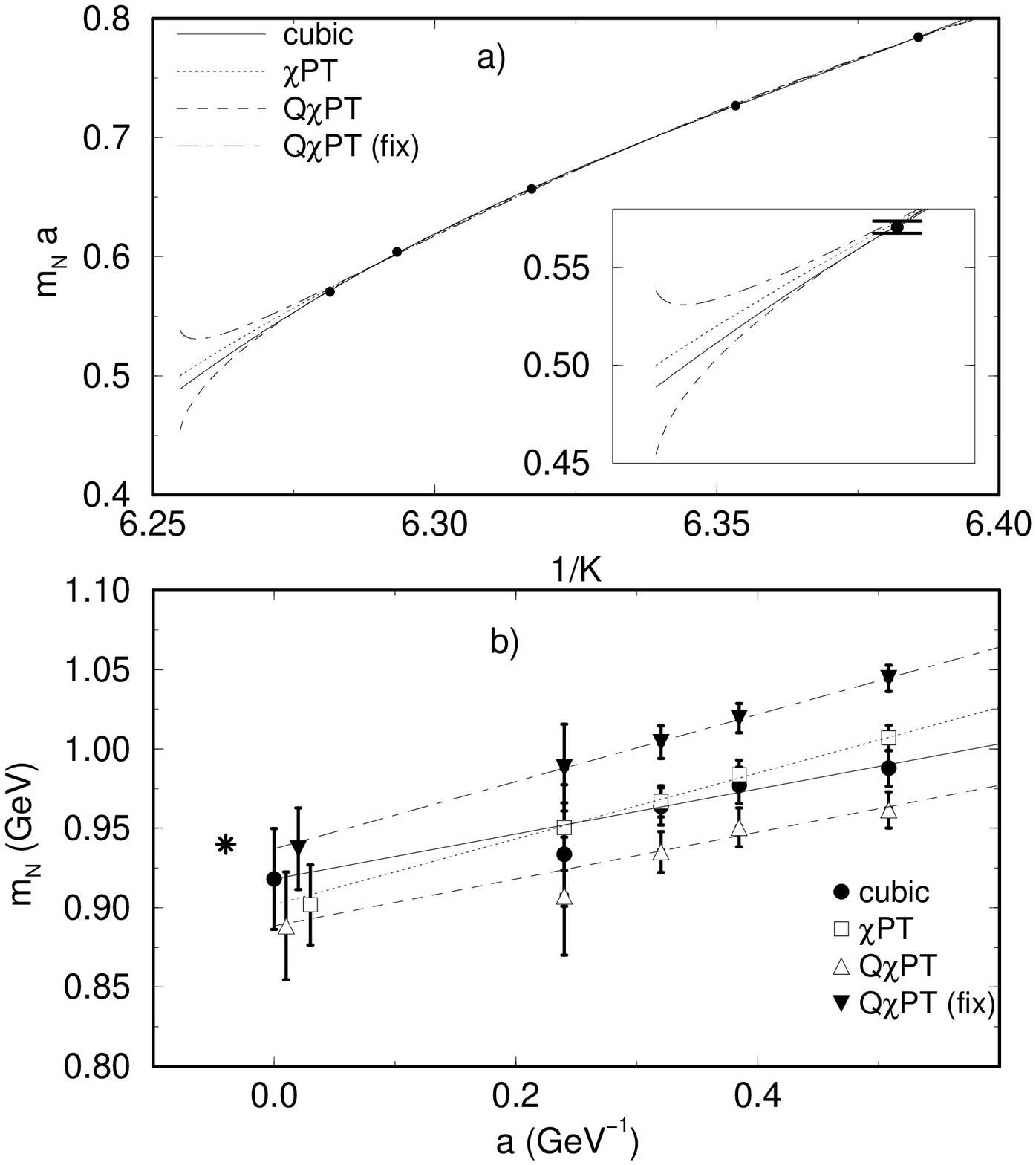}}
\caption{(A) Chiral fit to $M_{nucleon}$ data by the CP-PACS collaboration 
using the Wilson gauge and fermion actions. The data show negative 
curvature, though more points are needed to resolve between the cubic
fit $M_N = c_0 + c_1 M_\pi + c_2 M_\pi^2 + c_3 M_\pi^3 $, and $\chi$PT
($c_1=0)$, Q$\chi$PT ($c_3=0)$, $\chi$PT (fix) ($c_1=-0.53)$ fits. (B)
Continuum extrapolation of the nucleon mass obtained using the different 
chiral expressions are shown in (A).
The figures are reproduced from \cite{HM97yoshie}. }
\label{f:hmnucfit}
\end{figure}

\begin{figure} 
\hbox{\epsfxsize=\hsize\epsfbox{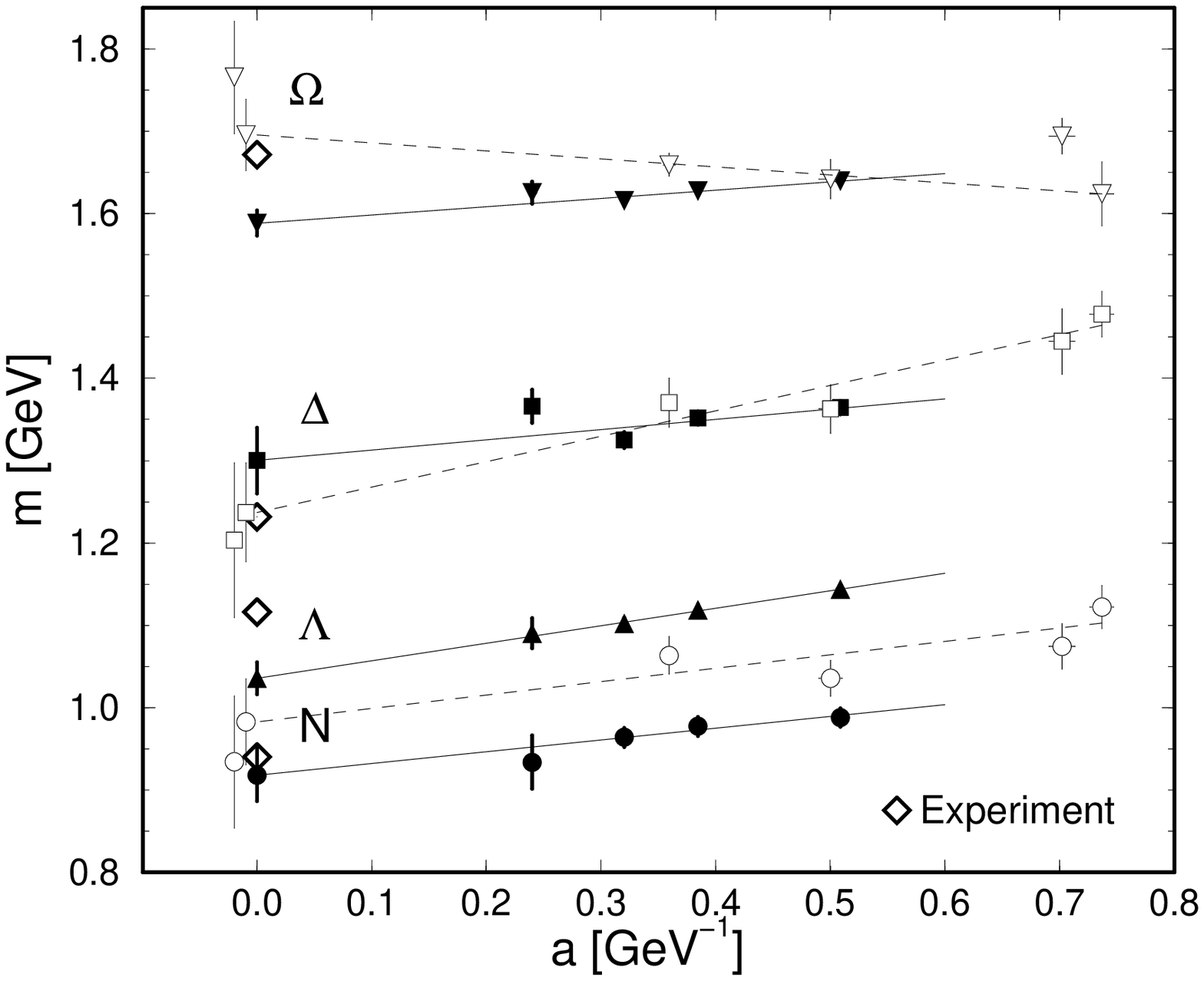}}
\caption{Continuum extrapolation of the nucleon, $\Lambda$, $\Delta$, 
and $\Omega$ masses obtained by the CP-PACS collaboration. The scale
$a$ is set by $M_\rho$.  The previous best results by the GF11
collaboration (unfilled symbols) are also shown. Experimental values
are shown by the symbol diamond. Figure reproduced from
\cite{HM97yoshie}. }
\label{f:hmbaryfit}
\end{figure}

The state-of-the-art Wilson fermion data have been obtained by the
CP-PACS collaboration. Data have been obtained at four values of the
lattice spacing so a reliable extrapolation to $a=0$ can be made.
Unfortunately, the data, in particular the baryon mass-splittings have
not yet been fully analyzed so some of the observations presented here
should be considered preliminary.  Fig.~\ref{f:hmnucfit}, taken from
Yoshie's review at LATTICE 97~\cite{HM97yoshie}, is CP-PACS's version
of the chiral fits.  They find that in the range $m_s/4-m_s$ only the
nucleon and the $\Lambda$ particles show any significant curvature.
Even though data at more values of the quark mass are required to
resolve the form of the leading higher order correction, the
conclusion of their analyses is that the negative curvature found in
their fits to nucleon data is sufficiently large to change the
quenched estimate of $M_N/M_\rho$ to almost perfect agreement with the
experimental value $\sim 1.22$. (This negative curvature was not
seen/resolved in calculations reported before 1997, consequently a
higher value in the range $1.3-1.4$ was usually reported.)  Their
extrapolation to $a=0$ of results with different chiral fits is shown
in Fig.~\ref{f:hmbaryfit}.

Second, they confirm that the decuplet masses are linear in the
quark masses, and the splittings, even after extrapolation to $a=0$,
come out too small with either $m_s(M_K)$ or $m_s(M_\phi)$.  Their 
full analyses on splittings in the baryon octet will appear soon. 

\begin{figure} 
\hbox{\epsfxsize=\hsize\epsfbox{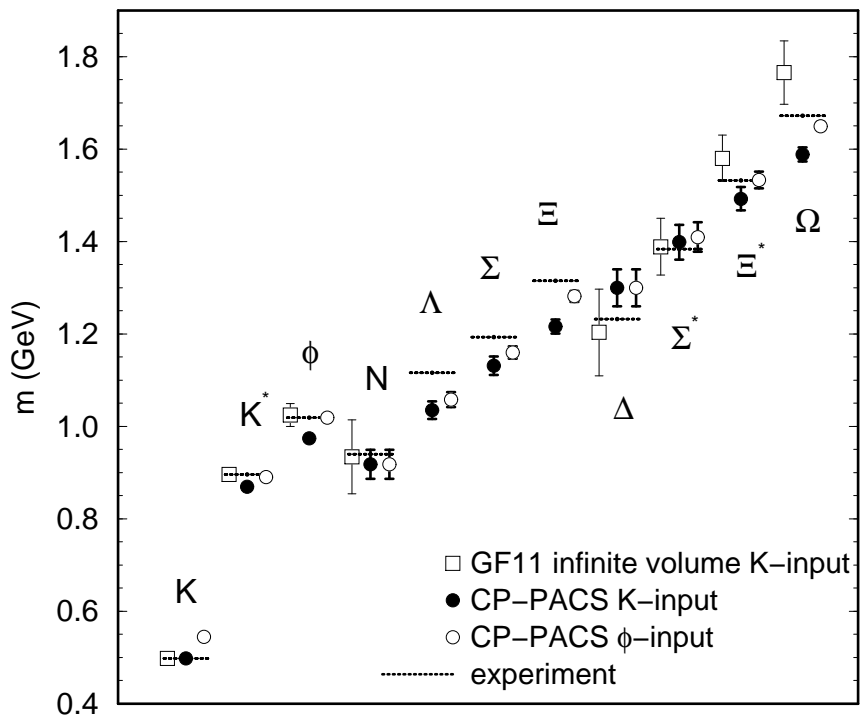}}
\caption{State-of-the-art results from the CP-PACS
collaboration for the meson and baryon masses after $a=0$ extrapolation and 
using Wilson fermions. The scale $a$ is set using $M_\rho$ and 
$\mbar$ using $M_\pi$. Results are shown for two ways of setting $m_s$ -- $M_K$ and 
$M_\phi$ and it is clear that neither give correct splittings in the baryon 
octet and decuplet. The data are reproduced from \cite{HM97yoshie}.}
\label{f:hmWILSON}
\end{figure}

A final summary of the quenched results taken from \cite{HM97yoshie}
using the Wilson action is given in Fig.~\ref{f:hmWILSON}. This figure
compares the recent CP-PACS results extrapolated to $a=0$ with the
previous best calculation by the GF11 collaboration. The main lesson
to be learned from this plot is that there does not seem to be a value
of the strange quark mass that allows a matching of all the meson and
baryon states. I shall come back to this point when discussing the
extraction of quark masses. The obvious question is -- is this
deviation an artifact of the quenched approximation? My view is that
we still do not have full control over the quenched data, and
therefore cannot point to quenching as the sole culprit. For example,
the definition of $m_q$ itself has corrections of $O(m_qa)$,
incorporating which would change the nature of the chiral fits. To
fully understand the discretization errors we need to wait for
similarly good quality data with the staggered and $O(a)$ improved
clover formulations before claiming definitive quenched values.

As of this writing dynamical simulations have begun in earnest, 
but precise quantitative results at a number of values of 
$a$ and with different actions are lacking. Thus, I feel it is 
too early to draw any conclusions from them. For a status report
on $n_f=2$ results see the recent reviews by Yoshie and Gottlieb
\cite{HM97yoshie,HM96gottlieb}.

\subsection{Decay Constants}

In addition to the extraction of masses from the rate of exponential decay
of the 2-point correlation function, the amplitudes $
\ME{n}{O}{0}$ give decay constants. For example, the axial
current matrix element $ \ME{\pi}{A_\mu(x)}{0} = i f_\pi p_\mu
\exp(ip \cdot x)$ gives the pion decay constant. Some of the recent 
results for decay constants have been reviewed by Prof. Martinelli at 
this school, and other recent reviews can be found in 
\cite{flynn96rev,onogi97rev}.   The technical points I
would like to explain here are the reasons why the errors in decay
constants are much larger than those in the determination of the
masses. First, from Eq.~\ref{eq:corrfuncdef} one notes that for a
$1\%$ error in the determination of the mass, the fractional error in
the decay amplitude is $\approx \sqrt{\exp(0.01 m \tau)} - 1 \approx
0.005m \tau$.  Since fits typically have $m\tau \approx 5$ for $1/a
\sim 2$ GeV lattices, the error in the decay constant is $\gsim 3\%$.
Second, for any given configuration, the correlation function is
highly correlated in $\tau$. Plotting $\log(\Gamma(\tau))$ for each
individual configuration one notes that the dominant fluctuation is in
the overall normalization, $i.e.$, the amplitude.  A very rapid growth
in these fluctuations with decreasing $m_q$ is, in fact, the signal
for exceptional configuration as discussed in
Section~\ref{s:wilsonfermions}.  These two sources of error combine to
produce a statistical error in decay constants that is much larger
than in the extraction of $M_{eff}$.  To this one has to add a systematic 
uncertainty coming from the renormalization constants for the local currents. 

\subsection{Lattice Calculations of the Glueball Spectrum}
\label{ss:glueballs}

The existence of glueballs, hadronic states composed mainly of gluons,
is a major untested prediction of QCD.  Assuming QCD is the correct
theory, we can determine the quantum numbers of a vast number of
possible glueball states, but cannot yet calculate their masses, or
their mixing with $q \bar q, q \bar qq \bar q, \ldots $ states, or
understand in detail the production and decay mechanisms.  Various
models like bag models, flux tube model, sum rules \etc\ give
estimates of masses that differ significantly from each other and the
reliability of these calculations is poor. A somewhat dated summary of
the theoretical expectations, compiled by Burnett and Sharpe, can be
found in \cite{sharpebarnett90}. The lack of knowledge of properties
of glueballs does not help the experimentalists who have to isolate
these states from the myriad of meson states in the $1 - 2.5 \ GeV$
region.  Clearly, LQCD calculations can play a vital role by
calculating the spectrum and the mixing with $q \bar q $ mesons.


The first goal of LQCD has been to calculate the quenched spectrum in
a world in which the mixing with quark states is turned off.  I shall
concentrate on just the $0^{++}$ and $2^{++}$ states as these have
been studied most extensively.  The glueball operators $\CO$ are
constructed out of Wilson loops.  Consider a $n \times m$ Wilson loop
$W^{n,m}_{xy}$ lying in the $xy$ plane. Then the rotationally
invariant sum $W^{n,m}_{xy} + W^{n,m}_{yz} + W^{n,m}_{zx}$ projects
onto $0^{++}$ state, while $W^{n,m}_{xy} - W^{n,m}_{yz}$ and its
permutations project onto $2^{++}$. The signal in the 2-point
correlator can be improved by taking a suitable linear combination of
different sized loops that maximize the overlap with physical
glueball states.  One such strategy is discussed in
\cite{gballs91LANL}. Unfortunately, the statistical signal in glueball
correlation functions is much weaker than that for meson and baryon
correlation functions. Even with a sample of $O(10^5)$ independent
configurations, the signal does not extend beyond $\tau \sim 10$ or
$M\tau \sim 6$ and one barely has a ``plateau'' in the effective mass
plot.  The reason for this poor signal in glueball correlation
functions is that they measure the fluctuations in the vacuum
corresponding to the creation and propagation of a virtual glueball,
whereas in meson and baryon correlators the quarks in the propagators
are added as external sources.

\begin{figure} 
\hbox{\epsfxsize=\hsize\epsfbox{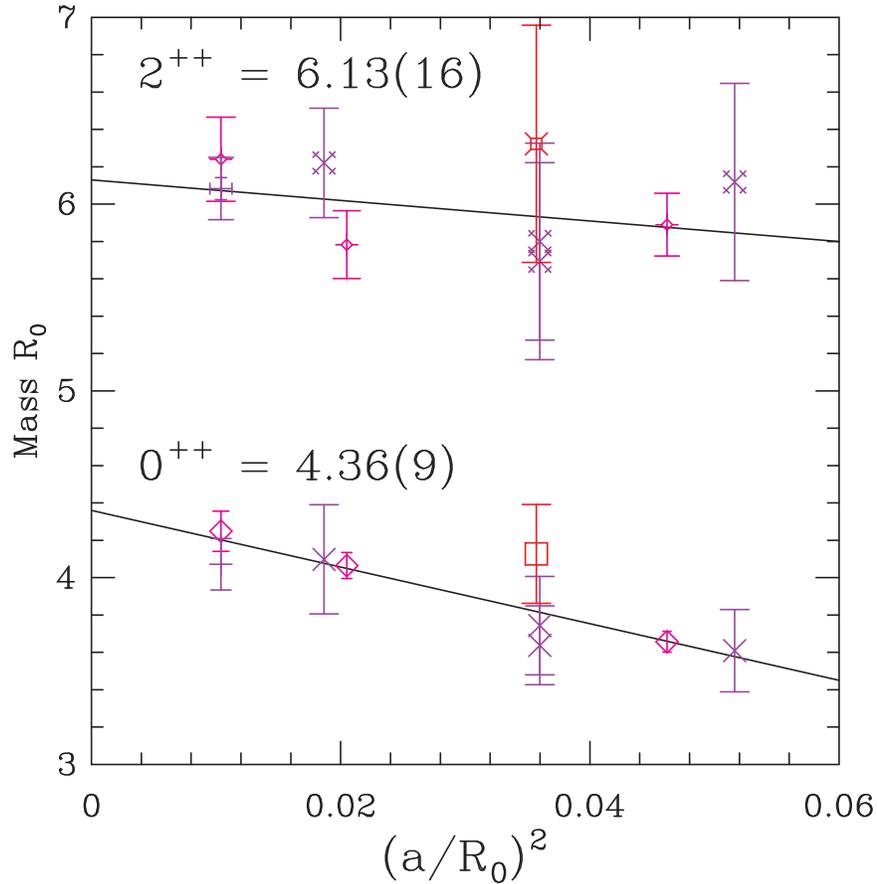}}
\caption{Extrapolation of $0^{++}$ and $2^{++}$ glueball mass data to $a=0$. The data are 
from UKQCD-WUPPERTAL collaboration (crosses), GF11 collaboration
(diamond), and Los Alamos collaboration (squares). The lattice scale
is set using $R_0$ which is related to the string tension by $1/R_0 =
\sqrt{\sigma}/1.18$. }
\label{f:GballsR0}
\end{figure}

The more recent, high statistics results for $M_{0^{++}}$ and
$M_{2^{++}}$ states have been taken from
Refs.~\cite{gballs91LANL,gballs89michaelteper,gballs93bali,gballs94GF11}.
The data and the extrapolation to $a=0$ are shown in
Fig.~\ref{f:GballsR0}. In this plot I have chosen to set the lattice
scale using $R_0 = 1.18 / \sqrt{\sigma}$, where $\sigma$ is the string
tension \cite{gballs98CM}. The figure makes it clear that the
lattice data from the various collaborations are completely
consistent, though the statistical quality of the GF11 data
\cite{gballs94GF11} is much higher.  A fit in $a^2$, the leading
discretization error, gives $M_{0^{++}} R_0 = 4.36(9) $ and
$M_{2^{++}} R_0 = 6.13(16) $ for the continuum limit of the quenched
theory. To obtain masses in physical units one needs to choose a value
for $R_0$ or equivalently $\sigma$. This brings us back to the overall
scale uncertainty inherent in all quenched calculations.  There are
two questions relevant to this analysis (i) what value should one
choose for $\sigma$, and (ii) by how much would the results change if
one had used a different quantity like $M_\rho$ or $\Lambda_{QCD}$ to
set the scale?  The two most quoted results are the following. The
UKQCD-WUPPERTAL collaboration use $1/R_0 = 373$ MeV based on choosing
$\sqrt{\sigma} = 440$ MeV and get $M_{0^{++}} = 1626(34)$ and
$M_{0^{++}} = 2286(60)$ MeV \cite{gballs98CM}. The updated GF11
collaboration estimates are $M_{0^{++}} = 1648(58)$ and $M_{2^{++}} =
2273(116)$ GeV using $\Lambda_{QCD}^{(0)} = 235(6) $ MeV which is
based on $M_\rho$ to set the scale \cite{gballs98GF11}. Thus, the
estimates of the two groups are consistent and agree with the fit
shown in Fig.~\ref{f:GballsR0}.  Therefore, in my view, based on these
fits, the current combined lattice estimate is $M_{0^{++}} =
1640(30)(160) $ MeV and $M_{2^{++}} = 2280(60)(230)$ MeV, where I have
added a second systematic error of $10\%$ to take into account the
scale uncertainty inherent to quenched simulations.

\subsection{Improved action Results on Anisotropic lattices}
\label{ss:glueballanisotropic}

To alleviate the problem that the signal in glueball 2-point
correlation functions falls rapidly, Peardon and Morningstar have
carried out simulations on anisotropic hypercubic lattices, $ a_t <<
a_s$ \cite{GB97Morningstar}.  The advantage of this trick is the 
following. Assume that there exists an approximate plateau in
$M_{eff}(\tau)$ between $\tau = 2 - 4$ in an isotropic lattice
simulation before the signal degrades to noise. Such a plateau is hard
to infer from three time-slices. On the other hand for $a_s / a_t = 5$
lattices, one has roughly 11 times-slices to play with, thereby
greatly increasing the confidence in the estimate. Using anisotropic
lattices does introduce an additional parameter into the analyses,
$a_S/a_T$ or the velocity of light. This they fix non-perturbatively
using the measured potential in spatial and temporal directions.

\begin{figure} 
\hbox{\epsfxsize=\hsize\epsfbox{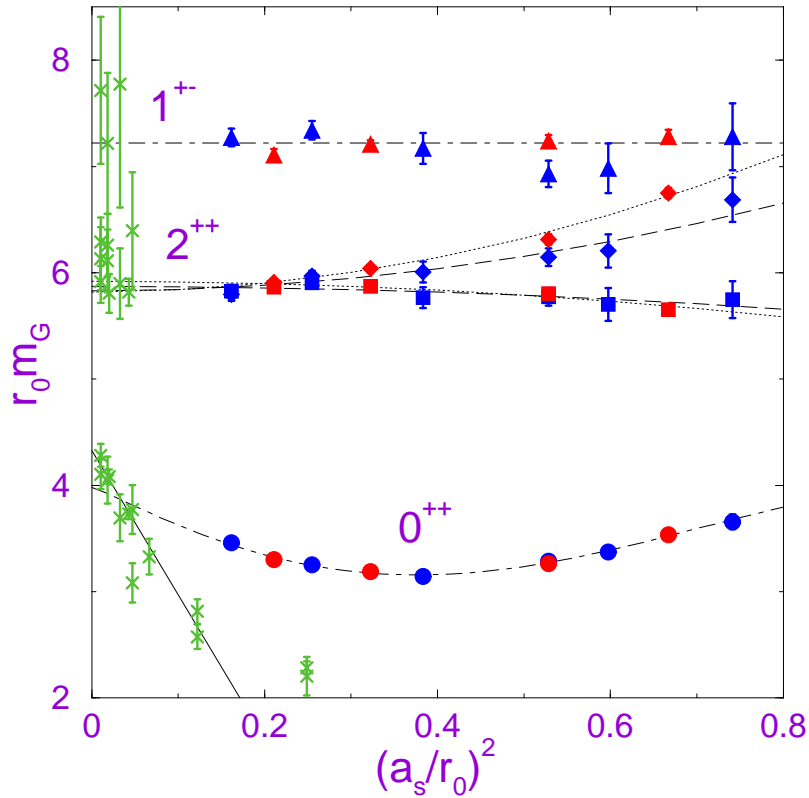}}
\caption{ Glueball mass estimates in terms of $R_0$ against the lattice
spacing $(a_s/R_0)^2$.  Results from the $\xi=5$ simulations for the
lattice irreps $A_1^{++}$ ($0^{++}$), $E^{++}$, $T_2^{++}$ ($2^{++}$),
and $T_1^{+-}$ ($1^{+-}$) are labeled $\circ, \Box, \Diamond$, and
$\triangle$ (blue), respectively.  The corresponding symbols in red
indicate the results from the $\xi=3$ simulations.  Data from Wilson
action simulations taken from
Refs.~\protect{\cite{gballs89michaelteper,gballs93bali,gballs94GF11}}
are shown using green crosses.  The dashed, dotted, and dash-dotted
curves indicate extrapolations to the continuum limit obtained by
fitting to the $\xi=3$ data, the $\xi=5$ data, and all data,
respectively.  The solid line indicates the extrapolation of the
Wilson action data to the continuum limit. }
\label{f:GBMorningstar}
\end{figure}

The second improvement they make is to tune the action.  They
transcribe the tadpole improved L\"uscher-Weisz glue action,
Eq.~\ref{eq:LWTIaction}, on to anisotropic lattices and include the
plaquette in both the fundamental and adjoint coupling space. The
motivation for the latter is to avoid the phase transition in the
fundamental-adjoint space as discussed in
Section~\ref{s:phasetransitions}.  Their new data
~\cite{GB98Morningstar} confirms that including a negative adjoint
coupling does improve the scaling in $M_{0^{++}}$.

Their results, prior to those including a negative adjoint coupling,
are shown in Fig.~\ref{f:GBMorningstar}. The advantage of improved
action and anisotropic lattices is clearly visible. First, the
calculations can be performed on much coarser lattices. Second, the
discretization errors are much smaller. Lastly, the values after
extrapolation to $a=0$ are consistent with conventional methods.
Their estimates $M_{0^{++}} R_0 = 3.98(15) $ and $M_{2^{++}} R_0 =
5.85(2) $, when combined with their estimate $R_0 = 410(20)$ MeV, are
consistent with the values from conventional simulations given in the
previous section. Thus, their results provide another confirmation of
the quenched estimates.

Let me end this discussion of gluey states with pointers to other
interesting calculations. Results for other spin-parity glueball
states can be found in
\cite{gballs93bali,GB97Morningstar,GBhyb97Peardon}; and for hybrid states in
\cite{GBhyb97Peardon,GBhyb96UKQCD,GBhyb97Manke,GBhyb97Collins}.

\subsection{Mixing between scalar quarkonia and glueballs}

The phenomenologically important question is whether the lattice
estimates can help pin down which of the two experimental candidates,
$f_0(1500)$ or $f_0(1710)$, is the scalar glueball. The answer, using
just the above quenched estimates, is no. The reason, in addition to
the fact that the above estimates overlap with both candidates, is
that effects of mixing with scalar quarkonia have not been
included. Lee and Weingarten have initiated a study of mixing by
calculating the two-pint functions $\VEV{\CS(\tau) \CS(0)}$ and
$\VEV{G(\tau) \CS(0)}$ where $\CS$ is the scalar density $(\bar u u +
\bar d d)/\sqrt{2}$ or $\bar s s$ and $G$ is the $0^{++}$ glueball
operator \cite{gballs98GF11}. The statistical quality of these very
hard calculations is not yet good. Lee and Weingarten, therefore,
diagonalize an approximate $3 \times 3$ matrix that is obtained mainly
from a phenomenological analyses. They then find that mixing {\it
raises} $M_{0^{++}}$ by $\sim 64 $ MeV, $i.e.$ $M_{0^{++}}$ changes
from $1640 \to 1704 $ MeV. Also, the associated wavefunctions of the
mixed states explain some of the unexpected features of decays, like
why the $K \bar K$ decay mode of $f_0(1500)$ is suppressed and why
$\Gamma(J/\psi \to \gamma f_0(1710)) > \Gamma(J/\psi \to \gamma
f_0(1390)) \gg \Gamma(J/\psi \to \gamma f_0(1500)) $.  Consequently,
Lee and Weingarten predict that $f_0(1710)$ is dominantly ($\sim
75\%)$ a glueball.

In obtaining these predictions, they neglected mixing with other
(multi-particle) states and decay widths in the mixing matrix and
quenching errors in the lattice calculations.  These, Lee and
Weingarten claim on the basis of phenomenological arguments, are small
and do not change their central conclusion. Hopefully, future LQCD
calculations will provide a much better quantitative validation of
their prediction. 

\subsection{Mass Inequalities}

In addition to getting hard numbers, LQCD (actually Euclidean field
theory) has also prompted the derivation of rigorous
mass-inequalities. The first such was derived by Weingarten showing
that the pion is the lightest meson state of non-zero isospin in QCD
\cite{MI83Weingarten}.  Consider the 2-point correlation function for
mesons, $i.e.$ $O = \bar u \Gamma d $ where $\Gamma$ can be any one of
the 16 elements of the Clifford algebra. Then
\begin{eqnarray}
\label{eq:pioncf}
\Gamma_\Gamma(\tau) \ &=& \ \ME{0}{\bar d \Gamma u(\tau,\vec y) \ \bar u \Gamma d(0,\vec x)}{0} \nonumber \\
                   &=& \ \VEV{{\rm Tr}S_F^d(0,\vec x;\tau,\vec y) \Gamma S_F^u(\tau,\vec y;0,\vec x) 
                      \Gamma } \nonumber \\
                   &=& \ \VEV{{\rm Tr}\Gamma  \gamma_5 S_F^{d\dagger}(\tau,\vec y;0,\vec x) 
			              \gamma_5\Gamma   S_F^u(\tau,\vec y;0,\vec x)  } .
\end{eqnarray}
For the special case of the pion, $\Gamma \gamma_5 =1$, consequently
$\Gamma_\pi(\tau)$ has the maximum value on each
configuration. Furthermore, for $m_u = m_d$ it is the absolute square
of the propagator. The other spin-parity states involve $\gamma$
matrices. Since these have both plus and minus signs, the various
spinor terms in the trace have cancellations.  The second property
needed to translate this property of correlation functions into an
inequality on masses is that the integration measure for QCD is
positive, $i.e.$ each configuration contributes with the same
sign. Thus, $\Gamma_\pi(\tau) \geq
\Gamma_\Gamma(\tau)$ for all $\tau$. Since, $\Gamma(\tau) \sim e^{-m
\tau}$, the inequality implies that the pion is the lightest meson.

Concurrently, Witten and Nussinov derived further inequalities. 
Examples of these are 

\begin{itemize}
\item
The electromagnetic mass shift in the pion is
positive~\cite{MI83Witten}. Specifically what is shown is that
$\VEV{V^3_\mu(k)V^3_\mu(-k) - A^3_\mu(k)A^3_\mu(-k) } \ge 0$ for any
Euclidean $k$.

\item
For mesons with valence flavors $A$ and $B$ the inequality $2M_{A \bar
B} \ge M_{A \bar A} + M_{B \bar B} $ holds in the limit that annihilation
into gluonic states can be neglected in the flavor singlet correlators
~\cite{MI83Witten,MI83Nussinov}.

\end{itemize}

Recently West has applied/extended these methods to glueballs ~\cite{MI83West}. 
He shows that $0^{++}$ is the lightest glueball state.

%% file: chap-alpha.tex
\section{The strong coupling constant $\alpha_s$}
\label{s:alpha}

Spectroscopy studies, in addition to testing QCD, allow the
determination of some of the fundamental parameters of the SM, $i.e$,
the coupling $\alpha_s$ and the quark masses.  To determine
$\alpha_s$, one needs to know not only its strength but also the
energy scale $Q$ at which the coupling is measured. The calculation of
$\alpha_s$, in general, proceeds as follows.  One chooses a quantity
$\VEV{\CO}$ that can be measured very accurately, either in high energy
experiments or via non-perturbative methods on the lattice, and for
which a perturbative expansion is known, $i.e.$,
\begin{equation}
\VEV{\CO} = \sum_n A_n \alpha_s^n(Q)  \ ,
\label{eq:defnalpha}
\end{equation}
where the $A_n$ are the perturbative coefficients. Inverting this
relation gives $\alpha_s(Q)$.

One promising lattice approach for calculating $\alpha_s$ has been
pioneered by the NRQCD collaboration \cite{alpha97NRQCD}.  They
examine several short-distance $\VEV{\CO}$, like the expectation value
of small Wilson loops, that can be calculated very accurately on the
lattice and for which perturbative expansions are known to at least
$O(\alpha_s^2)$.  The scale $Q$ characterizing these lattice
observable is typically of the form $ Q = C/a$ where $a$ is the
lattice spacing and $C$ is a constant that needs to be fixed.  They
determine $C$ relevant to $\VEV{\CO}$ by evaluating the mean momentum
flow in the perturbative expressions for these quantities.  The scale
$1/a$ is fixed using level splittings in quarkonia.  In particular
they use the spin-averaged S-P and the 1S-2S splittings in the Upsilon
system. The advantage of using these splittings is that they show
negligible sensitivity to the precise tuning of the bottom quark mass,
and the corrections due to light quark loops are expected to be small.

\begin{figure} 
\hbox{\epsfxsize=\hsize\epsfbox{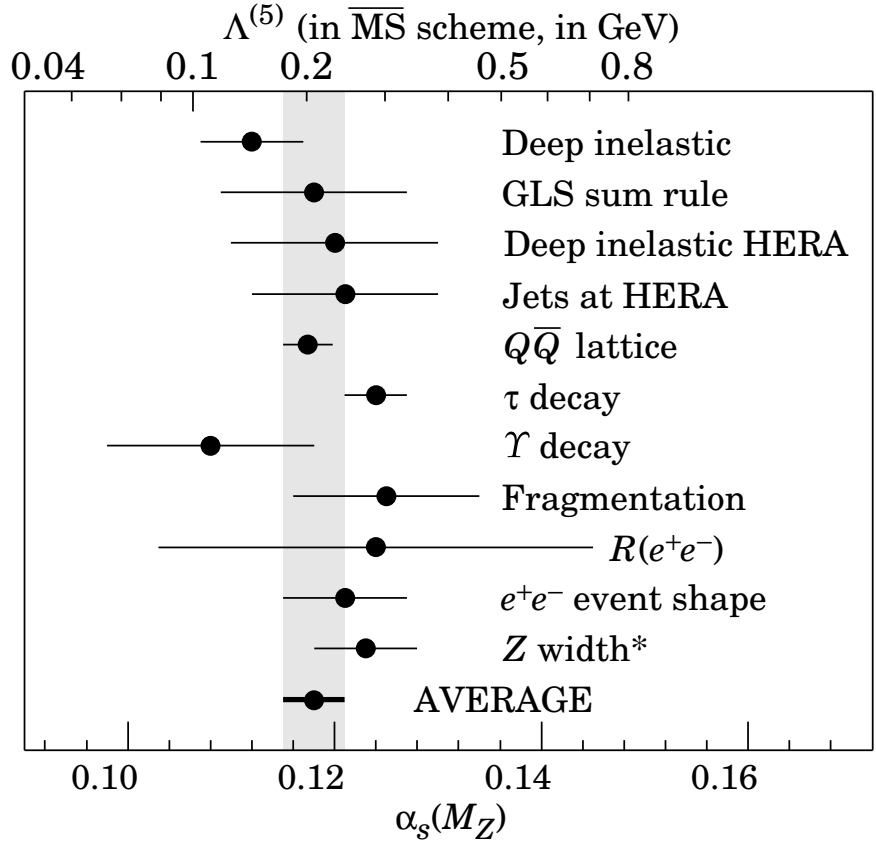}}
\caption{Summary of the values of $\alpha_s(M_Z)$ and $\Lambda^{(5)}$
(the QCD scale for 5 active quark flavors) extracted from various
processes. These are ordered from top to bottom by increasing energy
scale of measurements.  This figure is reproduced from the 1996 Review
of Particle Physics, updated to show the latest lattice result,
$\alpha_{\overline{MS}}(M_Z) = 0.1174 \pm 0.0024$. It is a remarkable 
confirmation of QCD that the $\alpha_s$ extracted from processes at 
different scales has the same value.}
\label{f:alphaworld}
\end{figure}

Finally, to convert from $\alpha_s$ in the lattice scheme to
$\alpha_{\MSbar}$ they use the two loop relation, and include
reasonable estimates of the $\alpha_s^3$ term in the error
analysis. Their final estimate
\begin{equation}
\alphamsbar (M_Z) = 0.1174(24) 
\end{equation}
includes quenching uncertainties. This remarkably precise result is
consistent with the world average of experimental determinations,
$\alpha_s = 0.118(3)$, as shown in Fig.~\ref{f:alphaworld}.

An alternate method for measuring $\alpha_s$ has been carried out by
the ALPHA collaboration \cite{alpha97ALPHA}. In their method, called
the Schr\"odinger functional approach, $\alpha_s$ is defined by
measuring the response of the lattice system to an external
field. They fix the lattice scale $a$ using the force at some fixed
distance ($R_0=0.5$ fermi). They evolve their coupling to a very high
scale using a non-perturbative renormalization group stepping
procedure.  At this high scale they define $\Lambda^{(0)}_{\bar{MS}}$
using the two loop relation and include the three loop uncertainty,
which is small, in error estimates. They have completed the
calculation for the quenched theory with the result
$\Lambda^{(0)}_{\bar{MS}} = 251(21)$ MeV, and are now doing the
$n_f=2$ simulations \cite{alpha97ALPHA}. The corresponding value
$\alphamsbar(90 \GeV) = 0.0818(10)$ is lower than in nature. This
difference is attributed to having neglected sea quark effects. A
similar low value is obtained in the NRQCD analyses if no corrections
are made for the quenched approximation. For details of the
Schr\"odinger functional method see the lectures by Prof. L\"uscher at
this school.

%% file: chap-mq.tex
\section{Quark Masses}
\label{s:mq}

Quarks are not asymptotic states of nature, so their masses cannot be
measured in experiments. They are generally defined in some
perturbative scheme like $\bar{MS}$ and at some scale $\mu$. Their
values remain amongst the least well known of the Standard Model
parameters. In this section I will discuss their extraction from
lattice data.

The extraction of the up, down, and strange quark masses from the
hadron spectrum is done by inverting Eqs.~\ref{eq:chiralforms} once the
coefficients $A_\pi$ $etc.$ are determined from fits to the lattice
data. As I have mentioned before, current lattice simulations neglect
iso-spin breaking effects, so we can only determine $\mbar=(m_u+m_d)/2$ and $m_s$.
I will discuss these first and then briefly mention the extraction of
the heavy quark masses $m_c$ and $m_b$.

In the quenched approximation, the most straightforward way to
estimate $\mbar$ is to extrapolate $M_\pi^2/X^2$ to the physical
value. Here $X$ is a dimensionful quantity like $f_\pi$ or $M_\rho$ or
$M_N$ or $M_\Delta$ that does not depend on $m_s$, and de facto sets
the lattice scale $1/a$.  Of the above quantities, using $M_\rho$ is
the most controversial as it decays via strong interactions and has a
large width, $170$ MeV. On the other hand the lattice data for
$M_\rho$ is most readily available and has good signal. Also, in the
quenched approximation the $\rho$ does not decay. Using $M_\rho$ to
set $a$ in QQCD versus some other quantity can therefore give an
overall scale uncertainty. The spread in $1/a$ extracted from
different quantities is found to be $\approx 10\%$, and this
uncertainty, inherent to the quenched approximation, will be treated
as an overall systematic error.  Similarly, one can set $m_s$ by
extrapolating the ratio $\CM_s/X$, where $\CM_s$ is any state that
depends on $m_s$ like $M_K$, or $M_{K^*}$, or $M_\phi$ or any of the
baryons containing a strange quark.  The world data show that
different choices of $\CM_s$ give rise to a $\approx 20\%$ spread in
$m_s$.  It is important to note that if one assumes the relation
$M_{\CP\CS}^2 = B_{\CP\CS} m_q$ (which LQCD has not validated as true for the
whole range $[\mbar,m_s]$), and uses $M_K$ to fix $m_s$, then the
ratio $m_s/\mbar$ has to be $25.9$ if $\mbar$ is set using $M_\pi^2$.

In the full theory $X$ depends on $m_s$, and similarly $\CM_s$ depends
on $\mbar$, through sea quark effects. Thus, $\mbar$, $a$, and $m_s$
have to be fixed by solving the coupled set of equations.

\begin{figure}[t]
\hbox{\hskip15bp\epsfxsize=0.9\hsize\epsfbox{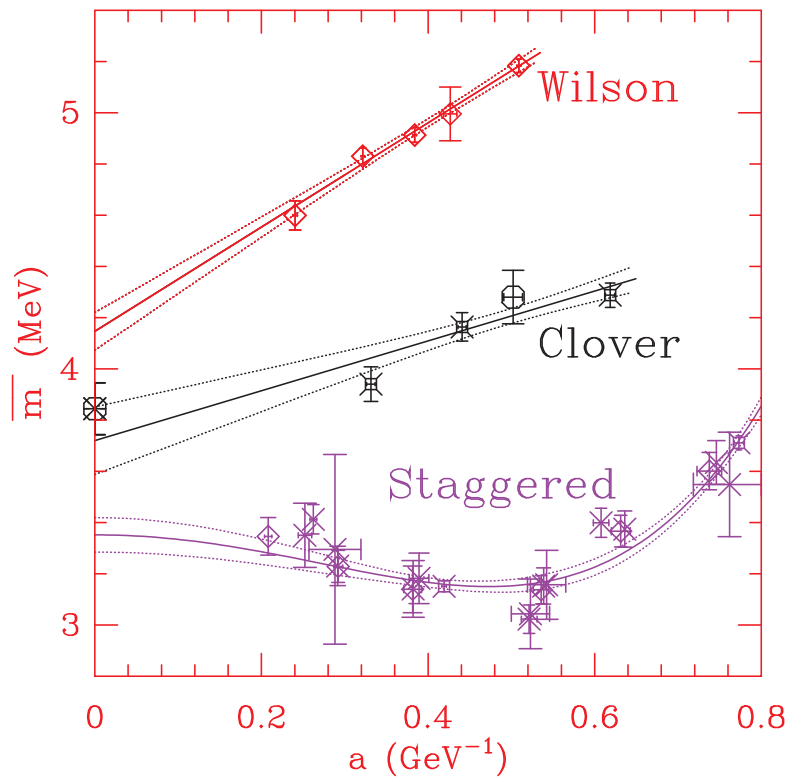}}
\vskip \baselineskip
\figcaption{The behavior of $m_u + m_d$ as a function of the lattice 
spacing $a$ for three discretization schemes -- Wilson, clover, and
staggered.  Linear extrapolations to $a=0$ are shown for Wilson and
clover formulations, while for staggered both $O(a^2)$ and $O(a^4)$
terms are incorporated in the fit. The theoretically expected form for
discretization errors in the clover formulation is $O(\alpha_s
a)$. The extrapolated value using this form is shown by the symbol
burst.}
\label{f:mbar}
\end{figure}

Simulations of the hadron spectrum have been done by a number of
collaborations around the world, and there exists a large amount of
data for the masses of hadrons as a function of the quark masses and
lattice spacing.  (Almost all lattice collaborations, for example the
APE, Columbia University, CP-PACS, Fermilab, GF11, HEMCGC, JLQCD, Los
Alamos, SCRI, staggered, UKQCD, and the Wuppertal University
collaborations, have contributed to these
calculations~\cite{RgTb96Mq}.)  In early analyses the
estimates of quark masses varied by almost a factor of two between
different formulations of fermions, and the pattern of the
discretization and quenching errors was not obvious from the work of
any one collaboration.  Clarification of this issue has been obtained
by putting together all the world data for Wilson, clover, and
staggered fermions \cite{RgTb96Mq,Mq97fermilab}. The final picture for $\mbar$,
based on the most recent, high statistics, and large lattices data is
shown in Fig.~\ref{f:mbar} for the three fermion formulations.  This
figure shows clearly that the factor of $\sim 2 $ difference at $1/a
\sim 2$ GeV shrinks to $\sim 15\%$ after extrapolation to $a= 0$,
$i.e.$ most of the difference can be explained as due to
discretization errors.

The behavior of $m_s$ versus $a$ is very similar.  A summary of the
quenched results in MeV at scale $\mu=2\ \GeV$, based on an analysis
of the 1997 world data \cite{Mq97rev}, is
\smallskip
\begin{center}
\setlength\tabcolsep{0.5cm}
\begin{tabular}{|l|c|c|c|}
\myhline
                 &  Wilson    & TI Clover   &  Staggered    \\
\myhline
$ \mbar(M_\pi)$  &  $4.1(1)$  & $3.8(1)$    & $ 3.5(1)$     \\
$ m_s(M_K)    $  &  $107(2)$  & $ 99(3)$    & $ 91(2)$      \\
$ m_s(M_\phi) $  &  $139(11)$ & $117(8)$    & $ 109(5)$  .  \\
\myhline
\end{tabular}
\label{tab:mqfinal}
\end{center}
\smallskip
These estimates were obtained using the following {\it Ansat\"ze}.
\begin{itemize}
\item
Chiral fits were made keeping only the lowest order terms shown in
Eqs.~\ref{eq:chiralforms}. 
\item
$M_\pi^2 / M_\rho$ is used to fix $\mbar$.
\item
Results for $m_s$ are given for two choices, $M_K^2 / M_\rho$ and
$M_\phi / M_\rho$.  (Using $M_{K^*} / M_\rho$ gives values almost
identical to those with $M_{\phi} / M_\rho$. This implies that for the
choice $m_s(M_\phi)$ one could equivalently have set the scale using
$M_{K^*}$ instead of $M_\rho$.)
\item
Since linear chiral fits have been used, $m_s(M_K) \equiv 25.9 \mbar$. 
\item
The matching factor, $Z_m$, between lattice and \MSbar\ schemes is the tadpole 
improved 1-loop result. 
\item
The extrapolation to $a=0$ is done using the lowest order
discretization error, $i.e.$ $O(a)$ and $O(\alpha_s a)$ for Wilson and
clover, and a quadratic in $a^2$ for staggered.
\end{itemize}

The final difference in estimates between Wilson, tadpole improved
clover (TI Clover), and staggered results is $\approx 15\%$ as shown
in Fig.~\ref{f:mbar}.  This could be due to the neglected higher order
discretization errors and/or due to the difference between
non-perturbative and 1-loop estimates of $Z$'s. Similarly, the
$\approx 20\%$ variation in $m_s$ with the state used to extract it,
$M_K$ versus $M_\phi$ (or equivalently $M_{K^*}$) could be due to the
quenched approximation or partly an artifact of keeping only the lowest
order discretization correction in the extrapolations. To disentangle these
discretization and quenching errors requires precise unquenched data
with roughly physical values for light quarks.

For best estimates of quenched results I average the data and use the
spread as the error.  To these, I add a second systematic error of
$10\%$ due to the uncertainty in the determination of the scale $1/a$,
which partly reflects the uncertainty due to the quenched
approximation.  The  results, in the $\MSbar$ scheme and evaluated at
$2 $ GeV, are~\cite{Mq97rev}
\begin{eqnarray}
\mbar &=& 3.8(4)(4)   \ \MeV  \nonumber\\
m_s   &=& 110(20)(11) \ \MeV  \,.
\label{eq:mqfinal}
\end{eqnarray}

\subsection{Light Quark Masses from Ward Identities}
\label{ss:mqWI}

A second method for extracting quark masses is 
based on axial and vector current Ward identities
\bea
Z_A \partial_\mu A_\mu &=& (m_1 + m_2)_R Z_P \CP \ , \nonumber \\
Z_V \partial_\mu V_\mu &=& (m_1 - m_2)_R Z_S \CS \ ,
\label{eq:mqwardA}
\eea
where $1,2$ are the flavor indices and $\CP$ and $\CS$ are the 
lattice pseudoscalar and scalar densities.  From these one can
construct ratios of 2-point correlation functions
\bea
\VEV{\partial_\mu A_\mu(\tau) \ \CP}   &=& 
	(m_1 + m_2)_R {Z_P \over Z_A} \VEV{\CP(\tau) \ \CP(0)} \nonumber \\
\VEV{\partial_\mu V_\mu(\tau) \ \CS(0)}&=& 
	(m_1 - m_2)_R {Z_S \over Z_V} \VEV{\CS(\tau) \ \CS(0)} \ , 
\label{eq:mqwardB}
\eea
which give the renormalized quark masses once the various
renormalization constants $Z_i$ are known. Since the Ward identities
are operator identities, Eq.~\ref{eq:mqwardB} holds for all $\tau$
except at very short times where there is influence of contact
terms. Thus it is not necessary for the signal to persist to
asymptotic $\tau$ as needed in spectroscopy method to isolate the
lowest state.

\begin{figure}[t]
\hbox{\hskip15bp\epsfxsize=0.9\hsize\epsfbox{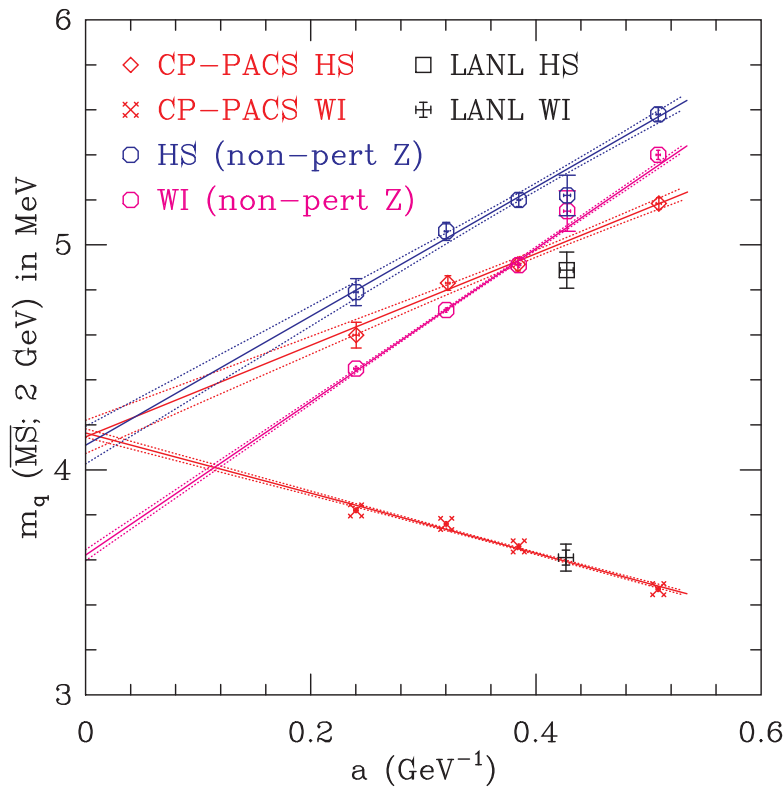}}
\vskip \baselineskip
\figcaption{The behavior of $\mbar=(m_u + m_d)/2$ as a function of the 
lattice spacing $a$ for Wilson fermions. Data from the hadron
spectroscopy and Ward Identity method are shown for two ways of
estimating the renormalization constants, $i.e.$ one loop perturbative
and non-perturbative.}
\label{f:Wmbar}
\end{figure}

The two estimates of quark masses, from spectroscopy and from the Ward
identities, should agree in the continuum limit.  A comparison of the
two data sets for the Wilson action and using 1-loop perturbative
$Z$'s is shown in Fig.~\ref{f:Wmbar}. The two estimates are
significantly different at finite $a$, however, a linear extrapolation
works for both sets of data, and the extrapolated values are in
surprisingly good agreement.  One could therefore attribute the
differences at finite $a$ to discretization errors. 

The recent non-perturbative calculations of the $Z$'s \cite{Mq98APE}
challenge this nice picture.  Fig.~\ref{f:Wmbar} also shows the same
data analyzed using these non-perturbative $Z$'s. At fixed $a$ the
non-perturbative $Z$'s increase both the hadron spectroscopy and WI results to roughly
the same value. This suggests that most of the difference is due to
the $O(\alpha_s^2)$ corrections to the 1-loop $Z$'s and not due to the
$O(a)$ discretization errors as mentioned before.  However, the
non-perturbative calculation of $Z$'s themselves have $O(a)$ errors
that have not been resolved in the results of \cite{Mq98APE}. As a
result we do not have a clear separation between the effects of the
$O(a)$ discretization errors and the $O(\alpha_s^2)$ errors in these 
$Z$'s.  

The bottom line is that, even though we have not resolved the two
kinds of errors, and while both the value at fixed $a$ and the slope
in $a$ increases significantly on changing from 1-loop to
non-perturbative $Z$'s, the extrapolated value is roughly unchanged.

To conclude, I believe that we have finally obtained values for light
quark masses with reliable estimates of statistical and all systematic
errors other than quenching.  Calculations are already underway to
significantly increase the number of data points with $n_f= 2$ and $
4$ flavors of dynamical quarks to reduce the quenching
uncertainty. Also, new ways to calculate the renormalization constants
non-perturbatively are being developed. Therefore, I expect that in
the next few years there will be a significant increase in the
precision of both QQCD and full QCD estimates of light quark masses.

\subsection{Schrodinger Functional Approach}

In the above discussion the term non-perturbative $Z$'s referred to
only the lattice scheme. In the required matching constants one still
has to use perturbative expressions for the renormalization constants
in the continuum scheme and make a good estimate of the matching
point.  The ALPHA Collaboration has proposed a method for removing
this last vestige of dependence on perturbation
theory~\cite{Mq97ALPHA}. The details of their method have been
discussed by Prof. L\"uscher at this school, so I will just summarize
the key points.

\begin{description}

\item{(i)}
Calculate the quark mass
using the axial WI method at some infrared lattice scale. At the same time 
calculate $Z_P$ and $Z_A$ non-perturbatively in the Schr\"odinger functional
scheme. 
\item{(ii)}
Evolve this lattice result to some very high scale using the
step scaling function calculated non-perturbatively. 

\item{(iii)} At this high scale, where $\alpha_s$ is small, 
define the renormalization group independent mass 
\be
\label{eq:RGindependentM}
\widehat m \ = \ \lim_{a \to 0} m_R \big( 2 \beta_0 g^2 \big)^{-\gamma_0/2 \beta_0}
\ee
This $\widehat m$ is scheme independent, thus subsequent conversion to
a running mass in some scheme and at some desired scale can be done in
the continuum using the most accurate versions (4-loop) of the scale
evolution equations \cite{97Vermaseren}.

\end{description}

To reiterate, the advantage of this method is that the lattice
calculations directly predict the scale and scheme invariant quantity
$\widehat m$.  The calculations by the ALPHA Collaboration are still
in progress and at this time estimates for $\widehat m$ have not been
released. It will be interesting to see how they compare with
conventional hadron spectroscopy or WI methods discussed above.  My
prejudice is that they will be very similar.

\subsection{Role of quark masses in CP violation}

A very opportune application of the precise determination of the light
quark masses is in the Standard Model prediction of direct CP
violation in kaon decays characterized by the ratio
$\epsilon'/\epsilon$ (see lectures by Prof. Buras and \cite{Buras96}).
The reason is that $\epsilon'/\epsilon$ is, to a good approximation,
proportional to $1/(m_d+m_s)^2$ \cite{Buras96eps}, 
\begin{equation} 
\epsilon'/\epsilon = A \bigg\{c_0 + \big[c_6 B_6^{1/2} + c_8 B_8^{3/2} \big] 
\big[ {158\MeV \over (m_s + m_d)} \big]^2 \bigg\} \ ,
\label{eq:masterEE}
\end{equation}
where all quantities are to be evaluated at the scale $m_c = 1.3
\GeV$. Eq.~\ref{eq:masterEE} highlights the dependence on the light
quark masses and the bag parameters $B_6^{1/2}$ and $B_8^{3/2}$ which
LQCD is also expected to provide.  The coefficients $c_0, c_6,$ and
$c_8$ have been calculated to next-to-next-to-leading order, and with
$M_{top}$ now known the main uncertainty in their extraction comes
from the value of $\Lambda_{QCD}$.  The quantity $A = |V_{ub}||V_{cb}|
\sin \delta$ in the parameterization of Eq.~\ref{eq:CKMstandard},
requires input of other SM parameters like $f_B, |V_{ub}|, |V_{cb}|,
B_K,
\sqrt{B_{B_d}}f_{B_d}$.  Using the central values quoted by Buras at 
scale $m_c = 1.3 \GeV$ \cite{Buras96eps} one gets $A = 1.29\times
10^{-4}$, $c_0 = - 1.4$, $c_6 = 7.9$, $c_8 = - 4.0$. Unfortunately,
the errors in these quantities make the resulting theoretical
prediction, $\epsilon'/\epsilon = 3.6(3.4)\times10^{-4}$
\cite{Buras96eps}, very loose.

Current experimental estimates are
$7.4(5.9)\times10^{-4}$ from Fermilab E731 \cite{epsE731} and
$23(7)\times10^{-4}$ from CERN NA31 \cite{epsNA31}.  
The new generation of experiments, Fermilab E832, CERN NA48, and
DA$\Phi$NE KLOE, will reduce the uncertainty to $\approx 1
\times10^{-4}$. First results from these experiments (see lectures by 
L. Fayard) should be available in the next couple of years. Thus, it
is very important to tighten the theoretical prediction.  Clearly,
lower estimates of quark masses suggested by LQCD would significantly
increase the size of direct CP violation, and make its experimental
determination and validation much easier.

\subsection{Heavy quark masses $m_c$ and $m_b$}

The masses of the heavy quarks, $m_c$ and $m_b$, have been
calculated using the charmonium and upsilon spectroscopy. For heavy
quarks there exist two different definitions of the quark mass $--$
the pole and $\bar{MS}$ masses.  The pole mass is defined to be
$m^{(pole)} = (M_{onia} - E_{binding})/2$, where $M_{onia}$ is taken
from experimental data and $E_{binding}$ is calculated on the lattice.
This calculation has been done using two approaches, NRQCD
\cite{Mb94NRQCD} and heavy quark effective theory (HQET) \cite{Mb97HQET},
and the results are consistent. The $\bar{MS}$ mass, which has the
advantage of being free of renormalon ambiguities that are inherent in
$m^{(pole)}$, is obtained from $m^{(pole)}$ using perturbation theory.
The present results are 
\begin{eqnarray}
  m_b^{\bar{MS}}(m_b^{\bar{MS}}) &=& 4.15(5)(20)\ \GeV              \quad{\rm (APE)} \nonumber\\
                                 &=& 4.16(15)\ \GeV    \phantom{(0)} \quad{\rm (NRQCD) } \,.
\end{eqnarray} 
These estimates are consistent
with determinations from other non-lattice methods.  

The agreement is not as good for various lattice estimates of $m_c ^{\bar{MS}}$. 
Current estimates lie between $1.2-1.5$ GeV. For example, three different 
methods investigated  by the APE collaboration \cite{Mc94APE,Mq98APE}, Fermilab 
collaboration \cite{Mc95Aida,Mc97Kronfeld}, and by Bochkarev and Forcrand 
\cite{Mc96Forcrand} who evaluate the 
correlation functions arising in QCD sum-rules on the lattice, give 
\begin{eqnarray}
m_c^{\bar{MS}}(m_c^{\bar{MS}}) &=& 1.525(40)(125) \GeV \qquad {\rm APE} \nonumber\\
			       &=& 1.33(8)        \GeV \qquad \phantom{(0125)} {\rm Fermilab}  \nonumber\\
			       &=& 1.22(5)        \GeV \qquad \phantom{(0125)} {\rm B\&F} \ . 
\end{eqnarray}
At this stage I do not consider the difference significant as the
various systematic errors have not been resolved equally well in all
the calculations.